\documentclass[11pt,twoside]{report}
\usepackage{natbib,accents,graphicx,epsf}
\def\@doubleleading{1.6} 
\def\@tightleading{1.1} 
\renewcommand\baselinestretch{\@tightleading} 

\def\eps@scaling{1.0} 
\newcommand\epsscale[1]{\gdef\eps@scaling{#1}} 
\newcommand\plotone[1]{ 
 \centering 
 \leavevmode 
 \includegraphics[width={\eps@scaling\columnwidth}]{#1} 
} 

\def\mycaption#1{
 \vskip 0.1truecm 
 \rightskip=3truepc
 \leftskip=3truepc
 \baselineskip=13pt
 \noindent{\small#1}
}

\def\aap  {A\&A}
\def\apj  {ApJ}
\def\apjl {ApJ}
\def\cqg  {Class. Quant. Grav.}
\def\mnras{MNRAS}
\def\nat  {Nature}
\def\pasj {PASJ}
\def\pl   {Phys.~Lett.}
\def\prd  {Phys.~Rev.~D.}
\def\prl  {Phys.~Rev.~Lett.}
\def\ptp  {Prog.~Theor.~Phys.}

\def\simgt{\lower.5ex\hbox{$\; \buildrel > \over \sim \;$}}
\def\simlt{\lower.5ex\hbox{$\; \buildrel < \over \sim \;$}}

\begin{document}
\begin{titlepage}
%%%%%%%%%%%%%%%%%%%%%%%%%%%%%%%%%%%%%%%%%%%%%%%%%%%%%%%%%%%%%%%%%%%
\begin{center}
 {\large THESIS}\\
 \vspace{2cm} 
 {\Large
   {\bf The Pursuit of Non-Gaussian Fluctuations\\
   \vspace{2mm}
        in the Cosmic Microwave Background}
 }
\end{center}
%%%%%%%%%%%%%%%%%%%%%%%%%%%%%%%%%%%%%%%%%%%%%%%%%%%%%%%%%%%%%%%%%%%
\vspace{3cm}
%%%%%%%%%%%%%%%%%%%%%%%%%%%%%%%%%%%%%%%%%%%%%%%%%%%%%%%%%%%%%%%%%%% 
\begin{center}
 {\large A dissertation submitted to\\
         Tohoku University\\
         in partial fulfillment of requirements for the degree of\\
         Doctor of Philosophy\\
         in\\
         Science}
\end{center} 
%%%%%%%%%%%%%%%%%%%%%%%%%%%%%%%%%%%%%%%%%%%%%%%%%%%%%%%%%%%%%%%%%%%
\vspace{3cm}
%%%%%%%%%%%%%%%%%%%%%%%%%%%%%%%%%%%%%%%%%%%%%%%%%%%%%%%%%%%%%%%%%%%
\begin{center} 
 {\Large {\bf Eiichiro Komatsu}\\
 \vspace{8mm}
 Astronomical Institute, Tohoku University}
\end{center}
%%%%%%%%%%%%%%%%%%%%%%%%%%%%%%%%%%%%%%%%%%%%%%%%%%%%%%%%%%%%%%%%%%%
\end{titlepage}
%%%%%%%%%%%%%%%%%%%%%%%%%%%%%%%%%%%%%%%%%%%%%%%%%%%%%%%%%%%%%%%%%%%

%%%%% abstract, table of contents, and acknowledgments %%%%% 
\pagenumbering{roman}

%%%%%%%%%%%%%%%%%%%%%%%%%%%%%%%%%%%%%%%%%%%%%%%%%%%%%%%%%%%%%%%%%%%
%
%  Abstract
%
%     1st draft:  06/18/2001
%     revision:   07/18/2001
%     final:      08/05/2001
%
%%%%%%%%%%%%%%%%%%%%%%%%%%%%%%%%%%%%%%%%%%%%%%%%%%%%%%%%%%%%%%%%%%%
\chapter*{Abstract}

%%%%%%%%%%%%%%%%%%%%%%%%%%%%%%%%%%%%%%%%%%%%%%%%%%%%%%%%%%%%%%%%%%%

%%%%% what do I in this thesis? %%%%%

We present theoretical and observational studies of non-Gaussian
fluctuations in the cosmic microwave background (CMB) radiation 
anisotropy.
We use the angular bispectrum and trispectrum, the 
harmonic transform of the angular three- and four-point correlation functions.
If the primordial fluctuations are non-Gaussian, then this non-Gaussianity 
will be apparent in the CMB sky.

%%%%% theory of the angular bispectrum %%%%%

Non-linearity in inflation produces the primordial non-Gaussianity. 
We predict the primary angular bispectrum from inflation 
down to arcminutes scales, and forecast how well we can measure 
the primordial non-Gaussian signal.
In addition to that, secondary anisotropy sources in the low-redshift
universe also produce non-Gaussianity, so do foreground emissions from
extragalactic or interstellar microwave sources.
We study how well we can measure these non-Gaussian signals, including the 
primordial signal, separately.
We find that when we can compute the predicted form of the bispectrum, 
it becomes a ``matched filter'' for finding non-Gaussianity in the data, 
being very powerful tool of measuring weak non-Gaussian signals and of
discriminating between different non-Gaussian components.
We find that slow-roll inflation produces too small bispectrum
to be detected by any experiments; thus, any detection strongly constrains 
this class of models.
We also find that the secondary bispectrum from coupling between the 
Sunyaev--Zel'dovich effect and the weak lensing effect, and
the foreground bispectrum from extragalactic point sources,
give detectable non-Gaussian signals on small angular scales.

%%%%% measurement of the angular bispectrum on the DMR map %%%%%

We test Gaussianity of the {\it COBE} DMR sky maps, by measuring all 
the modes of the angular bispectrum down to the DMR beam size.
We compare the data with the simulated Gaussian realizations, finding 
no significant signal of the bispectrum on the mode-by-mode basis. 
We also find that the previously reported detection of the bispectrum 
is consistent with a statistical fluctuation.
By fitting the theoretical prediction to the data for the primary 
bispectrum, we put a constraint on non-linearity in inflation. 
Simultaneously fitting the foreground bispectra, which are estimated from 
interstellar dust and synchrotron template maps, shows that neither 
dust nor synchrotron emissions contribute significantly to the bispectrum 
at high Galactic latitude.
We thus conclude that the angular bispectrum finds no significant 
non-Gaussian signals in the DMR data.

%%%%% measurement of the angular trispectrum on the DMR map %%%%%

We present the first measurement of the angular 
trispectrum on the DMR sky maps, further testing Gaussianity 
of the DMR data.
By applying the same method as used for the bispectrum to the DMR data, 
we find no significant non-Gaussian signals in the trispectrum.
Therefore, the angular bispectrum and trispectrum show that the 
DMR sky map is comfortably consistent with Gaussianity.

%%%%% what about the future? %%%%%

The methods that we have developed in this thesis can readily be 
applied to the {\it MAP} data, and will enable us to pursue non-Gaussian
CMB fluctuations with the unprecedented sensitivity.
We show that high-sensitivity measurement of the CMB bispectrum and 
trispectrum will probe the physics of the early universe as well as 
the astrophysics in the low-redshift universe, independently of the CMB
power spectrum.

   \tableofcontents
   \listoffigures
   \listoftables

%%%%%%%%%%%%%%%%%%%%%%%%%%%%%%%%%%%%%%%%%%%%%%%%%%%%%%%%%%%%%%%%%%%
%
%  Acknowledgments
%
%     1st draft:  07/15/2001
%     revision:   07/17/2001
%     final:      08/05/2001
%
%%%%%%%%%%%%%%%%%%%%%%%%%%%%%%%%%%%%%%%%%%%%%%%%%%%%%%%%%%%%%%%%%%%
\chapter*{Acknowledgments}

%%%%%%%%%%%%%%%%%%%%%%%%%%%%%%%%%%%%%%%%%%%%%%%%%%%%%%%%%%%%%%%%%%%

%%%% advisors and collaborators %%%%%

%
% Futamase sensei
%
I would like to thank Toshifumi Futamase for his support for my 
undergraduate and graduate studies.
He has taught me about inflation, generalized gravity theory, 
and quantum field theory in curved spacetime, which have always
fascinated me, and continue to do so.
In addition, he has opened up the CMB world to me, which has been and 
continues to be my main research field.
I will never forget his kind and constant support for my student life;
especially, for sending me to Princeton University, which has benefited 
my research life more than ever.

%
% David
%
I would like to thank David N. Spergel for his genuine support and 
constant encouragement during the last two years of my graduate student 
life in Princeton University.
He is not only a CMB professional, but also a superb theoretical 
astrophysicist.
He has always intrigued and stimulated me through
invaluable discussions on various topics.
I greatly appreciate his suggestion that I work on the CMB
non-Gaussianity; I am really enjoying the pursuit of CMB non-Gaussianity, 
hence the title of this thesis.
I also wish to thank him for involving me in the {\it MAP} project.
Joining the {\it MAP} project, has been my dream since I was an 
undergraduate student.
I also appreciate his remarkably generous support for assisting my wife 
and myself in adjusting to life in the United States.

%
% Ben, Tony, and Kris
%
The work in chapter~\ref{chap:obs_bl} and \ref{chap:obs_tl} has been
accomplished in collaboration with Benjamin D. Wandelt, Anthony J. 
Banday, and Krzysztof M. G\'orski.
Ben's outstanding knowledge of mathematics and statistics 
has helped me in completing the work in many ways.
I also appreciate his constant encouragement.
The collaboration started when Ben and I attended the CMB meeting
held at the Institute for Theoretical Physics (ITP), the University of
California, Santa Barbara.
I am very grateful for this chance meeting and for the work which we 
have done together.

%
% Licia
%
Licia Verde has taught me about the bispectrum.
Through many discussions on the higher-order moment statistics with her,
I have been able to improve the work in chapter~\ref{chap:theory_bl} 
substantially.
Licia also attended the ITP CMB meeting, after which she came to 
Princeton University (good for me!).
I appreciate the meeting which provided me with the opportunity to meet
these distinguished people.

%
% Inoue san
%
The discussions on global topology of the universe with 
Taro K. Inoue have benefited the work in 
chapter~\ref{chap:obs_tl} and appendix~\ref{app:CH}.
His remarkable, well-organized Ph.D. thesis \citep{Ino01b} has enabled
me to calculate the angular trispectrum in a closed compact hyperbolic
universe, even though I am just an amateur in topology.

%%%%% people outside Tohoku University %%%%%

In addition to the supervisors, collaborators, and colleagues 
who have directly contributed to this thesis work, my research life 
has been supported by many many generous, warm-hearted people, 
without whom I could not have accomplished the thesis.
I would like to thank them here.

%
% Sugiyama san
%
Naoshi Sugiyama has helped me in learning the basic physics of CMB.
I have attended his lectures on CMB twice: the first was when I was a 
first-year graduate student beginning serious work on CMB, and the
second was when I was a fourth-year student finishing one of the thesis 
projects. 
The lectures have triggered my interest in CMB explosively,
and encouraged me to pursue the CMB studies more and more.
I also sincerely appreciate his generous efforts which have been 
indispensable for me to be accepted at Princeton University.

%
% Suto san
%
Yasushi Suto has (implicitly) taught me how to pursue the research.  
I have learned about the Sunyaev--Zel'dovich (SZ) effect through 
the observational project (the SZ project) led by him
\citep{Kom99,Kom01c}.
This project has affected my research life dramatically, and 
broadened my field-of-view by many orders of magnitude.
I also appreciate his generous recommendation for me to be accepted at 
Princeton University.

%
% Hattori san
%
Makoto Hattori has assisted me with my undergraduate and graduate studies.
I still remember very clearly many of our discussions which took place at 
midnight; they were always a lot of fun.
He has taught me the physics and X-ray properties of clusters of
galaxies, which eventually led me to study the clusters with the 
SZ effect.
Also, I would like to thank him for involving me in the SZ project;
this involvement made my research world inflate exponentially.

%
% Matsuo san
%
Hiroshi Matsuo taught me the basics of radio observations 
through the SZ project.
The observations with him at the Nobeyama Radio Observatory were 
so joyful, as have been many interesting and stimulating discussions
--- which often took place with drink.
The tips on radio observations that I have learned from him are 
invaluable treasures for my research life.

%
% Kitayama san
%
I have been benefited by the collaboration with Tetsu Kitayama 
through the SZ project as well as the project on the CMB fluctuations 
induced by the cluster SZ effect \citep{KK99}.
I took advantage of collaborating with him to learn about 
the statistical treatment of the clusters of galaxies.
I also appreciate his warm friendship and encouragement.

%
% Uros
%
Through the collaboration with Uro$\check{\rm s}$ Seljak \citep{KSel01}, 
I have learned about the physics of dark matter halos.
He has expanded my understanding of clusters of galaxies substantially.
The discussions with him have always stimulated me, and also 
encouraged me very much.
I really appreciate his constant warm-hearted encouragement.

%
% Nagamine san, and others
%
I am indebted to Kentaro Nagamine for his generous help with my early 
life in the United States, without which I could not have survived 
until now, seriously.
I am grateful to all the people and colleagues in the Department of 
Astrophysical Sciences, Princeton University, for the warm hospitality 
and the exciting academic environment.

%
% Lila
%
I would like to thank my English tutor, Ms. Lila Lustberg, 
not only for teaching me
English, but also for wonderful friendship with my wife and myself.
I also sincerely appreciate her constant encouragement.

%%%%% Tohoku University fellows %%%%%

I have spent six years in T\^ohoku University as an undergraduate 
and a graduate student.
Throughout my school life, I have been benefited by many wonderful fellows
in the Astronomical Institute.
I would like to thank them here.

%
% Itoh
%
I would like to thank Yousuke Itoh for friendship throughout our 
undergraduate and graduate school
days, and his remarkably generous help with the paperwork necessary 
to continue my research abroad.
Yousuke and I have attended the weekly cosmology seminar
guided by our common supervisor, Futamase-sensei, when we were
undergraduate students.
Futamase-sensei's seminar has been known as one of the ``hardest'' 
student seminars in our institute, and it is really true.
(One day I suffered from appendicitis when preparing for the seminar!)
Yousuke is genuinely smart, and has always influenced me through the seminar.
This seminar was literally my starting point as a cosmologist, 
and if he were not attending, I could not have made it.

%
% Takada san and anyone else
%
I am deeply appreciative of the tremendous amount of discussions that I had 
with Masahiro Takada.
He and I have studied CMB together through the other seminar guided by 
Futamase-sensei, and he has helped with my understanding of CMB
significantly. 
I am grateful to the colleagues in our outstanding cosmology group:
Shijun Yoshida, Takashi Hamana, Etienne Pointecouteau, Keiichi Umetsu,
Jun'ichi Sato, Izumi Ohta, Yoshihiro Hamaji, and Nobuhiro Okabe,
as well as to the fellows in the same year:
Hiroshi Akitaya, Ken'ichiro Asai, Yuji Ikeda, Motoki Kino, and 
Naohiro Yamazaki, and to all the people in the Astronomical Institute.
Especially, I would like to thank Takashi Murayama 
and Takahiro Morishima for their remarkably generous efforts to 
maintain our outstanding computer environment in the institute.

%
% JSPS
%
I acknowledge financial support from the Japan Society for the 
Promotion of Sciences, which has enabled me to study abroad and 
to concentrate on the research activity without being concerned about
the living expenses.
This financial support has been a crucial factor for me to accomplish 
the thesis work.

%%%%% family and midori %%%%%

I would like to send my best thanks to my mother, Hideko, father,
Hidenori, and two sisters, Natsuko and Mikiko, for their every support 
for my long student life as well as for their understanding of my 
pursuing academic research.
Finally, I would like to thank my wife, Midori, who is my deepest love
of all in the universe.
She is the center of my universe.

\clearpage

\pagestyle{headings}

%%%%% main text %%%%%
\pagenumbering{arabic}

%%%%%%%%%%%%%%%%%%%%%%%%%%%%%%%%%%%%%%%%%%%%%%%%%%%%%%%%%%%%%%%%%%%
%
%  Introduction
%
%     1st draft:  06/28/2001
%     revision:   07/25/2001
%     final:      08/05/2001
%
%%%%%%%%%%%%%%%%%%%%%%%%%%%%%%%%%%%%%%%%%%%%%%%%%%%%%%%%%%%%%%%%%%%
\chapter{Introduction}
\label{chap:intro}

%%%%%%%%%%%%%%%%%%%%%%%%%%%%%%%%%%%%%%%%%%%%%%%%%%%%%%%%%%%%%%%%%%%
\section{Why Pursue Non-Gaussianity in CMB?}
\label{sec:why}

Modern understanding to emergence of inhomogeneity in the universe
through the cosmic history is outstanding; 
the inhomogeneity is {\it quantum} in origin. 
Then, it becomes classical through its evolution, producing 
fluctuations in the {\it cosmic microwave background} (CMB) radiation, 
and creating complex structures, such as galaxies, seen in the present 
universe.

A theory of early universe, the cosmic {\it inflation}
\citep{Guth81,Sato81,AS82,Lin82}, has predicted the emergence of the 
quantum fluctuations in early universe.
Inflation not only resolves several serious issues in the old Big-Bang
cosmology, but also gives a mechanism to produce inhomogeneity 
in the universe, and makes specific testable predictions for
the global structure and the inhomogeneity in the universe
\citep{GP82,Haw82,Sta82,BST83}.
The predictions may concisely be summarized as follows:
%%%%%%%%%%%%%%%%%%%%%%%%%%%%%%%%%%%%%%%%%%%%%%%%%%%%%%%%%%%%%%%%%%%
\begin{itemize}
 \item[(a)] The observable universe is spatially flat. 
 \item[(b)] The observable universe is homogeneous and isotropic 
	    on large angular scales, apart from tiny fluctuations.
 \item[(c)] The primordial inhomogeneity has a specific spatial pattern,
	    so-called the scale-invariant fluctuation.
 \item[(d)] Statistics of the primordial inhomogeneity obey 
           {\it Gaussian statistics}.
\end{itemize}
%%%%%%%%%%%%%%%%%%%%%%%%%%%%%%%%%%%%%%%%%%%%%%%%%%%%%%%%%%%%%%%%%%%
All these predictions but (d) have passed challenging observations.
Among those observations, the most firm evidence supporting the 
predictions has come from measurement of the angular distribution of CMB.
Being the oldest observable object in the universe, CMB is the best clue 
to early universe.

Observed isotropy of CMB with 0.001\% accuracy\footnote
{Apart from anisotropies due to the local motion of the Earth, 
which is of order 0.1\% \citep{Smo91}.}
supports the prediction (b) \citep{Mat90,Smo92}.
The measured angular distribution of the CMB anisotropy, 
0.001\% inhomogeneity, supports the primordial fluctuation distribution 
being consistent with the prediction (c) in the flat universe,
the prediction (a) \citep{Ben96,TOCO99,Boom00,Maxima00}.

Using various statistical techniques, many authors have attempted to test the 
prediction (d), the Gaussianity of the primordial inhomogeneity, 
using the CMB anisotropy on large angular scales ($\sim 7^\circ$) 
\citep{Kog96b,FMG98,PVF98,BT99,BZG00,MHL00,Mag00,SM00,Bar00}, 
on intermediate scales ($\sim 1^\circ$) 
\citep{Park01}, and on small scales ($\sim 10'$) \citep{Wu01}.

In contrast to the predictions (a)--(c) for which the observations
suggest no controversy, previous work on searching for non-Gaussian
CMB anisotropies has come to different conclusions from one another:
some do claim detection \citep{FMG98,PVF98,Mag00}, 
and the others do not \citep{Kog96b,SM00,Bar00,Park01,Wu01}.
Furthermore, \citet{BT99} and \citet{BZG00} claim the non-Gaussian signal
of \citet{FMG98} to be non-cosmological in origin.
\citet{MHL00} revise the conclusion of \citet{PVF98} by addressing
ambiguity in their method, from which the conclusion crucially depends 
upon an orientation of the data on the sky, and show no evidence for 
the non-Gaussianity.
The existence of non-Gaussianity in CMB is controversial.

Since non-Gaussianity has infinite degrees of freedom, testing a 
Gaussian hypothesis is difficult; one statistical method
showing CMB {\it consistent} with Gaussian does not mean CMB 
being {\it really} Gaussian.
In this sense, different statistical methods can come to different conclusions.
In addition to the difficulty, cosmological non-Gaussianity is hard to measure.
Instrumental and environmental effects in observations easily produce 
spurious non-Gaussian signals.
Astronomical microwave sources such as interstellar dust emissions 
also  produce strong non-Gaussian fluctuations on the sky.
Hence, we must be as careful as possible when searching for cosmological 
non-Gaussianity.

Pursuit of cosmological non-Gaussianity in CMB is a challenging test of 
inflation, the origin of inhomogeneity in the universe.
In this dissertation, we present theoretical and observational studies 
of the CMB non-Gaussianity.
Our primary goal is to test inflation with non-Gaussian 
CMB fluctuations.

As a statistical tool of searching for non-Gaussianity, we use 
the angular three- and four-point correlation functions in harmonic space,
the {\it angular bispectrum} and {\it trispectrum}, which are sensitive to 
weakly non-Gaussian fluctuations.
An advantage of these angular $n$-point harmonic spectra over other 
statistics is that they are predictable from not only inflation, 
but also secondary sources in the low-redshift universe 
(the Sunyaev--Zel'dovich effect, weak gravitational lensing effect, 
and so on), extragalactic foreground emissions, global topology of 
the universe, and so on.
When we fit those predictions to the measured spectra, 
the predictions become ``matched filters'' for detecting weak
non-Gaussianity in the data, and much more powerful than just
a null test of Gaussianity.

This dissertation is organized as follows.
The rest of this chapter will overview what the CMB sky looks like, and 
the previous study of CMB non-Gaussianity, followed by a 
heuristic description of non-Gaussian fluctuation production in inflation.

In chapter~\ref{chap:inflation}, we go through primordial fluctuation 
generation mechanism in inflation, and investigate how non-linear curvature 
perturbations are generated and propagated through the CMB anisotropy.
We present a possible quantum-to-classical transition mechanism of the 
quantum fluctuations on super horizon scales.

In chapter~\ref{chap:spectrum}, we study statistical properties
of the angular $n$-point harmonic spectrum for $n=2$ (power spectrum),
3 (bispectrum), and 4 (trispectrum).
We present practical methods of measuring the angular bispectrum and
trispectrum from observational data with many pixels.

In chapter~\ref{chap:theory_bl}, we make theoretical predictions
for the CMB angular bispectrum, which include primary contribution from 
inflation, secondary contribution from the Sunyaev--Zel'dovich effect
and the weak-lensing effect, and foreground contribution from
extragalactic radio and infrared astronomical sources. 
We estimate signal-to-noise ratios of detecting each contribution with
CMB experiments, further discussing how well we can 
measure each contribution separately.

In chapter~\ref{chap:obs_bl}, we measure the angular bispectrum 
on the {\it COBE} DMR four-year data, testing Gaussianity of the data.
By fitting the theoretical prediction for the primary bispectrum to the 
data, we constrain non-linearity in inflation;
also fitting foreground bispectra from interstellar microwave emissions
takes into account the effect of non-Gaussian interstellar dust and 
synchrotron emissions at high Galactic latitude.

In chapter~\ref{chap:obs_tl}, we measure the angular trispectrum 
on the DMR data, further testing Gaussianity of the data.

%%%%%%%%%%%%%%%%%%%%%%%%%%%%%%%%%%%%%%%%%%%%%%%%%%%%%%%%%%%%%%%%%%%
\section{Inhomogeneity in Microwave Sky}
\label{sec:CMB}

The cosmic microwave background radiation (CMB) is the isotropic 
microwave radiation filling the sky. 
The temperature is precisely measured to be 2.73~K \citep{Mat90}, and 
the peak intensity at $\nu=160~{\rm GHz}$ ($\lambda=1.9~{\rm mm}$) is 
$370~{\rm MJy~str^{-1}}$.
What if subtracting this mean radiation from the sky, what are we left with,
completely dark, literally nothing?

\subsection{{\it COBE} DMR sky maps}

The Differential Microwave Radiometer (DMR) mounted on the 
{\it Cosmic Background Explorer} ({\it COBE}), the satellite for 
full sky measurement of CMB launched by NASA in 1989, has revealed 
that what we are left with is the tiny inhomogeneity; the r.m.s. 
amplitude is about 30~$\mu$K \citep{Smo92}, 0.001\% of the mean temperature.

Various statistical analyses on the DMR sky maps have shown the angular 
distribution of the CMB anisotropy remarkably consistent with 
the scale-invariant fluctuation, the prediction of inflation 
\citep{Ben96,Gor96,Hin96,Wri96}.
The scale-invariant fluctuation implies that the r.m.s. amplitude of the 
CMB anisotropy is nearly independent of angular scales. 
DMR has measured 35~$\mu$K r.m.s. fluctuations on $7^\circ$ scale, 
29~$\mu$K on $10^\circ$ scale \citep{Ban97}.

\begin{figure}
 \begin{center}
  \leavevmode \epsfxsize=8cm \epsfbox{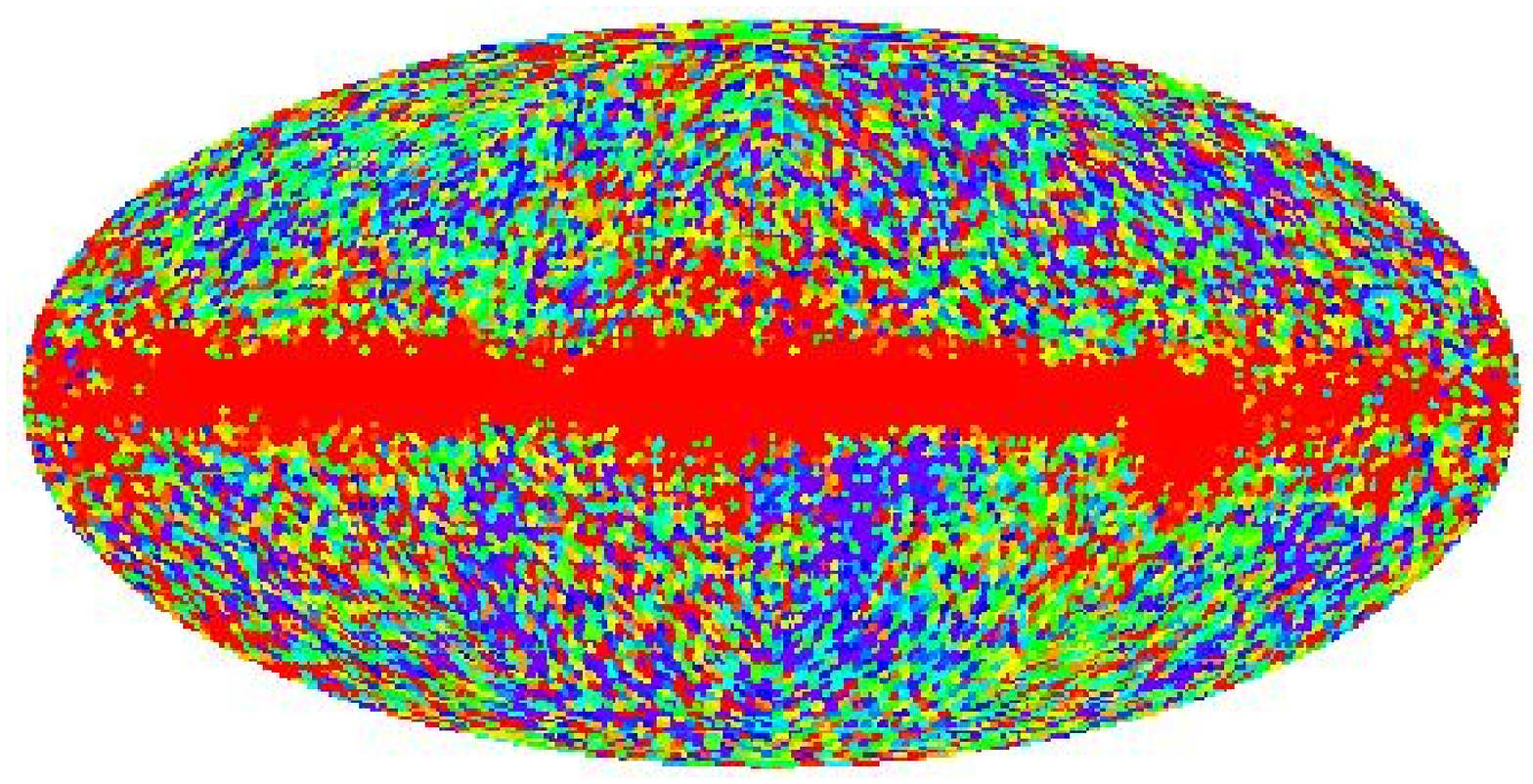}
  \leavevmode \epsfxsize=8cm \epsfbox{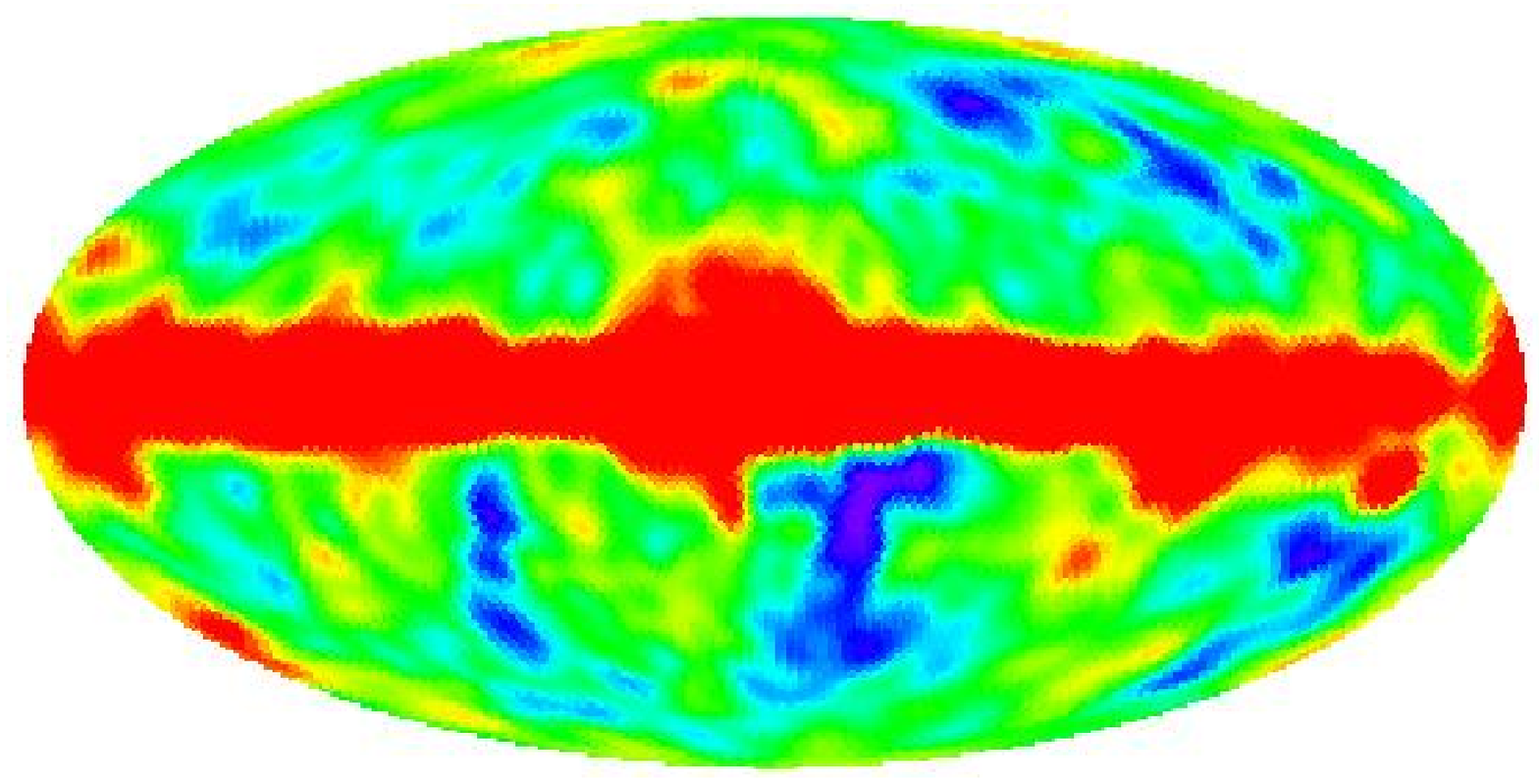}
  \vspace{5mm}
  \leavevmode \epsfxsize=9cm \epsfbox{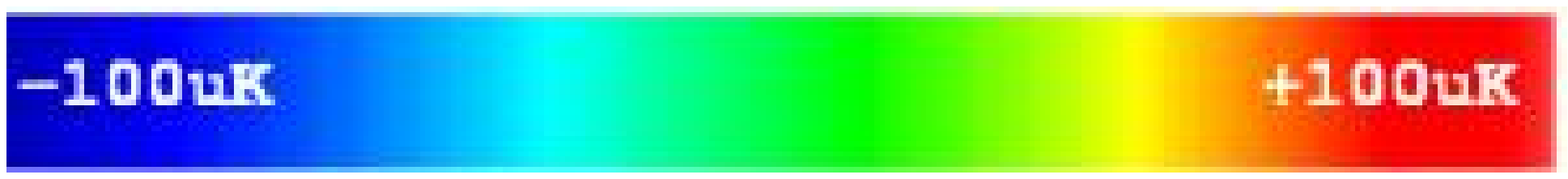}
 \end{center}
 \caption
 {{\it COBE} DMR Sky Map} 
 \mycaption
 {{\it COBE} DMR four-year $53+90$~GHz sky map in
 Galactic projection, using the HEALPix pixelization \citep{GHW98} with 
 $1.\hspace{-4pt}^\circ 83$ pixel size, leaving 12,288 pixels.
 The left is the raw map, while the right map has been smoothed 
 with a $7^\circ$~FWHM Gaussian.}
\label{fig:cobemap}
\end{figure}

DMR comprises 3 dual-horn antennas working at 31.5, 53, and 90~GHz.
The combination of 53 and 90~GHz maps gives the most sensitive sky map 
to CMB, while 31.5~GHz map is twice as noisier as the other channels. 
Figure~\ref{fig:cobemap} shows the combined DMR full sky map.
The left shows the raw map in which the instrumental noise dominates
appearance of the map; the right shows the smoothed map
with the DMR beam in which the instrumental noise is filtered out, giving
better appearance of CMB signals.
The mean signal-to-noise ratio of hot and cold spots in the smoothed map 
is 2, while a few prominent spots have 3 to 4.
Hence, we cannot say much about structures of the CMB anisotropy relying 
on the map basis; however, we can do say on the statistical basis.

\begin{figure}
 \plotone{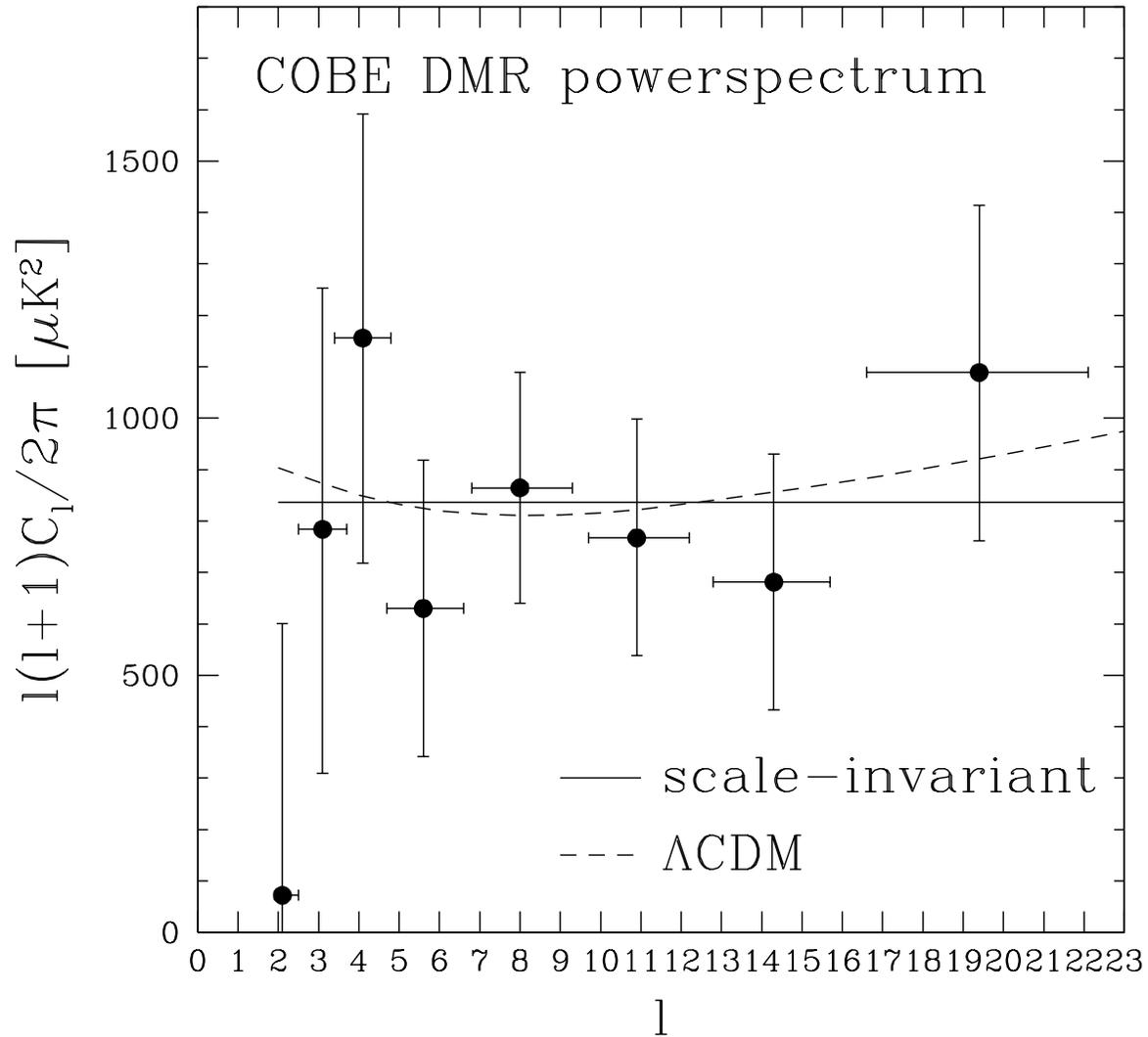}
 \caption
 {{\it COBE} DMR Angular Power Spectrum} 
 \mycaption
 {The CMB angular power spectrum, $C_l$, 
 measured on the {\it COBE} DMR
 four-year map. 
 The plotted quantity, $l(l+1)C_l/(2\pi)$, represents 
 the r.m.s. amplitude of fluctuations at a given angular scale, $l$.
 The data points (filled circles) are uncorrelated with each 
 other \citep{TH97}.
 The solid line shows the scale-invariant power spectrum, 
 $l(l+1)C_l=\mbox{constant}$,
 while the dashed
 line shows a $\Lambda$CDM spectrum.}
\label{fig:cobecl}
\end{figure}

Statistically, structures in the DMR map are inconsistent with pure 
instrumental noise; on the contrary, the structure has a distinct 
angular correlation pattern represented by the scale-invariant fluctuation.
To quantify this, it is useful to calculate the 
{\it angular power spectrum}, $C_l$, the harmonic transform of the angular
two-point correlation function, which measures how much fluctuation 
power exists on a given angular scale, $\theta\sim \pi/l$.
Figure~\ref{fig:cobecl} plots the measured $C_l$ on the DMR map.
What is actually plotted is $l(l+1)C_l/2\pi$, roughly mean squares
of fluctuations at $l$.  
The scale-invariant fluctuation implies 
$C_l\propto \left[l(l+1)\right]^{-1}$ \citep{P82}, and hence 
$l(l+1)C_l$ remains constant (solid line), so do the data points in the 
figure; the data points fit the inflation's prediction well.
The dashed line shows a more accurate prediction, taking into account
the effects of general relativistic photon-baryon fluid dynamics 
before the decoupling as well as of time evolution of gravitational 
potential field after the decoupling. 
The agreement with the data becomes better, further confirming
that the DMR angular power spectrum is consistent with inflation.

%%%%%%%%%%%%%%%%%%%%%%%%%%%%%%%%%%%%%%%%%%%%%%%%%%%%%%%%%%%%%%%%%%%%%%
%%%%%%%%%%%%%%%%%%%%%%%%%%%%%%%%%%%%%%%%%%%%%%%%%%%%%%%%%%%%%%%%%%%%%%

%%%%%%%%%%%%%%%%%%%%%%%%%%%%%%%%%%%%%%%%%%%%%%%%%%%%%%%%%%%%%%%%%%%%%%
%%%%%%%%%%%%%%%%%%%%%%%%%%%%%%%%%%%%%%%%%%%%%%%%%%%%%%%%%%%%%%%%%%%%%%

Gaussianity of the DMR data has been tested with various 
statistical methods 
\citep{Kog96b,FMG98,PVF98,BT99,BZG00,MHL00,Mag00,SM00,Bar00}.
\citet{FMG98} and \citet{Mag00} claim positive 
detection of non-Gaussian signals using the angular bispectrum, 
and \citet{PVF98} claim detection using the wavelet analysis.
The latter non-Gaussian signal has appeared to be less significant 
than they claim, as it disappears when the DMR map is rotated by 
$180^\circ$ \citep{MHL00,Bar00}.
For the former two bispectrum analyses, 
\citet{BT99} and \citet{BZG00} claim that the \citet{FMG98}'s signal 
is non-cosmological, but their claims do not account for the 
\citet{Mag00}'s signal.
In this thesis, we will argue that the reported non-Gaussian signals 
are not a matter of origin, but statistical fluctuations.

\subsection{Post-{\it COBE} era}

After the discovery of {\it COBE}, pursuit of the CMB anisotropy has 
been oriented toward measurement of the angular
power spectrum, $C_l$, on smaller angular scales, i.e., larger $l$.
Particularly, many efforts have been made to measure $C_l$ 
at $l\sim 200$ ($\theta\sim 1^\circ$), 
where inflation predicts a prominent peak in $l(l+1)C_l$ as a 
consequence of flatness of the universe, the prediction of
inflation \citep{KSS94}.

\begin{figure}
 \plotone{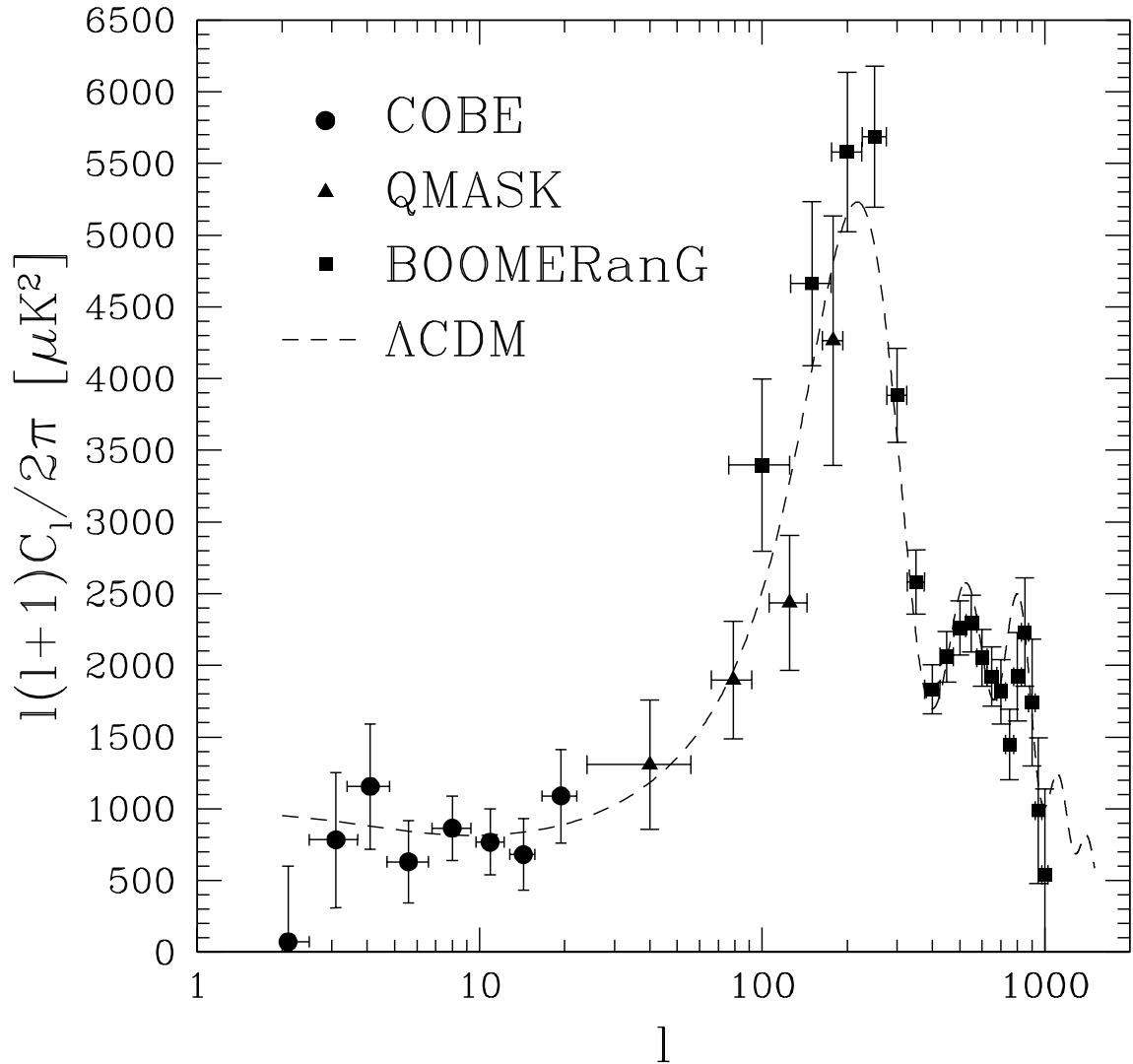}
 \caption
 {{\it COBE} DMR, {\it QMASK}, and {\it BOOMERanG} Angular Power Spectra} 
 \mycaption
 {The CMB angular power spectrum, measured by the three experiments 
 probing different angular scales.
 The circles are the {\it COBE} DMR data
 (full sky coverage with a Galactic cut, $7^\circ$ angular resolution), 
 the triangles are the {\it QMASK} data 
 (648-square-degree sky coverage, $0.\hspace{-3.pt}^\circ 68$ angular 
 resolution), and the squares are the {\it BOOMERanG} data
 (1,800-square-degree sky coverage, 
 $0.\hspace{-3.pt}^\circ 17$ angular resolution).
 The dashed line shows a prediction of inflation, a $\Lambda$CDM
 spectrum with $\Omega_{\rm m}=0.3$, $\Omega_{\Lambda}=0.7$, 
 $\Omega_{\rm b}=0.04$, $h=0.7$, and $n=0.95$.}
\label{fig:newcl}
\end{figure}

By early 2000, there has been strong evidence for the peak \citep{TOCO99}; 
in the end of 2000, {\it BOOMERanG} and {\it MAXIMA}, balloon-borne 
CMB experiments, have detected the peak \citep{Boom00,Maxima00},
further supporting inflation.
Figure~\ref{fig:newcl} compares the data from three experiments, 
which probe different angular scales from one another, 
with a prediction from inflation.
The agreement between the data and the prediction is outstanding.

%%%%%%%%%%%%%%%%%%%%%%%%%%%%%%%%%%%%%%%%%%%%%%%%%%%%%%%%%%%%%%%%%%%%%%
%%%%%%%%%%%%%%%%%%%%%%%%%%%%%%%%%%%%%%%%%%%%%%%%%%%%%%%%%%%%%%%%%%%%%%

Measurement of $C_l$ so far, however, has assumed CMB Gaussian.
If CMB is not Gaussian, then the measured $C_l$ is biased.
Moreover, when we fit the measured $C_l$ to a theoretical $C_l$,
we need to know the covariance matrix of $C_l$.
Since the covariance matrix of $C_l$ is the four-point harmonic spectrum,
the trispectrum, we have to investigate non-Gaussian signals 
in the trispectrum to construct the covariance matrix accurately.

Even if non-Gaussianity is small, we have to take it into account
in analyzing $C_l$; the next generation satellite experiments,
{\it MAP} and {\it Planck}, will measure $C_l$ with 1\% or better
accuracy, and we will use the measured $C_l$ to determine many of
cosmological parameters with 10\% or better accuracy.
Unless non-Gaussian effect is much smaller than the observational
uncertainty (who knows?), 
we have to take the effect into account, to achieve the 
accurate measurement of the parameters.

\begin{figure}
 \begin{center}
  \leavevmode \epsfxsize=9cm \epsfbox{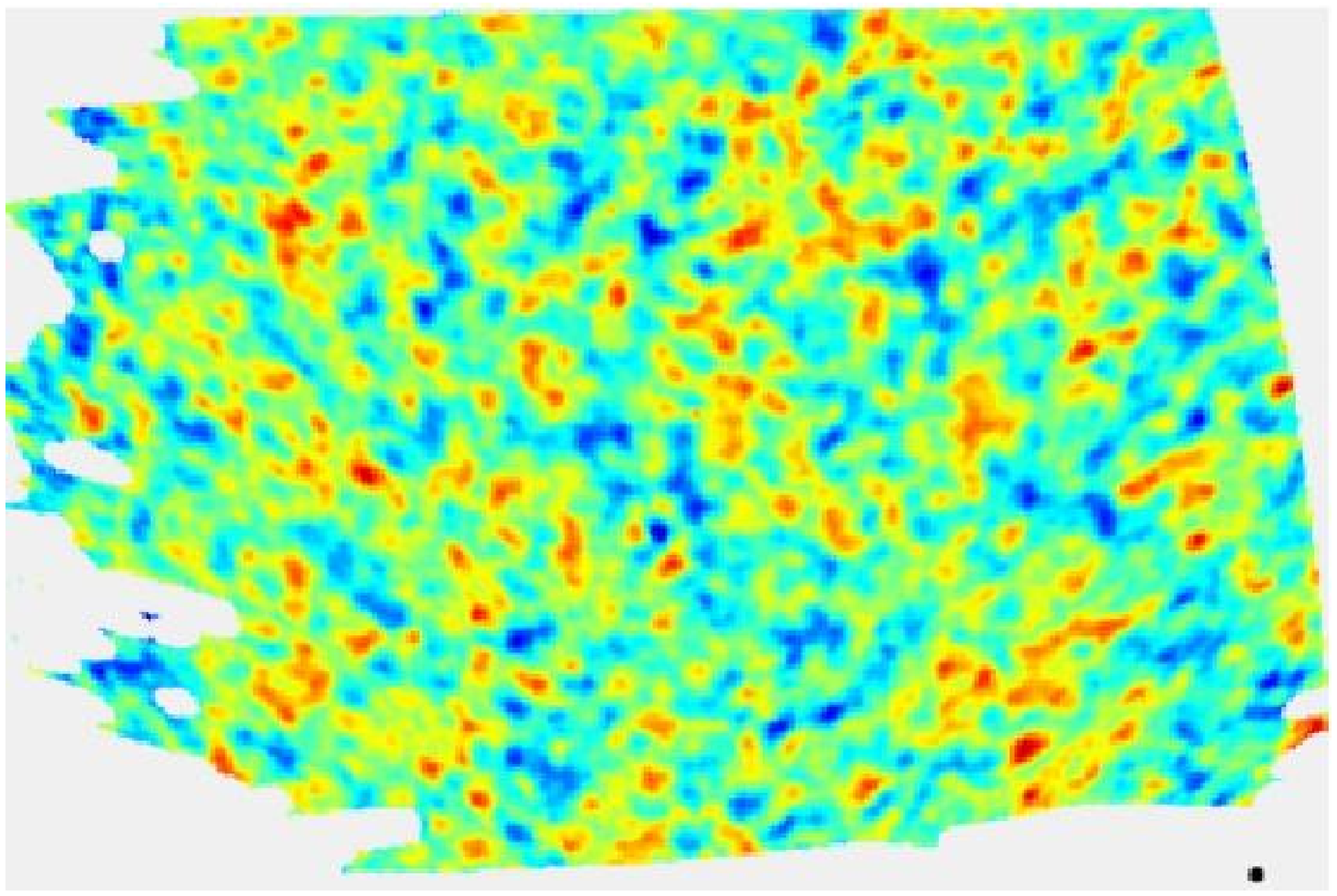}
  \hspace{5mm}
  \leavevmode \epsfxsize=5.4cm \epsfbox{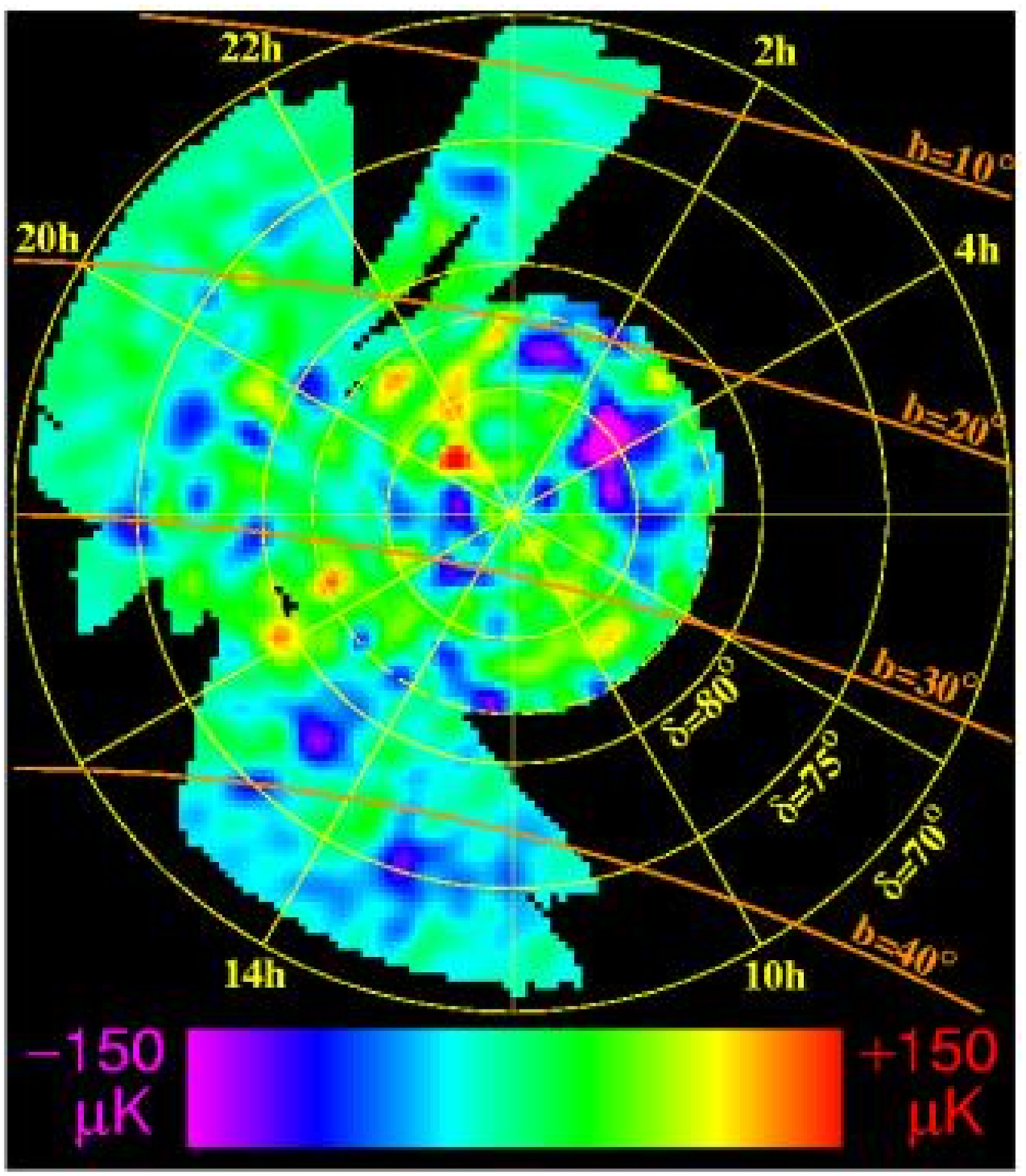}
 \end{center}
 \caption
 {Sub-degree CMB Maps: {\it BOOMERanG} and {\it QMASK}} 
 \mycaption
 {The left is the {\it BOOMERanG} sky map, which covers
 3\% of the sky, 1,800 square degrees, with $10'$ 
 ($0.\hspace{-3.pt}^\circ 17$) angular resolution \citep{Boom00}.
 The beam size is indicated by the filled circle at the bottom-right corner.
 The right is the {\it QMASK} sky map, which covers
 648 square degrees of the sky with $0.\hspace{-3.pt}^\circ 68$ 
 angular resolution \citep{Xu01}.
 Both maps have been scaled to match each other in size.
 The {\it QMASK} map is one-third of the {\it BOOMERanG} map.}
\label{fig:boomqmask}
\end{figure}

The state-of-the-art balloon-borne experiments provide not only 
$C_l$, but also high resolution, high signal-to-noise ratio CMB maps.
The left of figure~\ref{fig:boomqmask} shows the {\it BOOMERanG} 
sky map, which covers roughly 3\% of the sky, 1,800 square degrees, 
with $10'$ ($0.\hspace{-3.pt}^\circ 17$) angular resolution.
The high signal-to-noise ratios in the map show the observed structures 
in the map not instrumental noise, but CMB.

%%%%%%%%%%%%%%%%%%%%%%%%%%%%%%%%%%%%%%%%%%%%%%%%%%%%%%%%%%%%%%%%%%%%%%
%%%%%%%%%%%%%%%%%%%%%%%%%%%%%%%%%%%%%%%%%%%%%%%%%%%%%%%%%%%%%%%%%%%%%%

Visually, the CMB anisotropy in the map looks very much Gaussian,
implying non-Gaussianity is weak, if any.
As yet the {\it BOOMERanG} team has not performed Gaussianity tests
on the data; however, the {\it MAXIMA} team has done on their data.
The {\it MAXIMA} map covers 124 square degrees of the sky with 
$10'$ angular resolution, leaving 5,972 pixels.
They have found that the one-point probability density distribution 
function and the 
Minkowski functionals measured on the map are consistent with Gaussianity.

The genus statistic has been applied to the {\it QMASK} map
\citep{Xu01}, a high signal-to-noise ratio CMB map combining 
the balloon-borne {\it QMAP} \citep{Qmap1,Qmap2,Qmap3} and 
the ground-based {\it Saskatoon} experiments \citep{Sask97}, 
showing the map consistent with Gaussian \citep{Park01}.
The {\it QMASK} map is shown on the right of figure~\ref{fig:boomqmask},
where the size of the map has been scaled to match
the {\it BOOMERanG} map.
The {\it QMASK} map covers 648 square degrees of the sky,
one-third of the {\it BOOMERanG} map, with $0.\hspace{-3.pt}^\circ 68$ 
angular resolution, leaving 6,495 pixels.

The number of independent pixels that has been used for Gaussianity 
tests on the DMR (3,881), {\it MAXIMA} (5,972), and {\it QMASK} 
(6,495) data is similar to each other; although their angular scales 
are quite different, they have similar power of testing Gaussianity.
We will have a significant progress when the {\it BOOMERanG} data,
in which the number of pixels is 57,103, test the Gaussianity.

The just-launched CMB satellite, the {\it Microwave Anisotropy Probe} 
({\it MAP}), will increase our power of pursuing non-Gaussianity 
substantially. 
It will survey the full sky with the angular resolution better than 
$14'$, and have $\simgt 10^6$ pixels.
The statistical methods developed in this thesis can readily be applied to
the {\it MAP} data, enabling us to test the Gaussianity with 
unprecedented sensitivity.

%%%%%%%%%%%%%%%%%%%%%%%%%%%%%%%%%%%%%%%%%%%%%%%%%%%%%%%%%%%%%%%%%%%
\section{Non-Gaussian Fluctuations in Inflation}
\label{sec:nongaus}

\subsection{Adiabatic production of non-Gaussian fluctuations}

While inflation predicts Gaussian CMB fluctuations to very good accuracy,
strictly speaking, non-linearity in inflation produces weakly 
non-Gaussian fluctuations, which propagates through CMB.
Although the exact treatment is complicated, we present a basic
idea behind it concisely here \citep{KS01b}.

The curvature perturbations, $\Phi$, generate the CMB anisotropy, 
$\Delta T/T$.
The linear perturbation theory gives a linear relation between
$\Phi$ and $\Delta T/T$,
%%%%%%%%%%%%%%%%%%%%%%%%%%%%%%%%%%%%%%%%%%%%%%%%%%%%%%%%%%%%%%%%%%%%
\begin{equation}
  \frac{{\Delta T}}T\sim g_{\rm T}\Phi,
\end{equation}
%%%%%%%%%%%%%%%%%%%%%%%%%%%%%%%%%%%%%%%%%%%%%%%%%%%%%%%%%%%%%%%%%%%%
where $g_{\rm T}$ is the radiation transfer function.
For temperature fluctuations on super-horizon scales at the decoupling
epoch, the Sachs-Wolfe effect \citep{SW67} dominates, and 
$g_{\rm T}=-1/3$ for adiabatic fluctuations.
On sub-horizon scales, $g_{\rm T}$ oscillates (acoustic oscillation),
and we need to solve the Boltzmann photon transport equations
coupled with the Einstein equations for $g_{\rm T}$.
It follows from the relation, $\Delta T\propto \Phi$, 
that $\Delta T$ is Gaussian, if $\Phi$ is Gaussian.
As we will see, non-linearity in inflation makes $\Phi$ weakly 
non-Gaussian.

Even if $\Phi$ is Gaussian, $\Delta T/T$ can be non-Gaussian. 
According to the general relativistic cosmological perturbation theory,
there is a non-linear relation between $\Delta T/T$ and $\Phi$: 
%%%%%%%%%%%%%%%%%%%%%%%%%%%%%%%%%%%%%%%%%%%%%%%%%%%%%%%%%%%%%%%%%%%%
\begin{equation}
 \label{eq:T-Phi}
  \frac{{\Delta T}}T\sim g_{\rm T}\left(\Phi+f_\Phi \Phi^2\right).
\end{equation}
%%%%%%%%%%%%%%%%%%%%%%%%%%%%%%%%%%%%%%%%%%%%%%%%%%%%%%%%%%%%%%%%%%%%
Here, the second term with a coefficient of order unity, 
$f_\Phi\sim {\cal O}(1)$, is the higher-order correction arising from 
the second-order perturbation theory \citep{PC96}.
It produces non-Gaussian fluctuations; thus, even if $\Phi$ is 
Gaussian, $\Delta T$ becomes weakly non-Gaussian.

Is $\Phi$ Gaussian?
Non-linearity in inflation makes $\Phi$ weakly non-Gaussian. 
By expanding the fluctuation dynamics in inflation up to the 
second order, we obtain a non-linear relation between $\Phi$ and inflaton
fluctuations, $\delta\phi$:
%%%%%%%%%%%%%%%%%%%%%%%%%%%%%%%%%%%%%%%%%%%%%%%%%%%%%%%%%%%%%%%%%%%%
\begin{equation}
 \label{eq:Phi-inflaton}
  \Phi\sim m_{\rm pl}^{-1}g_{\Phi}
           \left(\delta\phi+m_{\rm pl}^{-1}f_{\delta\phi}\delta\phi^2\right).
\end{equation}
%%%%%%%%%%%%%%%%%%%%%%%%%%%%%%%%%%%%%%%%%%%%%%%%%%%%%%%%%%%%%%%%%%%%
\citet{SB90} show that this relation is a non-linear solution 
for curvature perturbations on super horizon scales;
the solution gives $g_{\Phi}\sim {\cal O}(10)$ and 
$f_{\delta\phi}\sim {\cal O}(10^{-1})$ for a class of slowly-rolling
single-field inflation models.

\begin{figure}
 \begin{center}
  \leavevmode\epsfxsize=8cm \epsfbox{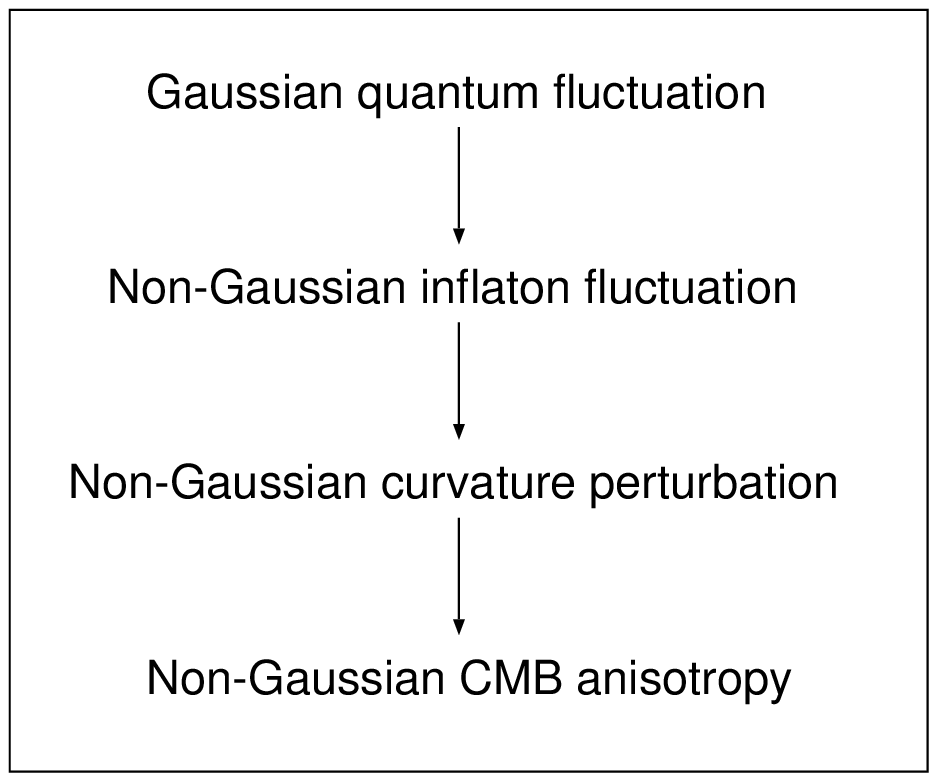}
 \end{center}
 \caption
 {Three Steps for Non-Gaussian CMB Anisotropy} 
 \mycaption
 {Adiabatic generation of non-Gaussian CMB anisotropies.
 First, inflation generates Gaussian quantum fluctuations, which
 become non-Gaussian inflaton fluctuations through a non-linear
 coupling between them (Eq.(\ref{eq:inflaton-eta})).
 Then, the inflaton fluctuations become more non-Gaussian curvature 
 perturbations through a non-linear relation between them
 (Eq.(\ref{eq:Phi-inflaton})).
 Finally, the curvature perturbations become more non-Gaussian CMB 
 anisotropies through non-linear gravitational effects (Eq.(\ref{eq:T-Phi})).}
\label{fig:chart}
\end{figure}

Quantum fluctuations produce Gaussian $\delta\phi$.
If the dynamics of $\delta\phi$ is simple enough to keep itself 
Gaussian throughout the evolution, then we can stop our consideration here;
however, it is not necessarily true. 
For example, non-trivial interaction terms in the equation of 
motion for inflaton fields \citep{FRS93}, or a non-linear coupling between 
long-wavelength classical fluctuations and short-wavelength quantum
fluctuations in the context of stochastic inflation \citep{Sta86,Gan94}, 
can make $\delta\phi$ weakly non-Gaussian, resulting in a non-linear
relation between $\delta\phi$ and a Gaussian field, $\eta$,
%%%%%%%%%%%%%%%%%%%%%%%%%%%%%%%%%%%%%%%%%%%%%%%%%%%%%%%%%%%%%%%%%%%%
\begin{equation}
 \label{eq:inflaton-eta}
  \delta\phi\sim g_{\delta\phi}
                 \left(\eta+m_{\rm pl}^{-1}f_{\eta}\eta^2\right),
\end{equation}
%%%%%%%%%%%%%%%%%%%%%%%%%%%%%%%%%%%%%%%%%%%%%%%%%%%%%%%%%%%%%%%%%%%%
where $\eta$ represents initially produced quantum
fluctuations, and $g_{\delta\phi}\sim 1$ and $f_{\eta}\sim {\cal O}(10^{-1})$.
Figure~\ref{fig:chart} summarizes the above three steps 
in the opposite order.

%%%%%%%%%%%%%%%%%%%%%%%%%%%%%%%%%%%%%%%%%%%%%%%%%%%%%%%%%%%%%%%%%%%%%%
%%%%%%%%%%%%%%%%%%%%%%%%%%%%%%%%%%%%%%%%%%%%%%%%%%%%%%%%%%%%%%%%%%%%%%

Collecting all the above contributions, we obtain a non-linear
relationship between $\Delta T/T$ and $\Phi$,
%%%%%%%%%%%%%%%%%%%%%%%%%%%%%%%%%%%%%%%%%%%%%%%%%%%%%%%%%%%%%%%%%%%%
\begin{equation}
 \label{eq:T-Phi_L}
  \frac{{\Delta T}}T\sim 
  g_{\rm T}\left[\Phi_{\rm L}
            +\left(f_\Phi
                  +g_\Phi^{-1}f_{\delta\phi}
                  +g_\Phi^{-1}g_{\delta\phi}^{-1}f_{\eta}
             \right)\Phi_{\rm L}^2\right],
\end{equation}
%%%%%%%%%%%%%%%%%%%%%%%%%%%%%%%%%%%%%%%%%%%%%%%%%%%%%%%%%%%%%%%%%%%%
where $\Phi_{\rm L}\equiv g_\Phi g_{\delta\phi}m_{\rm pl}^{-1}\eta 
\sim 10m_{\rm pl}^{-1}\eta$ 
is an auxiliary Gaussian curvature perturbation.
It may be useful to define a {\it non-linear coupling parameter},
$f_{\rm NL}= f_\Phi +g_\Phi^{-1}f_{\delta\phi}+
g_\Phi^{-1}g_{\delta\phi}^{-1}f_{\eta}$.
The first term in $f_{\rm NL}$, the second order gravity effect
$\sim {\cal O}(1)$, is dominant compared with the other two terms
$\sim {\cal O}(10^{-2})$, 
non-linearity in slow-roll inflation.
Note that $f_{\rm NL}$ corresponds to $-\Phi_3/2$ in \citet{Gan94} and
$-\alpha_\Phi$ in \citet{VWHK00}.
Using $f_{\rm NL}$, we rewrite equation~(\ref{eq:T-Phi_L}) as
$\Delta T({\mathbf x})/T\sim g_{\rm T}\Phi({\mathbf x})$, where
%%%%%%%%%%%%%%%%%%%%%%%%%%%%%%%%%%%%%%%%%%%%%%%%%%%%%%%%%%%%%%%%%%%%
\begin{equation}
 \label{eq:Phi}
  \Phi({\mathbf x})
  = \Phi_{\rm L}({\mathbf x})
   +f_{\rm NL}\left[
          \Phi^2_{\rm L}({\mathbf x})-\left<\Phi^2_{\rm L}({\mathbf x})\right>
          \right],
\end{equation}
%%%%%%%%%%%%%%%%%%%%%%%%%%%%%%%%%%%%%%%%%%%%%%%%%%%%%%%%%%%%%%%%%%%%
the angular bracket denoting the statistical ensemble average.

\begin{figure}
 \plotone{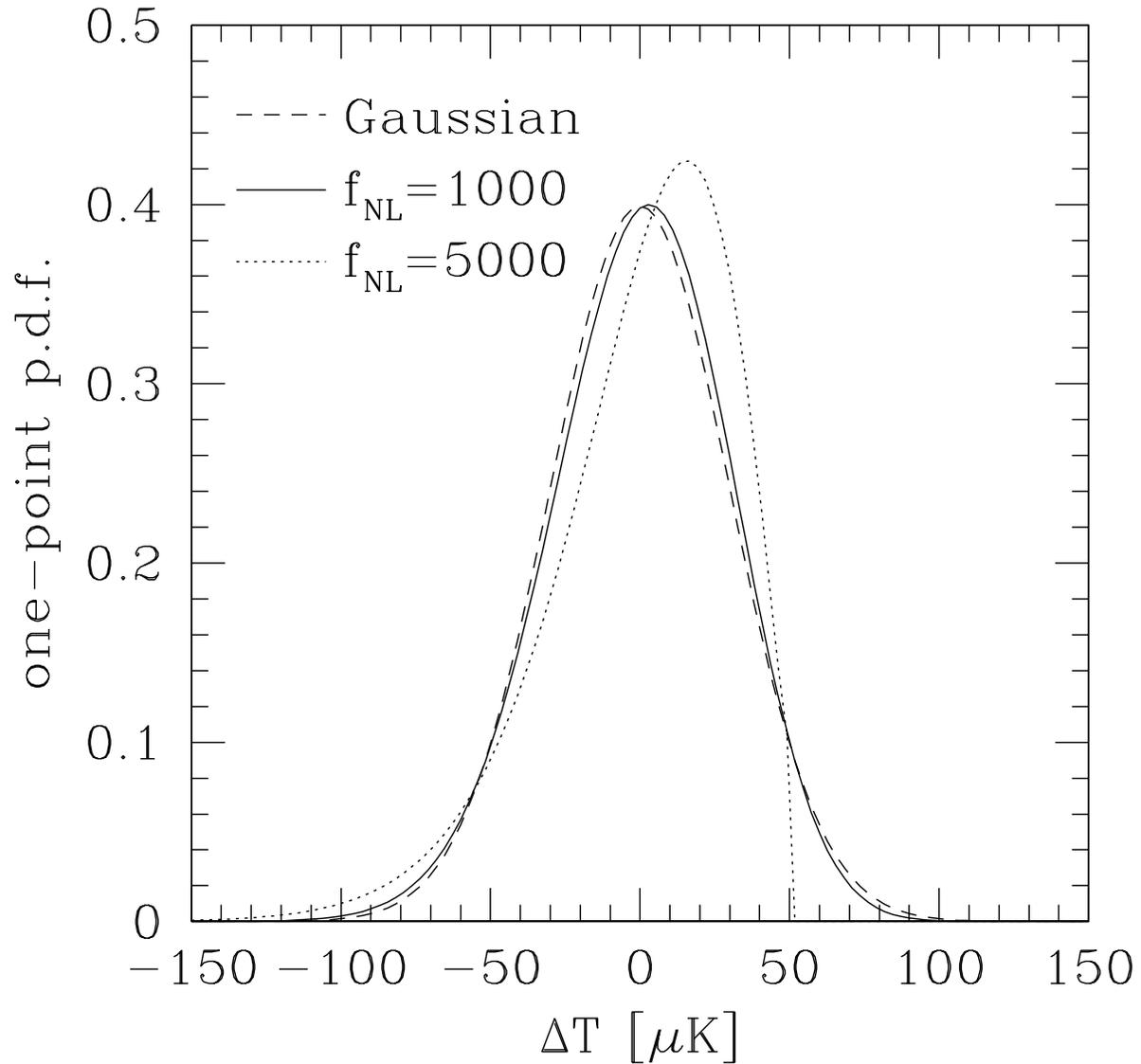}
 \caption
 {Adiabatic Non-Gaussianity: One-point p.d.f} 
 \mycaption
 {One-point probability density distribution function (p.d.f) of the CMB
 anisotropy, $\Delta T/T$, comparing a Gaussian p.d.f with 
 non-Gaussian p.d.f's of non-linear adiabatic fluctuations produced in inflation.
 The dashed line plots Gaussian distribution.
 The solid line plots non-Gaussian distribution for $f_{\rm NL}=1000$,
 the dotted line for $f_{\rm NL}=5000$.
 The larger $f_{\rm NL}$ is, the more negatively skewed p.d.f becomes.
 If $f_{\rm NL}<0$, then p.d.f becomes positively skewed.}
\label{fig:pdf}
\end{figure}

To see intuitively what non-Gaussian fluctuations that we 
have considered here look like, in figure~\ref{fig:pdf} we plot one-point 
probability density distribution function (p.d.f) of the CMB anisotropy.
We compare Gaussian p.d.f with non-Gaussian p.d.f of adiabatic 
fluctuations produced in inflation.
The dashed line is Gaussian distribution, i.e., no non-linear
perturbations are included ($f_{\rm NL}=0$).
The solid line includes non-linear coupling of order $f_{\rm NL}=1000$,
while the dotted line of order $f_{\rm NL}=5000$.
It follows from the figure that positive $f_{\rm NL}$ gives
negatively skewed p.d.f; negative $f_{\rm NL}$ gives positively skewed p.d.f.
The larger $\left|f_{\rm NL}\right|$ is, the more skewed p.d.f becomes.

%%%%%%%%%%%%%%%%%%%%%%%%%%%%%%%%%%%%%%%%%%%%%%%%%%%%%%%%%%%%%%%%%%%%%%
%%%%%%%%%%%%%%%%%%%%%%%%%%%%%%%%%%%%%%%%%%%%%%%%%%%%%%%%%%%%%%%%%%%%%%

We find that $f_{\rm NL}=1000$ gives virtually identical p.d.f to the 
Gaussian p.d.f.
In chapter~\ref{chap:theory_bl}, we will show that even with the ideal CMB 
experiment, we can measure $\left|f_{\rm NL}\right|$ no smaller than 60, if we use the 
skewness of the one-point p.d.f.
If we use the bispectrum, however, we can measure 
as small $\left|f_{\rm NL}\right|$ as 3.
The bispectrum is thus much more sensitive to the adiabatic non-Gaussianity
than the skewness of the one-point p.d.f is.
In chapter~\ref{chap:obs_bl}, we will show that the angular bispectrum
measured on the {\it COBE} DMR sky map constrains 
$\left|f_{\rm NL}\right|<1.6\times 10^3$.
The next generation satellite experiments, {\it MAP} and {\it Planck},
will improve the constraint substantially.

Since the minimum $\left|f_{\rm NL}\right|$ detectable with CMB 
experiments is 3, the second order gravity effect, 
$f_{\rm NL}\sim {\cal O}(1)$, will not produce detectable 
non-Gaussianity in CMB, nor will
slow-roll inflation, $f_{\rm NL}\sim {\cal O}(10^{-2})$.

Yet, inflation may not be so simple. 
Any significant deviation from slow-roll, or features in an inflaton 
potential \citep{KBHP91}, could produce a bigger $\left|f_{\rm NL}\right|$, 
bigger non-Gaussianity.
While we have restricted ourselves to adiabatic fluctuations,
the deviation from Gaussianity can be more 
significant if inflation produces non-negligible isocurvature fluctuations
\citep{LM97,BZ97,P97}. 
Measurement of non-Gaussian CMB anisotropies thus potentially constrains
non-linearity, ``slow-rollness'', and ``adiabaticity'' in inflation.
In the next subsection, we concisely describe a possible mechanism to
produce isocurvature fluctuations in inflation.

\subsection{Isocurvature fluctuations}

Isocurvature fluctuations do not perturb spatial curvature at the 
initial fluctuation-generation epoch.
In inflation, in addition to a scalar field responsible for 
adiabatic fluctuations, 
another scalar field, $\sigma$, may produce 
isocurvature density fluctuations with amplitude of 
$\delta\rho_{\sigma}/\rho_{\sigma}\sim H^2/(d\sigma/dt)$,
where $H$ is the Hubble parameter during inflation.
This formula assumes that $\sigma$ rolls down on its potential very slowly.
In some cases, this fluctuation amplitude is about the same as
adiabatic density fluctuations generated by a  
scalar field, $\phi$, which drives inflation: 
$\delta\rho_\sigma/\rho_\sigma\sim \delta\rho_\phi/\rho_\phi\sim 
H^2/(d\phi/dt)$.
This happens when both fluctuations are produced in a similar way,
through the quantum-fluctuation production in inflation.

Even if $\delta\rho/\rho$ is similar to each other, 
the energy density, $\rho$,  can be significantly different.
Since $\phi$ drives inflation, its energy density, $\rho_\phi$,
dominates the total energy density of the universe during inflation:
$\rho_\phi\gg \rho_\sigma$; thus, it gives $\delta\rho_\sigma\sim 
\delta\rho_\phi\left(\rho_\sigma/\rho_\phi\right)\ll \delta\rho_\phi$.
Then, the density fluctuations generate the curvature perturbations, $\Phi$.
Since $\delta\rho_\sigma\ll \delta\rho_\phi$, $\sigma$
makes negligible contribution to $\Phi$ compared with $\phi$, i.e., 
$\delta\rho_\sigma$ does not generate the curvature perturbations, being 
an isocurvature mode.
In this model, $\delta\rho_\sigma$ is Gaussian, as 
the quantum fluctuations have produced it linearly.

If $\sigma$ moves fast, then the quantum fluctuations produce 
$\delta\rho_\sigma$ non-linearly; we have
non-Gaussian density fluctuations.
\citet{LM97} have proposed a massive-free field oscillating about
its potential minimum as a possible non-Gaussian isocurvature-fluctuation
production mechanism in inflation.
The idea is as follows.
When a field rolls down on a potential, $V(\sigma)$, very slowly, 
quantum fluctuations of $\sigma$, $\delta\sigma$, produce
the energy density fluctuations of
$\delta\rho_\sigma\sim (dV/d\sigma)\delta\sigma$; thus,
$\delta\rho_\sigma$ is linear in $\delta\sigma$, being Gaussian.
In contrast, when a field oscillates rapidly about $\sigma=0$, 
there is no mean field; for example, a massive-free scalar field
with a potential $V(\sigma)=m^2\sigma^2/2$ for $m\simgt H$ 
produces the density fluctuations of
%%%%%%%%%%%%%%%%%%%%%%%%%%%%%%%%%%%%%%%%%%%%%%%%%%%%%%%%%%%%%%%%%%%%
\begin{equation}
 \delta\rho_\sigma\sim m^2\sigma\delta\sigma + m^2(\delta\sigma)^2
= m^2(\delta\sigma)^2.
\end{equation}
%%%%%%%%%%%%%%%%%%%%%%%%%%%%%%%%%%%%%%%%%%%%%%%%%%%%%%%%%%%%%%%%%%%%
Here, $m\simgt H$ ensures that $\sigma$ has rolled down to $\sigma=0$
quickly, and oscillates.
Hence, $\delta\rho_\sigma$ is {\it quadratic} in $\delta\sigma$, being
non-Gaussian.

After the initial generation of isocurvature fluctuations, 
$\sigma$ may produce the curvature perturbations through the evolution.
If $\sigma$ does not decay, or decays only very slowly, the energy 
density decreases as $a^{-3}$.
On the other hand, the radiation energy density that is produced
during the reheating phase by a decaying scalar field $\phi$ that has driven 
inflation decreases as $a^{-4}$;
thus, at some point in the cosmic evolution, 
the $\sigma$-field energy density dominates the 
universe, producing the curvature perturbations, 
%%%%%%%%%%%%%%%%%%%%%%%%%%%%%%%%%%%%%%%%%%%%%%%%%%%%%%%%%%%%%%%%%%%%
\begin{equation}
 \Phi({\mathbf x})= \eta^2({\mathbf x}) - \left<\eta^2({\mathbf x})\right>,
\end{equation}
%%%%%%%%%%%%%%%%%%%%%%%%%%%%%%%%%%%%%%%%%%%%%%%%%%%%%%%%%%%%%%%%%%%%
and hence the CMB anisotropies, $\Delta T/T\sim g_{\rm T}\Phi$.
Here, $\eta$ is a Gaussian fluctuation field which is related to 
$\delta\sigma$, and $g_{\rm T}$ is the isocurvature 
radiation transfer function.
The Sachs--Wolfe effect gives $g_{\rm T}= -2$.

The CMB experiments show that isocurvature fluctuations do not 
contribute to the curvature 
perturbations very much; on the contrary, their contribution 
is negligible compared with adiabatic contribution.
Figure~\ref{fig:newcl} compares a prediction for the CMB angular 
power spectrum from adiabatic fluctuations with the data.
The agreement is very good, and there is no need to invoke
isocurvature fluctuations. 
Moreover, the isocurvature fluctuations predict a very different
form of the power spectrum; thus, the data have excluded possibility 
of the isocurvature fluctuations dominating the observed CMB 
power spectrum at high significance.

Yet, there could exist isocurvature fluctuations in inflation.
Generally speaking, if there are many scalar fields, there must
exist isocurvature fluctuations.
It is rather unusual to assume {\it only one} scalar field during 
inflation, for currently viable theories of the high energy particle 
physics predict existence of many kinds of scalar fields
in a very high energy regime.
There is, however, little hope to detect their signatures in the 
CMB power spectrum, as they are so weak compared with adiabatic 
fluctuations.
Instead, searching for non-Gaussian signals in CMB is a promising
strategy to look for some of those isocurvature fluctuations which
are generally much more non-Gaussian than the adiabatic fluctuations.

\begin{figure}
 \plotone{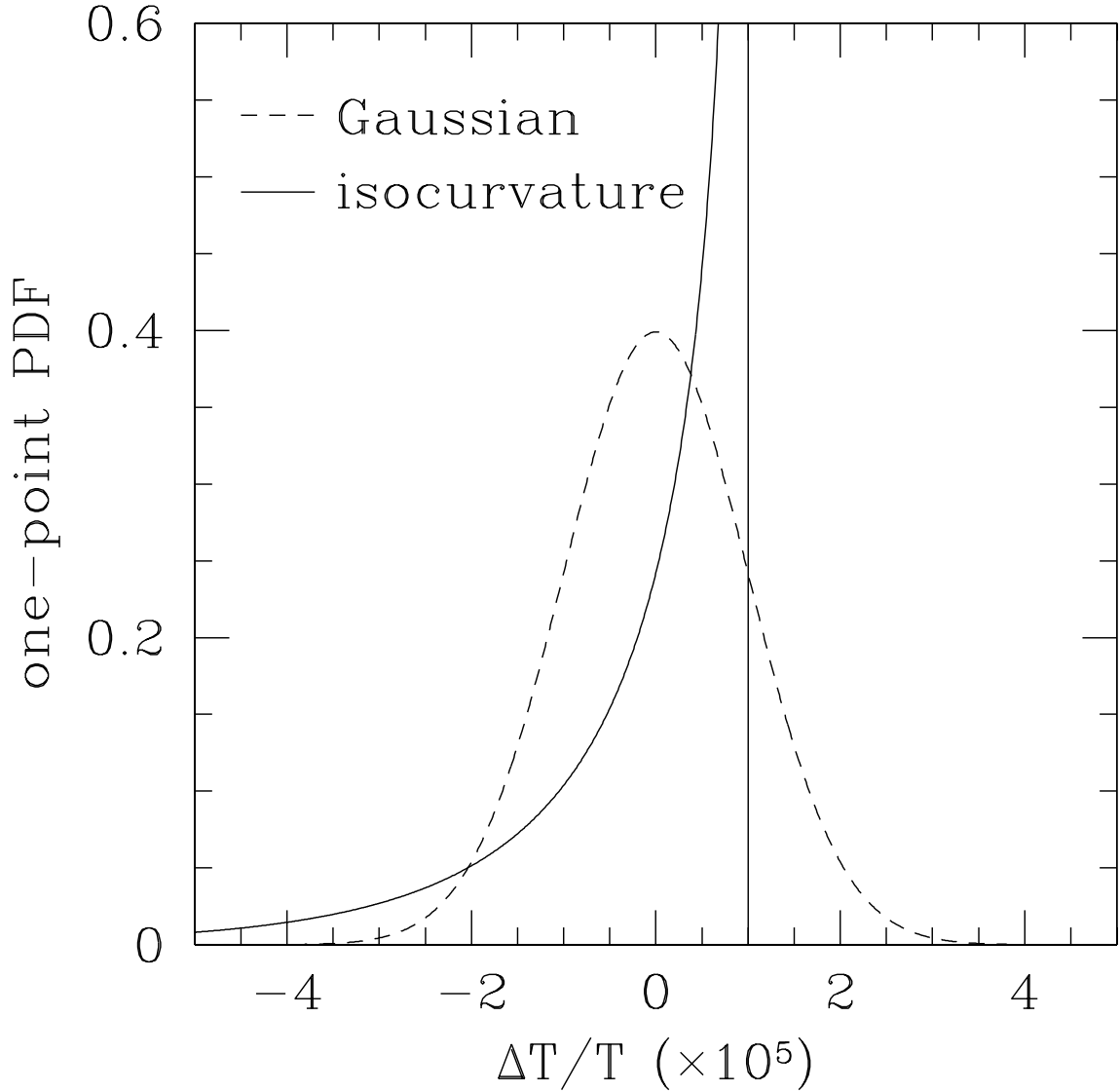}
 \caption
 {Isocurvature Non-Gaussianity: One-point p.d.f} 
 \mycaption
 {One-point probability density distribution function (p.d.f) of the CMB
 temperature anisotropy, $\Delta T/T$, comparing a Gaussian p.d.f with 
 a non-Gaussian p.d.f of isocurvature fluctuations produced in inflation
 \citep{LM97}.
 The dashed line plots Gaussian distribution, while 
 the solid line plots the non-Gaussian distribution.
 Note that we have assumed no beam smoothing; the large beam smoothing
 makes the non-Gaussian distribution similar to the Gaussian distribution 
 \citep{NSM00}.}
\label{fig:pdf_iso}
\end{figure}

Since $\Delta T/T$ is quadratic in a Gaussian variable, 
one-point p.d.f of $\Delta T/T$ is the $\chi^2$ distribution with
one degree of freedom.
Figure~\ref{fig:pdf_iso} plots the one-point p.d.f of the isocurvature
model (solid line) in comparison with Gaussian p.d.f (dashed line).
The predicted p.d.f is highly non-Gaussian.
If we assume the isocurvature CMB fluctuations dominating the universe,
then the predicted non-Gaussian p.d.f may look too non-Gaussian to be 
consistent with observations; however, the {\it COBE} DMR data
do not exclude this model on the basis of the non-Gaussianity
because of the large beam-smoothing effect \citep{NSM00}.
The bigger the beam is, the closer the smoothed $\chi^2$ distribution is
to Gaussian distribution \citep{NSM00}.
The CMB experiments probing much smaller angular scales than DMR will test the
isocurvature non-Gaussian models.

%%%%%%%%%%%%%%%%%%%%%%%%%%%%%%%%%%%%%%%%%%%%%%%%%%%%%%%%%%%%%%%%%%%%%%
%%%%%%%%%%%%%%%%%%%%%%%%%%%%%%%%%%%%%%%%%%%%%%%%%%%%%%%%%%%%%%%%%%%%%%

While we do not explore the isocurvature fluctuations
so extensively, we present in appendix~\ref{app:iso} an analytic 
prediction for the CMB angular bispectrum generated from the 
isocurvature fluctuations that we have described in this section.
This formula may be used to fit the measured bispectrum;
by doing so, we can constrain the model independently of the angular
power spectrum.
Although the one-point p.d.f of the isocurvature fluctuations 
is very similar to Gaussian distribution for large-beam CMB experiments,
the bispectrum may still be powerful enough to detect the non-Gaussian
signals.

%%%%%%%%%%%%%%%%%%%%%%%%%%%%%%%%%%%%%%%%%%%%%%%%%%%%%%%%%%%%%%%%%%%
%
%  Perturbation Theory in Inflation
%
%     1st draft:  07/11/2001
%     final:      08/05/2001
%
%%%%%%%%%%%%%%%%%%%%%%%%%%%%%%%%%%%%%%%%%%%%%%%%%%%%%%%%%%%%%%%%%%%
\chapter{Perturbation Theory in Inflation}
\label{chap:inflation}

%%%%%%%%%%%%%%%%%%%%%%%%%%%%%%%%%%%%%%%%%%%%%%%%%%%%%%%%%%%%%%%%%%%
\section{Inflation---Overview}
\label{sec:overview}

During inflation, the universe expands exponentially.
It implies the Hubble parameter, $H(t)=d\ln a/dt$, the expansion 
rate of the universe, being nearly constant in time, and the 
expansion scale factor, $a(t)$, given by
%%%%%%%%%%%%%%%%%%%%%%%%%%%%%%%%%%%%%%%%%%%%%%%%%%%%%%%%%%%%%%%%%%%
\begin{equation}
 \label{eq:accel}
  a(t)= a(t_0)\exp\left(\int_{t_0}^t H(t') dt'\right)
  \approx
  a(t_0)\exp\left[H(t)\left(t-t_0\right)\right].
\end{equation}
%%%%%%%%%%%%%%%%%%%%%%%%%%%%%%%%%%%%%%%%%%%%%%%%%%%%%%%%%%%%%%%%%%%
The exponential expansion drives the observable universe spatially flat,
for as the universe expands rapidly, a small section on a surface of 
a three-sphere of the universe approaches flat (we live on the section).
Thus, inflation predicts flatness of the universe, and 
recent CMB experiments have confirmed the prediction 
\citep{TOCO99,Boom00,Maxima00}.

What makes the exponential expansion possible?
One finds that neither matter nor radiation can make it; 
on the contrary, their energy density, $\rho$, and pressure, $p$, 
make the universe decelerate.
Since the universe accelerates only when $\rho+3p<0$,
one needs a negative pressure component dominating the universe.
How can it be possible?

A spatially homogeneous scalar field, $\phi$, with a potential, $V(\phi)$, 
provides negative pressure, making the exponential expansion possible.
The energy density is $\rho_\phi= \frac12(d\phi/dt)^2+V(\phi)$, while
the pressure is $p_\phi= \frac12(d\phi/dt)^2-V(\phi)$, giving 
%%%%%%%%%%%%%%%%%%%%%%%%%%%%%%%%%%%%%%%%%%%%%%%%%%%%%%%%%%%%%%%%%%%
\begin{equation}
 \label{eq:energycondition}
  \rho_\phi+3p_\phi=2\left[(d\phi/dt)^2-V(\phi)\right].
\end{equation}
%%%%%%%%%%%%%%%%%%%%%%%%%%%%%%%%%%%%%%%%%%%%%%%%%%%%%%%%%%%%%%%%%%%
Hence, one finds that $(d\phi/dt)^2 < V(\phi)$ suffices to accelerate the 
universe.
This {\it slowly-rolling} scalar field is a key ingredient of inflation;
by assuming a slowly-rolling scalar field dominant in early universe,
the universe expands exponentially.

\begin{figure}
 \begin{center}
  \leavevmode\epsfxsize=9cm \epsfbox{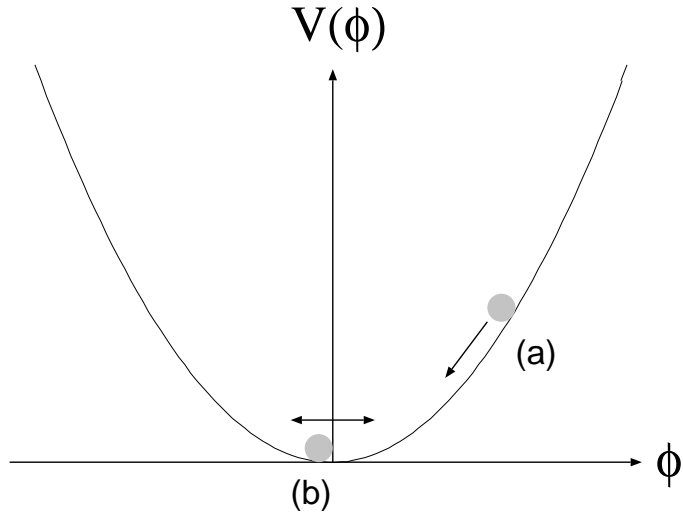}
 \end{center}
 \caption
 {Classical Evolution of Scalar Field} 
 \mycaption
 {Classical evolution of a scalar field, $\phi$, in a potential,
 $V(\phi)$.
 (a) $\phi$ rolls down on $V(\phi)$ slowly, driving the universe
 to expand exponentially.
 (b) $\phi$ oscillates rapidly about $\phi=0$, terminating inflation;
 it then decays into particles and radiation because of interactions, 
 reheating the universe.}
\label{fig:potential}
\end{figure}

While what is $\phi$ and how it comes to dominate the universe 
are still in debate, a simple model sketched in figure~\ref{fig:potential} 
works well.
In the phase (a), $\phi$ rolls down on $V(\phi)$ slowly, driving the 
universe to expand exponentially.
In the phase (b), $\phi$ oscillates rapidly, terminating inflation.
After inflation ends, interactions of $\phi$ with other particles 
lead $\phi$ to decay with a decay rate of $\Gamma_\phi$, producing 
particles and radiation. 
This is called a {\it reheating} phase of the universe,
as $\phi$ converts its energy density into heat by the particle
production.
A reheating temperature amounts to on the order of 
$(\Gamma_\phi m_{\rm pl})^{1/2}$.
While a precise value of reheating temperature depends upon models,
it is typically on the order of 
$10^{14-16}~{\rm GeV}\sim 10^{27-29}~{\rm K}$.
Note that the smallness of the observed CMB anisotropy implies 
that $\phi$ is coupled to other particles only very weakly, 
i.e., $\Gamma_\phi<H$, giving lower reheating temperature.
After the reheating, radiation dominates the universe, and 
the Big-bang scenario describes the rest of the cosmic history.

A class of inflation models with the potential sketched in 
figure~\ref{fig:potential} is called the {\it chaotic inflation},
for which $V(\phi)\propto \phi^n$ \citep{Lin83}.
Until now, this model has remained the most successful realization of
inflation with broad applications \citep{Lin90}.

%%%%%%%%%%%%%%%%%%%%%%%%%%%%%%%%%%%%%%%%%%%%%%%%%%%%%%%%%%%%%%%%%%%%%%
%%%%%%%%%%%%%%%%%%%%%%%%%%%%%%%%%%%%%%%%%%%%%%%%%%%%%%%%%%%%%%%%%%%%%%

%%%%%%%%%%%%%%%%%%%%%%%%%%%%%%%%%%%%%%%%%%%%%%%%%%%%%%%%%%%%%%%%%%%
\section{Quantum Fluctuations}
\label{sec:quantum}

Inflation predicts emergence of quantum fluctuations in early universe.
As soon as the fluctuations emerge from a vacuum, the exponential 
expansion stretches the proper wavelength of the fluctuations out of the 
Hubble-horizon scale, $H^{-1}$.
After leaving the horizon, the fluctuation amplitude does not change in time; 
on the contrary, it stays constant in time with characteristic 
r.m.s. amplitude, $\left|\phi\right|_{\rm rms}\sim H/(2\pi)$.

\begin{figure}
 \plotone{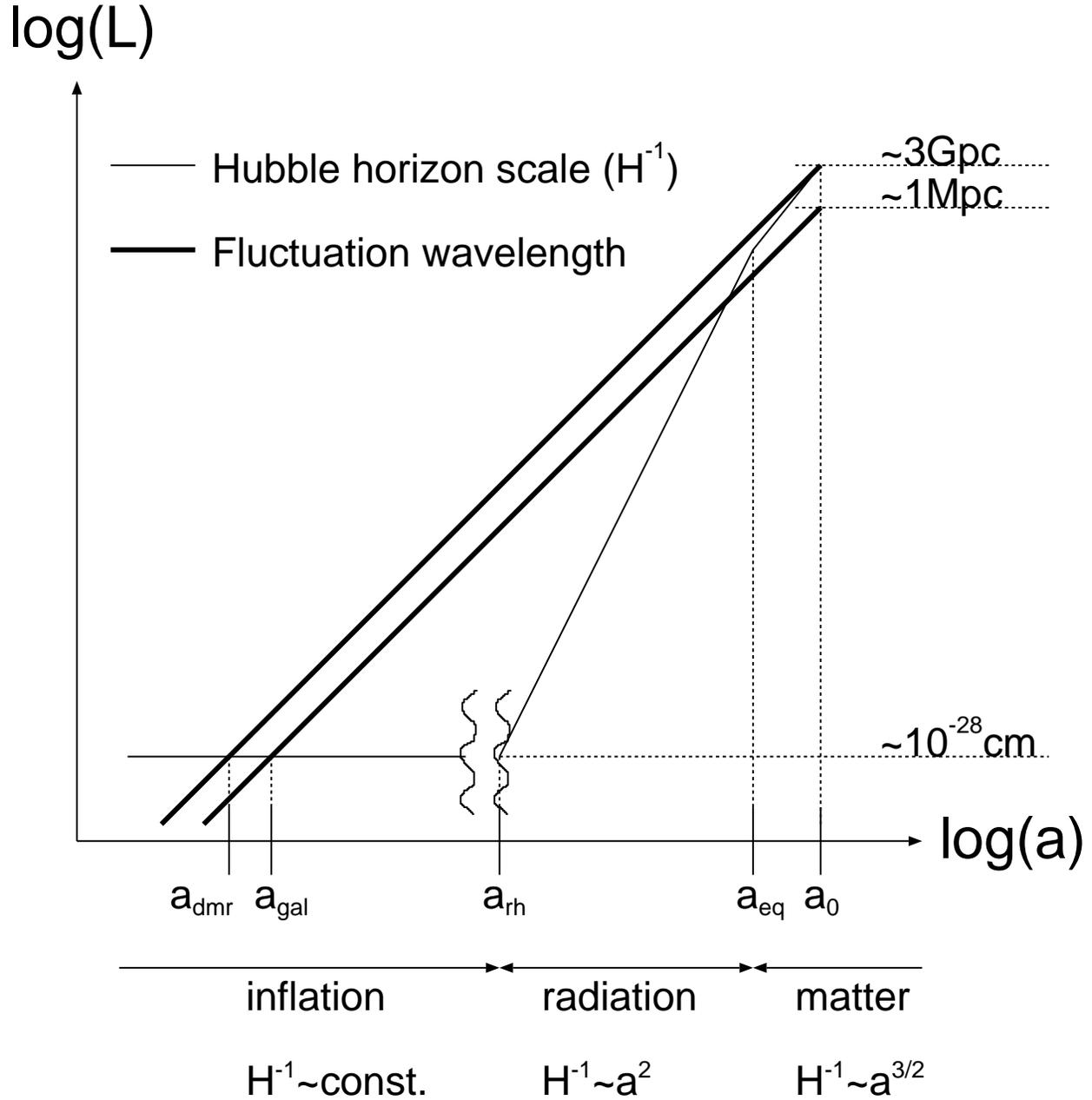}
 \caption
 {Physical Length Scales in Cosmic Evolution} 
 \mycaption
 {Representative physical length-scale evolution in the cosmic history.
 The thick lines draw the evolution of fluctuation wavelengths,
 $2\pi a/k$,
 while the thin line draws the evolution of the Hubble-horizon scale,
 $H^{-1}(a)$, where $a$ is the expansion scale factor.
 $H^{-1}(a)$ stays constant during inflation,
 grows as $a^2$ in the radiation era, and grows as $a^{3/2}$ in the
 matter era.
 There are characteristic scale factors:
 $a_0$ is the present-day, $a_{\rm eq}$ is the matter-radiation
 equality, and $a_{\rm rh}$ is the reheating epoch.
 The DMR-scale fluctuations leave the horizon
 at $a_{\rm dmr}$, while the galaxy-scale fluctuations leave
 at $a_{\rm gal}$.
 These scale factors are 
 related to each other roughly as follows:
 $a_0/a_{\rm eq}\sim 10^4$, $a_{\rm eq}/a_{\rm rh}\sim 10^{24}$, 
 $a_{\rm rh}/a_{\rm gal}\sim 10^{25}$ ($e^{58}$), and 
 $a_{\rm rh}/a_{\rm dmr}\sim 10^{28}$ ($e^{64}$).}
\label{fig:inflation}
\end{figure}

After inflation, as the universe decelerates, the fluctuations reenter 
the Hubble horizon, seeding matter and radiation fluctuations in the 
universe.
Figure~\ref{fig:inflation} summarizes the evolution of characteristic
length scales: the Hubble-horizon scale ($H^{-1}$), 
the {\it COBE} DMR-scale fluctuation wavelength, 
and the galaxy-scale fluctuation wavelength.

We estimate $H^{-1}$ during inflation as follows.
The scalar-field fluctuations produce CMB fluctuations of order $H/m_{\rm pl}$.
Using the DMR measurement \citep{Smo92}, 
$\Delta T/T\sim 10^{-5}$, we obtain $H\sim 10^{-5}m_{\rm pl}$, or  
$H^{-1}\sim 10^5m_{\rm pl}^{-1}\sim 10^{-28}~{\rm cm}$. 
Since $H^{-1}$ stays nearly constant in time during inflation, this value
represents the horizon scale throughout inflation approximately.
After inflation, $H^{-1}$ grows as $H^{-1}(a)\propto a^2$ in the 
radiation era, and $\propto a^{3/2}$ in the matter era.

DMR probes a present-day fluctuation wavelength on the order of 
$3~{\rm Gpc}\sim 10^{28}~{\rm cm}$.
By comparing the reheating temperature, $\sim 10^{27-29}~{\rm K}$, 
with the present-day CMB temperature, $2.73~{\rm K}$,
one finds that the universe has expanded by a factor of 
$a_0/a_{\rm rh}\sim 3\times 10^{26-28}$ since the reheating; thus, 
the DMR scale corresponds to a proper wavelength of $0.3-30$~cm 
at the reheating epoch (the number could be more uncertain).
Here, $a_0$ is the present-day scale factor, while $a_{\rm rh}$ is the 
reheating epoch.

The galaxy-scale fluctuations have the linear comoving wavelength
on the order of $1~{\rm Mpc}\sim 3\times 10^{24}~{\rm cm}$.
The galaxy-scale fluctuations have left the horizon later than 
the DMR-scale fluctuations:
$a_{\rm rh}/a_{\rm gal}\sim 10^{25}$, while
$a_{\rm rh}/a_{\rm dmr}\sim 10^{28}$.
Here, $a_{\rm gal}$ and $a_{\rm dmr}$ are the scale factors 
at which the galaxy- and DMR-scale fluctuations leave the horizon, 
respectively.

These ratios are often calculated with $e$-folding numbers, 
$N\equiv \ln (a_{\rm rh}/a)$.
For the galaxy- and DMR-scale fluctuations, we have 
$N_{\rm gal}= \ln(a_{\rm rh}/a_{\rm gal})\sim 58$, 
and $N_{\rm dmr}= \ln(a_{\rm rh}/a_{\rm dmr})\sim 64$.
Moreover, using equation~(\ref{eq:accel}), we obtain
$t_{\rm gal}-t_{\rm dmr}=H^{-1}(N_{\rm dmr}-N_{\rm gal})\sim 
10^{-38}~{\rm s}$; thus, inflation generates the fluctuations
on the DMR scales down to the galaxy scales almost instantaneously.

%%%%%%%%%%%%%%%%%%%%%%%%%%%%%%%%%%%%%%%%%%%%%%%%%%%%%%%%%%%%%%%%%%%%%%
%%%%%%%%%%%%%%%%%%%%%%%%%%%%%%%%%%%%%%%%%%%%%%%%%%%%%%%%%%%%%%%%%%%%%%

\subsection{Quantization in de Sitter spacetime}

A basic idea behind quantum-fluctuation generation in inflation is
well described by the second quantization of a massive-free scalar field in 
unperturbed de Sitter spacetime, for which $a(t)=a_0e^{H(t-t_0)}$ with
$H$ independent of $t$.
In this system, the problem is exactly solvable, and
finite mass captures an essential point of generating 
a ``tilted'' fluctuation spectrum, as we will show in the next subsection.
In this subsection, we describe the quantization procedure in 
de Sitter spacetime, following \citet{BD82}.

Quantization in curved spacetime is generally complicated, as
there is no unique vacuum state to define a ground state of quanta, 
even for inertial observers who detect no particles in
the Minkowski vacuum (the vacuum state for the quantum field theory
in Minkowski spacetime).
Fortunately, in the spatially-flat Robertson-Walker metric, 
there is a prescription for quantization, largely because of the metric being 
{\it conformal} to the Minkowski metric,
$g_{\mu\nu}=a^2\eta_{\mu\nu}$, or more specifically
%%%%%%%%%%%%%%%%%%%%%%%%%%%%%%%%%%%%%%%%%%%%%%%%%%%%%%%%%%%%%%%%%%%%%%%
\begin{equation}
 \label{eq:conformalmetric}
  ds^2 = - dt^2 + a^2(t)\delta_{ij}dx^idx^j
       = a^2(\tau)\left(- d\tau^2 + \delta_{ij}dx^idx^j\right),
\end{equation}
%%%%%%%%%%%%%%%%%%%%%%%%%%%%%%%%%%%%%%%%%%%%%%%%%%%%%%%%%%%%%%%%%%%%%%%
where $\tau\equiv \int^t a^{-1}(t')dt'$ is called the conformal time.
In this metric, there is a reasonable vacuum state for 
particles whose comoving frequencies, $\omega_k$, are higher than 
the conformal expansion rate, $d\ln a/d\tau$.
We should differ the conformal expansion rate,  $d\ln a/d\tau$, from
the expansion rate, $H=d\ln a/dt$. 
They are related through $d\ln a/d\tau= aH$.
In de Sitter spacetime, we obtain
%%%%%%%%%%%%%%%%%%%%%%%%%%%%%%%%%%%%%%%%%%%%%%%%%%%%%%%%%%%%%%%%%%%%%%%
\begin{equation}
 \label{eq:conformaltime}
  \tau = \tau_0 + \frac{1-\exp\left[-H(t-t_0)\right]}{a_0H}
       = \left(\tau_0+\frac1{a_0H}\right)
        - \frac1{a(\tau)H}.
\end{equation}
%%%%%%%%%%%%%%%%%%%%%%%%%%%%%%%%%%%%%%%%%%%%%%%%%%%%%%%%%%%%%%%%%%%%%%%
For simplicity, we set the first term vanishing, so that
$\tau=-\left[a(\tau)H\right]^{-1}$; thus, $\tau$ lies in
$-\infty < \tau < 0$ for $-\infty < t < \infty$ ($0<a<\infty$).
From now on, we will use dots for conformal-time derivatives:
$\dot{x}\equiv \partial x/\partial\tau$.

We start quantizing a scalar field, $\phi({\mathbf x},\tau)$, 
by expanding it into creation, 
$\hat{a}^\dag_{\mathbf k}$, and annihilation, $\hat{a}_{\mathbf k}$, 
operators, which satisfy the commutation relation,
$\left[\hat{a}_{\mathbf k},\hat{a}^\dag_{{\mathbf k}'}\right]
=\delta^{(3)}({\mathbf k}-{\mathbf k}')$.
We have
%%%%%%%%%%%%%%%%%%%%%%%%%%%%%%%%%%%%%%%%%%%%%%%%%%%%%%%%%%%%%%%%%%%%%
\begin{equation}
 \label{eq:expand}
  \phi({\mathbf x},\tau)
  =
  \int \frac{d^3{\mathbf k}}{(2\pi)^{3/2}} 
  \left[
  \hat{a}_{\mathbf k}
  \varphi_k(\tau)e^{i{\mathbf k}\cdot{\mathbf x}}
  +
  \hat{a}^\dag_{\mathbf k}
  \varphi^*_k(\tau)e^{-i{\mathbf k}\cdot{\mathbf x}}
  \right].
\end{equation}
%%%%%%%%%%%%%%%%%%%%%%%%%%%%%%%%%%%%%%%%%%%%%%%%%%%%%%%%%%%%%%%%%%%%%
The canonical commutation relation between $\phi$ and the conjugate
momentum, $\pi_\phi = a^2(\tau)\dot{\phi}$:
%%%%%%%%%%%%%%%%%%%%%%%%%%%%%%%%%%%%%%%%%%%%%%%%%%%%%%%%%%%%%%%%%%%%%
\begin{equation}
 \label{eq:comm}
  \left[\phi({\mathbf x},\tau),\pi_\phi({\mathbf x}',\tau)\right]
  = a^2(\tau)\left[\phi({\mathbf x},\tau),\dot{\phi}({\mathbf x}',\tau)\right]
  =i\delta^{(3)}({\mathbf x}-{\mathbf x}'),
\end{equation}
%%%%%%%%%%%%%%%%%%%%%%%%%%%%%%%%%%%%%%%%%%%%%%%%%%%%%%%%%%%%%%%%%%%%%
gives a normalization condition on $\varphi_k(\tau)$,
$a^2\left(
\varphi_k\dot{\varphi}^*_k-\varphi^*_k\dot{\varphi}_k\right)=i$.

The normalization condition motivates our using a new mode function 
given by $\chi_k\equiv a\varphi_k$, which satisfies a new 
normalization condition, 
$\chi_k\dot{\chi}^*_k-\chi^*_k\dot{\chi}_k=i$.
If $\chi_k$ has a positive frequency mode with respect to
the conformal timelike Killing vector
($\partial/\partial\tau$ for our metric),
i.e., $\dot{\chi}_k = -i\omega_k \chi_k$, then the condition gives
$\chi_k(\tau)= (2\omega_k)^{-1/2}e^{-i\omega_k\tau}$,
a ground state in the Minkowski vacuum.
Since $\chi_k$ gives the closest analogy to the Minkowski vacuum state, 
we will use $\chi_k$ more frequently than $\varphi_k$.

The Klein--Gordon equation for a massive-free scalar field, 
$g^{\alpha\beta}\phi_{,\alpha;\beta}=m^2\phi^2$, gives
equation of motion for $\chi_k(\tau)$,
%%%%%%%%%%%%%%%%%%%%%%%%%%%%%%%%%%%%%%%%%%%%%%%%%%%%%%%%%%%%%%%%%%%%%%
\begin{equation}
 \label{eq:KG}
  \ddot{\chi}_k(\tau) + 
  \left[k^2 + m^2_\chi(\tau)\right]\chi_k(\tau) = 0,
\end{equation}
%%%%%%%%%%%%%%%%%%%%%%%%%%%%%%%%%%%%%%%%%%%%%%%%%%%%%%%%%%%%%%%%%%%%%%
where $m^2_\chi(\tau)$ is the time-dependent effective mass,
%%%%%%%%%%%%%%%%%%%%%%%%%%%%%%%%%%%%%%%%%%%%%%%%%%%%%%%%%%%%%%%%%%%%%%
\begin{equation}
 \label{eq:effmass}
  m_\chi^2(\tau) 
  \equiv (m^2-2H^2)a^2(\tau)
  = m^2a^2(\tau)-\frac2{\tau^2}.
\end{equation}
%%%%%%%%%%%%%%%%%%%%%%%%%%%%%%%%%%%%%%%%%%%%%%%%%%%%%%%%%%%%%%%%%%%%%%
We thus find that the Hubble parameter effectively reduces $m^2$ by $2H^2$.
This time-dependent, negative contribution to $m_\chi^2(\tau)$
is the effect of de Sitter spacetime, which is not Minkowski but curved.

Fortunately, there is an exact solution to the Klein--Gordon 
equation~(\ref{eq:KG}):
%%%%%%%%%%%%%%%%%%%%%%%%%%%%%%%%%%%%%%%%%%%%%%%%%%%%%%%%%%%%%%%%%%%%%%
\begin{equation}
 \label{eq:chisolution}
  \chi_k(\tau)
  =
  \sqrt{-\tau}
  \left[C_1 H^{(1)}_\nu\left(-k\tau\right)
       +C_2 H^{(2)}_{\nu}\left(-k\tau\right)\right],
\end{equation}
%%%%%%%%%%%%%%%%%%%%%%%%%%%%%%%%%%%%%%%%%%%%%%%%%%%%%%%%%%%%%%%%%%%%%%
where $C_1$ and $C_2$ are integration constants, 
and $\nu^2 = 9/4 - {m^2}/{H^2}$.
$H^{(1)}_\nu(x)$ is a Hankel function of the first kind;
$H^{(2)}_\nu(x)=\left[H^{(1)}_\nu(x)\right]^*$.
We have negative sign in front of $\tau$ to recall that
$\tau$ lies in $-\infty < \tau < 0$.

How do we determine the integration constants, $C_1$ and $C_2$?
In other words, how do we normalize our mode function properly?
Since we know how to quantize a scalar field in the Minkowski vacuum, 
we should find a mode function that matches the Minkowski positive 
frequency mode, $\chi_k(\tau)= (2\omega_k)^{-1/2}e^{-i\omega_k\tau}$;
however, we cannot find an unique positive frequency mode valid 
throughout inflation, as the time-dependent spacetime creates particles.

Instead, we define a vacuum state in the {\it in} state, the remote past,
$\tau\rightarrow -\infty$.
Using an asymptotic form of the Hankel function,
%%%%%%%%%%%%%%%%%%%%%%%%%%%%%%%%%%%%%%%%%%%%%%%%%%%%%%%%%%%%%%%%%%%%%%
\begin{equation}
 \label{eq:hankel}
  H^{(1)}_\nu(x\gg 1)\approx \sqrt{\frac2{\pi x}}
  \exp\left[i\left(x-\nu\frac{\pi}2-\frac{\pi}4\right)\right],
\end{equation}
%%%%%%%%%%%%%%%%%%%%%%%%%%%%%%%%%%%%%%%%%%%%%%%%%%%%%%%%%%%%%%%%%%%%%%
we obtain $\chi_k$ in the in state ($\tau\rightarrow -\infty$)
from equation~(\ref{eq:chisolution}),
%%%%%%%%%%%%%%%%%%%%%%%%%%%%%%%%%%%%%%%%%%%%%%%%%%%%%%%%%%%%%%%%%%%%%%
\begin{equation}
 \label{eq:solution_limit}
  \chi_{k}(\tau\rightarrow-\infty)
  \longrightarrow
  \sqrt{\frac2{\pi k}}
  \left(C_1 e^{-ik\tau} + C_2 e^{ik\tau}\right).
\end{equation}
%%%%%%%%%%%%%%%%%%%%%%%%%%%%%%%%%%%%%%%%%%%%%%%%%%%%%%%%%%%%%%%%%%%%%%
Here, we have neglected the contribution from $m^2/H^2$ compared with
$-k\tau$ in the exponent.
The second term has a negative frequency, so that $C_2=0$.
The first term with $C_1=\sqrt{\pi}/2$ gives
$\chi_k(\tau)= (2k)^{-1/2}e^{-ik\tau}$, 
the Minkowski positive frequency mode with $\omega_k=k$, and thus
in the in state, the solution describes a ground state of
a massless field in the Minkowski vacuum.

Using the solution for $\chi_{k}(\tau)$, we obtain a solution for 
$\varphi_k(\tau)$,
%%%%%%%%%%%%%%%%%%%%%%%%%%%%%%%%%%%%%%%%%%%%%%%%%%%%%%%%%%%%%%%%%%%%%%
\begin{equation}
 \label{eq:modesolution}
  \varphi_k(\tau)
  =
  \frac{\sqrt{-\pi\tau}}{2a(\tau)}H^{(1)}_\nu\left(-k\tau\right),
\end{equation}
%%%%%%%%%%%%%%%%%%%%%%%%%%%%%%%%%%%%%%%%%%%%%%%%%%%%%%%%%%%%%%%%%%%%%%
and $\phi({\mathbf x},\tau)$ becomes
%%%%%%%%%%%%%%%%%%%%%%%%%%%%%%%%%%%%%%%%%%%%%%%%%%%%%%%%%%%%%%%%%%%%%
\begin{equation}
 \label{eq:expand*}
  \phi({\mathbf x},\tau)
  =
  \frac{\sqrt{-\pi\tau}}{2a(\tau)}
  \int \frac{d^3{\mathbf k}}{(2\pi)^{3/2}} 
  \left[
  \hat{a}_{\mathbf k}
  H^{(1)}_\nu\left(-k\tau\right)
  e^{i{\mathbf k}\cdot{\mathbf x}}
  +
  \hat{a}^\dag_{\mathbf k}
  H^{(2)}_\nu\left(-k\tau\right)
  e^{-i{\mathbf k}\cdot{\mathbf x}}
  \right].
\end{equation}
%%%%%%%%%%%%%%%%%%%%%%%%%%%%%%%%%%%%%%%%%%%%%%%%%%%%%%%%%%%%%%%%%%%%%
Since all the $k$ modes in the integral are independent of each other, 
the nearly infinite sum of those modes makes $\phi$ obey
Gaussian statistics almost exactly,
because of the central limit theorem;
thus, the two-point statistics specify all the statistical properties
of $\phi$.
This is a generic property of the ground-state quantum fluctuations.

The annihilation operator, $\hat{a}_{\mathbf k}$, annihilates the
vacuum state defined in the in state:
$\left.\left.\hat{a}_{\mathbf k}\right|0_{\rm in}\right>= 0$.
In this vacuum, we calculate amplitude of ground-state $\phi$ fluctuations as
%%%%%%%%%%%%%%%%%%%%%%%%%%%%%%%%%%%%%%%%%%%%%%%%%%%%%%%%%%%%%%%%%%%%%
\begin{equation}
 \label{eq:v_fluctuation}
  \left<0_{\rm in}\left|
		   \phi^\dag({\mathbf x},\tau)\phi({\mathbf x},\tau)
		 \right|0_{\rm in}\right>
  =
  \int_0^\infty \frac{k^2dk}{2\pi^2} \left|\varphi_k(\tau)\right|^2
  =
  \frac{-\tau}{8\pi a^2(\tau)}
  \int_0^\infty k^2dk \left|H_\nu^{(1)}(-k\tau)\right|^2.
\end{equation}
%%%%%%%%%%%%%%%%%%%%%%%%%%%%%%%%%%%%%%%%%%%%%%%%%%%%%%%%%%%%%%%%%%%%%
Since we probe a limited range of $k$ observationally,  
we also use the fluctuation spectrum in a logarithmic $k$ range,
$\Delta^2(k)$,
which represents variance of fluctuations at a given comoving 
wavelength $2\pi k^{-1}$,
%%%%%%%%%%%%%%%%%%%%%%%%%%%%%%%%%%%%%%%%%%%%%%%%%%%%%%%%%%%%%%%%%%%%%
\begin{equation}
 \label{eq:powphi}
  \Delta^2_\phi(k)
  \equiv \frac{k^3}{2\pi^2}\left|\varphi_k(\tau)\right|^2
  =
  \frac{-k^3\tau\left|H_\nu^{(1)}(-k\tau)\right|^2}
  {8\pi a^2(\tau)}
  =
  \frac{H^2}{8\pi}(-k\tau)^3\left|H_\nu^{(1)}(-k\tau)\right|^2,
\end{equation}
%%%%%%%%%%%%%%%%%%%%%%%%%%%%%%%%%%%%%%%%%%%%%%%%%%%%%%%%%%%%%%%%%%%%%
where we have used $a(\tau)= -(H\tau)^{-1}$ in the last equality.
Let us recall that $\nu^2 = 9/4 - {m^2}/{H^2}$, and $\tau$
lies in $-\infty<\tau<0$.

Here, we have the quantization of $\phi$ completed, and formally 
calculated the fluctuation spectrum.
These results are exact, and valid on all scales. 
In the next subsection, we study the solution on super-horizon scales,
where inflation produces observationally relevant fluctuations. 

\subsection{Scale-invariant fluctuations on super-horizon scales}

Equation~(\ref{eq:expand*}) describes a quantum massive-free 
scalar field, $\phi$, in the unperturbed de Sitter spacetime on all scales.
As the universe expands exponentially, 
$-k\tau=k/(aH)$ quickly becomes very small;
the mode leaves the Hubble-horizon scale, $H^{-1}$.
Figure~\ref{fig:inflation} shows that the fluctuations on 
the observationally relevant scales should have left the horizon during 
inflation.
Hence, the behavior of $\phi$ on super-horizon scales is practically
important.
In this subsection, we study the $\phi$ fluctuation spectrum
on super-horizon scales.

In equation~(\ref{eq:powphi}), using an asymptotic form of 
the Hankel function,
%%%%%%%%%%%%%%%%%%%%%%%%%%%%%%%%%%%%%%%%%%%%%%%%%%%%%%%%%%%%%%%%%%%%%%
\begin{equation}
 \label{eq:hankel_asym}
  H^{(1)}_\nu(x\ll 1)\approx 
  -i\frac{\Gamma(\nu)}{\pi}\left(\frac{x}2\right)^{-\nu},
\end{equation}
%%%%%%%%%%%%%%%%%%%%%%%%%%%%%%%%%%%%%%%%%%%%%%%%%%%%%%%%%%%%%%%%%%%%%%
we obtain the $\phi$ fluctuation spectrum on super-horizon scales,
%%%%%%%%%%%%%%%%%%%%%%%%%%%%%%%%%%%%%%%%%%%%%%%%%%%%%%%%%%%%%%%%%%%%%%
\begin{equation}
 \Delta^2_\phi(k)
  \approx
  \left( \frac{H}{2\pi} \right)^2
  2^{2\nu-3}  \left[\frac{\Gamma(\nu)}{\Gamma(3/2)}\right]^2
  \left(\frac{k}{aH}\right)^{3-2\nu},
\end{equation}
%%%%%%%%%%%%%%%%%%%%%%%%%%%%%%%%%%%%%%%%%%%%%%%%%%%%%%%%%%%%%%%%%%%%%%
where $\nu^2=9/4-m^2/H^2$.
We have used $\tau= -(aH)^{-1}$ and 
$\pi= \Gamma^2(1/2) = 4\Gamma^2(3/2)$.

One finds that $\nu=3/2$ is a special point, for which the spectrum
is independent of $k$, i.e., scale invariant, 
$\Delta^2_\phi(k)=H^2/(2\pi)^2$.
This happens when we assume $m^2/H^2\ll 1$, and thus
$\nu= 3/2-m^2/(3H^2) +{\cal O}(m^4/H^4)$, which gives
%%%%%%%%%%%%%%%%%%%%%%%%%%%%%%%%%%%%%%%%%%%%%%%%%%%%%%%%%%%%%%%%%%%%%%
\begin{equation}
 \label{eq:specphi}
 \Delta^2_\phi(k)
  \approx
  \left( \frac{H}{2\pi} \right)^2
  \left(\frac{k}{aH}\right)^{2m^2/(3H^2)},
\end{equation}
%%%%%%%%%%%%%%%%%%%%%%%%%%%%%%%%%%%%%%%%%%%%%%%%%%%%%%%%%%%%%%%%%%%%%%
or the spectral index,
%%%%%%%%%%%%%%%%%%%%%%%%%%%%%%%%%%%%%%%%%%%%%%%%%%%%%%%%%%%%%%%%%%%%%%
\begin{equation}
 \label{eq:index}
 \frac{d\ln\Delta^2_\phi}{d\ln k}
 =
 \frac{2m^2}{3H^2}.
\end{equation}
%%%%%%%%%%%%%%%%%%%%%%%%%%%%%%%%%%%%%%%%%%%%%%%%%%%%%%%%%%%%%%%%%%%%%%
Since $m^2\ll H^2$, the spectrum is almost scale invariant, giving
the $\phi$ fluctuations characteristic r.m.s. amplitude, 
$\left|\phi\right|_{\rm rms}=H/(2\pi)$.
Finite mass makes the spectrum slightly ``blue'', the power of $k$ 
being positive.

The above assumption, $m^2\ll H^2$, offers long-lasting inflation
that makes the observable universe flat and homogeneous; otherwise,
$\phi$ rolls down to a potential minimum too quickly, terminating 
inflation too early.
In the inflationary regime, the Friedmann equation gives
%%%%%%%%%%%%%%%%%%%%%%%%%%%%%%%%%%%%%%%%%%%%%%%%%%%%%%%%%%%%%%%%%%%%%%
\begin{equation}
 \frac{m^2}{H^2}
 =
 \frac{3}{4\pi}\frac{m_{\rm pl}^2}{\phi^2}\sim 10^{-2}.
\end{equation}
%%%%%%%%%%%%%%%%%%%%%%%%%%%%%%%%%%%%%%%%%%%%%%%%%%%%%%%%%%%%%%%%%%%%%%
Hence, for a massive-free scalar field to drive inflation, 
the mass cannot be comparable to the Hubble parameter,
and the fluctuation spectrum is almost exactly scale invariant.

\begin{figure}
 \begin{center}
  \leavevmode\epsfxsize=10cm \epsfbox{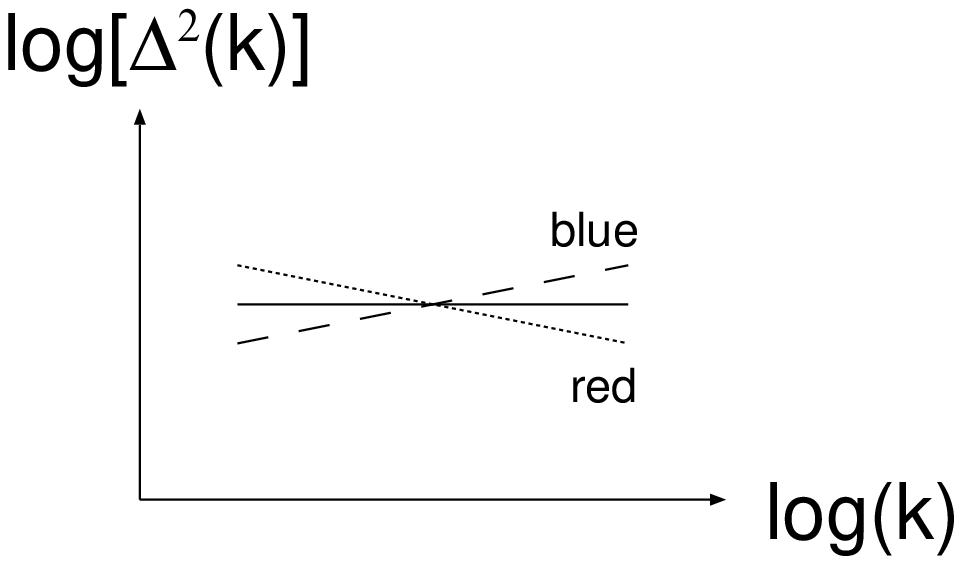}
 \end{center}
 \caption
 {Color of Spectrum} 
 \mycaption
 {A sketch of the fluctuation spectrum, $\Delta^2(k)$, which 
 represents the fluctuation power in a logarithmic $k$ range. 
 The solid line plots a scale-invariant spectrum, 
 the dotted line plots a ``red'' spectrum whose spectral index is 
 negative, and the dashed line plots a ``blue'' spectrum
 whose spectral index is positive.}
\label{fig:rb}
\end{figure}

The argument until now has assumed the exact de Sitter
spacetime in which $H$ is constant in time, and neglected 
perturbations in the metric.
As a result, we have obtained a blue spectrum whose spectral index
is $0< 2m^2/3H^2 \ll 1$.
In realistic inflation models, however, none of the above assumptions
apply: $H$ decreases slowly in time, and the metric is perturbed.
In the next subsection, we will show that both the effects give 
a tilted ``red'' spectrum, for which the power of $k$ is negative
of order $-10^{-2}$.
Figure~\ref{fig:rb} sketches what blue, scale-invariant, and red spectra
look like.

%%%%%%%%%%%%%%%%%%%%%%%%%%%%%%%%%%%%%%%%%%%%%%%%%%%%%%%%%%%%%%%%%%%%%%
%%%%%%%%%%%%%%%%%%%%%%%%%%%%%%%%%%%%%%%%%%%%%%%%%%%%%%%%%%%%%%%%%%%%%%

\subsection{Tilted ``red'' spectrum}

If $H$ decreases in time, and the metric is linearly perturbed,
a scalar field acquires a negative effective mass-squared, $\Delta m^2<0$.
It also modifies the mass of $\chi_k$, $m_{\chi}(\tau)$
(Eq.(\ref{eq:effmass})), as
%%%%%%%%%%%%%%%%%%%%%%%%%%%%%%%%%%%%%%%%%%%%%%%%%%%%%%%%%%%%%%%%%%%%%%
\begin{equation}
 \label{eq:effmass*}
  m_\chi^2(\tau) \longrightarrow \left(m^2+\Delta m^2\right)a^2(\tau)
  -\frac2{\tau^2}. 
\end{equation}
%%%%%%%%%%%%%%%%%%%%%%%%%%%%%%%%%%%%%%%%%%%%%%%%%%%%%%%%%%%%%%%%%%%%%%
We thus expect that $\Delta m^2$ modifies the fluctuation spectral
index, for the mass determines the spectral index through 
equation~(\ref{eq:index}).
This parameterization, $\Delta m^2$, may be useful to understand
what physical effect is responsible for the spectral index.

How large is $\Delta m^2$?
The derivation of $\Delta m^2$ is quite involved, but
\citet{MFB92} show that the exact form of $m_\chi^2(\tau)$
%%%%%%%%%%%%%%%%%%%%%%%%%%%%%%%%%%%%%%%%%%%%%%%%%%%%%%%%%%%%%%%%%%%%%%
\begin{equation}
 \label{eq:effmass**}
  m_\chi^2(\tau) 
  = -\frac{H}{\dot{\phi}}\frac{d^2({\dot{\phi}}/{H})}{d\tau^2}
  \approx 
  \left(\frac{d^2V}{d\phi^2}+9\frac{dH}{dt}\right)a^2(\tau)-\frac2{\tau^2}
  =m^2_{\rm eff}\hspace{0.4pt}a^2(\tau)-\frac2{\tau^2}.
\end{equation}
%%%%%%%%%%%%%%%%%%%%%%%%%%%%%%%%%%%%%%%%%%%%%%%%%%%%%%%%%%%%%%%%%%%%%%
Here, since $H$ and $\phi$ change in time $t$ 
only very slowly, we have retained the terms on the order of
$-dH/dt= 4\pi m_{\rm pl}^{-2}(d\phi/dt)^2$ or larger.
This approximation is called the slow-roll approximation, and
we evaluate the approximation explicitly in appendix~\ref{app:slowroll}.  
For a massive-free field, $V(\phi)=m^2\phi^2/2$, by comparing
equation~(\ref{eq:effmass**}) with (\ref{eq:effmass*}), we find
$\Delta m^2= 9dH/dt < 0$.
One can show that the 3 of the 9 comes from the effect of $H$ 
changing in time, and the 6 comes from the effect of the metric
perturbations.

By using the first-order slow-roll approximation, we obtain a relation between
$dH/dt$ and the potential slope, $dV/d\phi$,
$dH/dt\approx -\frac1{6V}(dV/d\phi)^2 = -\frac13m^2$.
It thus follows that the total effective mass-squared becomes negative, 
$m^2_{\rm eff}= m^2+\Delta m^2 \approx -2m^2 < 0$, and
we have a ``red'' spectrum or a negative spectral index,
%%%%%%%%%%%%%%%%%%%%%%%%%%%%%%%%%%%%%%%%%%%%%%%%%%%%%%%%%%%%%%%%%%%%%%
\begin{equation}
 \label{eq:index*}
 \frac{d\ln\Delta^2_\phi}{d\ln k}
 =
 \frac{2m^2_{\rm eff}}{3H^2}
 \approx
 -\frac{4m^2}{3H^2} < 0.
\end{equation}
%%%%%%%%%%%%%%%%%%%%%%%%%%%%%%%%%%%%%%%%%%%%%%%%%%%%%%%%%%%%%%%%%%%%%%

Let us summarize what has made the spectral index negative. 
The spectral index is determined by the mass, or the effective
mass, of a scalar field.
In inflation, the intrinsic mass, $m^2$, is over-compensated by
the induced mass from gravitational effects, $\Delta m^2$, which
is negative: $m^2+\Delta m^2<0$. 
As a result, the fluctuation spectrum becomes ``red''.
Actually, any power-law potential of the form 
$V(\phi)=A\phi^n$ with $A$ positive gives a negative 
$m_{\rm eff}^2$ through $m^2_{\rm eff}=-(1+n/2)nA\phi^{n-2}$.
Here, we have used $dH/dt\approx -\frac1{6V}(dV/d\phi)^2=
-\frac16n^2A\phi^{n-2}$.

For a generic scalar field with an arbitrary potential, we find
%%%%%%%%%%%%%%%%%%%%%%%%%%%%%%%%%%%%%%%%%%%%%%%%%%%%%%%%%%%%%%%%%%%%%%
\begin{equation}
 \label{eq:effmass***}
  m^2_{\rm eff}=
  \frac{d^2V}{d\phi^2}+9\frac{dH}{dt}
  =
  6\frac{dH}{dt} - 3H\frac{d^2\phi/dt^2}{d\phi/dt},
\end{equation}
%%%%%%%%%%%%%%%%%%%%%%%%%%%%%%%%%%%%%%%%%%%%%%%%%%%%%%%%%%%%%%%%%%%%%%
and the spectral index
%%%%%%%%%%%%%%%%%%%%%%%%%%%%%%%%%%%%%%%%%%%%%%%%%%%%%%%%%%%%%%%%%%%%%%
\begin{equation}
 \label{eq:liddle}
  \frac{d\ln \Delta^2_\phi}{d\ln k}
  =
  4\frac{dH/dt}{H^2}-2\frac{d^2\phi/dt^2}{H(d\phi/dt)},
\end{equation}
%%%%%%%%%%%%%%%%%%%%%%%%%%%%%%%%%%%%%%%%%%%%%%%%%%%%%%%%%%%%%%%%%%%%%%
which agrees with \citet{LL92}.

CMB experiments have shown that a scale-invariant fluctuation spectrum 
fits the data well \citep{Boom00,Maxima00}.
Combining all the CMB experiments to date,
\citet{WTZ01} show that a slightly red spectrum,
the power of $k$ being $\sim -0.07$, fits the data even better,
while the error of the fit is still of order $0.1$.
Accurate measurement of the spectral index constrains 
the shape of $V(\phi)$ through equation~(\ref{eq:liddle}), as
$V(\phi)$ determines the time variation of $H$ and $\phi$;
thus, it potentially discriminates between different inflation models.

Although the CMB experiments almost exclude possibility of 
the massive-free scalar field with sizable $m^2/H^2\sim 1$ 
dominating the matter and radiation fluctuations in the universe
(in terms of the spectrum index),
the field may produce some of the fluctuations that are non-Gaussian.
In this case, $\phi$ does not drive inflation, so that $m^2$ can be
comparable to $H^2$; the field rolls down to a potential minimum
quickly, and oscillates about the minimum, producing 
non-Gaussian isocurvature density fluctuations, 
$\delta\rho\propto \phi^2$ \citep{LM97}.
Even if the isocurvature fluctuations are sub-dominant in the universe, 
they could produce non-Gaussian temperature fluctuations in CMB. 
Measuring non-Gaussianity in CMB thus potentially 
probes particle physics in inflation.

\subsection{Emergence of classical fluctuations}

Until now, we have considered generation of quantum 
fluctuations in inflation, and derived a fluctuation spectrum on 
super-horizon scales.
We then expect that the fluctuations seed observed CMB anisotropies 
and large-scale structures in the universe; but, how?
Inflation generates {\it quantum} fluctuations, not {\it classical}
fluctuations.
How can the quantum fluctuations make classical objects like
galaxies seen today?

Generation of classical fluctuations, or quantum-to-classical transition
of fluctuations, during inflation, has been in debate
\citep{CH95,Matacz97a,Matacz97b,KPS98}.
In this subsection, we describe a possible mechanism to produce
classical fluctuations in inflation.
Our approach is partly close to \citet{KPS98}.

A basic idea behind our classical-to-quantum transition mechanism is to 
approximate a field, $\phi$ (Eq.(\ref{eq:expand})), with the sum of the 
long-wavelength (super-horizon) modes and the short-wavelength
(sub-horizon) modes.
This mode separation is unique, as the comoving Hubble-horizon scale, 
$(aH)^{-1}$, has been a characteristic length scale 
in the solutions for mode functions, $\varphi_k(\tau)$ 
(Eq.(\ref{eq:modesolution})).

In the super-horizon limit, $k\ll aH$, equations~(\ref{eq:modesolution})
and (\ref{eq:hankel_asym}) give 
$\varphi_{k\ll aH}^* = -\varphi_{k\ll aH}$.
Using this in equation~(\ref{eq:expand}), we obtain
%%%%%%%%%%%%%%%%%%%%%%%%%%%%%%%%%%%%%%%%%%%%%%%%%%%%%%%%%%%%%%%%%%%%%
\begin{eqnarray}
 \nonumber
  \phi({\mathbf x},\tau)
  &\approx&
  \int_{k<aH}d^3{\mathbf k}
  \left(\hat{a}_{\mathbf k}-\hat{a}^\dag_{-{\mathbf k}}\right)
  \varphi_{k\ll aH}(\tau)e^{i{\mathbf k}\cdot{\mathbf x}}\\
 \label{eq:classic}
  & &+
  a^{-1}(\tau)
  \int_{k>aH} \frac{d^3{\mathbf k}}{(2\pi)^{3/2}\sqrt{2k}} 
  \left(
  \hat{a}_{\mathbf k}
  e^{i{\mathbf k}\cdot{\mathbf x}-ik\tau}
  +
  \hat{a}^\dag_{\mathbf k}
  e^{-i{\mathbf k}\cdot{\mathbf x}+ik\tau}
  \right).
\end{eqnarray}
%%%%%%%%%%%%%%%%%%%%%%%%%%%%%%%%%%%%%%%%%%%%%%%%%%%%%%%%%%%%%%%%%%%%%
We name the first term $\phi_{\rm CL}$, the second term $\phi_{\rm QM}$.
The corresponding conjugate momenta are 
$\pi_{\rm CL}= a^2\dot{\phi}_{\rm CL}$ and
$\pi_{\rm QM}= a^2\dot{\phi}_{\rm QM}$, respectively.
The first term, $\phi_{\rm CL}$, becomes just the Fourier transform,
while the second term, $\phi_{\rm QM}$, becomes an ordinary 
ground state in the Minkowski vacuum except for $a^{-1}(\tau)$ in the 
front.
To see what happens to these terms in the context of 
the quantum field theory, we calculate the canonical 
commutation relations of $\phi_{\rm CL}$ and $\phi_{\rm QM}$.

For $\phi_{\rm QM}$, we find
%%%%%%%%%%%%%%%%%%%%%%%%%%%%%%%%%%%%%%%%%%%%%%%%%%%%%%%%%%%%%%%%%%%%%
\begin{equation}
 \left[\phi_{\rm QM}({\mathbf x},\tau),\pi_{\rm QM}({\mathbf x}',\tau)\right]
  = i\int_{k>aH}\frac{d^3{\mathbf k}}{(2\pi)^3}
  e^{i{\mathbf k}\cdot({\mathbf x}-{\mathbf x}')}
  = i\delta^{(3)}({\mathbf x}-{\mathbf x}')
\end{equation}
%%%%%%%%%%%%%%%%%%%%%%%%%%%%%%%%%%%%%%%%%%%%%%%%%%%%%%%%%%%%%%%%%%%%%
for $\left|{\mathbf x}-{\mathbf x}'\right| < (aH)^{-1}$;
thus, $\phi_{\rm QM}$ is a quantum field inside the horizon,
and $a\phi_{\rm QM}$ is identical to a quantum field in the Minkowski vacuum.
For $\phi_{\rm CL}$, we find the commutation relation vanishing,
%%%%%%%%%%%%%%%%%%%%%%%%%%%%%%%%%%%%%%%%%%%%%%%%%%%%%%%%%%%%%%%%%%%%%
\begin{equation}
 \left[\phi_{\rm CL}({\mathbf x},\tau),\pi_{\rm CL}({\mathbf x}',\tau)\right]
  = 0;
\end{equation}
%%%%%%%%%%%%%%%%%%%%%%%%%%%%%%%%%%%%%%%%%%%%%%%%%%%%%%%%%%%%%%%%%%%%%
thus, $\phi_{\rm CL}$ is no longer a quantum field, but 
a {\it classical} field.
In other words, once smoothing out the fluctuations inside the 
horizon, we are left with the classical fluctuations.
Notice that there has been no explicit decoherence mechanism in the 
system.
The exact solution to the Klein--Gordon equation in the exact
de Sitter spacetime naturally yields the classical fluctuations 
on super-horizon scales.

Without smoothing out the sub-horizon fluctuations, however, 
$\phi$ remains quantum.
The commutation relation of $\phi$ (Eq.(\ref{eq:comm}))
is exact as long as all the $k$ modes equally contribute to $\phi$.
If nothing happens to the sub-horizon fluctuations, then
the quantum, not classical, fluctuations will reenter the horizon 
after inflation.
Yet, practically speaking, this argument may not be relevant for 
our actual observations because of the following reason.
Consider a fluctuation wavelength just leaving the horizon at the 
end of inflation, at which the proper horizon size is 
$H^{-1}\sim 10^{-28}~{\rm cm}$. 
As the universe has expanded by a factor of order $10^{27}$ since 
the end of inflation, we may find the fluctuation wavelength to be 
$\sim 0.1~{\rm cm}$ in the present universe.
Now we ask: ``does this-size fluctuation affects classicality of 
galaxy-scale fluctuations?''
We may answer ``no'', if we assume that these 
substantially different-scale fluctuations have undergone 
different physics; if so, quantum coherence should have disappeared.

Even right after inflation, the super-horizon modes and the sub-horizon
modes have undergone different physics.
At the end of inflation, the reheating begins, and scalar fields
decay into particles and radiation, thermalizing the universe.
By causality, the thermalization process occurs only inside the horizon;
thus, the reheating affects the sub-horizon fluctuations differently
from the super-horizon fluctuations, breaking quantum nature of $\phi$.
In other words, in equation~(\ref{eq:classic}), the second term
may have disappeared during the reheating,
while the first term may remain unaffected.
As a result, the reheating effectively smoothes out the sub-horizon
scale fluctuations, and makes a quantum-to-classical transition possible.

%%%%%%%%%%%%%%%%%%%%%%%%%%%%%%%%%%%%%%%%%%%%%%%%%%%%%%%%%%%%%%%%%%%
\section{Linear Perturbation Theory in Inflation}
\label{sec:linear}

In the previous section, we have followed generation of scalar-field 
fluctuations in the unperturbed de Sitter spacetime,
i.e., no perturbations in the metric.
Scalar-field fluctuations, however, perturb the stress-energy tensor, 
and produce metric perturbations.
Since the metric perturbations regulate the matter and radiation fluctuations 
that we observe today, we must include the metric perturbations in the 
analysis, and follow the evolution.
In this section, we explore the {\it linear} perturbation theory in 
inflation, which includes perturbations to the metric and a scalar field.
Our notation 
follows \citet{Bardeen80}.

We use a linearly perturbed conformal Robertson--Walker metric of the form,
%%%%%%%%%%%%%%%%%%%%%%%%%%%%%%%%%%%%%%%%%%%%%%%%%%%%%%%%%%%%%%%%%%%%%%%
\begin{equation}
 \label{eq:metric}
  ds^2  = a^2(\tau)\left\{
		    -(1+2AQ) d\tau^2 
		    -2 B Q_i d\tau dx^i 
		    + \left[\left(1+2 H_{\rm L}Q\right)\delta_{ij} 
		       + 2 H_{\rm T}Q_{ij}\right]dx^idx^j
		  \right\}.
\end{equation}
%%%%%%%%%%%%%%%%%%%%%%%%%%%%%%%%%%%%%%%%%%%%%%%%%%%%%%%%%%%%%%%%%%%%%%%
Here, all the metric perturbations, $A$, $B$, $H_{\rm L}$, and $H_{\rm T}$,
are $\ll 1$, and functions of $\tau$.
The spatial coordinate dependence of the perturbations is described
by the scalar harmonic eigenfunctions, $Q$, $Q_i$, and $Q_{ij}$, that 
satisfy
$\delta^{ij}Q_{,ij} = -k^2 Q$, $Q_i= -k^{-1}Q_{,i}$, and
$Q_{ij} =  k^{-2}Q_{,ij} + \frac13\delta_{ij}Q$.
Note that $Q_{ij}$ is traceless: $\delta^{ij}Q_{ij}=0$.
\citet{KS84} use different symbols, $Y$, $Y_i$, and $Y_{ij}$, for 
$Q$, $Q_i$, and $Q_{ij}$, respectively.

The four metric-perturbation variables are not entirely free, but
some of which should be fixed to fix our coordinate system before
we analyze the perturbations.
The choice of coordinate system is often called the choice of {\it gauge},
or the {\it gauge transformation}; we will describe it later.

\subsection{Fluid representation of scalar field}

The metric perturbations enter into the stress-energy tensor 
perturbations, $\delta T^{\mu}_{\nu}$.
We expand a scalar field into its homogeneous mean field, $\phi(\tau)$, and 
fluctuations about the mean, $\delta\phi(\tau) Q({\mathbf x})$.
The energy density and pressure fluctuations are given by 
%%%%%%%%%%%%%%%%%%%%%%%%%%%%%%%%%%%%%%%%%%%%%%%%%%%%%%%%%%%%%%%%%%%%%%%
\begin{eqnarray}
 \delta\rho_{\phi}Q &\equiv& -\delta T^0_0 
  = \left[a^{-2}\left(\dot{\phi}\delta\dot{\phi} - A\dot{\phi}^2\right)
   + V_{,\phi}\delta\phi\right]Q, \\
 \delta p_{\phi}Q &\equiv& \frac{\delta T^k_k}3 
  = \left[a^{-2}\left(\dot{\phi}\delta\dot{\phi} - A\dot{\phi}^2\right)
   - V_{,\phi}\delta\phi\right]Q.
\end{eqnarray}
%%%%%%%%%%%%%%%%%%%%%%%%%%%%%%%%%%%%%%%%%%%%%%%%%%%%%%%%%%%%%%%%%%%%%%% 
The energy flux, $T^0_i$, gives the velocity field, $v_\phi Q_i$,
%%%%%%%%%%%%%%%%%%%%%%%%%%%%%%%%%%%%%%%%%%%%%%%%%%%%%%%%%%%%%%%%%%%%%%%
\begin{equation}
 \left(\rho_\phi+p_\phi\right)\left(v_\phi-B\right)Q_i
  \equiv T^0_i
  = \left(\frac{\dot{\phi}}{a^2}k\delta\phi\right) Q_i.
\end{equation}
%%%%%%%%%%%%%%%%%%%%%%%%%%%%%%%%%%%%%%%%%%%%%%%%%%%%%%%%%%%%%%%%%%%%%%%
Using $\rho_\phi+p_\phi= a^{-2}\dot{\phi}^2$, we obtain 
$v_\phi -B = k\dot{\phi}^{-1}\delta\phi$; thus, 
$\delta\phi$ is directly responsible for the fluid's peculiar motion.
The anisotropic stress, $T^i_j-p_\phi\delta^i_j$, is a second-order 
perturbation variable for a scalar field, being negligible.

When we choose our coordinate system so as $B\equiv v_\phi$ 
(a fluid element is comoving with the origin of the spatial coordinate), 
we have $\delta\phi$ vanishing, $\delta\phi\equiv 0$.
This coordinate is called the comoving gauge, and we write
the scalar-field fluctuations in this gauge as $\delta\phi_{\rm com}\equiv 0$.

Since we have only one degree of freedom,
a scalar field, in the system, $\delta\rho_\phi$, $\delta p_\phi$, 
and $v_\phi$ are not independent of each other. 
Nevertheless, this fluid representation is useful, as the cosmological
linear perturbation theory has been developed as the general relativistic
fluid dynamics.
We can plague these fluid variables into 
the well-established general relativistic fluid equations, and 
see what happens to the metric perturbations.
While we do not use those fluid equations explicitly in the following,
but solve equation of motion for a scalar field 
(Klein--Gordon equation) directly, the fluid equations give the same
answer.

\subsection{Gauge-invariant perturbations}

In the previous subsection, we have seen that scalar-field fluctuations
vanish in the comoving gauge in which $B\equiv v_\phi$; 
thus, a choice of gauge defines perturbations.
For the scalar-type perturbations that we are considering, 
the gauge transformation is
%%%%%%%%%%%%%%%%%%%%%%%%%%%%%%%%%%%%%%%%%%%%%%%%%%%%%%%%%%%%%%%%%%%%%%%
\begin{eqnarray}
 \tau &\longrightarrow& \tau' = \tau + T(\tau)Q({\mathbf x}),\\
 x^i  &\longrightarrow&  x'{}^i = x^i + L(\tau)Q_i({\mathbf x}),
\end{eqnarray}
%%%%%%%%%%%%%%%%%%%%%%%%%%%%%%%%%%%%%%%%%%%%%%%%%%%%%%%%%%%%%%%%%%%%%%%
where $T$ and $L$ are $\ll 1$.
Accordingly, scalar-field fluctuations, $\delta\phi$, transform as
%%%%%%%%%%%%%%%%%%%%%%%%%%%%%%%%%%%%%%%%%%%%%%%%%%%%%%%%%%%%%%%%%%%%%%%
\begin{equation}
 \label{eq:phitrans}
 \delta\phi(\tau) \longrightarrow 
  \widetilde{\delta\phi}(\tau') = 
  \delta\phi(\tau) - \dot{\phi}(\tau)T(\tau).
\end{equation}
%%%%%%%%%%%%%%%%%%%%%%%%%%%%%%%%%%%%%%%%%%%%%%%%%%%%%%%%%%%%%%%%%%%%%%%
Hence, if we choose $T=\dot{\phi}^{-1}\delta\phi$, then we obtain
$\widetilde{\delta\phi}=0$.
This choice of $T$ defines the comoving gauge, $\delta\phi_{\rm com}=0$.
In this way, we find different values for the perturbation variables 
in different gauges.

So, what gauge should we use?
Unfortunately, there is no answer to the question:
``what gauge {\it should} we use?''.
Although there is no best gauge in the world,
depending on a problem that we intend to solve, we may find that 
one gauge is more {\it useful} than the other, or vice versa.
As long as we fix the gauge uniquely, and understand what
gauge we are working on clearly, no problems occur.

In practice, however, problems occur when 
one author understands its own gauge, but does not understand
the other author's gauge.
Since there is no best gauge in the world, different authors may
use different gauges, and may disagree with each other
because of their misunderstanding of the gauges.
In other words, one author's calculation on amplitude of $\delta\phi$
may disagree with the other's calculation, if they are using different 
gauges.
The author using the comoving gauge sees $\delta\phi=0$, but
others may see $\delta\phi\neq 0$.

One way to overcome this undesirable property is to make 
perturbation variables {\it invariant} under the 
gauge transformation, and let them represent {\it gauge-invariant} 
perturbations.
As an example, consider a new perturbation variable \citep{MFB92},
%%%%%%%%%%%%%%%%%%%%%%%%%%%%%%%%%%%%%%%%%%%%%%%%%%%%%%%%%%%%%%%%%%%%%%%
\begin{equation}
 \label{eq:u}
 u\equiv \delta\phi 
  - \frac{\dot{\phi}}{aH}\left(H_{\rm L}+\frac13H_{\rm T}\right).
\end{equation}
%%%%%%%%%%%%%%%%%%%%%%%%%%%%%%%%%%%%%%%%%%%%%%%%%%%%%%%%%%%%%%%%%%%%%%%
One can prove this variable gauge invariant, $\widetilde{u}=u$, using
equation~(\ref{eq:phitrans}) and  
%%%%%%%%%%%%%%%%%%%%%%%%%%%%%%%%%%%%%%%%%%%%%%%%%%%%%%%%%%%%%%%%%%%%%%%
\begin{equation}
 \widetilde{H}_{\rm L}
  +\frac13\widetilde{H}_{\rm T}
  =
  H_{\rm L} + \frac13H_{\rm T} - aHT.
\end{equation}
%%%%%%%%%%%%%%%%%%%%%%%%%%%%%%%%%%%%%%%%%%%%%%%%%%%%%%%%%%%%%%%%%%%%%%%
Actually, $H_{\rm L}+\frac13H_{\rm T}$ represents perturbations in
the intrinsic spatial curvature, ${\cal R}$, as it is the scalar 
potential of the 3-dimension Ricci scalar: $\delta^{(3)}R= a^{-2}k^2{\cal R}Q$,
where ${\cal R}\equiv H_{\rm L}+\frac13H_{\rm T}$.
While $u$ reduces to $\delta\phi$ in the spatially flat gauge
(${\cal R}\equiv 0$), or to $-(\dot{\phi}/aH){\cal R}$ in the 
comoving gauge ($\delta\phi\equiv 0$), its value is invariant 
under any gauge transformation.
Any authors should agree upon the value of $u$.

For the physical interpretation of $u$, we may name $u$ 
``scalar-field fluctuations in the spatially flat
gauge'' or ``intrinsic spatial curvature perturbations in the comoving
gauge''.
Either name describes the physical meaning of $u$ correctly.
The physical meaning of $u$ depends upon what gauge we are using;
however, the most important point is that the value of $u$ is 
independent of a gauge choice. 
In this sense, $u$ can be a ``common language'' among different authors.

\citet{BST83} use a similar gauge-invariant variable to $u$, 
%%%%%%%%%%%%%%%%%%%%%%%%%%%%%%%%%%%%%%%%%%%%%%%%%%%%%%%%%%%%%%%%%%%%%%%
\begin{equation}
 \zeta\equiv -\frac{aH}{\dot{\phi}}u = {\cal R}-\frac{aH}{\dot{\phi}}
  \delta\phi,
\end{equation} 
%%%%%%%%%%%%%%%%%%%%%%%%%%%%%%%%%%%%%%%%%%%%%%%%%%%%%%%%%%%%%%%%%%%%%%%
that reduces to ${\cal R}$ in the comoving gauge, or to 
$-(aH/\dot{\phi})\delta\phi$ in the spatially flat gauge.
This variable helps our perturbation analysis not only because of
being gauge invariant, but also being {\it conserved} on super-horizon
scales throughout the cosmic evolution.
We will show this property in the next subsection.

Using gauge invariance of $u$ or $\zeta$, we obtain a relation between 
$\delta\phi$ in the spatially flat gauge, 
$\delta\phi_{\rm flat}$, and ${\cal R}$ in the comoving gauge, 
${\cal R}_{\rm com}$, as 
%%%%%%%%%%%%%%%%%%%%%%%%%%%%%%%%%%%%%%%%%%%%%%%%%%%%%%%%%%%%%%%%%%%%%%%
\begin{equation}
 \label{eq:linearrelation}
 {\cal R}_{\rm com} = -\frac{aH}{\dot{\phi}}\delta\phi_{\rm flat}.
\end{equation}
%%%%%%%%%%%%%%%%%%%%%%%%%%%%%%%%%%%%%%%%%%%%%%%%%%%%%%%%%%%%%%%%%%%%%%%
It is derived from $u_{\rm com}=u_{\rm flat}$, or 
$\zeta_{\rm com}=\zeta_{\rm flat}$.
As we have seen in the previous section, $\delta\phi$ obeys Gaussian
statistics to very good accuracy because of the central limit theorem,
that is, $\delta\phi$ is the sum of the nearly infinite number of independent 
modes (Eq.(\ref{eq:expand*})).
Since ${\cal R}_{\rm com}$ is linearly related to 
$\delta\phi_{\rm flat}$, ${\cal R}_{\rm com}$ also obeys Gaussian
statistics in the linear order; however, as we will show in the next
section, non-linear correction to this linear relation makes 
${\cal R}_{\rm com}$ weakly non-Gaussian.

The spatial curvature perturbation, ${\cal R}$, is more relevant for
the structure formation in the universe than the scalar-field
fluctuation, $\delta\phi$, itself, as ${\cal R}$ regulates the matter density
and velocity perturbations through the Poisson equation.
Actually, ${\cal R}$ reduces to the Newtonian potential inside the horizon. 
Since the quantum fluctuations generate $\delta\phi$, we expect it to generate
 ${\cal R}$ through ${\cal R}_{\rm com}=-(aH/\dot{\phi})\delta\phi_{\rm flat}$.
This is naively true, but may sound tricky. 
In the next subsection, we will show how to calculate ${\cal R}$
generated in inflation more rigorously.

\subsection{Generation of spatial curvature perturbations}

To calculate the intrinsic spatial curvature perturbation, 
${\cal R}= H_{\rm L}+\frac13H_{\rm T}$, that is generated in inflation,
we need to track its evolution equation, and figure out how it is 
related to $\delta\phi$.
We will show in this subsection that ${\cal R}$ is actually more than
related to $\delta\phi$; it is almost {\it equivalent} to $\delta\phi$.
We can track the evolution of $\delta\phi$ and ${\cal R}$ 
simultaneously, using the gauge-invariant variable, $u$ (Eq.(\ref{eq:u})).

\citet{MFB92} show that 
$au= a\delta\phi- (\dot{\phi}/H){\cal R}$ obeys the same 
Klein--Gordon equation as we have used in the previous section 
(Eq.(\ref{eq:KG})),
%%%%%%%%%%%%%%%%%%%%%%%%%%%%%%%%%%%%%%%%%%%%%%%%%%%%%%%%%%%%%%%%%%%
\begin{equation}
 \ddot{\chi}_k 
  + \left[k^2+m^2_\chi(\tau)\right]\chi_k =0,
\end{equation}
%%%%%%%%%%%%%%%%%%%%%%%%%%%%%%%%%%%%%%%%%%%%%%%%%%%%%%%%%%%%%%%%%%%
where $\chi_k$ is the mode function that expands 
$au$ (see Eq.(\ref{eq:expand})).
It thus follows that $a\delta\phi$ and $(\dot{\phi}/H){\cal R}$
obey the same equation, and our argument on the quantum-fluctuation
generation during inflation in the previous section applies to 
$u$ as well.
$u$ being quantum fluctuations means that it also obeys Gaussian 
statistics very well because of the central limit theorem 
(see~Eq.(\ref{eq:expand*}) and the text after equation).

We consider the Klein--Gordon equation for $au$ 
on super-horizon scales.
As equation~(\ref{eq:expand}), we expand $au$
into the mode functions, $\chi_k= a\varphi_k$,
where $\chi_k$ and $\varphi_k$ are exactly the same functions that we
have used in the previous section.

We give a slightly different expression for the solution,
emphasizing its time dependence on super-horizon scales.  
Taking the long-wavelength limit, $k^2\ll m_\chi^2(\tau)$, and using 
the exact form of $m^2_\chi(\tau)$ (Eq.(\ref{eq:effmass**}))
\citep{MFB92}, we obtain the Klein--Gordon equation on super-horizon scales,
%%%%%%%%%%%%%%%%%%%%%%%%%%%%%%%%%%%%%%%%%%%%%%%%%%%%%%%%%%%%%%%%%%%
\begin{equation}
 \ddot{\chi}_k 
  -\frac{H}{\dot{\phi}}\frac{d^2({\dot{\phi}}/{H})}{d\tau^2}\chi_k
  =0.
\end{equation}
%%%%%%%%%%%%%%%%%%%%%%%%%%%%%%%%%%%%%%%%%%%%%%%%%%%%%%%%%%%%%%%%%%%
There is an exact solution to this equation,
%%%%%%%%%%%%%%%%%%%%%%%%%%%%%%%%%%%%%%%%%%%%%%%%%%%%%%%%%%%%%%%%%%%
\begin{equation}
 \frac{H}{\dot{\phi}}{\chi}_k =
  \frac{aH}{\dot{\phi}}{\varphi}_k =
  C_1 + C_2 \int \frac{H^2}{\dot{\phi}^2} d\tau,
\end{equation}
%%%%%%%%%%%%%%%%%%%%%%%%%%%%%%%%%%%%%%%%%%%%%%%%%%%%%%%%%%%%%%%%%%%
where $C_1$ and $C_2$ are integration constants independent of $\tau$.
The second term is a decaying mode as
$\int \dot{\phi}^{-2}H^2 d\tau = \int a^{-3} (d\phi/dt)^{-2} H^2 dt$,
and thus $(H/\dot{\phi}){\chi}_k$ remains constant in time
on super-horizon scales.
This implies that $\zeta=-(aH/\dot{\phi})u$ also remains constant 
in time on super-horizon scales.
Note that $\zeta$ obeys Gaussian statistics in the linear order, as
it is related to a Gaussian variable, $u$, linearly;
however, as we will show in the next
section, non-linear correction to this linear relation makes 
$\zeta$ weakly non-Gaussian.
This statement is equivalent to that we have made on 
${\cal R}_{\rm com}$.

The solution obtained here for $\zeta$ is valid throughout the cosmic
history regardless of whether a scalar field, radiation, or matter
dominates the universe; thus, once created and leaving
the Hubble horizon during inflation, $\zeta$ remains constant in time 
throughout the subsequent cosmic evolution until reentering the horizon.
The amplitude of $\zeta$, i.e., $C_1$, is fixed by the 
quantum-fluctuation amplitude derived in the previous 
section~(Eq.(\ref{eq:specphi})),
%%%%%%%%%%%%%%%%%%%%%%%%%%%%%%%%%%%%%%%%%%%%%%%%%%%%%%%%%%%%%%%%%%%
\begin{equation}
 \frac{k^3}{2\pi^2}\left|C_1\right|^2 
  = \left(\frac{aH}{\dot{\phi}}\right)^2
  \Delta_\phi^2(k)
  \approx \left(\frac{aH^2}{2\pi\dot{\phi}}\right)^2
  = \left[\frac{H^2}{2\pi (d\phi/dt)}\right]^2.
\end{equation}
%%%%%%%%%%%%%%%%%%%%%%%%%%%%%%%%%%%%%%%%%%%%%%%%%%%%%%%%%%%%%%%%%%%
This is the spectrum of $\zeta$, $\Delta_\zeta^2(k)$, 
on super-horizon scales.
While we have neglected the $k$ dependence of the spectrum here, 
the spectral index of $\zeta$ is the same as of $\phi$
(Eq.(\ref{eq:liddle})),
%%%%%%%%%%%%%%%%%%%%%%%%%%%%%%%%%%%%%%%%%%%%%%%%%%%%%%%%%%%%%%%%%%%%%%
\begin{equation}
 \label{eq:indzeta}
  \frac{d\ln \Delta^2_\zeta}{d\ln k}
  =
  4\frac{dH/dt}{H^2}-2\frac{d^2\phi/dt^2}{H(d\phi/dt)}.
\end{equation}
%%%%%%%%%%%%%%%%%%%%%%%%%%%%%%%%%%%%%%%%%%%%%%%%%%%%%%%%%%%%%%%%%%%%%%

$\Delta_\zeta^2(k)$ gives the {\it primordial curvature-perturbation spectrum}.
This is very important prediction of inflation, as
it directly predicts the observables such as
the CMB anisotropy spectrum and the matter fluctuation spectrum.
Strictly speaking, $\Delta_\zeta^2(k)$ reduces to 
the curvature-perturbation spectrum in the comoving gauge.

To summarize, the quantum fluctuations generate the gauge-invariant
perturbation, $u$, that reduces to either $\delta\phi_{\rm flat}$ or 
$(\dot{\phi}/aH){\cal R}_{\rm com}$ depending on which gauge we use, 
either the spatially flat gauge or the comoving gauge.
Hence, $\delta\phi_{\rm flat}$ and $(\dot{\phi}/aH){\cal R}_{\rm com}$ 
are essentially equivalent to each other.
The benefit of $u$ is that it relates these two variables unambiguously,
simplifying the transformation between 
$\delta\phi_{\rm flat}$ and ${\cal R}_{\rm com}$.
This is a virtue of the linear perturbation theory; we do not have
this simplification when dealing with non-linear perturbations for which
we have to find non-linear transformation  between
$\delta\phi_{\rm flat}$ and ${\cal R}_{\rm com}$.
The non-linear transformation actually makes ${\cal R}_{\rm com}$
weakly non-Gaussian, even if $\delta\phi_{\rm flat}$ is exactly Gaussian.
We will see this in the next section.

Here, we have the generation of the primordial spatial curvature 
perturbations completed.
In the next subsection, we will derive the CMB anisotropy spectrum.

\subsection{Generation of primary CMB anisotropy}

The metric perturbations perturb CMB, producing the CMB anisotropy
on the sky.
Among the metric perturbation variables, the curvature perturbations
play a central role in producing the CMB anisotropy.

As we have shown in the previous subsection, 
the gauge-invariant perturbation, $\zeta$, does not change in time 
on super-horizon scales throughout the cosmic evolution regardless 
of whether a scalar field, radiation, or matter dominates the universe.
The intrinsic spatial curvature perturbation, ${\cal R}$, 
however, does change when equation of state of the universe,
$w\equiv p/\rho$, changes.
Since $\zeta$ remains constant, it may be useful to write the evolution
of ${\cal R}$ in terms of $\zeta$ and $w$; however,
${\cal R}$ is {\it not} gauge invariant itself, but $\zeta$ is gauge
invariant, so that the relation between ${\cal R}$ and $\zeta$ may
look misleading.

\citet{Bardeen80} has introduced another gauge-invariant variable,
$\Phi$ (or $\Phi_{\rm H}$ in the original notation),
which reduces to ${\cal R}$ in the zero-shear gauge, or the Newtonian
gauge, in which $B\equiv 0\equiv H_{\rm T}$.
$\Phi$ is given by
%%%%%%%%%%%%%%%%%%%%%%%%%%%%%%%%%%%%%%%%%%%%%%%%%%%%%%%%%%%%%%%%%%%
\begin{equation}
 \Phi\equiv {\cal R}-\frac{aH}k\left(-B+\frac{\dot{H}_{\rm T}}{k}\right).
\end{equation}
%%%%%%%%%%%%%%%%%%%%%%%%%%%%%%%%%%%%%%%%%%%%%%%%%%%%%%%%%%%%%%%%%%%
Here, the terms in the parenthesis represent the shear, or the anisotropic
expansion rate, of the $\tau={\rm constant}$ hypersurfaces.
While $\Phi$ represents the curvature perturbations in the zero-shear gauge,
it also represents the shear in the spatially flat gauge in which 
${\cal R}\equiv 0$.
Using $\Phi$, we may write $\zeta$ as
%%%%%%%%%%%%%%%%%%%%%%%%%%%%%%%%%%%%%%%%%%%%%%%%%%%%%%%%%%%%%%%%%%%
\begin{equation}
 \label{eq:zetaphi}
 \zeta= {\cal R} - \frac{aH}{\dot{\phi}}\delta\phi 
      = \Phi - \frac{aH}k \left(v_\phi-\frac{\dot{H}_{\rm T}}{k}\right),
\end{equation}
%%%%%%%%%%%%%%%%%%%%%%%%%%%%%%%%%%%%%%%%%%%%%%%%%%%%%%%%%%%%%%%%%%%
where the terms in the parenthesis represent the 
gauge-invariant fluid velocity.

Why use $\Phi$?
We use $\Phi$ because it gives the closest analogy to the Newtonian
potential, for $\Phi$ reduces to ${\cal R}$ in the zero-shear gauge
(or the Newtonian gauge) in which the metric (Eq.(\ref{eq:metric})) 
becomes just like the Newtonian limit of the general relativity.
It thus gives a natural connection to the ordinary Newtonian analysis.

The gauge-invariant velocity term, $v-k^{-1}\dot{H}_{\rm T}$,
differs $\zeta$ from $\Phi$.
In other words, the velocity and $\Phi$ share the value of $\zeta$.
Since a fraction of sharing depends upon equation of state of the universe,
$w=p/\rho$, the velocity and $\Phi$ change as $w$ changes.
$\zeta$ is independent of $w$.

The general relativistic cosmological linear perturbation theory 
gives the evolution of $\Phi$ on super-horizon scales \citep{KS84},
%%%%%%%%%%%%%%%%%%%%%%%%%%%%%%%%%%%%%%%%%%%%%%%%%%%%%%%%%%%%%%%%%%%
\begin{equation}
 \Phi
  = \frac{3+3w}{5+3w}\zeta,
\end{equation}
%%%%%%%%%%%%%%%%%%%%%%%%%%%%%%%%%%%%%%%%%%%%%%%%%%%%%%%%%%%%%%%%%%%
for adiabatic fluctuations, and hence $\Phi=\frac23\zeta$ in the radiation era
($w=1/3$), and $\Phi=\frac35\zeta$ in the matter era ($w=0$).
$\Phi$ then perturbs CMB through the so-called (static) Sachs--Wolfe 
effect \citep{SW67}.

\begin{figure}
 \begin{center}
  \leavevmode\epsfxsize=10cm \epsfbox{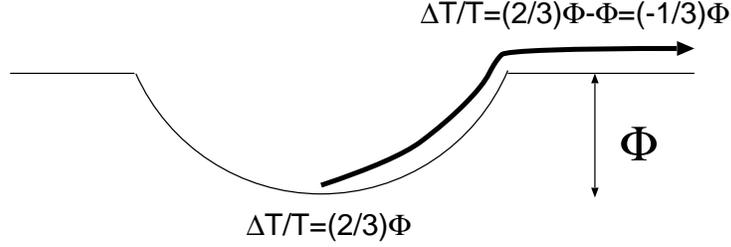}
 \end{center}
 \caption
 {The Static Sachs--Wolfe Effect} 
 \mycaption
 {The static Sachs--Wolfe effect predicts that CMB that resides in a 
 $\Phi$ potential well has an initial adiabatic temperature fluctuation 
 of $\Delta T/T= \frac23\Phi$ in the matter era.
 It further receives an additional fluctuation of $-\Phi$ when 
 climbing up the potential at the decoupling epoch.
 In total, we observe $\Delta T/T= -\frac13\Phi$.}
\label{fig:sw}
\end{figure}

The Sachs--Wolfe effect predicts that CMB that resides 
in a $\Phi$ potential well initially has an initial adiabatic 
temperature fluctuation of $\Delta T/T= [2/3(1+w)]\Phi$,
and it further receives an additional fluctuation of $-\Phi$ when 
climbing up the potential at the decoupling epoch.
In total, the CMB temperature fluctuations that we observe today amount to
%%%%%%%%%%%%%%%%%%%%%%%%%%%%%%%%%%%%%%%%%%%%%%%%%%%%%%%%%%%%%%%%%%%
\begin{equation}
 \label{eq:sweffect}
 \frac{\Delta T}T 
  = \frac{2}{3(1+w)}\Phi-\Phi
  = -\frac{1+3w}{3+3w}\Phi
  = -\frac{1+3w}{5+3w}\zeta.
\end{equation}
%%%%%%%%%%%%%%%%%%%%%%%%%%%%%%%%%%%%%%%%%%%%%%%%%%%%%%%%%%%%%%%%%%%
Figure~\ref{fig:sw} sketches the static Sachs--Wolfe effect.

%%%%%%%%%%%%%%%%%%%%%%%%%%%%%%%%%%%%%%%%%%%%%%%%%%%%%%%%%%%%%%%%%%%%%%
%%%%%%%%%%%%%%%%%%%%%%%%%%%%%%%%%%%%%%%%%%%%%%%%%%%%%%%%%%%%%%%%%%%%%%

For isocurvature fluctuations, initial temperature fluctuations
in a potential well are given by $-\Phi$ in both the radiation era
and the matter era; thus, total temperature fluctuations amount to
$\Delta T/T= -\Phi-\Phi= -2\Phi$.
By definition of the isocurvature fluctuations, $\Phi$ is initially
zero, but there exist non-vanishing initial entropy fluctuations.  
As the universe evolves, the entropy fluctuations create $\Phi$,
and hence the temperature fluctuations.

At the decoupling epoch, the universe has already been in the matter era
in which $w=0$,
so that we observe adiabatic temperature fluctuations of
$\Delta T/T= -\frac13\Phi= -\frac15\zeta$,
and the CMB fluctuation spectrum of the Sachs--Wolfe effect,
$\Delta_{\rm SW}^2(k)$, is
%%%%%%%%%%%%%%%%%%%%%%%%%%%%%%%%%%%%%%%%%%%%%%%%%%%%%%%%%%%%%%%%%%%
\begin{equation}
 \Delta_{\rm SW}^2(k)
  = \frac19\Delta_{\Phi}^2(k)
  = \frac1{25}\Delta_{\zeta}^2(k)
  = \left[\frac{H^2}{10\pi (d\phi/dt)}\right]^2,
\end{equation}
%%%%%%%%%%%%%%%%%%%%%%%%%%%%%%%%%%%%%%%%%%%%%%%%%%%%%%%%%%%%%%%%%%%
where $H$ is the Hubble parameter during inflation.
While we have not shown the $k$ dependence of the spectrum here,
the spectral index is given by equation~(\ref{eq:indzeta}).
By projecting the 3-dimension CMB fluctuation spectrum,
$\Delta_{\rm SW}^2(k)$, on the sky, we obtain
the angular power spectrum, $C_l$ \citep{BE87},
%%%%%%%%%%%%%%%%%%%%%%%%%%%%%%%%%%%%%%%%%%%%%%%%%%%%%%%%%%%%%%%%%%%
\begin{equation}
 \label{eq:sw}
 C_l^{\rm SW} = 4\pi \int_0^\infty \frac{dk}k~
  \Delta_{\rm SW}^2(k) 
  j^2_l\left[k(\tau_0-\tau_{\rm dec})\right]
  =
  C^{\rm SW}_2
  \frac{\Gamma\left[(9-n)/2\right]\Gamma\left[l+(n-1)/2\right]}
       {\Gamma\left[(n+3)/2\right]\Gamma\left[l+(5-n)/2\right]},
\end{equation}
%%%%%%%%%%%%%%%%%%%%%%%%%%%%%%%%%%%%%%%%%%%%%%%%%%%%%%%%%%%%%%%%%%%
where $\tau_0$ and $\tau_{\rm dec}$ denote the present day and the 
decoupling epoch, respectively, and 
$n\equiv 1+\left[d\ln \Delta^2(k)/d\ln k\right]$ 
is a spectral index which is conventionally used in the literature.
If the spectrum is exactly scale invariant, $n=1$,
then we obtain $C_l^{\rm SW}=\left[l(l+1)\right]^{-1}6C_2^{\rm SW}$.

Equation~(\ref{eq:sw}) provides a simple yet good fit to the CMB 
power spectrum measured by {\it COBE} DMR.
The spectrum comprises two parameters, $C_2^{\rm SW}$ and $n$.
\citet{Ben96} find $n=1.2\pm 0.3$ and
$C_2^{\rm SW}=7.9^{+2.0}_{-1.4}\times 10^{-11}$.
When fixing $n=1$,  they find
$C_2^{\rm SW|n=1}=(9.3\pm 0.8)\times 10^{-11}$.
The measured $C_l$ is thus consistent with CMB being scale invariant,
supporting inflation.
Moreover, it implies that $H^2/(d\phi/dt)\sim 
\left(C_2^{\rm SW}\right)^{1/2}\sim 10^{-5}$, constraining amplitude
of the Hubble parameter during inflation.

On the angular scales smaller than the DMR angular scales, the Sachs--Wolfe
approximation breaks down, and the acoustic physics in the photon-baryon
fluid system modifies the primordial radiation spectrum
\citep{PY70,BE87}.
To calculate the modification, we have to solve the Boltzmann
photon transfer equation together with the Einstein equations.
The modification is often described by the {\it radiation transfer
function}, $g_{{\rm T}l}(k)$, which can be calculated numerically with
the Boltzmann code such as {\sf CMBFAST} \citep{SZ96}.
Using $g_{{\rm T}l}(k)$, we write the CMB power spectrum, $C_l$, as
%%%%%%%%%%%%%%%%%%%%%%%%%%%%%%%%%%%%%%%%%%%%%%%%%%%%%%%%%%%%%%%%%%%
\begin{equation}
 C_l= 4\pi \int_0^\infty \frac{dk}k~
  \Delta_{\Phi}^2(k)
  g^2_{{\rm T}l}(k).
\end{equation}
%%%%%%%%%%%%%%%%%%%%%%%%%%%%%%%%%%%%%%%%%%%%%%%%%%%%%%%%%%%%%%%%%%%
Note that for the static Sachs--Wolfe effect,
adiabatic fluctuations give 
$g_{{\rm T}l}(k)=-\frac13j_l\left[k(\tau_0-\tau_{\rm dec})\right]$,
while isocurvature fluctuations give 
$g_{{\rm T}l}(k)=-2j_l\left[k(\tau_0-\tau_{\rm dec})\right]$.
Here, we have used $\Delta_{\Phi}^2(k)$ rather than 
$\Delta_{\rm SW}^2(k)$ or $\Delta_{\zeta}^2(k)$, following the literature.
The literature often uses the $\Phi$ power spectrum, $P_\Phi(k)$, to replace
$\Delta_{\Phi}^2(k)$; the relation is 
$\Delta_{\Phi}^2(k)= (2\pi^2)^{-1} k^3P_\Phi(k)$. 
$\Delta_{\Phi}^2(k)$ is called the dimensionless power spectrum.

If $\Phi$ were exactly Gaussian, then $C_l$ would specify all the 
statistical properties of $\Phi$, which are equivalent to 
those of $\zeta$.
Since $\zeta$ is related to a Gaussian variable, $u$, through
$\zeta=-(aH/\dot{\phi})u$, in the linear order $\zeta$ also obeys 
Gaussian statistics; however, the relation between $\zeta$ and $u$
becomes {\it non-linear} when we take into account non-linear
perturbations.
As a result, $\zeta$, and hence $\Phi$, becomes non-Gaussian even if
$u$ is exactly Gaussian, yielding non-Gaussian CMB anisotropies.
In the next section, we will analyze non-linear perturbations in inflation.

Using the second-order gravitational perturbation theory, \citet{PC96} 
derive the second-order correction to the relation between 
$\Delta T$ and $\Phi$ (Eq.(\ref{eq:sweffect})).
It gives $\Delta T/T= -\frac13\Phi + {\cal O}(1)\Phi^2$; thus, 
even if $\Phi$ is Gaussian, $\Delta T$ becomes weakly non-Gaussian.

%%%%%%%%%%%%%%%%%%%%%%%%%%%%%%%%%%%%%%%%%%%%%%%%%%%%%%%%%%%%%%%%%%%
\section{Non-linear Perturbations in Inflation}
\label{sec:nonlinear}

In the previous section, we have shown that the quantum fluctuations
generate the gauge-invariant perturbation, 
$u=\delta\phi - (\dot{\phi}/aH){\cal R}$, and
$u$ obeys Gaussian statistics very well because
of the central limit theorem.
Another gauge-invariant variable,
$\zeta= -(aH/\dot{\phi})u = {\cal R} - (aH/\dot{\phi})\delta\phi$,
which remains constant in time outside the horizon,
also obeys Gaussian statistics in the linear order, as 
it is related to a Gaussian variable, $u$, linearly.

In the non-linear order, however, the situation may change.
In the relation between $\zeta$ and $u$,
the factor in front of $u$, $aH/\dot{\phi}$, is also a function of $\phi$,
and it may produce additional fluctuations like 
$[\partial(aH/\dot{\phi})/\partial\phi]\delta\phi$.
Suppose that ${\cal R}$ in the comoving gauge 
($\delta\phi_{\rm com}\equiv 0$), 
${\cal R}_{\rm com}$, is an arbitrary function of a scalar field:
${\cal R}_{\rm com}= f(\phi)$. 
Note that ${\cal R}_{\rm com}$ is equivalent to $\zeta$.
By perturbing $\phi$ as $\phi=\phi_0+\delta\phi_{\rm flat}$, where
$\delta\phi_{\rm flat}$ is a scalar-field fluctuation in the
spatially flat gauge (${\cal R}_{\rm flat}\equiv 0$), we have
%%%%%%%%%%%%%%%%%%%%%%%%%%%%%%%%%%%%%%%%%%%%%%%%%%%%%%%%%%%%%%%%%%%
\begin{eqnarray}
 \nonumber
 {\cal R}_{\rm com}&=&
  f(\phi_0+\delta\phi_{\rm flat})\\
  &=&
  f(\phi_0) + 
  \left(\frac{\partial f}{\partial\phi}\right)\delta\phi_{\rm flat}
  +\frac12\left(\frac{\partial^2f}{\partial\phi^2}\right)
  \delta\phi^2_{\rm flat}
  +{\cal O}\left(\delta\phi_{\rm flat}^3\right).
\end{eqnarray}
%%%%%%%%%%%%%%%%%%%%%%%%%%%%%%%%%%%%%%%%%%%%%%%%%%%%%%%%%%%%%%%%%%%
By comparing this equation with the linear-perturbation result
(Eq.(\ref{eq:linearrelation})),
${\cal R}_{\rm com}= -(aH/\dot{\phi})\delta\phi_{\rm flat}$,
we find $f(\phi_0)=0$, $\partial f/\partial\phi= 
-(aH/\dot{\phi})$, and
%%%%%%%%%%%%%%%%%%%%%%%%%%%%%%%%%%%%%%%%%%%%%%%%%%%%%%%%%%%%%%%%%%%
\begin{equation}
 \label{eq:crude}
 {\cal R}_{\rm com}=
  -\frac{aH}{\dot{\phi}}\delta\phi_{\rm flat}
  -\frac12\frac{\partial}{\partial\phi}\left(\frac{aH}{\dot{\phi}}\right)
  \delta\phi^2_{\rm flat}
  +{\cal O}\left(\delta\phi_{\rm flat}^3\right);
\end{equation}
%%%%%%%%%%%%%%%%%%%%%%%%%%%%%%%%%%%%%%%%%%%%%%%%%%%%%%%%%%%%%%%%%%%
thus, even if $\delta\phi_{\rm flat}$ is exactly Gaussian, 
${\cal R}_{\rm com}$, and hence $\zeta$, becomes weakly non-Gaussian 
because of $\delta\phi_{\rm flat}^2$ or the higher-order terms.
While the treatment here may look rather crude, we will show in this 
section that the solution~(Eq.(\ref{eq:crude})) actually satisfies 
a more proper treatment of non-linear perturbations in inflation.

In the linear regime, we have the gauge-invariant perturbation
variable that characterizes the curvature perturbations as well as the 
scalar-field fluctuations, and the single equation that describes the 
perturbation evolution on all scales.
In the non-linear regime, however, we cannot make such 
great simplification. 
Since the Einstein equations are highly non-linear, 
fully analyzing non-linear problems is technically very difficult.
Hence, we need a certain approximation.

\subsection{Gradient expansion of Einstein equations}

In inflation, there is an useful scheme of approximation, the so-called
anti-Newtonian approximation \citep{Tom75,Tom82,TD92}, 
or later called the long-wavelength approximation \citep{KH98,ST98}
or the {\it gradient expansion method} 
\citep{SB90,SS92,NT96,NT98}.

The approximation neglects higher-order spatial derivatives in the Einstein
equations as well as in the equations of motion for matter fields, and
is equivalent to taking a long-wavelength limit of the system.
The equation system is further simplified if we set the shift vector
zero in the metric.
Once neglecting higher-order spatial derivatives and the shift vector,
one finds that the shear decays away very rapidly.

We use the metric of the form
%%%%%%%%%%%%%%%%%%%%%%%%%%%%%%%%%%%%%%%%%%%%%%%%%%%%%%%%%%%%%%%%%%%
\begin{equation}
 \label{eq:metric*}
  ds^2 = -N^2d\tau^2 + {}^{(3)}g_{ij}dx^idx^j,
\end{equation}
%%%%%%%%%%%%%%%%%%%%%%%%%%%%%%%%%%%%%%%%%%%%%%%%%%%%%%%%%%%%%%%%%%%
where $N$ is the Lapse function, and ${}^{(3)}g_{ij}$ describes the 3-metric. 
We have set the shift vector zero; it corresponds to $B\equiv 0$ in the
linearized metric~(\ref{eq:metric}). 
We perturb $N$ and ${}^{(3)}g_{ij}$ non-linearly.
While the full treatment of non-linear evolution of the system 
is highly complicated, by neglecting higher-order spatial derivatives,
we reduce the Einstein equations to rather simplified forms
\citep{SB90}:
%%%%%%%%%%%%%%%%%%%%%%%%%%%%%%%%%%%%%%%%%%%%%%%%%%%%%%%%%%%%%%%%%%%
\begin{eqnarray}
 \label{eq:00}
 H^2 - \frac13\sigma^{ij}\sigma_{ij} &=&
  \frac{8\pi G}3\left[\frac1{2N^2}\dot{\phi}^2+V(\phi)\right],\\
 \label{eq:0i}
 H_{,i}-\frac12\sigma^{k}_{i,k} &=& -4\pi G\frac{\dot{\phi}}N\phi_{,i},\\
 \label{eq:trace}
 \frac{2}N\dot{H}+3H^2+\sigma^{ij}\sigma_{ij}
  &=& -8\pi G\left[\frac1{2N^2}\dot{\phi}^2-V(\phi)\right],\\
 \label{eq:traceless}
 \frac1N\dot{\sigma}^i_j
  +3H\sigma^i_j&=& 0.
\end{eqnarray}
%%%%%%%%%%%%%%%%%%%%%%%%%%%%%%%%%%%%%%%%%%%%%%%%%%%%%%%%%%%%%%%%%%%

$H$ is the inhomogeneous Hubble parameter which 
defines the isotropic expansion rate of the $\tau={\rm constant}$
hypersurfaces,
%%%%%%%%%%%%%%%%%%%%%%%%%%%%%%%%%%%%%%%%%%%%%%%%%%%%%%%%%%%%%%%%%%%
\begin{equation}
 \label{eq:inhomogeneoushubble}
 H\equiv \frac1{6N}{}^{(3)}g^{ij}{}^{(3)}\dot{g}_{ij},
\end{equation}
%%%%%%%%%%%%%%%%%%%%%%%%%%%%%%%%%%%%%%%%%%%%%%%%%%%%%%%%%%%%%%%%%%%
or ${}^{(3)}\dot{g}_{ij}=2NH{}^{(3)}g_{ij}$.
By introducing the inhomogeneous scale factor, $a({\mathbf x},\tau)$, 
given by $H=\dot{a}/(Na)$, we can write the 3-metric as 
${}^{(3)}g_{ij}= a^2({\mathbf x},\tau)\gamma_{ij}({\mathbf x})$, 
where $\gamma_{ij}({\mathbf x})$ is a function of 
the spatial coordinate only.
This form of ${}^{(3)}g_{ij}$ satisfies 
equation~(\ref{eq:inhomogeneoushubble}). 
We have obtained the simplified form for the 3-metric because of setting
the shift vector zero.

$\sigma^i_j$ is the shear which quantifies the anisotropic expansion rate,
%%%%%%%%%%%%%%%%%%%%%%%%%%%%%%%%%%%%%%%%%%%%%%%%%%%%%%%%%%%%%%%%%%%
\begin{equation}
 \sigma^i_j\equiv \frac1{2N}{}^{(3)}\dot{g}_{ij}-H{}^{(3)}g_{ij}.
\end{equation}
%%%%%%%%%%%%%%%%%%%%%%%%%%%%%%%%%%%%%%%%%%%%%%%%%%%%%%%%%%%%%%%%%%%
It follows from the traceless-part equation~(\ref{eq:traceless}) and
$NH=\dot{a}/a$ that the shear decays rapidly as the universe expands:
$\sigma^i_j\propto a^{-3}$; thus, we neglect
the shear terms in the Einstein equations henceforth.

By neglecting the shear term in the trace-part equation~(\ref{eq:trace}),
and substituting the Friedmann equation~(\ref{eq:00}) for $H^2$,
we obtain $\dot{H} = -4\pi G N^{-1}{\dot{\phi}^2}$.
Comparing this equation with the momentum constraint equation~(\ref{eq:0i}) 
without the shear term, ${H}_{,i} = -4\pi G N^{-1}\dot{\phi}{\phi}_{,i}$,
we find $H({\mathbf x},\tau)= H\left(\phi({\mathbf x},\tau)\right)$, and
the scalar-field momentum
%%%%%%%%%%%%%%%%%%%%%%%%%%%%%%%%%%%%%%%%%%%%%%%%%%%%%%%%%%%%%%%%%%%
\begin{equation}
 \label{eq:0i*}
 \frac{\dot{\phi}}N = 
 -\frac1{4\pi G}\left(\frac{\partial H}{\partial\phi}\right).
\end{equation}
%%%%%%%%%%%%%%%%%%%%%%%%%%%%%%%%%%%%%%%%%%%%%%%%%%%%%%%%%%%%%%%%%%%
Substituting this for the kinetic term in the 
Friedmann equation~(\ref{eq:00}), and 
neglecting the shear term, we finally obtain a closed evolution 
equation for $H(\phi)$,
%%%%%%%%%%%%%%%%%%%%%%%%%%%%%%%%%%%%%%%%%%%%%%%%%%%%%%%%%%%%%%%%%%%
\begin{equation}
 \label{eq:00**}
 H^2(\phi)= \frac1{12\pi G}\left(\frac{\partial H}{\partial\phi}\right)^2
  + \frac{8\pi G}3V(\phi).
\end{equation}
%%%%%%%%%%%%%%%%%%%%%%%%%%%%%%%%%%%%%%%%%%%%%%%%%%%%%%%%%%%%%%%%%%%
From this equation, $H(\phi)$ may be solved as 
$H(\phi,I)$, where $I$ is an integration constant which
parameterizes the initial condition.
This equation fully describes the evolution of the system including 
non-linear perturbations.
Note that $H(\phi)$ depends upon the spatial coordinate through
$\phi=\phi({\mathbf x},\tau)$.

A perturbation to $H$ is given by
$\delta H=\left(\partial H/\partial\phi\right)_I\delta\phi
+\left(\partial H/\partial I\right)_\phi\delta I$.
For the latter term, by differentiating equation~(\ref{eq:00**})
with respect to $I$ for a fixed $\phi$, we find
$\left(\partial H/\partial I\right)_\phi\propto a^{-3}$; thus,
it decays very rapidly during inflation, giving
$\delta H=\left(\partial H/\partial\phi\right)_I\delta\phi$.
Hence, the comoving gauge, $\delta\phi\equiv 0$, coincides with
the constant Hubble parameter gauge, $\delta H\equiv 0$.
\citet{ST98} also observe this property from a different point of view,
and find that this property holds for multiple scalar-field system
as well.

\subsection{Generation of non-linear curvature perturbations}

Our goal in this subsection is to find a non-linear relation between 
${\cal R}_{\rm com}$ and $\delta\phi_{\rm flat}$, following \citet{SB90}.
In the absence of the shear, ${\cal R}$ obeys 
$\dot{\cal R}=\delta\left(NH\right)= 
\delta\left(\partial\ln a/\partial\tau\right)$, and hence
${\cal R}$ is equivalent to fluctuations in the inhomogeneous 
scale factor, $\delta\ln a$.

We calculate ${\cal R}_{\rm com}$ by perturbing 
$\ln a$ non-linearly in the comoving gauge.
Since $\phi$ is homogeneous in the comoving gauge, 
we choose $\phi$ as a time coordinate: $\tau\equiv \phi$.
We find the Lapse function 
%%%%%%%%%%%%%%%%%%%%%%%%%%%%%%%%%%%%%%%%%%%%%%%%%%%%%%%%%%%%%%%%%%%
\begin{equation}
 \label{eq:N}
  N= -\frac{4\pi G}{\partial H/\partial\phi}
\end{equation}
%%%%%%%%%%%%%%%%%%%%%%%%%%%%%%%%%%%%%%%%%%%%%%%%%%%%%%%%%%%%%%%%%%%
for this time coordinate from
equation~(\ref{eq:0i*}) with setting $\dot{\phi}\equiv 1$.

To calculate the scalar-field fluctuations in the spatially flat gauge, 
$\delta\phi_{\rm flat}$, we need to generate quantum fluctuations first;
however, since we are solving the equation system on super-horizon 
scales only, we cannot calculate quantum fluctuations of $\delta\phi$ 
within the current framework.
Instead, we assume that $\delta\phi$ on super-horizon scales 
is provided by the small-scale quantum fluctuations that are stretched
out of the horizon by inflationary expansion.
We use the linear perturbation theory to calculate the fluctuation
amplitude at the horizon crossing, and provide 
$\delta\phi$ as initially linear, Gaussian fluctuations.
We then calculate $\delta\phi_{\rm flat}$ on the $\ln a={\rm constant}$ 
hypersurfaces.
While this treatment may sacrifice a virtue of the current framework 
which does not assume linearity of perturbations,
non-linearity of the scalar-field fluctuations may also be incorporated into
the analysis with the so-called stochastic inflation approach 
\citep{Sta86,SB91}.
Using this, \citet{Gan94} show that the non-linearity makes 
$\delta\phi_{\rm flat}$ weakly non-Gaussian.

\begin{figure}
 \plotone{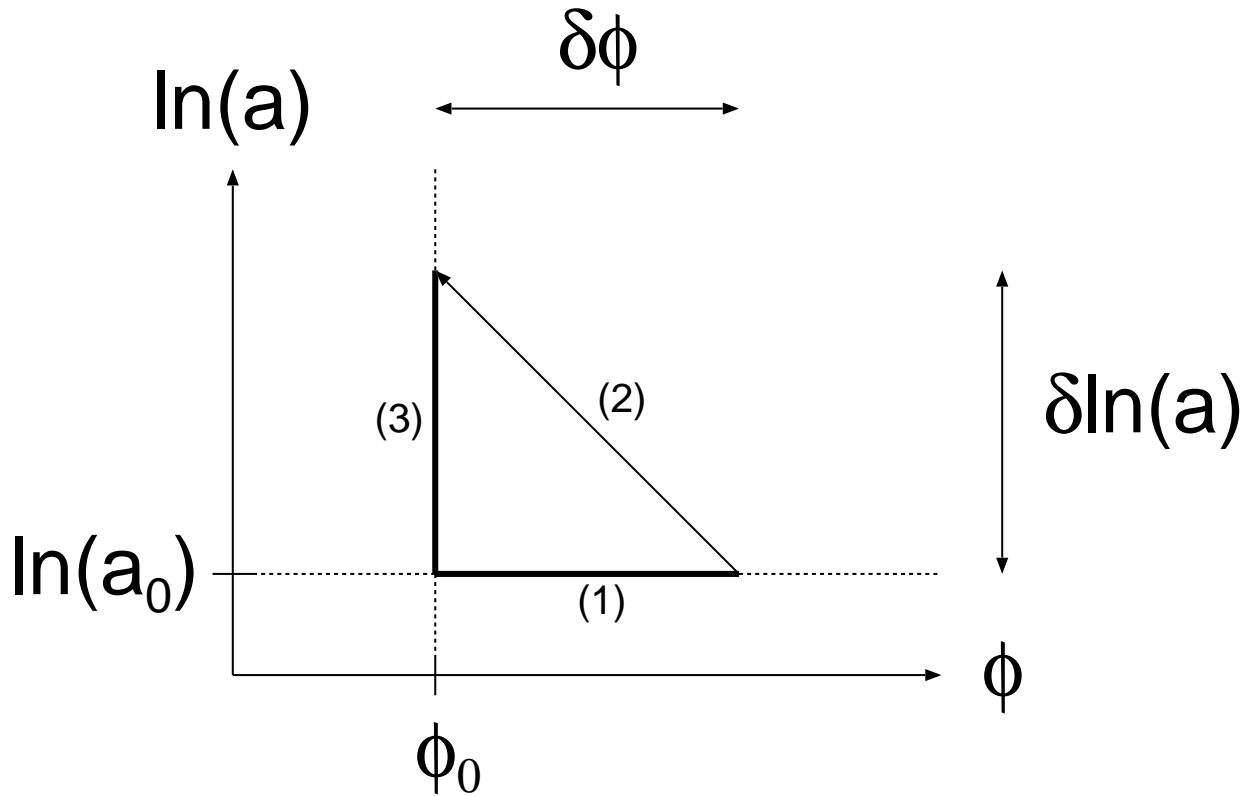}
 \caption
 {Hypersurface Transformation} 
 \mycaption
 {Hypersurface transformation of the scalar-field fluctuations 
 in the spatially flat
 gauge, $\delta\phi_{\rm flat}=\delta\phi(\ln a_0)$,  
 into the spatial curvature perturbations in the comoving gauge, 
 ${\cal R}_{\rm com}=\delta\ln a(\phi_0)$.
 (1) To represent $\delta\phi(\ln a_0)$, draw a short line from 
 $(\phi_0,\ln a_0)$ to $(\phi_0+\delta\phi,\ln a_0)$ 
 in parallel to the $\phi$ axis. 
 (2) By evolving $\ln a$, evolve this line until it coincides with 
 the $\phi=\phi_0$ line.
 (3) Measure the line length, $\delta\ln a(\phi_0)$.}
\label{fig:trans}
\end{figure}

After all, our goal is to relate $\delta\phi(\ln a)$ to
$\delta\ln a(\phi)$.
In other words, we transform the perturbations on 
the $\ln a= {\rm constant}$ hypersurfaces to the ones on the
$\phi= {\rm constant}$ hypersurfaces.
We do this as follows \citep{SB90},
and figure~(\ref{fig:trans}) shows the following process schematically.

%%%%%%%%%%%%%%%%%%%%%%%%%%%%%%%%%%%%%%%%%%%%%%%%%%%%%%%%%%%%%%%%%%%%%%
%%%%%%%%%%%%%%%%%%%%%%%%%%%%%%%%%%%%%%%%%%%%%%%%%%%%%%%%%%%%%%%%%%%%%%

First, imagine $\phi-\ln a$ plane on which we transform the perturbations.
We then mark a point on the plane with $(\phi_0,\ln a_0)$, and draw a 
short line from this point to $(\phi_0+\delta\phi,\ln a_0)$ 
in parallel to the $\phi$ axis.
This line represents $\delta\phi({\mathbf x},\ln a_0)\equiv 
\phi({\mathbf x},\ln a_0)-\phi_0$, i.e., $\phi$ perturbations on 
a $\ln a = \ln a_0$ hypersurface.
Next, using evolution equation of $\ln a$, 
we evolve the line until it coincides with the $\phi=\phi_0$ line, namely
%%%%%%%%%%%%%%%%%%%%%%%%%%%%%%%%%%%%%%%%%%%%%%%%%%%%%%%%%%%%%%%%%%%
\begin{equation}
 \ln a_0 = \ln \left[a({\mathbf x},\phi_0+ \delta\phi)\right]
  \longrightarrow \ln \left[a({\mathbf x},\phi_0)\right].
\end{equation}
%%%%%%%%%%%%%%%%%%%%%%%%%%%%%%%%%%%%%%%%%%%%%%%%%%%%%%%%%%%%%%%%%%%
The evolution equation of $\ln a$  is $d\ln a/d\tau = NH$.
By integrating this equation, we obtain
%%%%%%%%%%%%%%%%%%%%%%%%%%%%%%%%%%%%%%%%%%%%%%%%%%%%%%%%%%%%%%%%%%%
\begin{equation}
 \ln \left[a({\mathbf x},\phi_0)\right] - \ln a_0
 =
 \int_{\phi_0+\delta\phi({\mathbf x},\ln a_0)}^{\phi_0}d\phi~
 \frac{N(\phi)H(\phi)}{\dot{\phi}},
\end{equation}
%%%%%%%%%%%%%%%%%%%%%%%%%%%%%%%%%%%%%%%%%%%%%%%%%%%%%%%%%%%%%%%%%%%
where the Lapse function, $N$, is given by equation~(\ref{eq:N})
for $\phi$ being the time coordinate.
Finally, we measure the line length,
$\delta\ln\left[a({\mathbf x},\phi_0)\right]
\equiv \ln\left[a({\mathbf x},\phi_0)\right] - \ln a_0$, to obtain 
${\cal R}_{\rm com}$.

By noting $\delta\phi_{\rm flat}= \delta\phi({\mathbf x},\ln a_0)$, 
we find a {\it non-linear} relationship between ${\cal R}_{\rm com}$ and 
$\delta\phi_{\rm flat}$,
%%%%%%%%%%%%%%%%%%%%%%%%%%%%%%%%%%%%%%%%%%%%%%%%%%%%%%%%%%%%%%%%%%%
\begin{equation}
 \label{eq:rigorous}
  {\cal R}_{\rm com}
  =
  -\int_{\phi_0}^{\phi_0+\delta\phi_{\rm flat}}
  d\phi~\frac{N(\phi)H(\phi)}{\dot{\phi}}
  =
  4\pi G\int_{\phi_0}^{\phi_0+\delta\phi_{\rm flat}}d\phi
  \left[\frac{\partial\ln H}{\partial\phi}\right]^{-1}.
\end{equation}
%%%%%%%%%%%%%%%%%%%%%%%%%%%%%%%%%%%%%%%%%%%%%%%%%%%%%%%%%%%%%%%%%%%
This non-linear relation should be compared with the crude 
estimate from the linear analysis, equation~(\ref{eq:crude}).
By expanding the non-linear relation into Taylor series with respect
to $\delta\phi_{\rm flat}$, we find that the crude estimate
agrees with the non-linear result on the term-by-term basis.
Note that $a$ in equation~(\ref{eq:crude}) should be replaced by $N$,
as it has used the conformal time as the time coordinate, i.e.,
$N=a$.

\subsection{Generation of weakly non-Gaussian adiabatic fluctuations}

We have shown that non-linearity in inflation creates weakly non-Gaussian
curvature perturbations outside the horizon from Gaussian quantum
fluctuations inside the horizon.
By expanding the non-linear relation between ${\cal R}_{\rm com}$ and
$\delta\phi_{\rm flat}$ up to the second order, we obtain
a non-linear curvature perturbation,
${\cal R}_{\rm com}={\cal R}_{\rm com}^{\rm L}
+ {\cal R}_{\rm com}^{\rm NL}$, where
%%%%%%%%%%%%%%%%%%%%%%%%%%%%%%%%%%%%%%%%%%%%%%%%%%%%%%%%%%%%%%%%%%%
\begin{eqnarray}
  {\cal R}_{\rm com}^{\rm L}
   &\equiv& {4\pi G}
   \left(\frac{\partial\ln H}{\partial\phi}\right)^{-1}\delta\phi_{\rm flat},\\
 {\cal R}_{\rm com}^{\rm NL}
  &\equiv& -\frac1{8\pi G}\left(\frac{\partial^2\ln H}{\partial\phi^2}\right)
  \left({\cal R}_{\rm com}^{\rm L}\right)^2.
\end{eqnarray}
%%%%%%%%%%%%%%%%%%%%%%%%%%%%%%%%%%%%%%%%%%%%%%%%%%%%%%%%%%%%%%%%%%%
Using equation~(\ref{eq:zetaphi}), we obtain a non-linear
Newtonian potential, $\Phi=\Phi_{\rm L}+\Phi_{\rm NL}$, where
$\Phi_{\rm L}=\frac23{\cal R}_{\rm com}^{\rm L}$
and $\Phi_{\rm NL}=\frac23{\cal R}_{\rm com}^{\rm NL}$ in the 
radiation era, and
$\Phi_{\rm L}=\frac35{\cal R}_{\rm com}^{\rm L}$
and $\Phi_{\rm NL}=\frac35{\cal R}_{\rm com}^{\rm NL}$ in the 
matter era.
Note that these formulae should be modified when we include  
non-linear effects of the stochastic inflation approach \citep{Gan94},
which add extra terms to the formulae.

In the following chapters, we will focus on measuring 
$\Phi_{\rm NL}$ using non-Gaussian CMB temperature fluctuations.
We parameterize the amplitude of $\Phi_{\rm NL}$ with a {\it non-linear
coupling parameter}, $f_{\rm NL}$, as
%%%%%%%%%%%%%%%%%%%%%%%%%%%%%%%%%%%%%%%%%%%%%%%%%%%%%%%%%%%%%%%%%%%
\begin{equation}
 \Phi_{\rm NL}\left({\mathbf x}\right)
  = f_{\rm NL}\left[
	       \Phi_{\rm L}^2\left({\mathbf x}\right)
	     -\left<\Phi_{\rm L}^2\left({\mathbf x}\right)\right>\right].
\end{equation}
%%%%%%%%%%%%%%%%%%%%%%%%%%%%%%%%%%%%%%%%%%%%%%%%%%%%%%%%%%%%%%%%%%%
How big is $f_{\rm NL}$?
If a scalar field rolls down on a potential, $V(\phi)$, slowly,
then $\partial\ln H/\partial\phi\approx \frac12\partial\ln
V/\partial\phi$.
We thus obtain
%%%%%%%%%%%%%%%%%%%%%%%%%%%%%%%%%%%%%%%%%%%%%%%%%%%%%%%%%%%%%%%%%%%
\begin{equation}
 f_{\rm NL}=  
  -\frac5{24\pi G}\left(\frac{\partial^2\ln H}{\partial\phi^2}\right)
  \approx
  -\frac5{48\pi G}\left(\frac{\partial^2\ln V}{\partial\phi^2}\right).
\end{equation}
%%%%%%%%%%%%%%%%%%%%%%%%%%%%%%%%%%%%%%%%%%%%%%%%%%%%%%%%%%%%%%%%%%%
As an example, consider a power-law potential, $V(\phi)\propto \phi^n$.
We find 
%%%%%%%%%%%%%%%%%%%%%%%%%%%%%%%%%%%%%%%%%%%%%%%%%%%%%%%%%%%%%%%%%%%
\begin{equation}
 f_{\rm NL}=  \frac{5n}{48\pi}\frac{m_{\rm pl}^2}{\phi^2}
  \sim 10^{-2}.
\end{equation}
%%%%%%%%%%%%%%%%%%%%%%%%%%%%%%%%%%%%%%%%%%%%%%%%%%%%%%%%%%%%%%%%%%%
On the other hand, we will show in chapter~\ref{chap:theory_bl}
that $\left|f_{\rm NL}\right|$ should be larger than of order unity to produce
detectable non-Gaussian signals in CMB.
Even for the fastest motion of $\phi$ allowed in slowly-rolling 
single-field inflation, we find $\left|f_{\rm NL}\right|<5/2$.
Therefore, it is quite hard for this class of inflation models to
produce detectable CMB non-Gaussianity.
In other words, any detection of cosmological non-Gaussian signals in CMB
constrains slow-roll inflation models very strongly, or may favor
multiple scalar-field models or isocurvature fluctuations.

%%%%%%%%%%%%%%%%%%%%%%%%%%%%%%%%%%%%%%%%%%%%%%%%%%%%%%%%%%%%%%%%%%%
%
%  Angular n-point Harmonic Spectrum on the Sky
%
%     1st draft:  06/21/2001
%     revision:   07/18/2001
%     final:      08/05/2001
%
%%%%%%%%%%%%%%%%%%%%%%%%%%%%%%%%%%%%%%%%%%%%%%%%%%%%%%%%%%%%%%%%%%%
\chapter{Angular $n$-point Harmonic Spectrum on the Sky}
\label{chap:spectrum}

\begin{figure}
 \begin{center}
  \leavevmode\epsfxsize=9cm \epsfbox{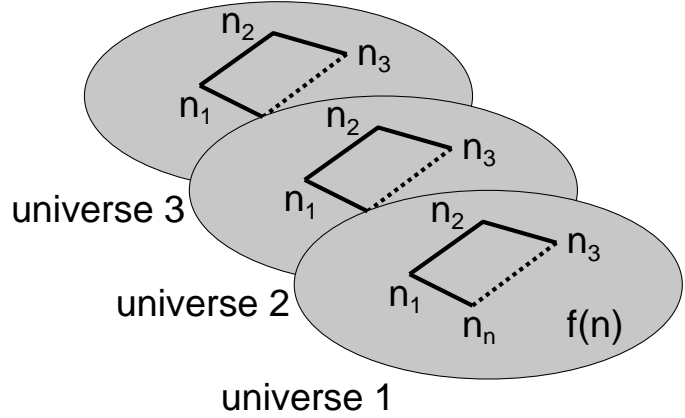}
 \end{center}
 \caption
 {Ensemble Average of Angular Correlation Function} 
 \mycaption
 {A schematic view of the ensemble average of the $n$-point angular 
 correlation function,  
 $f(\hat{\mathbf{n}}_1)f(\hat{\mathbf{n}}_2)f(\hat{\mathbf{n}}_3)
 \dots f(\hat{\mathbf{n}}_n)$. 
 We measure it on each universe, and then average it over many
 universes.}
\label{fig:ensemble}
\end{figure}

%%%%%%%%%%%%%%%%%%%%%%%%%%%%%%%%%%%%%%%%%%%%%%%%%%%%%%%%%%%%%%%%%%%
The angular $n$-point correlation function,
%%%%%%%%%%%%%%%%%%%%%%%%%%%%%%%%%%%%%%%%%%%%%%%%%%%%%%%%%%%%%%%%%%%
\begin{equation}
 \label{eq:n_corr}
  \left<f(\hat{\mathbf{n}}_1)f(\hat{\mathbf{n}}_2)
   \dots f(\hat{\mathbf{n}}_n)\right>,
\end{equation}
%%%%%%%%%%%%%%%%%%%%%%%%%%%%%%%%%%%%%%%%%%%%%%%%%%%%%%%%%%%%%%%%%%%
is a simple statistic characterizing a clustering pattern of
fluctuations on the sky, $f(\hat{\mathbf{n}})$.
Here, the bracket denotes the ensemble average, and
figure~\ref{fig:ensemble} sketches the meaning.
If the fluctuation is Gaussian, then the two-point correlation function
specifies all the statistical properties of $f(\hat{\mathbf{n}})$, for
the two-point correlation function is the only parameter in Gaussian 
distribution. 
If it is not Gaussian, then we need higher-order correlation 
functions to determine the statistical properties.

%%%%%%%%%%%%%%%%%%%%%%%%%%%%%%%%%%%%%%%%%%%%%%%%%%%%%%%%%%%%%%%%%%%%%%
%%%%%%%%%%%%%%%%%%%%%%%%%%%%%%%%%%%%%%%%%%%%%%%%%%%%%%%%%%%%%%%%%%%%%%

Yet simple, one disadvantage of the angular correlation function
is that data points of the correlation function at different angular
scales are generally not independent of each other, but correlated: 
two-point correlation at 1 degree is correlated with that at 2 degrees, 
and so on.
This property makes a detailed statistical analysis and interpretation 
of the data complicated.

Hence, one finds it more convenient to expand $f(\hat{\mathbf{n}})$
into spherical harmonics, the orthonormal basis on the sphere, as
%%%%%%%%%%%%%%%%%%%%%%%%%%%%%%%%%%%%%%%%%%%%%%%%%%%%%%%%%%%%%%%%%%%
\begin{equation}
 \label{eq:f_alm}
 f(\hat{\mathbf{n}}) 
 = \sum_{l=0}^\infty\sum_{m=-l}^{l} a_{lm}Y_{lm}(\hat{\mathbf{n}}),
\end{equation}
%%%%%%%%%%%%%%%%%%%%%%%%%%%%%%%%%%%%%%%%%%%%%%%%%%%%%%%%%%%%%%%%%%%
and then to consider the angular $n$-point harmonic spectrum, 
$\left<a_{l_1m_1}a_{l_2m_2}\dots a_{l_nm_n}\right>$.
While $a_{lm}$ for $m\neq 0$ is complex, reality of
$f(\hat{\mathbf{n}})$ gives $a_{l-m}=a_{lm}^*(-1)^m$, and thus the 
number of independent modes is not $4l+1$, but $2l+1$.

Especially, the angular two-, three-, and four-point harmonic spectra
are called the angular {\it power spectrum}, {\it bispectrum}, and 
{\it trispectrum}, respectively.
For a Gaussian field, the angular spectra at different angular scales, 
or at different $l$'s, are uncorrelated.
Even for a non-Gaussian field, 
they are reasonably uncorrelated as long as the non-Gaussianity is weak.
Moreover, since the spherical harmonic is orthogonal for different $l$'s, 
it highlights characteristic structures on the sky at a given $l$.
In other words, even if the angular correlation function is featureless,
the angular spectrum may have a distinct structure, 
for inflation predicts a prominent peak in the angular power spectrum, 
not in the angular correlation function.
In this chapter, we study statistical properties of the angular $n$-point 
harmonic spectra.

%%%%%%%%%%%%%%%%%%%%%%%%%%%%%%%%%%%%%%%%%%%%%%%%%%%%%%%%%%%%%%%%%%%
\section{Statistical Isotropy of the Universe}\label{sec:isotropy}

In reality, we cannot measure the ensemble average of the angular 
harmonic spectrum, but one realization such as 
$a_{l_1m_1}a_{l_2m_2}\dots a_{l_nm_n}$, which is so noisy that we want to 
average it somehow to reduce the noise.

We assume {\it statistical isotropy} of the universe from which
it follows that our sky is isotropic and has no preferred direction.
Isotropy of CMB justifies the assumption.
The assumption readily implies that one can average the spectrum over $m_i$
with an appropriate weight, as $m_i$ represent an azimuthal orientation 
on the sky.
The average over $m_i$ enables us to reduce statistical error of 
the measured harmonic spectra.

\begin{figure}
 \begin{center}
  \leavevmode\epsfxsize=9cm \epsfbox{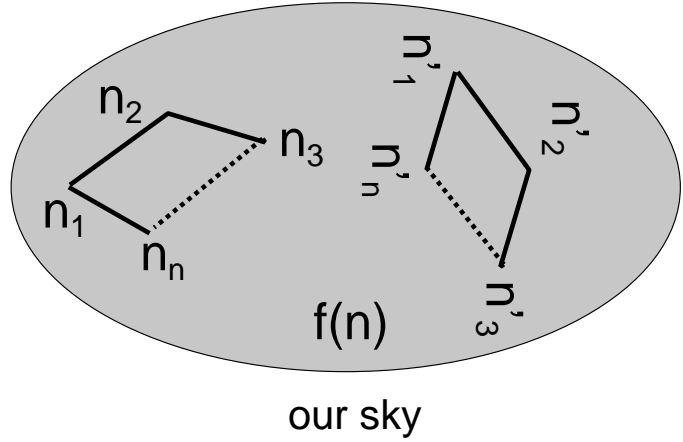}
 \end{center}
 \caption
 {Statistical Isotropy of Angular Correlation Function} 
 \mycaption{A schematic view of statistical isotropy of the angular 
 correlation function.
 As long as its configuration is preserved, we can average
 $f(\hat{\mathbf{n}}_1)\dots f(\hat{\mathbf{n}}_n)$ 
 over all possible orientations and 
 positions on the sky.}
\label{fig:rotinv}
\end{figure}

How can we find the weight?
One finds it as a solution to statistical isotropy, or 
{\it rotational invariance} of the angular correlation function on the sky,
%%%%%%%%%%%%%%%%%%%%%%%%%%%%%%%%%%%%%%%%%%%%%%%%%%%%%%%%%%%%%%%%%%%
\begin{equation}
 \label{eq:rotinv}
  \left<Df(\hat{\mathbf{n}}_1)Df(\hat{\mathbf{n}}_2)
   \dots Df(\hat{\mathbf{n}}_n)\right>
  = \left<f(\hat{\mathbf{n}}_1)f(\hat{\mathbf{n}}_2)
     \dots f(\hat{\mathbf{n}}_n)\right>,
\end{equation}
%%%%%%%%%%%%%%%%%%%%%%%%%%%%%%%%%%%%%%%%%%%%%%%%%%%%%%%%%%%%%%%%%%%
where $D=D(\alpha,\beta,\gamma)$ is a rotation matrix for the Euler 
angles $\alpha$, $\beta$, and $\gamma$.
Figure~\ref{fig:rotinv} sketches the meaning of statistical isotropy. 
Substituting equation~(\ref{eq:f_alm}) for $f(\hat{\mathbf{n}})$ 
in equation~(\ref{eq:rotinv}), we then need 
rotation of the spherical harmonic, $DY_{lm}(\hat{\mathbf{n}})$.
It is formally represented by the rotation matrix element, 
$D_{m'm}^{(l)}(\alpha,\beta,\gamma)$, as \citep{RBMW59}
%%%%%%%%%%%%%%%%%%%%%%%%%%%%%%%%%%%%%%%%%%%%%%%%%%%%%%%%%%%%%%%%%%%
\begin{equation}
 \label{eq:rotYlm}
  DY_{lm}(\hat{\mathbf{n}})= 
  \sum_{m'=-l}^{l} D_{m'm}^{(l)} Y_{lm'}(\hat{\mathbf{n}}).
\end{equation}
%%%%%%%%%%%%%%%%%%%%%%%%%%%%%%%%%%%%%%%%%%%%%%%%%%%%%%%%%%%%%%%%%%%
The matrix element, $D_{m'm}^{(l)}=\left<l,m'\left|D\right|l,m\right>$,
describes finite rotation of an initial state whose 
orbital angular momentum is represented by $l$ and $m$ into
a final state represented by $l$ and $m'$.
Finally, we obtain the statistical isotropy condition on the 
angular $n$-point harmonic spectrum:
%%%%%%%%%%%%%%%%%%%%%%%%%%%%%%%%%%%%%%%%%%%%%%%%%%%%%%%%%%%%%%%%%%%
\begin{equation}
 \label{eq:condition}
  \left<a_{l_1m_1}a_{l_2m_2}\dots a_{l_nm_n}\right>
  = \sum_{{\rm all}~m'} \left<a_{l_1m'_1}a_{l_2m'_2}
			       \dots a_{l_nm'_n}\right>
  D_{m_1'm_1}^{(l_1)}D_{m_2'm_2}^{(l_2)}\dots D_{m'_nm_n}^{(l_n)}.
\end{equation}
%%%%%%%%%%%%%%%%%%%%%%%%%%%%%%%%%%%%%%%%%%%%%%%%%%%%%%%%%%%%%%%%%%%

%%%%%%%%%%%%%%%%%%%%%%%%%%%%%%%%%%%%%%%%%%%%%%%%%%%%%%%%%%%%%%%%%%%%%%
%%%%%%%%%%%%%%%%%%%%%%%%%%%%%%%%%%%%%%%%%%%%%%%%%%%%%%%%%%%%%%%%%%%%%%

Using this equation, \citet{Hu01} has systematically evaluated appropriate 
weights for averaging the angular power spectrum ($n=2$), 
bispectrum ($n=3$), and trispectrum ($n=4$), over azimuthal angles.
Some of those may be found more intuitively; however, this method
allows us to find the weight for any higher-order harmonic spectrum.
In the following sections, we derive rotationally invariant, 
azimuthally averaged harmonic spectra for $n=2$, 3, and 4, and 
study their statistical properties.

%%%%%%%%%%%%%%%%%%%%%%%%%%%%%%%%%%%%%%%%%%%%%%%%%%%%%%%%%%%%%%%%%%%
\section{Angular Power Spectrum}\label{sec:powerspectrum}

The angular power spectrum measures how much fluctuations exist on a 
given angular scale.
For example, the variance of $a_{lm}$ for $l\ge 1$,
$\left<a_{lm}a_{lm}^*\right>$, measures the amplitude of fluctuations
at a given $l$.

Generally speaking, the covariance matrix of $a_{lm}$, 
$\left<a_{l_1m_1}a_{l_2m_2}^*\right>$, is not necessarily diagonal.
It is, however, actually diagonal once we assume full sky coverage and 
rotational invariance of the angular two-point correlation function, 
as we will show in this section.
The variance of $a_{lm}$ thus describes the two-point correlation completely.

Rotational invariance (Eq.(\ref{eq:condition})) requires 
%%%%%%%%%%%%%%%%%%%%%%%%%%%%%%%%%%%%%%%%%%%%%%%%%%%%%%%%%%%%%%%%%%%
\begin{equation}
 \label{eq:condition2}
  \left<a_{l_1m_1}a^*_{l_2m_2}\right>
  = \sum_{m_1'm_2'} \left<a_{l_1m'_1}a^*_{l_2m'_2}\right>
  D_{m_1'm_1}^{(l_1)}D_{m'_2m_2}^{(l_2)*}
\end{equation}
%%%%%%%%%%%%%%%%%%%%%%%%%%%%%%%%%%%%%%%%%%%%%%%%%%%%%%%%%%%%%%%%%%%
to be satisfied, where we have used the complex conjugate for simplifying 
calculations.
From this equation, we seek for a rotationally invariant representation 
of the angular power spectrum. 
Suppose that the covariance matrix of $a_{lm}$ is diagonal, i.e., 
$  \left<a_{l_1m_1}a^*_{l_2m_2}\right>= 
  \left<C_{l_1}\right> \delta_{l_1l_2} \delta_{m_1m_2}$.
Equation~(\ref{eq:condition2}) then reduces to
%%%%%%%%%%%%%%%%%%%%%%%%%%%%%%%%%%%%%%%%%%%%%%%%%%%%%%%%%%%%%%%%%%%
\begin{equation}
 \nonumber
  \left<a_{l_1m_1}a^*_{l_2m_2}\right>
  =
  \left<C_{l_1}\right> \delta_{l_1l_2}
  \sum_{m_1'}D_{m_1'm_1}^{(l_1)}D_{m'_1m_2}^{(l_1)*}
  =
  \left<C_{l_1}\right> \delta_{l_1l_2} \delta_{m_1m_2}.
\end{equation}
%%%%%%%%%%%%%%%%%%%%%%%%%%%%%%%%%%%%%%%%%%%%%%%%%%%%%%%%%%%%%%%%%%%
Thus, we have proven $\left<C_l\right>$ rotationally invariant.
Rotational invariance implies that the covariance matrix is diagonal.

\subsection{Estimator}

Observationally, the unbiased estimator of $\left<C_l\right>$ should be
%%%%%%%%%%%%%%%%%%%%%%%%%%%%%%%%%%%%%%%%%%%%%%%%%%%%%%%%%%%%%%%%%%%
\begin{eqnarray}
 \nonumber
  C_l&=& \frac1{2l+1}\sum_{m=-l}^{l} a_{lm}a_{lm}^*
     = \frac1{2l+1}\left(a_{l0}^2 + 2\sum_{m=1}^{l}a_{lm}a_{lm}^*\right)\\
 \label{eq:estimate2}
     &=& \frac1{2l+1}\left\{
		      a_{l0}^2 + 2\sum_{m=1}^{l}
     \left[\left(\Re{a_{lm}}\right)^2 + \left(\Im{a_{lm}}\right)^2\right]
     \right\}.
\end{eqnarray}
%%%%%%%%%%%%%%%%%%%%%%%%%%%%%%%%%%%%%%%%%%%%%%%%%%%%%%%%%%%%%%%%%%%
The second equality follows from $a_{l-m}=a_{lm}^*(-1)^m$, 
i.e., $a_{l-m}a_{l-m}^*=a_{lm}a_{lm}^*$, and hence
we average $2l+1$ independent samples for a given $l$.
It suggests that fractional statistical error of $C_l$ is 
reduced by $\sqrt{1/(2l+1)}$.
This property is the main motivation of our considering the azimuthally
averaged harmonic spectrum.

We find it useful to define an azimuthally averaged
harmonic transform, $e_{l}(\hat{\mathbf n})$, as 
%%%%%%%%%%%%%%%%%%%%%%%%%%%%%%%%%%%%%%%%%%%%%%%%%%%%%%%%%%%%%%%%%%%
\begin{equation}
 \label{eq:el}
   e_{l}(\hat{\mathbf n})
   \equiv
   \sqrt{\frac{4\pi}{2l+1}}
   \sum_{m=-l}^{l} a_{lm} Y_{lm}(\hat{\mathbf n}),
\end{equation}
%%%%%%%%%%%%%%%%%%%%%%%%%%%%%%%%%%%%%%%%%%%%%%%%%%%%%%%%%%%%%%%%%%
which is interpreted as a square-root of $C_l$ at a given position of 
the sky,
%%%%%%%%%%%%%%%%%%%%%%%%%%%%%%%%%%%%%%%%%%%%%%%%%%%%%%%%%%%%%%%%%%
\begin{equation}
  \label{eq:cl}
   \int\frac{d^2\hat{\mathbf n}}{4\pi}~
   e^2_{l}(\hat{\mathbf n})
   = C_l.
\end{equation}
%%%%%%%%%%%%%%%%%%%%%%%%%%%%%%%%%%%%%%%%%%%%%%%%%%%%%%%%%%%%%%%%%%
$e_{l}(\hat{\mathbf n})$ is particularly useful for measuring 
the angular bispectrum \citep{SG99,Kom01b} (chapter~\ref{chap:obs_bl}), 
trispectrum (chapter~\ref{chap:obs_tl}), and 
probably any higher-order harmonic spectra, because of being 
computationally very fast to calculate.
This is very important, as the new satellite experiments,
{\it MAP} and {\it Planck}, have more than millions of pixels, for which
we will crucially need a fast algorithm of measuring these 
higher-order harmonic spectra.

\subsection{Covariance matrix}

We derive the covariance matrix of $C_l$,
$\left<C_lC_{l'}\right>-\left<C_l\right>\left<C_{l'}\right>$, with
 the four-point function, the trispectrum.
Starting with
%%%%%%%%%%%%%%%%%%%%%%%%%%%%%%%%%%%%%%%%%%%%%%%%%%%%%%%%%%%%%%%%%%%
\begin{equation}
 \label{eq:error2}
  \left<C_lC_{l'}\right>
  =
  \frac1{(2l+1)(2l'+1)}
   \sum_{mm'}\left<a_{lm}a_{lm}^*a_{l'm'}a_{l'm'}^*\right>,
\end{equation}
%%%%%%%%%%%%%%%%%%%%%%%%%%%%%%%%%%%%%%%%%%%%%%%%%%%%%%%%%%%%%%%%%%%
we obtain the power spectrum covariance matrix
%%%%%%%%%%%%%%%%%%%%%%%%%%%%%%%%%%%%%%%%%%%%%%%%%%%%%%%%%%%%%%%%%%%
\begin{eqnarray}
 \nonumber
  \left<C_lC_{l'}\right> - \left<C_l\right>\left<C_{l'}\right>
  &=&
  \frac{2\left<C_l\right>^2}{2l+1}\delta_{ll'}
   + \frac1{(2l+1)(2l'+1)}
   \sum_{mm'}\left<a_{lm}a_{lm}^*a_{l'm'}a_{l'm'}^*\right>_{\rm c}\\
  \label{eq:error2*}
 &=&
  \frac{2\left<C_l\right>^2}{2l+1}\delta_{ll'}
   + \frac{(-1)^{l+l'}}{\sqrt{(2l+1)(2l'+1)}}
   \left<T^{ll}_{l'l'}(0)\right>_{\rm c},
\end{eqnarray}
%%%%%%%%%%%%%%%%%%%%%%%%%%%%%%%%%%%%%%%%%%%%%%%%%%%%%%%%%%%%%%%%%%%
where $\left<a_{lm}a_{lm}^*a_{l'm'}a_{l'm'}^*\right>_{\rm c}$
is the connected four-point harmonic spectrum, the connected trispectrum, 
which is exactly zero for a Gaussian field.
It follows from this equation that the covariance matrix of $C_l$ is 
exactly diagonal only when $a_{lm}$ is Gaussian.
$\left<T^{l_1l_2}_{l_3l_4}(L)\right>_{\rm c}$ is the ensemble average of
the angular averaged connected trispectrum, which we will 
define in \S~\ref{sec:trispectrum} (Eq.(\ref{eq:tl})).

Unfortunately, we cannot measure the connected $T^{ll}_{l'l'}(0)$
directly from the angular trispectrum (see \S~\ref{sec:trispectrum}).
We will thus never be sure if the power spectrum covariance is 
precisely diagonal, as long as we use the angular trispectrum.
We need the other statistics that can pick up information of the 
connected $T^{ll}_{l'l'}(0)$, even though they are indirect.
Otherwise, we need a model for the connected trispectrum, and
use the model to constrain the connected $T^{ll}_{l'l'}(0)$ from 
the other trispectrum configurations.
We will discuss this point in chapter~\ref{chap:obs_tl}.

There is no reason to assume the connected $T^{ll}_{l'l'}(0)$ small.
It is produced on large angular scales, 
if topology of the universe is closed hyperbolic \citep{Ino01b}.
In appendix~\ref{app:CH}, we derive an analytic prediction for the 
connected trispectrum produced in a closed hyperbolic universe.
On small angular scales, several authors have shown that 
the weak gravitational lensing effect produces the connected
trispectrum or four-point correlation function \citep{Ber97,ZS99,Zal00};
\citet{Hu01} finds that the 
induced off-diagonal terms are negligible compared with the diagonal 
terms out to $l\sim 2000$.

If the connected trispectrum is negligible, then we obtain
%%%%%%%%%%%%%%%%%%%%%%%%%%%%%%%%%%%%%%%%%%%%%%%%%%%%%%%%%%%%%%%%%%%
\begin{equation}
 \label{eq:error2**}
  \left<C_lC_{l'}\right> - \left<C_l\right>\left<C_{l'}\right>
  \approx
  \frac{2\left<C_l\right>^2}{2l+1}\delta_{ll'}.
\end{equation}
%%%%%%%%%%%%%%%%%%%%%%%%%%%%%%%%%%%%%%%%%%%%%%%%%%%%%%%%%%%%%%%%%%%
The fractional error of $C_l$ is thus proportional to 
$\sqrt{1/(2l+1)}$, as expected from our having $2l+1$ independent samples 
to average for a given $l$.
The exact form follows from $C_l$ being $\chi^2$ distribution
with $2l+1$ degrees of freedom when $a_{lm}$ is Gaussian.
If $a_{lm}$ is Gaussian, then its probability density distribution is
%%%%%%%%%%%%%%%%%%%%%%%%%%%%%%%%%%%%%%%%%%%%%%%%%%%%%%%%%%%%%%%%%%%
\begin{equation}
 \label{eq:probalm}
  P\left(a_{lm}\right)
  =
  \frac{\exp\left[-{a_{lm}^2}/(2\left<C_l\right>)\right]}
  {\sqrt{2\pi \left<C_l\right>}}.
\end{equation}
%%%%%%%%%%%%%%%%%%%%%%%%%%%%%%%%%%%%%%%%%%%%%%%%%%%%%%%%%%%%%%%%%%%
We use this distribution to generate Gaussian random realizations
of $a_{lm}$ for a given $\left<C_l\right>$.
First, we calculate $\left<C_l\right>$ with the {\sf CMBFAST} code \citep{SZ96}
for a set of cosmological parameters.
We then generate a realization of $a_{lm}$,
$a_{lm}= \epsilon\left<C_l\right>^{1/2}$, where $\epsilon$ is a
Gaussian random variable with the unit variance.

%%%%%%%%%%%%%%%%%%%%%%%%%%%%%%%%%%%%%%%%%%%%%%%%%%%%%%%%%%%%%%%%%%
\section{Angular Bispectrum}\label{sec:bispectrum}

\begin{figure}
 \begin{center}
  \leavevmode\epsfxsize=5cm \epsfbox{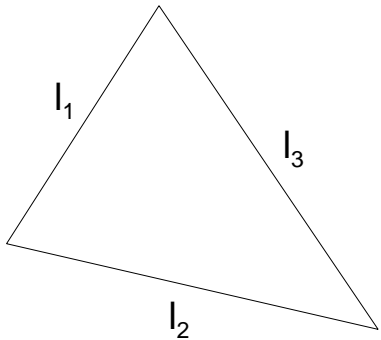}
 \end{center}
 \caption
 {Angular Bispectrum Configuration}
\label{fig:triangle}
\end{figure}

The angular bispectrum consists of three harmonic transforms, 
$a_{l_1m_1}a_{l_2m_2}a_{l_3m_3}$.
For Gaussian $a_{lm}$, the expectation value is exactly zero.
By imposing statistical isotropy upon the angular three-point
correlation function, one finds that the angular averaged bispectrum, 
$B_{l_1l_2l_3}$, given by
%%%%%%%%%%%%%%%%%%%%%%%%%%%%%%%%%%%%%%%%%%%%%%%%%%%%%%%%%%%%%%%%%%
\begin{equation}
  \label{eq:blll*}
  \left<a_{l_1m_1}a_{l_2m_2}a_{l_3m_3}\right>
  =
  \left<B_{l_1l_2l_3}\right>
  \left(\begin{array}{ccc}l_1&l_2&l_3\\m_1&m_2&m_3\end{array}\right)
\end{equation}
%%%%%%%%%%%%%%%%%%%%%%%%%%%%%%%%%%%%%%%%%%%%%%%%%%%%%%%%%%%%%%%%%%
satisfies rotational invariance (Eq.(\ref{eq:condition})).
Here, the matrix denotes the Wigner-$3j$ symbol 
(see appendix~\ref{app:wigner}).
Since $l_1$, $l_2$, and $l_3$ form a triangle, $B_{l_1l_2l_3}$ satisfies the 
triangle condition, $\left|l_i-l_j\right|\leq l_k \leq l_i+l_j$ 
for all permutations of indices. Parity invariance of 
the angular correlation function demands $l_1+l_2+l_3={\rm even}$.
Figure~\ref{fig:triangle} sketches a configuration of the angular bispectrum.

%%%%%%%%%%%%%%%%%%%%%%%%%%%%%%%%%%%%%%%%%%%%%%%%%%%%%%%%%%%%%%%%%%%%%%
%%%%%%%%%%%%%%%%%%%%%%%%%%%%%%%%%%%%%%%%%%%%%%%%%%%%%%%%%%%%%%%%%%%%%%

The Wigner-$3j$ symbol, which describes coupling of two angular momenta,
represents the azimuthal angle dependence of the angular bispectrum, 
for the bispectrum forms a triangle.
Suppose that two ``states'' with $(l_1,m_1)$ and $(l_2,m_2)$ angular momenta 
form a coupled state with $(l_3,m_3)$. 
They form a triangle whose orientation is represented by 
$m_1$, $m_2$, and $m_3$, with satisfying $m_1+m_2+m_3=0$. 
As we rotate the system, the Wigner-$3j$ symbol transforms $m$'s, 
yet preserving the configuration of the triangle.
Similarly, rotational invariance of the angular bispectrum demands
that the same triangle configuration give the same amplitude of
the bispectrum regardless of its orientation, and thus 
the Wigner-$3j$ symbol describes the azimuthal angle dependence.

The proof of $\left<B_{l_1l_2l_3}\right>$ to be rotationally invariant 
is as follows.
Substituting equation~(\ref{eq:blll*}) for the statistical isotropy
condition (Eq.(\ref{eq:condition})) for $n=3$, we obtain 
%%%%%%%%%%%%%%%%%%%%%%%%%%%%%%%%%%%%%%%%%%%%%%%%%%%%%%%%%%%%%%%%%%%
\begin{eqnarray}
 \nonumber
  & &
 \left<a_{l_1m_1}a_{l_2m_2}a_{l_3m_3}\right>\\
 \nonumber
  &=& 
  \sum_{{\rm all}~m'} \left<a_{l_1m_1'}a_{l_2m_2'}a_{l_3m_3'}\right>
  D_{m_1'm_1}^{(l_1)}D_{m_2'm_2}^{(l_2)}D_{m_3'm_3}^{(l_3)}\\
 \nonumber
 &=&
  \left<B_{l_1l_2l_3}\right>
 \sum_{{\rm all}~m'}
 \left(\begin{array}{ccc}l_1&l_2&l_3\\m_1'&m_2'&m_3'\end{array}\right)  \\
 \nonumber
 & &\times
 \sum_{LMM'}
 \left(\begin{array}{ccc}l_1&l_2&L\\m_1'&m_2'&M'\end{array}\right)
 \left(\begin{array}{ccc}l_1&l_2&L\\m_1&m_2&M\end{array}\right)
 (2L+1)
 D_{M'M}^{(L)*} D_{m_3'm_3}^{(l_3)}\\
 \nonumber
 &=&
  \left<B_{l_1l_2l_3}\right>
 \sum_{m_3'} \sum_{LMM'} \delta_{l_3L} \delta_{m_3'M'}
 \left(\begin{array}{ccc}l_1&l_2&L\\m_1&m_2&M\end{array}\right)
 D_{M'M}^{(L)*} D_{m_3'm_3}^{(l_3)}\\
 &=&
  \left<B_{l_1l_2l_3}\right>
 \left(\begin{array}{ccc}l_1&l_2&l_3\\m_1&m_2&m_3\end{array}\right).
\end{eqnarray}
%%%%%%%%%%%%%%%%%%%%%%%%%%%%%%%%%%%%%%%%%%%%%%%%%%%%%%%%%%%%%%%%%%%
In the second equality, we have reduced 
$D_{m_1'm_1}^{(l_1)}D_{m_2'm_2}^{(l_2)}$ to $D_{M'M}^{(L)*}$,
using equation~(\ref{eq:rotreduce}).
In the third equality, we have used the identity \citep{RBMW59},
%%%%%%%%%%%%%%%%%%%%%%%%%%%%%%%%%%%%%%%%%%%%%%%%%%%%%%%%%%%%%%%%%%%
\begin{equation}
 \label{eq:orthonormal}
 \sum_{m'_1m'_2}
  \left(\begin{array}{ccc}l_1&l_2&l_3\\m'_1&m'_2&m'_3\end{array}\right)
  \left(\begin{array}{ccc}l_1&l_2&L\\m'_1&m'_2&M'\end{array}\right)
  =
   \frac{\delta_{l_3L}\delta_{m'_3M'}}{2L+1}.
\end{equation}
%%%%%%%%%%%%%%%%%%%%%%%%%%%%%%%%%%%%%%%%%%%%%%%%%%%%%%%%%%%%%%%%%%%

\subsection{Estimator}

To obtain the unbiased estimator of the angular averaged bispectrum,
$B_{l_1l_2l_3}$, we invert equation~(\ref{eq:blll*}) with the 
identity~(\ref{eq:orthonormal}), and obtain 
%%%%%%%%%%%%%%%%%%%%%%%%%%%%%%%%%%%%%%%%%%%%%%%%%%%%%%%%%%%%%%%%%%
\begin{equation}
  \label{eq:best}
  B_{l_1l_2l_3}= \sum_{{\rm all}~m}
  \left(
  \begin{array}{ccc}
  l_1&l_2&l_3\\
  m_1&m_2&m_3
  \end{array}
  \right)
  a_{l_1m_1}a_{l_2m_2}a_{l_3m_3}.
\end{equation}
%%%%%%%%%%%%%%%%%%%%%%%%%%%%%%%%%%%%%%%%%%%%%%%%%%%%%%%%%%%%%%%%%%
We can rewrite this expression into a more computationally useful form.
Using the azimuthally averaged harmonic transform, 
$e_{l}(\hat{\mathbf n})$ (Eq.(\ref{eq:el})), and the identity \citep{RBMW59},
%%%%%%%%%%%%%%%%%%%%%%%%%%%%%%%%%%%%%%%%%%%%%%%%%%%%%%%%%%%%%%%%%%
\begin{eqnarray}
 \nonumber
  \left(\begin{array}{ccc}l_1&l_2&l_3\\m_1&m_2&m_3\end{array}\right)
  &=&
  \left(\begin{array}{ccc}l_1&l_2&l_3\\0&0&0\end{array}\right)^{-1}
  \sqrt{
  \frac{(4\pi)^3}{\left(2l_1+1\right)\left(2l_2+1\right)\left(2l_3+1\right)}
  }\\
 \label{eq:gaunt_spec}
 & &\times
  \int \frac{d^2\hat{\mathbf n}}{4\pi}~
  Y_{l_1m_1}(\hat{\mathbf n})
  Y_{l_2m_2}(\hat{\mathbf n})
  Y_{l_3m_3}(\hat{\mathbf n}),
\end{eqnarray}
%%%%%%%%%%%%%%%%%%%%%%%%%%%%%%%%%%%%%%%%%%%%%%%%%%%%%%%%%%%%%%%%%%
we rewrite equation~(\ref{eq:best}) as
%%%%%%%%%%%%%%%%%%%%%%%%%%%%%%%%%%%%%%%%%%%%%%%%%%%%%%%%%%%%%%%%%%
\begin{equation}
%  \label{eq:bobs}
  B_{l_1l_2l_3}=
  \left(\begin{array}{ccc}l_1&l_2&l_3\\0&0&0\end{array}\right)^{-1}
  \int \frac{d^2\hat{\mathbf n}}{4\pi}~
  e_{l_1}(\hat{\mathbf n})
  e_{l_2}(\hat{\mathbf n})
  e_{l_3}(\hat{\mathbf n}).
\end{equation}
%%%%%%%%%%%%%%%%%%%%%%%%%%%%%%%%%%%%%%%%%%%%%%%%%%%%%%%%%%%%%%%%%%
This expression is computationally efficient; we can quickly calculate 
$e_l(\hat{\mathbf n})$ with the spherical harmonic transform. 
Then, the average over the full sky, $\int d^2\hat{\mathbf n}/(4\pi)$,
is done by the sum over all pixels divided by the total number 
of pixels, $N^{-1}\sum_i^{N}$, if all the pixels have the equal area. 
Note that the integral over $\hat{\mathbf n}$ must be done on the 
full sky even when a sky-cut is applied, as 
$e_{l}(\hat{\mathbf n})$ already encapsulates information of 
partial sky coverage through $a_{lm}$, which may be measured 
on the incomplete sky.

\subsection{Covariance matrix}

We calculate the covariance matrix of $B_{l_1l_2l_3}$, provided that
non-Gaussianity is weak, $\left<B_{l_1l_2l_3}\right>\approx 0$.
Since the covariance matrix is a product of six $a_{lm}$'s,
we have $_6C_2\cdot{}_4C_2/3!= 15$ terms to evaluate, 
according to the Wick's theorem; however, using the identity
\citep{RBMW59},
%%%%%%%%%%%%%%%%%%%%%%%%%%%%%%%%%%%%%%%%%%%%%%%%%%%%%%%%%%%%%%%%%%
\begin{equation}
 \label{eq:helpidentity}
 (-1)^m 
 \left(\begin{array}{ccc}l&l&l'\\m&-m&0\end{array}\right)
 = \frac{(-1)^l}{\sqrt{2l+1}}\delta_{l'0},
\end{equation}
%%%%%%%%%%%%%%%%%%%%%%%%%%%%%%%%%%%%%%%%%%%%%%%%%%%%%%%%%%%%%%%%%%
and assuming none of $l$'s zero, we find only $3!=6$ terms that do not 
include $\left<a_{l_im_i}a_{l_jm_j}\right>$ but 
include only $\left<a_{l_im_i}a_{l_jm_j}^*\right>$ non-vanishing.
Evaluating these 6 terms, we obtain \citep{Luo94,heavens98,SG99,GM00}
%%%%%%%%%%%%%%%%%%%%%%%%%%%%%%%%%%%%%%%%%%%%%%%%%%%%%%%%%%%%%%%%%%
\begin{eqnarray}
 \nonumber
  & &\left<B_{l_1l_2l_3}B_{l_1'l_2'l_3'}\right>\\
  \nonumber
  &=&
  \sum_{{\rm all}~mm'}
  \left(\begin{array}{ccc}l_1&l_2&l_3\\m_1&m_2&m_3\end{array}\right)
  \left(\begin{array}{ccc}l_1'&l_2'&l_3'\\m_1'&m_2'&m_3'\end{array}\right)
  \left<a_{l_1m_1}a_{l_2m_2}a_{l_3m_3}
   a_{l_1'm_1'}^*a_{l_2'm_2'}^*a_{l_3'm_3'}^*\right>\\
 \nonumber
 &=&
 \left<C_{l_1}\right>\left<C_{l_2}\right>\left<C_{l_3}\right>
 \left[
  \delta_{l_1l_2l_3}^{l_1'l_2'l_3'} + \delta_{l_1l_2l_3}^{l_3'l_1'l_2'}
  + \delta_{l_1l_2l_3}^{l_2'l_3'l_1'} 
  + (-1)^{l_1+l_2+l_3} \left(\delta_{l_1l_2l_3}^{l_1'l_3'l_2'}
  + \delta_{l_1l_2l_3}^{l_2'l_1'l_3'} + \delta_{l_1l_2l_3}^{l_3'l_2'l_1'}
  \right)\right],\\
 \label{eq:error3}
\end{eqnarray}
%%%%%%%%%%%%%%%%%%%%%%%%%%%%%%%%%%%%%%%%%%%%%%%%%%%%%%%%%%%%%%%%%%
where $\delta_{l_1l_2l_3}^{l_1'l_2'l_3'}\equiv
\delta_{l_1l_1'}\delta_{l_2l_2'}\delta_{l_3l_3'}$, and so on.
Hence, the covariance matrix is diagonal in the weak 
non-Gaussian limit.
The diagonal terms for $l_i\neq 0$ and $l_1+l_2+l_3={\rm even}$ are
%%%%%%%%%%%%%%%%%%%%%%%%%%%%%%%%%%%%%%%%%%%%%%%%%%%%%%%%%%%%%%%%%%
\begin{equation}
 \label{eq:variance3}
  \left<B_{l_1l_2l_3}^2\right>
  =
  \left<C_{l_1}\right>\left<C_{l_2}\right>\left<C_{l_3}\right>
  \left(1+ 2\delta_{l_1l_2}\delta_{l_2l_3} +
   \delta_{l_1l_2}+\delta_{l_2l_3}+\delta_{l_3l_1}\right).
\end{equation}
%%%%%%%%%%%%%%%%%%%%%%%%%%%%%%%%%%%%%%%%%%%%%%%%%%%%%%%%%%%%%%%%%%
The variance is amplified by a factor of 2 or 6, when two or all 
$l$'s are same, respectively.

\begin{figure}
 \plotone{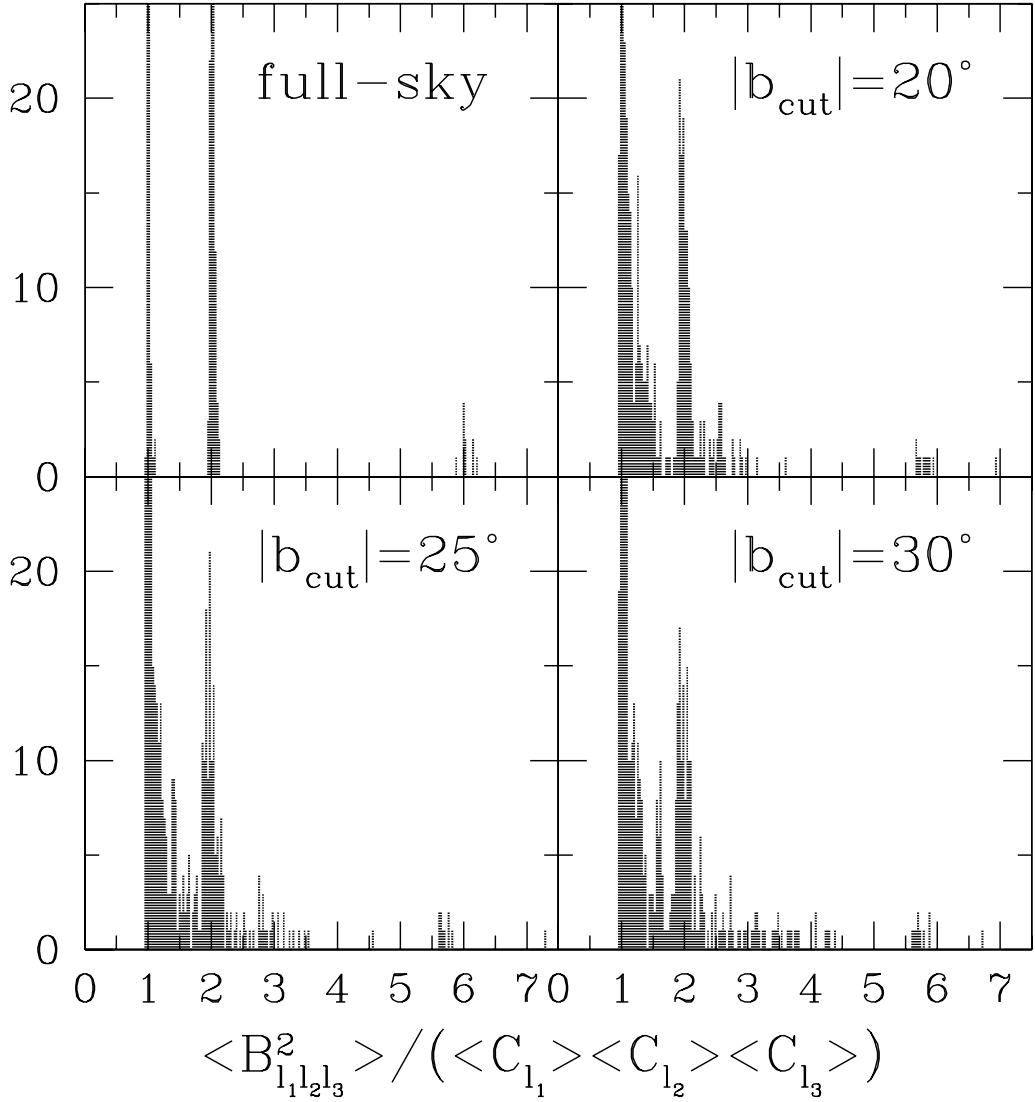}
 \caption
 {Variance of Angular Bispectrum} 
 \mycaption{Histograms of variance of the angular bispectrum
 for $l_1\leq l_2\leq l_3$  up to a maximum multipole of 20.
 There are 466 modes.
 These are derived from simulated realizations of a Gaussian sky.
 The top-left panel shows the case of full sky coverage, while
 the rest of panels show the cases of incomplete sky coverage.
 The top-right, bottom-left, and bottom-right panels use
 the $20^\circ$, $25^\circ$, and $30^\circ$ Galactic sky-cuts, respectively.}
\label{fig:variance_bare}
\end{figure}

We find that equation~(\ref{eq:variance3}) becomes not exact on 
the incomplete sky, where the variance distribution becomes more scattered.
Using simulated realizations of a Gaussian sky, we have measured
the variance on the full sky as well as on the incomplete sky for 
three different Galactic sky-cuts, $20^\circ$, $25^\circ$, and $30^\circ$.
Figure~\ref{fig:variance_bare} plots the results; 
we find that equation~(\ref{eq:variance3}) holds only approximately 
on the incomplete sky.

%%%%%%%%%%%%%%%%%%%%%%%%%%%%%%%%%%%%%%%%%%%%%%%%%%%%%%%%%%%%%%%%%%%%%%
%%%%%%%%%%%%%%%%%%%%%%%%%%%%%%%%%%%%%%%%%%%%%%%%%%%%%%%%%%%%%%%%%%%%%%

%%%%%%%%%%%%%%%%%%%%%%%%%%%%%%%%%%%%%%%%%%%%%%%%%%%%%%%%%%%%%%%%%%%
\section{Angular Trispectrum}\label{sec:trispectrum}

\begin{figure}
 \begin{center}
  \leavevmode\epsfxsize=7cm \epsfbox{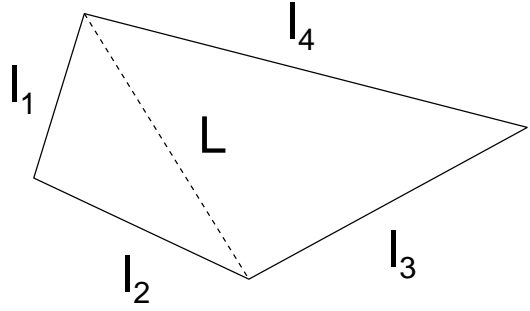}
 \end{center}
 \caption
 {Angular Trispectrum Configuration}
\label{fig:quad}
\end{figure}

The angular trispectrum consists of four harmonic transforms, 
$a_{l_1m_1}a_{l_2m_2}a_{l_3m_3}a_{l_4m_4}$.
\citet{Hu01} finds a rotationally invariant solution for
the angular trispectrum as
%%%%%%%%%%%%%%%%%%%%%%%%%%%%%%%%%%%%%%%%%%%%%%%%%%%%%%%%%%%%%%%%%%%
\begin{equation}
 \label{eq:tl}
 \left<a_{l_1m_1}a_{l_2m_2}a_{l_3m_3}a_{l_4m_4}\right>
  =
  \sum_{LM}
  \left(\begin{array}{ccc}l_1&l_2&L\\m_1&m_2&-M\end{array}\right)
  \left(\begin{array}{ccc}l_3&l_4&L\\m_3&m_4&M\end{array}\right)
  (-1)^M \left<T^{l_1l_2}_{l_3l_4}(L)\right>.
\end{equation}
%%%%%%%%%%%%%%%%%%%%%%%%%%%%%%%%%%%%%%%%%%%%%%%%%%%%%%%%%%%%%%%%%%%
One can prove this solution, $\left<T^{l_1l_2}_{l_3l_4}(L)\right>$, 
rotationally invariant by similar calculations to those proving the 
angular bispectrum to be so.
By construction, $l_1$, $l_2$, and $L$ form one triangle,
while $l_3$, $l_4$, and $L$ form the other triangle in a quadrilateral
with sides of $l_1$, $l_2$, $l_3$, and $l_4$.
$L$ represents a diagonal of the quadrilateral.
Figure~\ref{fig:quad} sketches a configuration of the angular 
trispectrum.
When we arrange $l_1$, $l_2$, $l_3$, and $l_4$ in order of 
$l_1\le l_2\le l_3\le l_4$, $L$ lies in 
$\max(l_2-l_1,l_4-l_3)\le L\le \min(l_1+l_2,l_3+l_4)$.
Parity invariance of the angular four-point correlation function 
demands $l_1+l_2+L=\mbox{even}$ and $l_3+l_4+L=\mbox{even}$.

%%%%%%%%%%%%%%%%%%%%%%%%%%%%%%%%%%%%%%%%%%%%%%%%%%%%%%%%%%%%%%%%%%%%%%
%%%%%%%%%%%%%%%%%%%%%%%%%%%%%%%%%%%%%%%%%%%%%%%%%%%%%%%%%%%%%%%%%%%%%%

The angular trispectrum generically consists of two parts.
One is the unconnected part, the contribution from Gaussian fields,
which is given by the angular power spectra \citep{Hu01},
%%%%%%%%%%%%%%%%%%%%%%%%%%%%%%%%%%%%%%%%%%%%%%%%%%%%%%%%%%%%%%%%%%
\begin{eqnarray}
 \nonumber
  & &
  \left<T^{l_1l_2}_{l_3l_4}(L)\right>_{\rm unconnected}\\
 \nonumber
  &=&
  (-1)^{l_1+l_3}\sqrt{(2l_1+1)(2l_3+1)}\left<C_{l_1}\right>\left<C_{l_3}\right>
  \delta_{l_1l_2}\delta_{l_3l_4}\delta_{L0} \\
 \label{eq:unconnected*}
  & & 
  + (2L+1)\left<C_{l_1}\right>\left<C_{l_2}\right>\left[(-1)^{l_2+l_3+L}
		       \delta_{l_1l_3}\delta_{l_2l_4}
		     +\delta_{l_1l_4}\delta_{l_2l_3}\right].
\end{eqnarray}
%%%%%%%%%%%%%%%%%%%%%%%%%%%%%%%%%%%%%%%%%%%%%%%%%%%%%%%%%%%%%%%%%%
For $l_1\le l_2\le l_3\le l_4$, the unconnected terms are 
non-zero only when $L=0$ or $l_1=l_2=l_3=l_4$.
We have numerically confirmed that our estimator given below 
(Eq.(\ref{eq:tobs*})) accurately reproduces the unconnected terms
(Eq.(\ref{eq:unconnected*})) on a simulated Gaussian sky.

The other is the connected part whose expectation value is 
exactly zero for Gaussian fields; thus, the connected part is sensitive 
to non-Gaussianity.
When none of $l$'s are same in $T^{l_1l_2}_{l_3l_4}(L)$,
one might expect the trispectrum to comprise the connected part only;
however, it is true only on the full sky.
The unconnected terms on the incomplete sky, which are often much bigger 
than the connected terms, leak the power to the other modes for which 
all $l$'s are different.
We should take this effect into account in the analysis.

\subsection{Estimator}

Inverting equation~(\ref{eq:tl}), we obtain the unbiased estimator of 
$T^{l_1l_2}_{l_3l_4}(L)$ \citep{Hu01},
%%%%%%%%%%%%%%%%%%%%%%%%%%%%%%%%%%%%%%%%%%%%%%%%%%%%%%%%%%%%%%%%%%%%%%
\begin{eqnarray}
 \nonumber
  T^{l_1l_2}_{l_3l_4}(L)
  &=& (2L+1)\sum_{{\rm all}~m}\sum_M(-1)^M
  \left(\begin{array}{ccc}l_1&l_2&L\\m_1&m_2&M\end{array}\right)
  \left(\begin{array}{ccc}l_3&l_4&L\\m_3&m_4&-M\end{array}\right)\\
 \label{eq:test}
 & &\times
  a_{l_1m_1}a_{l_2m_2}a_{l_3m_3}a_{l_4m_4}.
\end{eqnarray}
%%%%%%%%%%%%%%%%%%%%%%%%%%%%%%%%%%%%%%%%%%%%%%%%%%%%%%%%%%%%%%%%%%%%%%
Note that this expression includes both the connected and the
unconnected terms.

We find that this estimator has a special property for $L=0$
(which demands $l_1=l_2$ and $l_3=l_4$).
The trispectrum estimator for these configurations,
$T^{l_1l_1}_{l_3l_3}(0)$, reduces to a product of two power spectrum 
estimators, $C_{l_1}C_{l_3}$,
%%%%%%%%%%%%%%%%%%%%%%%%%%%%%%%%%%%%%%%%%%%%%%%%%%%%%%%%%%%%%%%%%%%%%%
\begin{eqnarray}
 \nonumber
  T^{l_1l_1}_{l_3l_3}(0)
  &=&
  \sum_{m_1m_3}
  \left(\begin{array}{ccc}l_1&l_1&0\\m_1&-m_1&0\end{array}\right)
  \left(\begin{array}{ccc}l_3&l_3&0\\m_3&-m_3&0\end{array}\right)
  a_{l_1m_1}a_{l_1-m_1}a_{l_3m_3}a_{l_3-m_3}\\
  &=& 
   \label{eq:special}
   (-1)^{l_1+l_3}\sqrt{(2l_1+1)(2l_3+1)}C_{l_1}C_{l_3},
\end{eqnarray}
%%%%%%%%%%%%%%%%%%%%%%%%%%%%%%%%%%%%%%%%%%%%%%%%%%%%%%%%%%%%%%%%%%
where $C_l=(2l+1)^{-1}\sum_ma_{lm}a_{lm}^*$.
We have used the identity, equation~(\ref{eq:helpidentity}),
and $a_{l-m}=(-1)^ma_{lm}^*$ in the second equality.
From this equation, one may assume that $T^{l_1l_1}_{l_3l_3}(0)$ 
coincides with the unconnected
terms for $L=0$ (see Eq.(\ref{eq:unconnected*})),
%%%%%%%%%%%%%%%%%%%%%%%%%%%%%%%%%%%%%%%%%%%%%%%%%%%%%%%%%%%%%%%%%%
\begin{equation}
 \label{eq:unco}
  \left<T^{l_1l_1}_{l_3l_3}(0)\right>_{\rm unconnected}
  =
  (-1)^{l_1+l_3}\sqrt{(2l_1+1)(2l_3+1)}\left<C_{l_1}\right>\left<C_{l_3}\right>
  + 2\left<C_{l_1}\right>^2 \delta_{l_1l_3}.
\end{equation}
%%%%%%%%%%%%%%%%%%%%%%%%%%%%%%%%%%%%%%%%%%%%%%%%%%%%%%%%%%%%%%%%%%
They are, however, different for non-Gaussian fields, 
because of the power spectrum covariance, equation~(\ref{eq:error2*}).
By taking the ensemble average of $T^{l_1l_1}_{l_3l_3}(0)$, and 
substituting equation~(\ref{eq:error2*}) for
$\left<C_{l_1}C_{l_3}\right>$, we find a rather trivial result:
%%%%%%%%%%%%%%%%%%%%%%%%%%%%%%%%%%%%%%%%%%%%%%%%%%%%%%%%%%%%%%%%%%
\begin{eqnarray}
 \nonumber
 \left<T^{l_1l_1}_{l_3l_3}(0)\right>
  &=&
  (-1)^{l_1+l_3}\sqrt{(2l_1+1)(2l_3+1)}\left<C_{l_1}C_{l_3}\right>\\
 &=&
  \nonumber
  (-1)^{l_1+l_3}\sqrt{(2l_1+1)(2l_3+1)}\left<C_{l_1}\right>\left<C_{l_3}\right>
  + 2\left<C_{l_1}\right>^2 \delta_{l_1l_3}
      +
      \left<T^{l_1l_1}_{l_3l_3}(0)\right>_{\rm c}\\   
  &=&
  \left<T^{l_1l_1}_{l_3l_3}(0)\right>_{\rm unconnected}
  +
  \left<T^{l_1l_1}_{l_3l_3}(0)\right>_{\rm c}.
\end{eqnarray}
%%%%%%%%%%%%%%%%%%%%%%%%%%%%%%%%%%%%%%%%%%%%%%%%%%%%%%%%%%%%%%%%%%
Hence, $T^{l_1l_1}_{l_3l_3}(0)$ contains information not only 
of the unconnected trispectrum, but also of the connected trispectrum.

Unfortunately, we cannot measure the connected part of 
$T^{l_1l_1}_{l_3l_3}(0)$ directly
from the angular trispectrum because of the following reason.
To measure the connected terms, we have to subtract the
unconnected terms from the measured trispectrum first.
Since we are never able to measure the ensemble average of the 
unconnected terms (Eq.(\ref{eq:unco})), 
we estimate them by using estimated power spectrum, $C_l$.
If we subtract the estimated unconnected terms,
$\propto C_{l_1}C_{l_3}$, from measured $T^{l_1l_1}_{l_3l_3}(0)$,
then it follows from equation~(\ref{eq:special}) that 
$T^{l_1l_1}_{l_3l_3}(0)$ vanishes {\it exactly}: 
$T^{l_1l_1}_{l_3l_3}(0)=0$; thus, 
$T^{l_1l_1}_{l_3l_3}(0)$ has no statistical power of measuring the 
connected terms.

For practical measurement of the angular trispectrum, we rewrite
the trispectrum estimator given by equation~(\ref{eq:test}) with
the azimuthally averaged harmonic transform, 
$e_{l}(\hat{\mathbf n})$ (Eq.(\ref{eq:el})).
We find that the following form is particularly computationally efficient:
%%%%%%%%%%%%%%%%%%%%%%%%%%%%%%%%%%%%%%%%%%%%%%%%%%%%%%%%%%%%%%%%%%
\begin{equation}
  \label{eq:tobs*}
   T^{l_1l_2}_{l_3l_4}(L)
   =
   \frac1{2L+1} \sum_{M=-L}^{L} t_{LM}^{l_1l_2*} t_{LM}^{l_3l_4},
\end{equation}
%%%%%%%%%%%%%%%%%%%%%%%%%%%%%%%%%%%%%%%%%%%%%%%%%%%%%%%%%%%%%%%%%%
where $t_{LM}^{l_1l_2}$ is given by
%%%%%%%%%%%%%%%%%%%%%%%%%%%%%%%%%%%%%%%%%%%%%%%%%%%%%%%%%%%%%%%%%%
\begin{equation}
 \label{eq:tLM}
  t_{LM}^{l_1l_2}
  \equiv
  \sqrt{\frac{2L+1}{4\pi}}  
  \left(
   \begin{array}{ccc}
    l_1 & l_2 & L \\ 0 & 0 & 0 
   \end{array}
 \right)^{-1} 
  \int d^2\hat{\mathbf n}
  \left[e_{l_1}(\hat{\mathbf n}) e_{l_2}(\hat{\mathbf n})\right]
  Y_{LM}^*(\hat{\mathbf n}).
\end{equation}
%%%%%%%%%%%%%%%%%%%%%%%%%%%%%%%%%%%%%%%%%%%%%%%%%%%%%%%%%%%%%%%%%%
Since $t_{LM}^{l_1l_2}$ is the harmonic transform on the full sky, 
we can calculate it quickly. 
This method makes measurement of the angular trispectrum computationally
feasible even for the {\it MAP} data in which we have more than millions of
pixels; thus, the methods developed here can be applied not only to the 
{\it COBE} DMR data, but also to the {\it MAP} data.

\subsection{Covariance matrix}

We calculate the covariance of the trispectrum in the weakly non-Gaussian
limit.
Since the trispectrum covariance comprises eight $a_{lm}$'s,
the total number of terms is $_8C_2\cdot{}_6C_2\cdot{}_4C_2/4!=105$
according to the Wick's theorem.
The full calculation will be a nightmare, for we have to deal with
%%%%%%%%%%%%%%%%%%%%%%%%%%%%%%%%%%%%%%%%%%%%%%%%%%%%%%%%%%%%%%%%%%
\begin{eqnarray}
 \nonumber
 & &\left<T^{l_1l_2}_{l_3l_4}(L)T^{l_1'l_2'}_{l_3'l_4'}(L')\right>\\
 \nonumber
 &=& (2L+1)(2L'+1)\sum_{{\rm all}~mm'}\sum_{MM'}(-1)^{M+M'}\\
 \nonumber
  & &\times
  \left(\begin{array}{ccc}l_1&l_2&L\\m_1&m_2&M\end{array}\right)
  \left(\begin{array}{ccc}l_3&l_4&L\\m_3&m_4&-M\end{array}\right)
  \left(\begin{array}{ccc}l_1'&l_2'&L'\\m_1'&m_2'&M'\end{array}\right)
  \left(\begin{array}{ccc}l_3'&l_4'&L'\\m_3'&m_4'&-M'\end{array}\right)\\
 \label{eq:error4}
  & &\times \left<
   a_{l_1m_1}a_{l_2m_2}a_{l_3m_3}a_{l_4m_4}
   a^*_{l_1'm_1'}a^*_{l_2'm_2'}a^*_{l_3'm_3'}a^*_{l_4'm_4'}
   \right>.
\end{eqnarray}
%%%%%%%%%%%%%%%%%%%%%%%%%%%%%%%%%%%%%%%%%%%%%%%%%%%%%%%%%%%%%%%%%%
We reduce this intricate expression to much more simplified
forms for some particular configurations. 
For $L,L'\neq 0$ terms, thanks to the identity 
(\ref{eq:helpidentity}), only $4!=24$ terms that do not include 
$\left<a_{l_im_i}a_{l_jm_j}\right>$ but include only 
$\left<a_{l_im_i}a_{l_jm_j}^*\right>$ are non-vanishing.
For $L=L'=0$ terms, the triangle conditions in a quadrilateral
demand $l_1=l_2$ and $l_3=l_4$ (see figure~\ref{fig:quad}).
As we have shown, these configurations have no statistical power of 
measuring the connected trispectrum of interest.
Hence, we evaluate $L,L'\neq 0$ terms in the following.

\begin{figure}
 \plotone{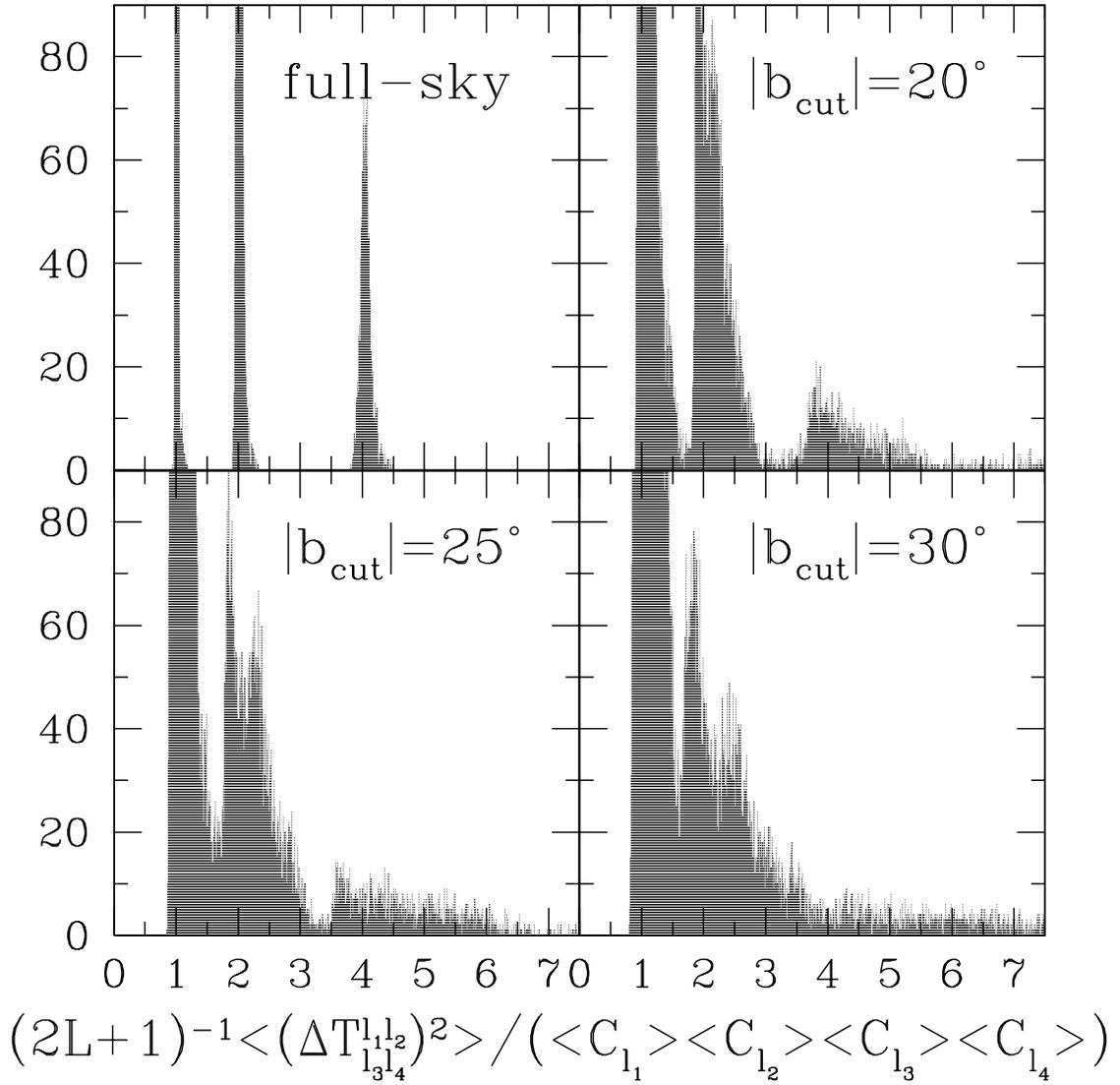}
 \caption
 {Variance of Angular Trispectrum I} 
 \mycaption{Histograms of variance of the angular trispectrum for
 $L\neq 0$ and $l_1\leq l_2<l_3\leq l_4$, for which the unconnected 
 terms vanish on the full sky.
 There are 16,554 modes, up to a maximum multipole of 20. 
 The meaning of the panels is the same as in figure~\ref{fig:variance_bare}.}
\label{fig:variance_tl_bare_c1}
\end{figure}

Evaluating 24 $L,L'\neq 0$ terms is still a headache; however, for  
$l_1\leq l_2 < l_3\leq l_4$, we have only 8 terms left:
%%%%%%%%%%%%%%%%%%%%%%%%%%%%%%%%%%%%%%%%%%%%%%%%%%%%%%%%%%%%%%%%%%
\begin{eqnarray}
 \nonumber
 & &\frac{\left<T^{l_1l_2}_{l_3l_4}(L)T^{l_1'l_2'}_{l_3'l_4'}(L')\right>}
 {(2L+1)\left<C_{l_1}\right>\left<C_{l_2}\right>\left<C_{l_3}\right>
  \left<C_{l_4}\right>}\\
 &=&
  \nonumber
  \delta_{LL'}
  \left[\delta^{l_1l_2l_3l_4}_{l_1'l_2'l_3'l_4'}
       +\delta^{l_1l_2l_3l_4}_{l_3'l_4'l_1'l_2'}
       + (-1)^{l_1+l_2+l_3+l_4}\left(
				\delta^{l_1l_2l_3l_4}_{l_2'l_1'l_4'l_3'}
				+\delta^{l_1l_2l_3l_4}_{l_4'l_3'l_2'l_1'}
                               \right)\right. \\
 \label{eq:error4*}
   & &\left. + (-1)^{l_1+l_2+L}\left(
		      \delta^{l_1l_2l_3l_4}_{l_2'l_1'l_3'l_4'}
		      +\delta^{l_1l_2l_3l_4}_{l_4'l_3'l_1'l_2'}
		    \right)
   + (-1)^{l_3+l_4+L}\left(
		    \delta^{l_1l_2l_3l_4}_{l_1'l_2'l_4'l_3'}
		    +\delta^{l_1l_2l_3l_4}_{l_3'l_4'l_2'l_1'}
		    \right)\right],
\end{eqnarray}
%%%%%%%%%%%%%%%%%%%%%%%%%%%%%%%%%%%%%%%%%%%%%%%%%%%%%%%%%%%%%%%%%
where $\delta_{l_1l_2l_3l_4}^{l_1'l_2'l_3'l_4'}\equiv
\delta_{l_1l_1'}\delta_{l_2l_2'}\delta_{l_3l_3'}\delta_{l_4l_4'}$, and so on.
Using parity invariance, $l_1+l_2+L={\rm even}$ and 
$l_3+l_4+L={\rm even}$, we find the covariance matrix diagonal. 
Thus, the diagonal terms for $L\neq 0$ and $l_1\leq l_2<l_3\leq l_4$ are
simplified very much as
%%%%%%%%%%%%%%%%%%%%%%%%%%%%%%%%%%%%%%%%%%%%%%%%%%%%%%%%%%%%%%%%%
\begin{equation}
 \label{eq:variance4}
  {\left<\left[T^{l_1l_2}_{l_3l_4}(L)\right]^2\right>} 
  = {(2L+1)\left<C_{l_1}\right>\left<C_{l_2}\right>\left<C_{l_3}\right>
  \left<C_{l_4}\right>}
  \left(
   1 + \delta_{l_1l_2} + \delta_{l_3l_4} + \delta_{l_1l_2}\delta_{l_3l_4}
   \right).
\end{equation}
%%%%%%%%%%%%%%%%%%%%%%%%%%%%%%%%%%%%%%%%%%%%%%%%%%%%%%%%%%%%%%%%%% 
This result is strictly correct only on the full sky;
the incomplete sky makes the variance distribution much more scattered.
Figure~\ref{fig:variance_tl_bare_c1} plots the variance on the full sky 
as well as on the incomplete sky.

%%%%%%%%%%%%%%%%%%%%%%%%%%%%%%%%%%%%%%%%%%%%%%%%%%%%%%%%%%%%%%%%%%%%%%
%%%%%%%%%%%%%%%%%%%%%%%%%%%%%%%%%%%%%%%%%%%%%%%%%%%%%%%%%%%%%%%%%%%%%%

\begin{figure}
 \plotone{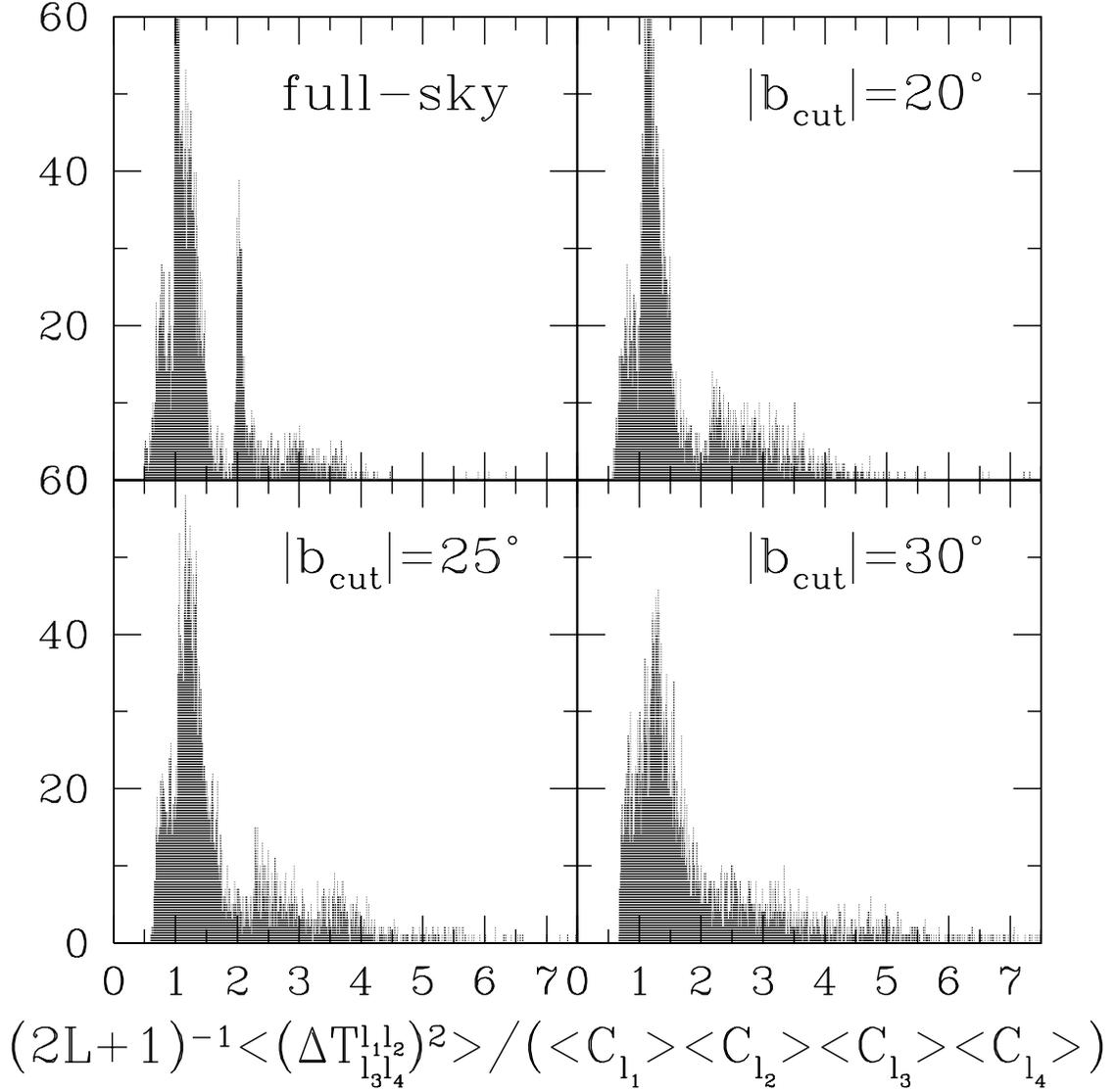}
 \caption
 {Variance of Angular Trispectrum II} 
 \mycaption{Histograms of variance of the angular trispectrum
 for $L\neq 0$, $l_2=l_3$, and $l_1\neq l_4$,
 for which the unconnected terms vanish on the full sky.
 There are 4,059 modes, up to a maximum multipole of 20.
 The meaning of the panels is the same as in figure~\ref{fig:variance_bare}.}
\label{fig:variance_tl_bare_c2}
\end{figure}

For the rest of configurations for which the unconnected terms vanish, 
$L\neq 0$, $l_2=l_3$, and $l_1\neq l_4$, 
the covariance matrix is no longer diagonal in $L,L'$ \citep{Hu01}.
Figure~\ref{fig:variance_tl_bare_c2} plots the numerically evaluated 
variance on the full sky as well as on the incomplete sky.
The variance divided by
${(2L+1)\left<C_{l_1}\right>\left<C_{l_2}\right>\left<C_{l_3}\right>
\left<C_{l_4}\right>}$ is no longer an integer, but
more scattered than that for $L\neq 0$ and $l_1\leq l_2<l_3\leq l_4$ 
even on the full sky.

%%%%%%%%%%%%%%%%%%%%%%%%%%%%%%%%%%%%%%%%%%%%%%%%%%%%%%%%%%%%%%%%%%%%%%
%%%%%%%%%%%%%%%%%%%%%%%%%%%%%%%%%%%%%%%%%%%%%%%%%%%%%%%%%%%%%%%%%%%%%%

%%%%%%%%%%%%%%%%%%%%%%%%%%%%%%%%%%%%%%%%%%%%%%%%%%%%%%%%%%%%%%%%%%%%%%
\section{Power Spectrum and Bispectrum on the Incomplete Sky}

Incomplete sky coverage destroys orthonormality of the 
spherical harmonics on the sky.
The degree to which orthonormality is broken is often
characterized by the coupling integral \citep{P80},
%%%%%%%%%%%%%%%%%%%%%%%%%%%%%%%%%%%%%%%%%%%%%%%%%%%%%%%%%%%%%%%%%%
\begin{equation}
 \label{eq:coupling}
  W_{ll'mm'}
  \equiv
  \int
  d^2\hat{\mathbf n}~
  W(\hat{\mathbf n})
  Y_{lm}^*\left(\hat{\mathbf n}\right)
  Y_{l'm'}\left(\hat{\mathbf n}\right)
  =
  \int_{\Omega_{\rm obs}} 
  d^2\hat{\mathbf n}~
  Y_{lm}^*\left(\hat{\mathbf n}\right)
  Y_{l'm'}\left(\hat{\mathbf n}\right),
\end{equation}
%%%%%%%%%%%%%%%%%%%%%%%%%%%%%%%%%%%%%%%%%%%%%%%%%%%%%%%%%%%%%%%%%%
where $W(\hat{\mathbf n})$ is zero in a cut region otherwise 1, and 
$\Omega_{\rm obs}$ denotes a solid angle of the observed sky.
When $W_{ll'mm'}\neq \delta_{ll'}\delta_{mm'}$,
the measured harmonic transform of the temperature 
anisotropy field, $a_{lm}$, becomes a 
{\it biased} estimator of the true harmonic transform, 
$a_{lm}^{\rm true}$, through
%%%%%%%%%%%%%%%%%%%%%%%%%%%%%%%%%%%%%%%%%%%%%%%%%%%%%%%%%%%%%%%%%%
\begin{equation}
 \label{eq:alm_obs}
  a_{lm}=
  \sum_{l'=0}^\infty\sum_{m'=-l'}^{l'}a_{l'm'}^{\rm true} W_{ll'mm'}.
\end{equation}
%%%%%%%%%%%%%%%%%%%%%%%%%%%%%%%%%%%%%%%%%%%%%%%%%%%%%%%%%%%%%%%%%%
Hence, we must correct our estimators of the power spectrum 
and the bispectrum for the bias arising from incomplete sky coverage.

First, we derive a relationship between the angular power spectrum
on the incomplete sky and that on the full sky. 
Taking the ensemble average of the estimator of the power spectrum,
the pseudo-$C_l$ \citep{WHG98,WHG00},
$C_l=(2l+1)^{-1}\sum_m\left|a_{lm}\right|^2$,
we have
%%%%%%%%%%%%%%%%%%%%%%%%%%%%%%%%%%%%%%%%%%%%%%%%%%%%%%%%%%%%%%%%%
\begin{eqnarray}
 \nonumber
 \left<C_l\right>
  &=& \frac1{2l+1}\sum_{l'}
  C_{l'}^{\rm true}
  \sum_{mm'}
  \left|W_{ll'mm'}\right|^2\\
 \nonumber
  &\approx&
  \frac1{2l+1}
  C_{l}^{\rm true}
  \sum_{m}
  \sum_{l'm'}
  \int 
  {d^2\hat{\mathbf n}}~
  W(\hat{\mathbf n})
  Y_{lm}^*\left(\hat{\mathbf n}\right)
  Y_{l'm'}\left(\hat{\mathbf n}\right)
    \int 
  {d^2\hat{\mathbf m}}~
  W(\hat{\mathbf m})
  Y_{lm}\left(\hat{\mathbf m}\right)
  Y_{l'm'}^*\left(\hat{\mathbf m}\right)\\
 \nonumber
  &=&
  \frac1{2l+1}
  C_{l}^{\rm true}
  \sum_m
  \int 
  {d^2\hat{\mathbf n}}~
  W(\hat{\mathbf n})
  Y_{lm}^*\left(\hat{\mathbf n}\right)
  \int 
  {d^2\hat{\mathbf m}}~
  W(\hat{\mathbf m})
  Y_{lm}\left(\hat{\mathbf m}\right)
  \delta^{(2)}\left(\hat{\mathbf n}-\hat{\mathbf m}\right)\\
 \nonumber
  &=&
  C_{l}^{\rm true}
  \int
  \frac{d^2\hat{\mathbf n}}{4\pi}~
  W(\hat{\mathbf n})
  P_{l}\left(1\right)\\
 &=&
  C_{l}^{\rm true}\frac{\Omega_{\rm obs}}{4\pi}.
\end{eqnarray}
%%%%%%%%%%%%%%%%%%%%%%%%%%%%%%%%%%%%%%%%%%%%%%%%%%%%%%%%%%%%%%%%%
In the second equality, we have taken $C_{l'}^{\rm true}$ out of the 
summation over $l'$,
as $\left|W_{ll'mm'}\right|^2$ peaks very sharply at $l=l'$, and
$C_{l'}^{\rm true}$ varies much more slowly than 
$\left|W_{ll'mm'}\right|^2$ in $l'$.
This approximation is good for nearly full sky coverage.
In the third equality, we have used 
$\sum_{l'm'}Y_{l'm'}\left(\hat{\mathbf n}\right)
Y_{l'm'}^*\left(\hat{\mathbf m}\right)=
\delta^{(2)}\left(\hat{\mathbf n}-\hat{\mathbf m}\right)$.
In the forth equality, we have used 
$\sum_m Y^*_{lm}\left(\hat{\mathbf n}\right)
Y_{lm}\left(\hat{\mathbf m}\right)= 
\frac{2l+1}{4\pi}P_l(\hat{\mathbf n}\cdot\hat{\mathbf m})$.
The result indicates that the bias amounts approximately to a 
fraction of the sky covered by observations.

Next, we derive a relationship between the angular bispectrum on the 
incomplete sky and that on the full sky. 
We begin with 
%%%%%%%%%%%%%%%%%%%%%%%%%%%%%%%%%%%%%%%%%%%%%%%%%%%%%%%%%%%%%%%%%
\begin{equation}
  \left<a_{l_1m_1}a_{l_2m_2}a_{l_3m_3}\right>
 = 
 \sum_{{\rm all}~l'm'}
 \left<a^{\rm true}_{l_1'm_1'}a^{\rm true}_{l_2'm_2'}
  a^{\rm true}_{l_3'm_3'}\right>
  W_{l_1l_1'm_1m_1'}W_{l_2l_2'm_2m_2'}W_{l_3l_3'm_3m_3'}.
\end{equation}
%%%%%%%%%%%%%%%%%%%%%%%%%%%%%%%%%%%%%%%%%%%%%%%%%%%%%%%%%%%%%%%%%
Rotational and parity invariance of the bispectrum 
implies the bispectrum given by
%%%%%%%%%%%%%%%%%%%%%%%%%%%%%%%%%%%%%%%%%%%%%%%%%%%%%%%%%%%%%%%%%
\begin{equation}
  \left<a_{l_1m_1}a_{l_2m_2}a_{l_3m_3}\right>
   =
   b_{l_1l_2l_3}
   \int
  d^2\hat{\mathbf n}~
  Y_{l_1m_1}^*\left(\hat{\mathbf n}\right)
  Y_{l_2m_2}^*\left(\hat{\mathbf n}\right)
  Y_{l_3m_3}^*\left(\hat{\mathbf n}\right),
\end{equation}
%%%%%%%%%%%%%%%%%%%%%%%%%%%%%%%%%%%%%%%%%%%%%%%%%%%%%%%%%%%%%%%%%
where $b_{l_1l_2l_3}$ is an arbitrary real symmetric function,
which is related to the angular averaged bispectrum, $B_{l_1l_2l_3}$.
When $b^{\rm true}_{l_1l_2l_3}$ varies much more slowly than 
the coupling integral, we obtain
%%%%%%%%%%%%%%%%%%%%%%%%%%%%%%%%%%%%%%%%%%%%%%%%%%%%%%%%%%%%%%%%%
\begin{eqnarray}
 \left<a_{l_1m_1}a_{l_2m_2}a_{l_3m_3}\right>
 &=&
  \nonumber
  \sum_{{\rm all}~l'}
  b_{l_1'l_2'l_3'}^{\rm true}
  \sum_{{\rm all}~m'}
  \int
  d^2\hat{\mathbf n}~
  Y_{l'_1m'_1}^*\left(\hat{\mathbf n}\right)
  Y_{l'_2m'_2}^*\left(\hat{\mathbf n}\right)
  Y_{l'_3m'_3}^*\left(\hat{\mathbf n}\right) \\
 \nonumber
 & &\times
 \int  d^2\hat{\mathbf n}_1~
  W(\hat{\mathbf n}_1)
  Y_{l_1'm_1'}\left(\hat{\mathbf n}_1\right)
  Y_{l_1m_1}^*\left(\hat{\mathbf n}_1\right) \\
 \nonumber
 & &\times
  \int  d^2\hat{\mathbf n}_2~
  W(\hat{\mathbf n}_2)
  Y_{l_2'm_2'}\left(\hat{\mathbf n}_2\right)
  Y_{l_2m_2}^*\left(\hat{\mathbf n}_2\right) \\
 \nonumber
 & &\times
  \int  d^2\hat{\mathbf n}_3~
  W(\hat{\mathbf n}_3)
  Y_{l_3'm_3'}\left(\hat{\mathbf n}_3\right)
  Y_{l_3m_3}^*\left(\hat{\mathbf n}_3\right)\\
 \label{eq:bl_obs}
  &\approx&
   b_{l_1l_2l_3}^{\rm true}
   \int  d^2\hat{\mathbf n}~
  W(\hat{\mathbf n})
  Y_{l_1m_1}^*\left(\hat{\mathbf n}\right)
  Y_{l_2m_2}^*\left(\hat{\mathbf n}\right)
  Y_{l_3m_3}^*\left(\hat{\mathbf n}\right).
\end{eqnarray}
%%%%%%%%%%%%%%%%%%%%%%%%%%%%%%%%%%%%%%%%%%%%%%%%%%%%%%%%%%%%%%%%%
Then, we calculate the angular averaged bispectrum, $B_{l_1l_2l_3}$
(Eq.(\ref{eq:blll*})). 
By convolving equation~(\ref{eq:best}) with the Wigner-3$j$ symbol
and using the identity~(\ref{eq:gaunt_spec}),
we obtain
%%%%%%%%%%%%%%%%%%%%%%%%%%%%%%%%%%%%%%%%%%%%%%%%%%%%%%%%%%%%%%%%%
\begin{eqnarray}
 \nonumber
 \left<B_{l_1l_2l_3}\right>
 &\approx&
 b_{l_1l_2l_3}^{\rm true}
 \sqrt{\frac{4\pi}{(2l_1+1)(2l_2+1)(2l_3+1)}}
   \left(\begin{array}{ccc}l_1&l_2&l_3\\0&0&0\end{array}\right)^{-1}\\
 \nonumber
  & &\times
 \sum_{{\rm all}~m}
  \int
  d^2\hat{\mathbf m}~
  Y_{l_1m_1}\left(\hat{\mathbf m}\right)
  Y_{l_2m_2}\left(\hat{\mathbf m}\right)
  Y_{l_3m_3}\left(\hat{\mathbf m}\right) \\
 & &\times 
 \nonumber
  \int
  d^2\hat{\mathbf n}~
  W(\hat{\mathbf n})
  Y_{l_1m_1}^*\left(\hat{\mathbf n}\right)
  Y_{l_2m_2}^*\left(\hat{\mathbf n}\right)
  Y_{l_3m_3}^*\left(\hat{\mathbf n}\right)\\
 \nonumber
 &=&
 b_{l_1l_2l_3}^{\rm true}
 \sqrt{\frac{(2l_1+1)(2l_2+1)(2l_3+1)}{4\pi}}
   \left(\begin{array}{ccc}l_1&l_2&l_3\\0&0&0\end{array}\right)^{-1}\\
 \nonumber
  & &\times
  \int  \frac{d^2\hat{\mathbf m}}{4\pi}
  \int  \frac{d^2\hat{\mathbf n}}{4\pi}~
  W(\hat{\mathbf n})
  P_{l_1}\left(\hat{\mathbf m}\cdot\hat{\mathbf n}\right)
  P_{l_2}\left(\hat{\mathbf m}\cdot\hat{\mathbf n}\right)
  P_{l_3}\left(\hat{\mathbf m}\cdot\hat{\mathbf n}\right)\\
 \nonumber
 &=&
  b_{l_1l_2l_3}^{\rm true}
  \sqrt{\frac{(2l_1+1)(2l_2+1)(2l_3+1)}{4\pi}}
	\left(\begin{array}{ccc}l_1&l_2&l_3\\0&0&0\end{array}\right)
	\frac{\Omega_{\rm obs}}{4\pi}\\
 &=&
  \label{eq:biasbl}
  B_{l_1l_2l_3}^{\rm true}\frac{\Omega_{\rm obs}}{4\pi},
\end{eqnarray}
%%%%%%%%%%%%%%%%%%%%%%%%%%%%%%%%%%%%%%%%%%%%%%%%%%%%%%%%%%%%%%%%%
where we have used the identity,
%%%%%%%%%%%%%%%%%%%%%%%%%%%%%%%%%%%%%%%%%%%%%%%%%%%%%%%%%%%%%%%%%
\begin{equation}
 \int_{-1}^{1}\frac{dx}2~
  P_{l_1}(x)P_{l_2}(x)P_{l_3}(x)
  =
  \left(\begin{array}{ccc}l_1&l_2&l_3\\0&0&0\end{array}\right)^2.
\end{equation}
%%%%%%%%%%%%%%%%%%%%%%%%%%%%%%%%%%%%%%%%%%%%%%%%%%%%%%%%%%%%%%%%%
Thus, the bias for the angular bispectrum on the incomplete sky is also 
approximately given by a fraction of the sky covered by observations.

%%%%%%%%%%%%%%%%%%%%%%%%%%%%%%%%%%%%%%%%%%%%%%%%%%%%%%%%%%%%%%%%%%%
%
%  Theoretical Predictions for the CMB Bispectrum
%
%     1st draft:  06/20/2001
%     final:      08/06/2001
%
%%%%%%%%%%%%%%%%%%%%%%%%%%%%%%%%%%%%%%%%%%%%%%%%%%%%%%%%%%%%%%%%%%%
\chapter{Theoretical Predictions for the CMB Bispectrum}
\label{chap:theory_bl}

%%%%%%%%%%%%%%%%%%%%%%%%%%%%%%%%%%%%%%%%%%%%%%%%%%%%%%%%%%%%%%%%%%%
In inflation, quantum fluctuations of scalar fields 
generate the matter and the radiation fluctuations 
in the universe (chapter~\ref{chap:inflation}).
In the stochastic inflationary scenario of \citet{Sta86},
quantum fluctuations decohere to generate classical fluctuations.
There are two potential sources of non-Gaussianity in this 
inflationary model:
(a) non-linear coupling between the classical inflaton field and 
the observed fluctuation field, and
(b) non-linear coupling between the quantum noise field and the
classical fluctuation field.
\citet{SB90,SB91} have studied the former; 
\citet{Gan94} have studied the latter.

\citet{CH95} and \citet{Matacz97a,Matacz97b} present
an alternative treatment of the decoherence process that leads to 
different results for the primordial density perturbations 
from \citet{Sta86}.
Matacz's treatment makes similar predictions for the level of 
non-Gaussianity to the Starobinsky's treatment \citep{Matacz97a,Matacz97b}.
These studies conclude that in a slow-roll regime, fluctuations are Gaussian;
however, features in a inflaton potential can produce 
significant non-Gaussianity \citep{KBHP91}.

Previous work on the primary non-Gaussianity
has focused on very large angular scales, the {\it COBE} scale, where 
the temperature fluctuations trace the primordial fluctuations.
For {\it MAP} and {\it Planck}; however, we need the 
full effect of the radiation transfer function.
In this chapter, we develop a formalism for doing this, and then present 
numerical results. 
Both the formalism and the numerical results are main results of this chapter.
We also discuss how well we can separate the primary bispectrum from 
various secondary bispectra.

This chapter is organized as follows. 
In \S~\ref{sec:formulation}, we define the angular bispectrum, the 
Gaunt integral, and a new quantity called the {\it reduced} bispectrum, 
which plays a fundamental role in estimating physical properties of the 
bispectrum.
In \S~\ref{sec:bl+s}, we formulate the primary bispectrum that uses the full 
radiation transfer function, and presents numerical results for the primary
bispectrum and skewness. 
In \S~\ref{sec:secondary}, we calculate secondary bispectra from
the coupling between the Sunyaev--Zel'dovich effect and the weak lensing
effect \citep{SG99,GS99,CH00}, and from extragalactic radio and 
infrared sources.
In \S~\ref{sec:measure},  we study how well we can measure each
bispectrum, and how well we can discriminate between those bispectra.
\S~\ref{sec:discussion} is devoted to further discussion and 
our conclusions in this chapter.

%%%%%%%%%%%%%%%%%%%%%%%%%%%%%%%%%%%%%%%%%%%%%%%%%%%%%%%%%%%%%%%%%%
\section{Reduced Bispectrum}
\label{sec:formulation}

We expand the observed CMB temperature fluctuation field, 
$\Delta T(\hat{\mathbf n})/T$, into the spherical harmonics,
%%%%%%%%%%%%%%%%%%%%%%%%%%%%%%%%%%%%%%%%%%%%%%%%%%%%%%%%%%%%%%%%%%
\begin{equation}
  a_{lm}= \int d^2\hat{\mathbf n}\frac{\Delta T(\hat{\mathbf n})}{T}
  Y_{lm}^*(\hat{\mathbf n}),
\end{equation}
%%%%%%%%%%%%%%%%%%%%%%%%%%%%%%%%%%%%%%%%%%%%%%%%%%%%%%%%%%%%%%%%%%
where the hats denote unit vectors. 
The CMB angular bispectrum is given by 
%%%%%%%%%%%%%%%%%%%%%%%%%%%%%%%%%%%%%%%%%%%%%%%%%%%%%%%%%%%%%%%%%%
\begin{equation}
  \label{eq:blllmmm}
  B_{l_1l_2l_3}^{m_1m_2m_3}\equiv 
  \left<a_{l_1m_1}a_{l_2m_2}a_{l_3m_3}\right>,
\end{equation}
%%%%%%%%%%%%%%%%%%%%%%%%%%%%%%%%%%%%%%%%%%%%%%%%%%%%%%%%%%%%%%%%%%
and the angular averaged bispectrum is (Eq.(\ref{eq:best}))
%%%%%%%%%%%%%%%%%%%%%%%%%%%%%%%%%%%%%%%%%%%%%%%%%%%%%%%%%%%%%%%%%%
\begin{equation}
  \label{eq:blll}
  B_{l_1l_2l_3}= \sum_{{\rm all}~m}
  \left(
  \begin{array}{ccc}
  l_1&l_2&l_3\\
  m_1&m_2&m_3
  \end{array}
  \right)
  B_{l_1l_2l_3}^{m_1m_2m_3}, 
\end{equation}
%%%%%%%%%%%%%%%%%%%%%%%%%%%%%%%%%%%%%%%%%%%%%%%%%%%%%%%%%%%%%%%%%%
where the matrix is the Wigner-$3j$ symbol.
The bispectrum, $B_{l_1l_2l_3}^{m_1m_2m_3}$,
satisfies the triangle conditions and parity invariance:
$m_1+m_2+m_3=0$, $l_1+l_2+l_3={\rm even}$, and 
$\left|l_i-l_j\right|\leq l_k \leq l_i+l_j$ for all permutations
of indices. 
It implies that $B_{l_1l_2l_3}^{m_1m_2m_3}$ consists of the Gaunt integral,  
${\cal G}_{l_1l_2l_3}^{m_1m_2m_3}$, defined by
%%%%%%%%%%%%%%%%%%%%%%%%%%%%%%%%%%%%%%%%%%%%%%%%%%%%%%%%%%%%%%%%%%
\begin{eqnarray}
  \nonumber
  {\cal G}_{l_1l_2l_3}^{m_1m_2m_3}
  &\equiv&
  \int d^2\hat{\mathbf n}
  Y_{l_1m_1}(\hat{\mathbf n})
  Y_{l_2m_2}(\hat{\mathbf n})
  Y_{l_3m_3}(\hat{\mathbf n})\\
  \label{eq:gaunt}
  &=&\sqrt{
   \frac{\left(2l_1+1\right)\left(2l_2+1\right)\left(2l_3+1\right)}
        {4\pi}
        }
  \left(
  \begin{array}{ccc}
  l_1 & l_2 & l_3 \\ 0 & 0 & 0 
  \end{array}
  \right)
  \left(
  \begin{array}{ccc}
  l_1 & l_2 & l_3 \\ m_1 & m_2 & m_3 
  \end{array}
  \right).
\end{eqnarray}
%%%%%%%%%%%%%%%%%%%%%%%%%%%%%%%%%%%%%%%%%%%%%%%%%%%%%%%%%%%%%%%%%%
${\cal G}_{l_1l_2l_3}^{m_1m_2m_3}$ is real, and satisfies 
all the conditions mentioned above.

Rotational invariance of the angular three-point correlation
function implies that $B_{l_1l_2l_3}$ is written as
%%%%%%%%%%%%%%%%%%%%%%%%%%%%%%%%%%%%%%%%%%%%%%%%%%%%%%%%%%%%%%%%%%
\begin{equation}
  \label{eq:func}
  B_{l_1l_2l_3}^{m_1m_2m_3}
  ={\cal G}_{l_1l_2l_3}^{m_1m_2m_3}b_{l_1l_2l_3}, 
\end{equation}
%%%%%%%%%%%%%%%%%%%%%%%%%%%%%%%%%%%%%%%%%%%%%%%%%%%%%%%%%%%%%%%%%%
where $b_{l_1l_2l_3}$ is an arbitrary real symmetric function 
of $l_1$, $l_2$, and $l_3$.
This form, equation (\ref{eq:func}), is necessary and
sufficient to construct generic $B_{l_1l_2l_3}^{m_1m_2m_3}$ under 
rotational invariance; thus, we will use $b_{l_1l_2l_3}$ more frequently than
$B_{l_1l_2l_3}^{m_1m_2m_3}$ in this chapter, and call this function
the {\it reduced} bispectrum, as $b_{l_1l_2l_3}$ 
contains all physical information in $B_{l_1l_2l_3}^{m_1m_2m_3}$.
Since the reduced bispectrum does not contain the Wigner-$3j$ symbol,
which merely ensures the triangle conditions and parity invariance, 
it is easier to calculate physical properties of the bispectrum.

We calculate the angular averaged bispectrum, $B_{l_1l_2l_3}$,
by substituting equation (\ref{eq:func}) into (\ref{eq:blll}),
%%%%%%%%%%%%%%%%%%%%%%%%%%%%%%%%%%%%%%%%%%%%%%%%%%%%%%%%%%%%%%%%%%
\begin{equation}
  \label{eq:wigner*}
  B_{l_1l_2l_3}
  =
  \sqrt{\frac{(2l_1+1)(2l_2+1)(2l_3+1)}{4\pi}}
  \left(
  \begin{array}{ccc}
  l_1&l_2&l_3\\
  0&0&0
  \end{array}
  \right)b_{l_1l_2l_3},
\end{equation}
%%%%%%%%%%%%%%%%%%%%%%%%%%%%%%%%%%%%%%%%%%%%%%%%%%%%%%%%%%%%%%%%%%
where we have used the identity,
%%%%%%%%%%%%%%%%%%%%%%%%%%%%%%%%%%%%%%%%%%%%%%%%%%%%%%%%%%%%%%%%%%
\begin{equation}
  \label{eq:wigner}
  \sum_{{\rm all}~m}
  \left(
  \begin{array}{ccc}
  l_1&l_2&l_3\\
  m_1&m_2&m_3
  \end{array}
  \right)
  {\cal G}_{l_1l_2l_3}^{m_1m_2m_3}
  =
  \sqrt{\frac{(2l_1+1)(2l_2+1)(2l_3+1)}{4\pi}}
  \left(
  \begin{array}{ccc}
  l_1&l_2&l_3\\
  0&0&0
  \end{array}
  \right).
\end{equation}
%%%%%%%%%%%%%%%%%%%%%%%%%%%%%%%%%%%%%%%%%%%%%%%%%%%%%%%%%%%%%%%%%%

Alternatively, one can define the bispectrum in the flat-sky
approximation, 
%%%%%%%%%%%%%%%%%%%%%%%%%%%%%%%%%%%%%%%%%%%%%%%%%%%%%%%%%%%%%%%%%%
\begin{equation}
 \label{eq:smallangle}
  \left<a({\mathbf l}_1)a({\mathbf l}_1)a({\mathbf l}_3)\right>
  =(2\pi)^2\delta^{(2)}\left({\mathbf l}_1+{\mathbf l}_2+{\mathbf l}_3\right)
  B({\mathbf l}_1,{\mathbf l}_2,{\mathbf l}_3),
\end{equation}
%%%%%%%%%%%%%%%%%%%%%%%%%%%%%%%%%%%%%%%%%%%%%%%%%%%%%%%%%%%%%%%%%%
where ${\mathbf l}$ is a two-dimensional wave vector on the sky.
This definition of $B({\mathbf l}_1,{\mathbf l}_2,{\mathbf l}_3)$ 
reduces to equation (\ref{eq:func}) with the correspondence, 
${\cal G}_{l_1l_2l_3}^{m_1m_2m_3}\rightarrow 
(2\pi)^2\delta^{(2)}\left({\mathbf l}_1+{\mathbf l}_2+{\mathbf l}_3\right)$, 
in the flat-sky limit \citep{Hu00}. 
Thus, we have
%%%%%%%%%%%%%%%%%%%%%%%%%%%%%%%%%%%%%%%%%%%%%%%%%%%%%%%%%%%%%%%%%%
\begin{equation}
 \label{eq:smallangle*}
  b_{l_1l_2l_3}\approx
  B({\mathbf l}_1,{\mathbf l}_2,{\mathbf l}_3)
  \qquad \mbox{(flat-sky approximation)}.
\end{equation}
%%%%%%%%%%%%%%%%%%%%%%%%%%%%%%%%%%%%%%%%%%%%%%%%%%%%%%%%%%%%%%%%%% 
This fact motivates our using the reduced bispectrum, 
$b_{l_1l_2l_3}$, rather than the angular averaged bispectrum, $B_{l_1l_2l_3}$.
Note that $b_{l_1l_2l_3}$ is similar to $\hat{B}_{l_1l_2l_3}$ defined by 
Magueijo \citep{Mag00}; the relation is 
$b_{l_1l_2l_3}=\sqrt{4\pi}\hat{B}_{l_1l_2l_3}$.

%%%%%%%%%%%%%%%%%%%%%%%%%%%%%%%%%%%%%%%%%%%%%%%%%%%%%%%%%%%%%%%%%%
\section{Primary Bispectrum and Skewness}
\label{sec:bl+s}

\subsection{Model of the primordial non-Gaussianity}

If primordial fluctuations are adiabatic scalar fluctuations, then
%%%%%%%%%%%%%%%%%%%%%%%%%%%%%%%%%%%%%%%%%%%%%%%%%%%%%%%%%%%%%%%%%%
\begin{equation}
  \label{eq:almphi}
  a_{lm}=4\pi(-i)^l
  \int\frac{d^3{\mathbf k}}{(2\pi)^3}\Phi({\mathbf k})g_{{\rm T}l}(k)
  Y_{lm}^*(\hat{\mathbf k}),
\end{equation}
%%%%%%%%%%%%%%%%%%%%%%%%%%%%%%%%%%%%%%%%%%%%%%%%%%%%%%%%%%%%%%%%%%
where $\Phi({\mathbf k})$ is the primordial curvature perturbation
in Fourier space, and $g_{{\rm T}l}(k)$ is the radiation transfer function.
$a_{lm}$ takes over the non-Gaussianity, if any, from $\Phi({\mathbf k})$.
Although equation (\ref{eq:almphi}) is valid only if the
universe is flat, it is straightforward to extend this
to an arbitrary geometry.
We can calculate the isocurvature fluctuations similarly by
using the entropy perturbation and the proper transfer function.

In this chapter, we explore the simplest weak non-linear coupling case, 
%%%%%%%%%%%%%%%%%%%%%%%%%%%%%%%%%%%%%%%%%%%%%%%%%%%%%%%%%%%%%%%%%%
\begin{equation}
  \label{eq:modelreal}
  \Phi({\mathbf x})
 =\Phi_{\rm L}({\mathbf x})
 +f_{\rm NL}\left[
              \Phi^2_{\rm L}({\mathbf x})-
	      \left<\Phi^2_{\rm L}({\mathbf x})\right>
        \right],
\end{equation}
%%%%%%%%%%%%%%%%%%%%%%%%%%%%%%%%%%%%%%%%%%%%%%%%%%%%%%%%%%%%%%%%%%
in real space, where $\Phi_{\rm L}({\mathbf x})$ denotes a linear Gaussian 
part of the perturbation, and $\left<\Phi({\mathbf x})\right>=0$ is guaranteed.
Henceforth, we call $f_{\rm NL}$ the {\it non-linear coupling parameter}.
This model is based upon slow-roll inflation;  
\citet{SB90,SB91} and \citet{Gan94} have found that $f_{\rm NL}$ is 
given by a certain combination of slope and curvature of an inflaton
potential ($\Phi_3=-2f_{\rm NL}$ in \citet{Gan94}). 
\citet{Gan94} have found that $\left|f_{\rm NL}\right|\sim 10^{-2}$
for quadratic and quartic potential models.
We have reproduced these results in chapter~\ref{chap:inflation}.

In Fourier space, we decompose $\Phi({\mathbf k})$ into two parts,
%%%%%%%%%%%%%%%%%%%%%%%%%%%%%%%%%%%%%%%%%%%%%%%%%%%%%%%%%%%%%%%%%%
\begin{equation}
  \label{eq:model}
  \Phi({\mathbf k})=\Phi_{\rm L}({\mathbf k})+\Phi_{\rm NL}({\mathbf k}),
\end{equation}
%%%%%%%%%%%%%%%%%%%%%%%%%%%%%%%%%%%%%%%%%%%%%%%%%%%%%%%%%%%%%%%%%%
and accordingly we have
%%%%%%%%%%%%%%%%%%%%%%%%%%%%%%%%%%%%%%%%%%%%%%%%%%%%%%%%%%%%%%%%%%
\begin{equation}
  \label{eq:model*}
  a_{lm}=a_{lm}^{\rm L}+a_{lm}^{\rm NL},
\end{equation}
%%%%%%%%%%%%%%%%%%%%%%%%%%%%%%%%%%%%%%%%%%%%%%%%%%%%%%%%%%%%%%%%%%
where $\Phi_{\rm NL}({\mathbf k})$ is a non-linear curvature
perturbation defined by
%%%%%%%%%%%%%%%%%%%%%%%%%%%%%%%%%%%%%%%%%%%%%%%%%%%%%%%%%%%%%%%%%%
\begin{equation}
  \label{eq:nonlinear}
  \Phi_{\rm NL}({\mathbf k})\equiv 
  f_{\rm NL}
  \left[
  \int \frac{d^3{\mathbf p}}{(2\pi)^3}
  \Phi_{\rm L}({\mathbf k}+{\mathbf p})\Phi^*_{\rm L}({\mathbf p})
  -(2\pi)^3\delta^{(3)}({\mathbf k})\left<\Phi^2_{\rm L}({\mathbf x})\right>
  \right].
\end{equation}
%%%%%%%%%%%%%%%%%%%%%%%%%%%%%%%%%%%%%%%%%%%%%%%%%%%%%%%%%%%%%%%%%%
One can confirm that $\left<\Phi({\mathbf k})\right>=0$ is satisfied.
In this model, a non-vanishing component of the 
$\Phi({\mathbf k})$-field bispectrum is
%%%%%%%%%%%%%%%%%%%%%%%%%%%%%%%%%%%%%%%%%%%%%%%%%%%%%%%%%%%%%%%%%%
\begin{equation}
  \label{eq:phispec}
  \left<\Phi_{\rm L}({\mathbf k}_1)
	\Phi_{\rm L}({\mathbf k}_2)
	\Phi_{\rm NL}({\mathbf k}_3)\right>
  = 2(2\pi)^3\delta^{(3)}({\mathbf k}_1+{\mathbf k}_2+{\mathbf k}_3)
    f_{\rm NL}P_\Phi(k_1)P_\Phi(k_2),
\end{equation}
%%%%%%%%%%%%%%%%%%%%%%%%%%%%%%%%%%%%%%%%%%%%%%%%%%%%%%%%%%%%%%%%%%	
where $P_\Phi(k)$ is the linear power spectrum given by 
%%%%%%%%%%%%%%%%%%%%%%%%%%%%%%%%%%%%%%%%%%%%%%%%%%%%%%%%%%%%%%%%%%
\begin{equation}
 \left<\Phi_{\rm L}({\mathbf k}_1)\Phi_{\rm L}({\mathbf k}_2)\right>
  =(2\pi)^3P_\Phi(k_1)\delta^{(3)}({\mathbf k}_1+{\mathbf k}_2).
\end{equation}
%%%%%%%%%%%%%%%%%%%%%%%%%%%%%%%%%%%%%%%%%%%%%%%%%%%%%%%%%%%%%%%%%%
We have also used 
%%%%%%%%%%%%%%%%%%%%%%%%%%%%%%%%%%%%%%%%%%%%%%%%%%%%%%%%%%%%%%%%%%
\begin{equation}
 \left<\Phi_{\rm L}({\mathbf k}+{\mathbf p})\Phi^*_{\rm L}({\mathbf p})\right>
  =(2\pi)^3P_\Phi(p)\delta^{(3)}({\mathbf k}),
\end{equation}
%%%%%%%%%%%%%%%%%%%%%%%%%%%%%%%%%%%%%%%%%%%%%%%%%%%%%%%%%%%%%%%%%%
and 
%%%%%%%%%%%%%%%%%%%%%%%%%%%%%%%%%%%%%%%%%%%%%%%%%%%%%%%%%%%%%%%%%%
\begin{equation}
 \left<\Phi^2_{\rm L}({\mathbf x})\right>
  =(2\pi)^{-3}\int d^3{\mathbf k} P_\Phi(k).
\end{equation}
%%%%%%%%%%%%%%%%%%%%%%%%%%%%%%%%%%%%%%%%%%%%%%%%%%%%%%%%%%%%%%%%%%

Substituting equation (\ref{eq:almphi}) into (\ref{eq:blllmmm}),
using equation (\ref{eq:phispec}) for the $\Phi({\mathbf k})$-field
bispectrum, and then integrating over angles 
$\hat{\mathbf k}_1$, $\hat{\mathbf k}_3$, and $\hat{\mathbf k}_3$, we obtain 
the primary CMB angular bispectrum,
%%%%%%%%%%%%%%%%%%%%%%%%%%%%%%%%%%%%%%%%%%%%%%%%%%%%%%%%%%%%%%%%%%
\begin{eqnarray}
 B_{l_1l_2l_3}^{m_1m_2m_3}
  \nonumber
  &=& 
  \left<a_{l_1m_1}^{\rm L}a_{l_2m_2}^{\rm L}a_{l_3m_3}^{\rm NL}\right>
  + \left<a_{l_1m_1}^{\rm L}a_{l_2m_2}^{\rm NL}a_{l_3m_3}^{\rm L}\right>
  + \left<a_{l_1m_1}^{\rm NL}a_{l_2m_2}^{\rm L}a_{l_3m_3}^{\rm L}\right>\\
 \nonumber
  &=& 2{\cal G}_{l_1l_2l_3}^{m_1m_2m_3}
	\int_0^\infty r^2 dr 
    \left[
          b^{\rm L}_{l_1}(r)b^{\rm L}_{l_2}(r)b^{\rm NL}_{l_3}(r)+
	  b^{\rm L}_{l_1}(r)b^{\rm NL}_{l_2}(r)b^{\rm L}_{l_3}(r)\right.\\
  \label{eq:almspec}
  & &\left.+
	  b^{\rm NL}_{l_1}(r)b^{\rm L}_{l_2}(r)b^{\rm L}_{l_3}(r)
    \right],
\end{eqnarray}
%%%%%%%%%%%%%%%%%%%%%%%%%%%%%%%%%%%%%%%%%%%%%%%%%%%%%%%%%%%%%%%%%%	
where 
%%%%%%%%%%%%%%%%%%%%%%%%%%%%%%%%%%%%%%%%%%%%%%%%%%%%%%%%%%%%%%%%%%
\begin{eqnarray}
  \label{eq:bLr}
  b^{\rm L}_{l}(r) &\equiv&
  \frac2{\pi}\int_0^\infty k^2 dk P_\Phi(k)g_{{\rm T}l}(k)j_l(kr),\\
  \label{eq:bNLr}
  b^{\rm NL}_{l}(r) &\equiv&
  \frac2{\pi}\int_0^\infty k^2 dk f_{\rm NL}g_{{\rm T}l}(k)j_l(kr).
\end{eqnarray}
%%%%%%%%%%%%%%%%%%%%%%%%%%%%%%%%%%%%%%%%%%%%%%%%%%%%%%%%%%%%%%%%%%	
Note that $b^{\rm L}_{l}(r)$ is dimensionless, 
while $b^{\rm NL}_{l}(r)$ has a dimension of $L^{-3}$.

One confirms that equation (\ref{eq:func}) holds; thus, the reduced 
bispectrum, $b_{l_1l_2l_3}$ (Eq.(\ref{eq:func})), for the primordial 
non-Gaussianity is
%%%%%%%%%%%%%%%%%%%%%%%%%%%%%%%%%%%%%%%%%%%%%%%%%%%%%%%%%%%%%%%%%%
\begin{eqnarray}
 \nonumber
  b_{l_1l_2l_3}^{\rm primary}
  &=& 2\int_0^\infty r^2 dr
    \left[
          b^{\rm L}_{l_1}(r)b^{\rm L}_{l_2}(r)b^{\rm NL}_{l_3}(r)+
	  b^{\rm L}_{l_1}(r)b^{\rm NL}_{l_2}(r)b^{\rm L}_{l_3}(r)\right.\\
 \label{eq:blprim}
  & &\left.+ b^{\rm NL}_{l_1}(r)b^{\rm L}_{l_2}(r)b^{\rm L}_{l_3}(r)
    \right].
\end{eqnarray}
%%%%%%%%%%%%%%%%%%%%%%%%%%%%%%%%%%%%%%%%%%%%%%%%%%%%%%%%%%%%%%%%%%	
We can fully specify $b_{l_1l_2l_3}^{\rm primary}$ by a single constant
parameter, $f_{\rm NL}$, for the CMB angular power spectrum, $C_l$, 
will precisely measure cosmological parameters \citep{BET97}.
We stress again that this is the special case in the slow-roll limit. 
If the slow-roll condition is not satisfied, 
then $f_{\rm NL}=f_{\rm NL}(k_1,k_2,k_3)$ at equation 
(\ref{eq:phispec}) \citep{Gan94}.
\citet{WK00} have developed a formula to 
compute $B_{l_1l_2l_3}$ from generic $\Phi({\mathbf k})$-field bispectrum.
Our formula (Eq.(\ref{eq:almspec})) agrees with theirs,
given our form of the $\Phi({\mathbf k})$-field bispectrum
(Eq.(\ref{eq:phispec})).

Even if inflation produces Gaussian fluctuations,
the general relativistic second-order 
perturbation theory produces $f_{\rm NL}\sim {\cal O}(1)$ \citep{PC96}.
For generic slow-roll models, these terms dominate the primary non-Gaussianity.

\subsection{Numerical results of the primary bispectrum}

We calculate the primary CMB bispectrum 
(Eqs.(\ref{eq:almspec})--(\ref{eq:blprim})) numerically as follows.
We compute the full radiation transfer function, $g_{{\rm T}l}(k)$,
with the {\sf CMBFAST} code \citep{SZ96}, assuming
a single power-law spectrum, $P_\Phi(k)\propto k^{n-4}$,
for the primordial curvature fluctuations.
After doing the integration over $k$ (Eqs.(\ref{eq:bLr}) and (\ref{eq:bNLr})) 
with the same algorithm as of {\sf CMBFAST},
we do the integration over $r$ (Eq.(\ref{eq:blprim})),
$r=c\left(\tau_0-\tau\right)$, where $\tau$ is the conformal time.
$\tau_0$ is the present-day value.
In our model, $c\tau_0=11.8\ {\rm Gpc}$, and 
the decoupling occurs at $c\tau_*=235\ {\rm Mpc}$ 
at which the differential visibility has a maximum.
Our $c\tau_0$ includes radiation effects on the expansion of the
universe; otherwise, $c\tau_0=12.0\ {\rm Gpc}$.
Since the most of the primary signal is generated at $\tau_*$,
we choose the $r$ integration boundary as 
$c\left(\tau_0-2\tau_*\right)\leq r\leq c\left(\tau_0-0.1\tau_*\right)$.
We use a step-size of $0.1c\tau_*$, as we have found that a step size of  
$0.01c\tau_*$ gives very similar results.
As a cosmological model, we use the scale-invariant standard 
CDM model with $\Omega_{\rm m}=1$, $\Omega_\Lambda=0$, 
$\Omega_{\rm b}=0.05$, $h=0.5$, and $n=1$, and with the power spectrum,
$P_\Phi(k)$, normalized to {\it COBE} \citep{BW97}.
Although this model is almost excluded by current observations,
it is still useful to depict basic effects of the transfer function 
on the bispectrum.

\begin{figure}
 \plotone{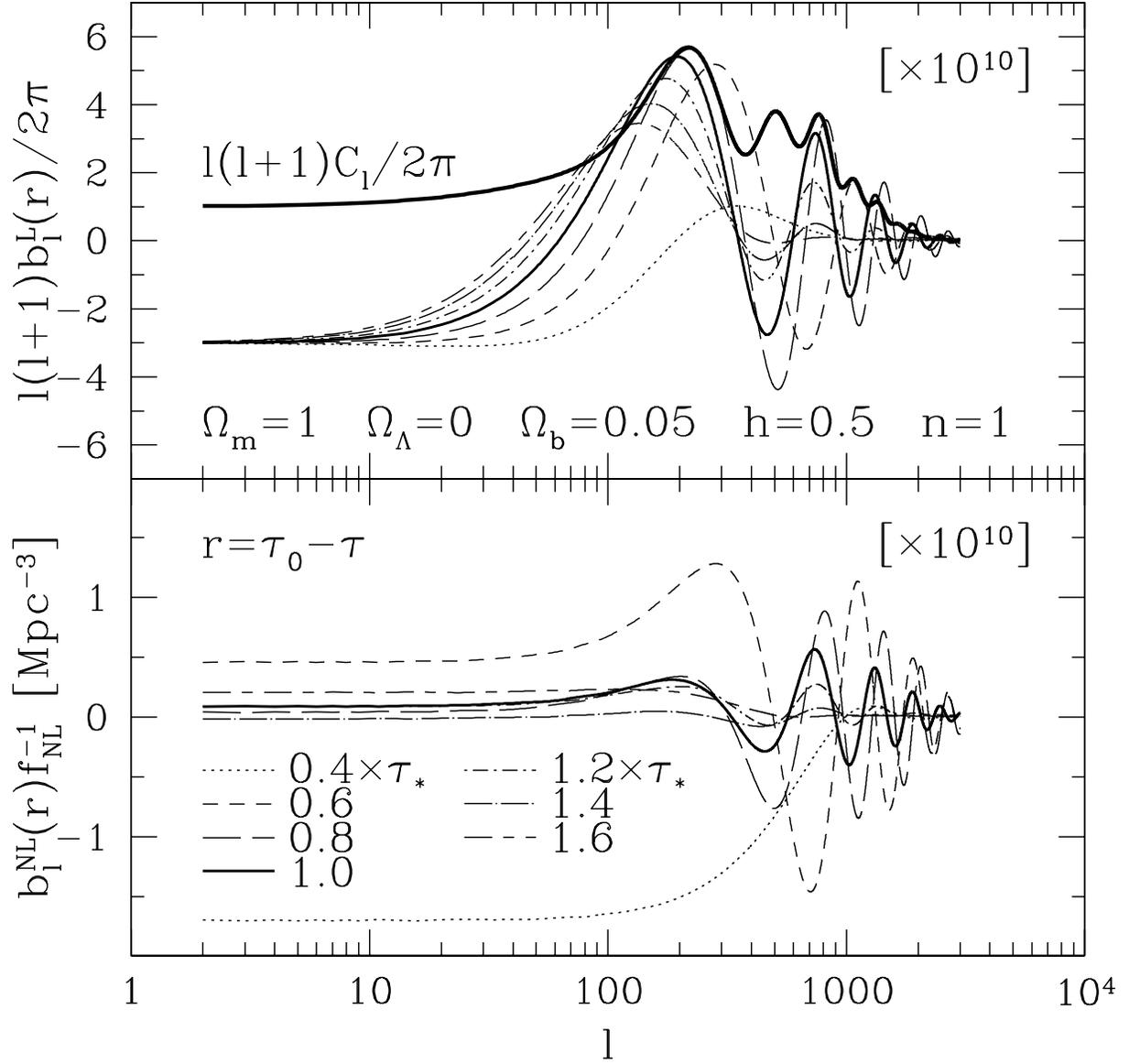}
 \caption{Components of Primary CMB Bispectrum}
 \mycaption{
 This figure shows $b_l^{\rm L}(r)$ (Eq.(\ref{eq:bLr})) and 
 $b_l^{\rm NL}(r)$ (Eq.(\ref{eq:bNLr})), the two terms in our calculation of 
 the primary CMB angular bispectrum, as a function of $r$. 
 Various lines in the top panel show 
 $\left[l(l+1)b_l^{\rm L}(r)/2\pi\right]\times 10^{10}$, where
 $r=c\left(\tau_0-\tau\right)$, at $\tau=0.4,0.6,0.8,1.0,1.2,1.4$, 
 and $1.6\times \tau_*$ (decoupling time);
 $\left[b_l^{\rm NL}(r)f^{-1}_{\rm NL}\right]\times 10^{10}$ 
 are shown in the bottom panel. 
 $\tau_0$ is the present-day conformal time.
 Note that $c\tau_0=11.8\ {\rm Gpc}$, and $c\tau_*=235\ {\rm Mpc}$ in our 
 cosmological model chosen here.
 The thickest solid line in the top panel is the CMB angular power spectrum,
 $\left[l(l+1)C_l/2\pi\right]\times 10^{10}$.  
 $C_l$ is shown for comparison.}
\label{fig:bl}
\end{figure}

Figure~\ref{fig:bl} shows $b_l^{\rm L}(r)$ (Eq.(\ref{eq:bLr})) and 
$b_l^{\rm NL}(r)$ (Eq.(\ref{eq:bNLr})) for several different values of $r$.
We find that $b^{\rm L}_l(r)$ and $C_l$ look very similar to each other in 
shape and amplitude at $l\simgt 100$, 
although the amplitude in the Sachs--Wolfe regime is different by a 
factor of $-3$. 
This is because $C_l\propto P_\Phi(k)g_{{\rm T}l}^2(k)$,
while $b_l^{\rm L}(r)\propto P_\Phi(k)g_{{\rm T}l}(k)$, where $g_{{\rm T}l}=-1/3$.
We also find that $b_l^{\rm L}(r)$ has a good phase coherence over 
wide range of $r$, while the phase of $b_l^{\rm NL}(r)$ in the high-$l$ regime 
oscillates rapidly as a function of $r$. 
This strongly damps the integrated result (Eq.(\ref{eq:almspec})) 
in the high-$l$ regime.
The main difference between $C_l$ and $b_l(r)$ is that
$b_l(r)$ changes the sign, while $C_l$ does not.

%%%%% Figure 1: bL & bNL %%%%%
%%%%%%%%%%%%%%%%%%%%%%%%%%%%%%%%%%%%%%%%%%%%%%%%%%%%%%%%%%%%%%%%%%
% Figure 1
%%%%%%%%%%%%%%%%%%%%%%%%%%%%%%%%%%%%%%%%%%%%%%%%%%%%%%%%%%%%%%%%%%

\begin{figure}
 \plotone{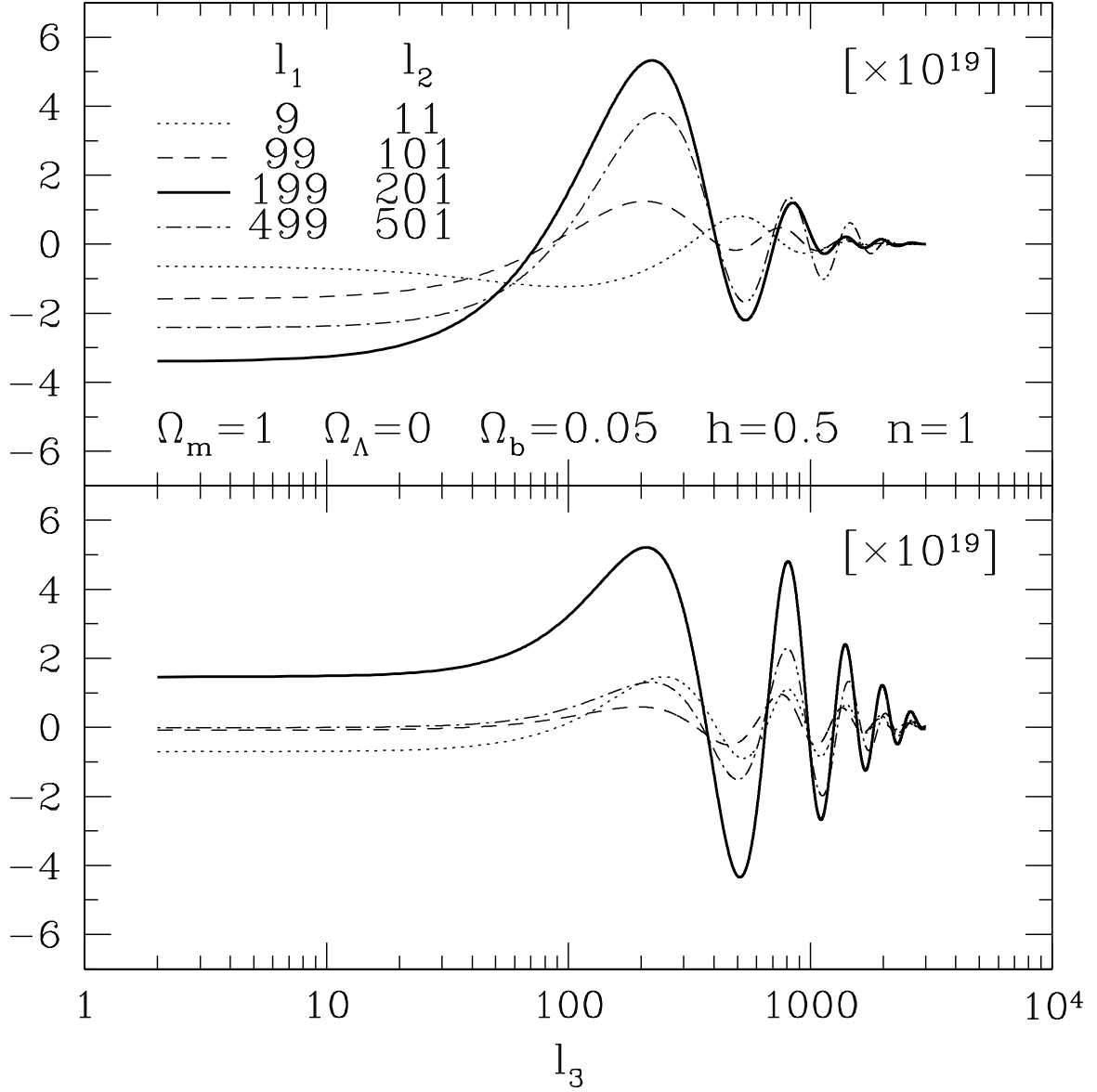}
 \caption{Primary CMB Bispectrum}
 \mycaption{The primary angular bispectrum (Eq.(\ref{eq:almspec})), 
 divided by
 the Gaunt integral, ${\cal G}_{l_1l_2l_3}^{m_1m_2m_3}$ (Eq.(\ref{eq:gaunt})).
 The bispectrum is plotted as a function of $l_3$ for 
 $(l_1,l_2)=$(9,11), (99,101), (199,201), and (499,501).
 Each panel plots a different 1-dimensional slice of the bispectrum. 
 The top panel plots 
 $l_2(l_2+1)l_3(l_3+1)\left<
 a_{l_1m_1}^{\rm NL} a_{l_2m_2}^{\rm L} a_{l_3m_3}^{\rm L}\right>f_{\rm NL}^{-1}
 \left({\cal G}_{l_1l_2l_3}^{m_1m_2m_3}\right)^{-1}/(2\pi)^2$,
 while the bottom panel plots
 $l_1(l_1+1)l_2(l_2+1)\left<
 a_{l_1m_1}^{\rm L} a_{l_2m_2}^{\rm L} a_{l_3m_3}^{\rm NL}\right>f_{\rm NL}^{-1}
 \left({\cal G}_{l_1l_2l_3}^{m_1m_2m_3}\right)^{-1}/(2\pi)^2$.
 Note that we have multiplied the bispectrum in each panel by a factor 
 of $10^{19}$.}
\label{fig:bispectrum}
\end{figure}

Looking at figure~\ref{fig:bl}, we find
$l^2b_l^{\rm L}\sim 2\times 10^{-9}$ and
$b_l^{\rm NL}f^{-1}_{\rm NL}\sim 10^{-10}\ {\rm Mpc^{-3}}$.
The most signal coming from the decoupling, 
the volume element at $\tau_*$ is 
$r_*^2\Delta r_*\sim (10^4)^2\times 10^2\ {\rm Mpc^3}$;
thus, we estimate an order of magnitude of the primary reduced
bispectrum (Eq.(\ref{eq:blprim})) as
%%%%%%%%%%%%%%%%%%%%%%%%%%%%%%%%%%%%%%%%%%%%%%%%%%%%%%%%%%%%%%%%%%
\begin{equation}
  \label{eq:orderest}
  b_{lll}^{\rm primary}\sim 
  l^{-4}
  \left[2 r_*^2\Delta r_*\left(l^2b_l^{\rm L}\right)^2
   b_l^{\rm NL}\times 3\right]
  \sim l^{-4}\times 2\times 10^{-17}f_{\rm NL}. 
\end{equation}
%%%%%%%%%%%%%%%%%%%%%%%%%%%%%%%%%%%%%%%%%%%%%%%%%%%%%%%%%%%%%%%%%%
Since $b_l^{\rm NL}f^{-1}_{\rm NL}\sim r_*^{-2}\delta(r-r_*)$ 
(see Eq.(\ref{eq:deltadelta})),
$r_*^2\Delta r_* b_l^{\rm NL}f^{-1}_{\rm NL}\sim 1$.
This rough estimate agrees with the numerical result below 
(figure~\ref{fig:bispectrum}).

Figure~\ref{fig:bispectrum} 
shows the integrated bispectrum (Eq.(\ref{eq:almspec}))
divided by the Gaunt integral, ${\cal G}_{l_1l_2l_3}^{m_1m_2m_3}$,
which is the reduced bispectrum, $b_{l_1l_2l_3}^{\rm primary}$.
While the bispectrum is a 3-d function, we show different 1-d slices of the 
bispectrum in this figure.
We plot 
$$l_2(l_2+1)l_3(l_3+1)
\left<a_{l_1m_1}^{\rm NL}a_{l_2m_2}^{\rm L}a_{l_3m_3}^{\rm L}\right>
\left({\cal G}_{l_1l_2l_3}^{m_1m_2m_3}\right)^{-1}/(2\pi)^2$$ as a 
function of $l_3$ in the top panel, while we plot
$$l_1(l_1+1)l_2(l_2+1)
\left<a_{l_1m_1}^{\rm L}a_{l_2m_2}^{\rm L}a_{l_3m_3}^{\rm NL}\right>
\left({\cal G}_{l_1l_2l_3}^{m_1m_2m_3}\right)^{-1}/(2\pi)^2$$
in the bottom panel.
We have multiplied each $b_{l}^{\rm L}(r)$ which contains $P_\Phi(k)$
by $l(l+1)/(2\pi)$ so that the Sachs--Wolfe plateau at $l_3\simlt 10$ 
is easily seen.
We have chosen $l_1$ and $l_2$ so as $(l_1,l_2)=(9,11),(99,101),(199,201)$,
and $(499,501)$.
We find that the $(l_1,l_2)=(199,201)$ mode,  
the first acoustic peak mode, has the largest signal in this 
family of parameters.
The top panel has a prominent first acoustic peak, and strongly damped
oscillations in the high-$l$ regime; the bottom panel also has a first peak, 
but damps more slowly.
Typical amplitude of the reduced bispectrum is
$l^4b^{\rm primary}_{lll}f^{-1}_{\rm NL}\sim 10^{-17}$, which agrees with
an order of magnitude estimate (Eq.(\ref{eq:orderest})).

%%%%% Figure 2: integrated bispectrum %%%%%
%%%%%%%%%%%%%%%%%%%%%%%%%%%%%%%%%%%%%%%%%%%%%%%%%%%%%%%%%%%%%%%%%%
% Figure 2
%%%%%%%%%%%%%%%%%%%%%%%%%%%%%%%%%%%%%%%%%%%%%%%%%%%%%%%%%%%%%%%%%%

Our formula (Eq.(\ref{eq:blprim})) and numerical results agree with 
\citet{Gan94} in the Sachs--Wolfe regime, where 
$g_{{\rm T}l}(k)\approx -j_l(kr_*)/3$, and
%%%%%%%%%%%%%%%%%%%%%%%%%%%%%%%%%%%%%%%%%%%%%%%%%%%%%%%%%%%%%%%%%%
\begin{equation}
  \label{eq:SWapp}
  b_{l_1l_2l_3}^{\rm primary}
  \approx
  -6f_{\rm NL}
  \left(C_{l_1}^{\rm SW}C_{l_2}^{\rm SW}+
        C_{l_1}^{\rm SW}C_{l_3}^{\rm SW}+
        C_{l_2}^{\rm SW}C_{l_3}^{\rm SW}\right)
  \qquad
  \mbox{(SW approximation)}.
\end{equation}
%%%%%%%%%%%%%%%%%%%%%%%%%%%%%%%%%%%%%%%%%%%%%%%%%%%%%%%%%%%%%%%%%%
Each term is in the same order as equation (\ref{eq:blprim}).
Here, $C_l^{\rm SW}$ is the CMB angular power spectrum in the Sachs--Wolfe
approximation,
%%%%%%%%%%%%%%%%%%%%%%%%%%%%%%%%%%%%%%%%%%%%%%%%%%%%%%%%%%%%%%%%%%
\begin{equation}
  \label{eq:clsw}
  C_l^{\rm SW}
  \equiv
  \frac2{9\pi}\int_0^\infty k^2 dk P_\Phi(k)j^2_l(kr_*).
\end{equation}
%%%%%%%%%%%%%%%%%%%%%%%%%%%%%%%%%%%%%%%%%%%%%%%%%%%%%%%%%%%%%%%%%%
In deriving equation (\ref{eq:SWapp}) from (\ref{eq:blprim}), 
we have approximated $b_l^{\rm NL}(r)$ (Eq.(\ref{eq:bNLr})) with
%%%%%%%%%%%%%%%%%%%%%%%%%%%%%%%%%%%%%%%%%%%%%%%%%%%%%%%%%%%%%%%%%%
\begin{equation}
  \label{eq:deltadelta}
  b_l^{\rm NL}(r)
  \approx
  \left(-\frac{f_{\rm NL}}3\right)
  \frac2{\pi}\int_0^\infty k^2 dk j_{l}(kr_*)j_l(kr)
  = -\frac{f_{\rm NL}}3 r_*^{-2}\delta(r-r_*).
\end{equation}
%%%%%%%%%%%%%%%%%%%%%%%%%%%%%%%%%%%%%%%%%%%%%%%%%%%%%%%%%%%%%%%%%%

The Sachs--Wolfe approximation (Eq.(\ref{eq:SWapp})) 
is valid only when $l_1$, $l_2$, and $l_3$ are all smaller than 
$\sim 10$, for which \citet{Gan94} gives $\sim -6\times 10^{-20}$ in
figure~\ref{fig:bispectrum}.
We stress again that the Sachs--Wolfe approximation 
gives a qualitatively different result from our full calculation 
(Eq.(\ref{eq:blprim})) at $l_i\simgt 10$. 
The full bispectrum does change the sign; the 
approximation never changes the sign because of using $C_l^{\rm SW}$.
The acoustic oscillation and the sign change 
are actually great advantages, when we try
to separate the primary bispectrum from various secondary bispectra.
We will study this point later.

\subsection{Primary skewness}

The skewness, $S_3$, given by
%%%%%%%%%%%%%%%%%%%%%%%%%%%%%%%%%%%%%%%%%%%%%%%%%%%%%%%%%%%%%%%%%%
\begin{equation}
  S_3\equiv \left<\left(\frac{\Delta T(\hat{\mathbf n})}{T}\right)^3\right>,
\end{equation}
%%%%%%%%%%%%%%%%%%%%%%%%%%%%%%%%%%%%%%%%%%%%%%%%%%%%%%%%%%%%%%%%%%
is the simplest statistic characterizing non-Gaussianity. 
We expand $S_3$ in terms of $B_{l_1l_2l_3}$ (Eq.(\ref{eq:blll})),
or $b_{l_1l_2l_3}$ (Eq.(\ref{eq:func})), as 
%%%%%%%%%%%%%%%%%%%%%%%%%%%%%%%%%%%%%%%%%%%%%%%%%%%%%%%%%%%%%%%%%%
\begin{eqnarray}
  \nonumber
  S_3
  &=&
  \frac1{4\pi}\sum_{l_1l_2l_3}
  \sqrt{
  \frac{\left(2l_1+1\right)\left(2l_2+1\right)\left(2l_3+1\right)}
        {4\pi}
        }
  \left(
  \begin{array}{ccc}
  l_1 & l_2 & l_3 \\ 0 & 0 & 0 
  \end{array}
  \right)
   B_{l_1l_2l_3}
   W_{l_1}W_{l_2}W_{l_3}\\
 \nonumber
  &=&
  \frac1{2\pi^2}\sum_{2\leq l_1l_2l_3}
  \left(l_1+\frac12\right)\left(l_2+\frac12\right)\left(l_3+\frac12\right)
  \left(
  \begin{array}{ccc}
  l_1 & l_2 & l_3 \\ 0 & 0 & 0 
  \end{array}
  \right)^2\\
  \label{eq:skewness}
   & &\times
   b_{l_1l_2l_3}
   W_{l_1}W_{l_2}W_{l_3},
\end{eqnarray}
%%%%%%%%%%%%%%%%%%%%%%%%%%%%%%%%%%%%%%%%%%%%%%%%%%%%%%%%%%%%%%%%%%
where $W_l$ is the experimental window function. 
We have used equation (\ref{eq:wigner*}) to replace 
$B_{l_1l_2l_3}$ by the reduced bispectrum, $b_{l_1l_2l_3}$, 
in the last equality. 
Since $l=0$ and $1$ modes are not observable, we have excluded them 
from the summation.
Throughout this chapter, we consider a single-beam window function,
$W_l=e^{-l(l+1)/(2\sigma_{\rm b}^2)}$, where 
$\sigma_{\rm b}={\rm FWHM}/\sqrt{8\ln2}$.
Since $\left(\begin{array}{ccc}l_1&l_2&l_3\\0&0&0
\end{array}\right)^2 b_{l_1l_2l_3}$ is symmetric under permutation of
indices, we change the way of summation as
%%%%%%%%%%%%%%%%%%%%%%%%%%%%%%%%%%%%%%%%%%%%%%%%%%%%%%%%%%%%%%%%%%
\begin{equation} 
  \label{eq:sumchange}
  \sum_{2\leq l_1l_2l_3}
  \longrightarrow
  6 \sum_{2\leq l_1\leq l_2\leq l_3}.
\end{equation}
%%%%%%%%%%%%%%%%%%%%%%%%%%%%%%%%%%%%%%%%%%%%%%%%%%%%%%%%%%%%%%%%%%
This reduces the number of summations by a factor of $\simeq 6$.
We will use this convention henceforth.

\begin{figure}
 \plotone{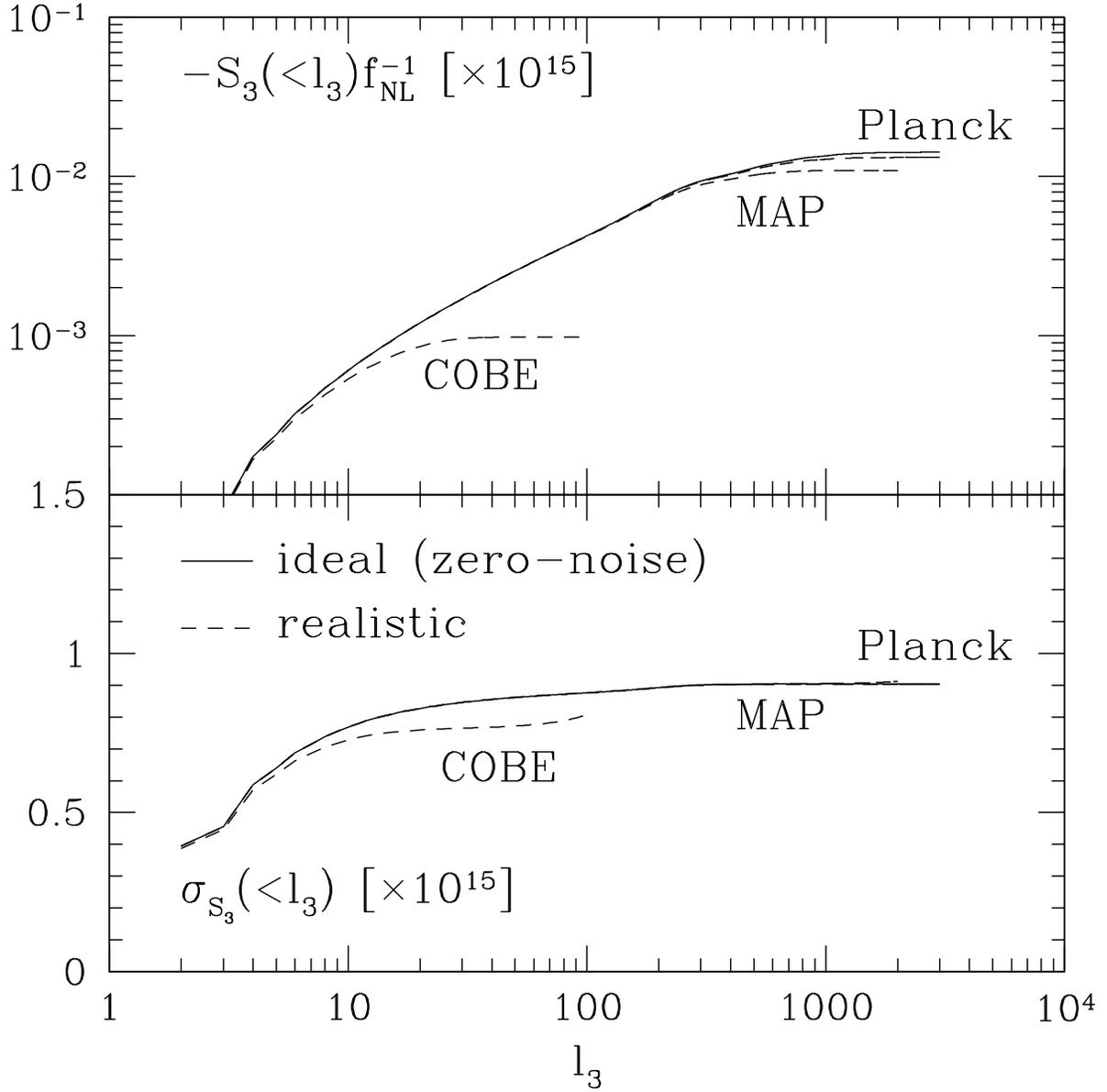}
 \caption{Primary Skewness}
 \mycaption{The top panel shows the primary CMB skewness 
 (Eq.(\ref{eq:skewness})) summed up to a certain $l_3$, 
 $-S_3(<l_3)f_{\rm NL}^{-1}\times 10^{15}$.
 The bottom panel shows the error of $S_3$ (Eq.(\ref{eq:skewvar})) 
 summed up to $l_3$, $\sigma_{S_3}(<l_3)\times 10^{15}$.
 The solid line represents the zero-noise ideal experiment, 
 while the dotted lines show
 {\it COBE}, {\it MAP}, and {\it Planck} experiments.}
\label{fig:skewness}
\end{figure}

The top panel of figure~\ref{fig:skewness} plots $S_3(<l_3)$, 
which is $S_3$ summed up to a certain $l_3$, for FWHM beam sizes 
of $7^\circ$, $13'$, and $5'\hspace{-2.5pt}.5$. 
These values correspond to {\it COBE}, {\it MAP}, and 
{\it Planck} beam sizes, respectively. 
Figure~\ref{fig:skewness} also plots the infinitesimally thin beam case.
We find that {\it MAP}, {\it Planck}, and the ideal experiments 
measure very similar $S_3$ to one another, despite the fact that 
{\it Planck} and the ideal experiments can use much more number of modes
than {\it MAP}. 
The reason is as follows.
Looking at equation (\ref{eq:skewness}), one finds that $S_3$
is a linear integral of $b_{l_1l_2l_3}$ over $l_i$; thus, integrating
oscillations in $b_{l_1l_2l_3}^{\rm primary}$ around zero 
(see figure~\ref{fig:bispectrum}) damps the non-Gaussian signal on small 
angular scales, $l\simgt 300$.
Since the Sachs--Wolfe effect, no oscillation, 
dominates the {\it COBE} scale anisotropy, the cancellation on the 
{\it COBE} scales affects $S_3$ less significantly than on the {\it MAP} 
and {\it Planck} scales. 
{\it Planck} suffers from severe cancellation in small angular scales:
{\it Planck} and the ideal experiments measure only 
the same amount of $S_3$ as {\it MAP} does.
As a result, measured $S_3$ almost saturates at the {\it MAP} resolution
scale, $l\sim 500$.

%%%%% Figure 3: skewness %%%%%
%%%%%%%%%%%%%%%%%%%%%%%%%%%%%%%%%%%%%%%%%%%%%%%%%%%%%%%%%%%%%%%%%%
% Figure 3
%%%%%%%%%%%%%%%%%%%%%%%%%%%%%%%%%%%%%%%%%%%%%%%%%%%%%%%%%%%%%%%%%%

We conclude this section by noting that 
when we can calculate the expected form of the bispectrum, then it becomes
a ``matched filter'' for detecting the non-Gaussianity in data,
and thus much more powerful tool than the skewness in which the information
is lost through the coarse-graining.

%%%%%%%%%%%%%%%%%%%%%%%%%%%%%%%%%%%%%%%%%%%%%%%%%%%%%%%%%%%%%%%%%%
\section{Secondary Sources of the CMB Bispectrum}
\label{sec:secondary}

Even if the CMB bispectrum were significantly detected in the CMB map, 
the origin would not necessarily be primordial, but rather
would be various secondary sources such as the Sunyaev--Zel'dovich (SZ) 
effect \citep{ZS69}, the weak lensing effect, and so on, or
foreground sources such as extragalactic radio sources.
To isolate the primordial origin from the others, we have to know the 
accurate form of bispectra produced by secondary and foreground sources.

\subsection{Coupling between the weak lensing and the 
Sunyaev--Zel'dovich effects}

The coupling between the SZ effect and the weak lensing
effect produces an observable effect in the bispectrum \citep{GS99,CH00}. 
We expand the CMB temperature field including the SZ and the lensing
effect as
%%%%%%%%%%%%%%%%%%%%%%%%%%%%%%%%%%%%%%%%%%%%%%%%%%%%%%%%%%%%%%%%%%
\begin{eqnarray}
 \nonumber
  \frac{\Delta T(\hat{\mathbf n})}T
  &=& \frac{\Delta T^{\rm P}\left(
                          \hat{\mathbf n}+\nabla\Theta(\hat{\mathbf n})
                    \right)}T
   +\frac{\Delta T^{\rm SZ}(\hat{\mathbf n})}T\\
 \label{eq:lenscouple}
  &\approx& \frac{\Delta T^{\rm P}(\hat{\mathbf n})}T
  +\nabla\left(\frac{\Delta T^{\rm P}(\hat{\mathbf n})}T\right)\cdot
   \nabla\Theta(\hat{\mathbf n})
  +\frac{\Delta T^{\rm SZ}(\hat{\mathbf n})}T,
\end{eqnarray}
%%%%%%%%%%%%%%%%%%%%%%%%%%%%%%%%%%%%%%%%%%%%%%%%%%%%%%%%%%%%%%%%%%
where P denotes the primary anisotropy, 
$\Theta(\hat{\mathbf n})$ is the lensing potential,
%%%%%%%%%%%%%%%%%%%%%%%%%%%%%%%%%%%%%%%%%%%%%%%%%%%%%%%%%%%%%%%%%%
\begin{equation}
 \Theta(\hat{\mathbf n})
  \equiv
  -2\int_0^{r_*} dr \frac{r_*-r}{rr_*}
  \Phi(r,\hat{\mathbf n}r),
\end{equation}
%%%%%%%%%%%%%%%%%%%%%%%%%%%%%%%%%%%%%%%%%%%%%%%%%%%%%%%%%%%%%%%%%%
and SZ denotes the SZ effect,
%%%%%%%%%%%%%%%%%%%%%%%%%%%%%%%%%%%%%%%%%%%%%%%%%%%%%%%%%%%%%%%%%%
\begin{equation}
  \label{eq:sz}
  \frac{\Delta T^{\rm SZ}(\hat{\mathbf n})}T
  =
  y(\hat{\mathbf n})j_\nu,
\end{equation}
%%%%%%%%%%%%%%%%%%%%%%%%%%%%%%%%%%%%%%%%%%%%%%%%%%%%%%%%%%%%%%%%%%
where $j_\nu$ is a spectral function of the SZ effect \citep{ZS69}.
$y(\hat{\mathbf n})$ is the Compton $y$-parameter given by
%%%%%%%%%%%%%%%%%%%%%%%%%%%%%%%%%%%%%%%%%%%%%%%%%%%%%%%%%%%%%%%%%%
\begin{equation}
  \label{eq:yparam}
  y(\hat{\mathbf n}) \equiv y_0\int \frac{dr}{r_*} 
  \frac{T_\rho(r,\hat{\mathbf n}r)}{\overline{T}_{\rho0}} a^{-2}(r),
\end{equation}
%%%%%%%%%%%%%%%%%%%%%%%%%%%%%%%%%%%%%%%%%%%%%%%%%%%%%%%%%%%%%%%%%%
where
%%%%%%%%%%%%%%%%%%%%%%%%%%%%%%%%%%%%%%%%%%%%%%%%%%%%%%%%%%%%%%%%%%
\begin{equation}
  \label{eq:y0}
  y_0\equiv \frac{\sigma_T \overline{\rho}_{\rm gas0} k_{\rm B} 
            \overline{T}_{\rho0} r_*}{\mu_e m_p m_e c^2} 
     = 4.3\times 10^{-4}\mu_e^{-1}\left(\Omega_{\rm b} h^2\right) 
       \left(\frac{k_{\rm B} \overline{T}_{\rho0}}{1~{\rm keV}}\right)
       \left(\frac{r_*}{10~{\rm Gpc}}\right).
\end{equation}
%%%%%%%%%%%%%%%%%%%%%%%%%%%%%%%%%%%%%%%%%%%%%%%%%%%%%%%%%%%%%%%%%%
$T_\rho\equiv \rho_{\rm gas} T_e/\overline{\rho}_{\rm gas}$ is the 
electron temperature weighted by the gas mass density, the overline denotes
the volume average, and the subscript 0 means the present epoch.
We adopt $\mu_e^{-1}=0.88$, where 
$\mu_e^{-1}\equiv n_e/(\rho_{\rm gas}/m_p)$ is the number of electrons 
per proton mass in the fully ionized medium.
Other quantities have their usual meanings.

Transforming equation (\ref{eq:lenscouple}) into harmonic space, we obtain 
%%%%%%%%%%%%%%%%%%%%%%%%%%%%%%%%%%%%%%%%%%%%%%%%%%%%%%%%%%%%%%%%%%
\begin{eqnarray}
  \nonumber
  a_{lm}
  &=&
  a_{lm}^{\rm P}
  +\sum_{l'm'}\sum_{l''m''}(-1)^m
  {\cal G}_{l l' l''}^{-m m' m''}\\
 \nonumber
  & &\times
  \frac{l'(l'+1)-l(l+1)+l''(l''+1)}2
  a_{l'm'}^{\rm P}\Theta_{l''m''}
  +a_{lm}^{\rm SZ}\\
 \nonumber
  &=&
  a_{lm}^{\rm P}
  +\sum_{l'm'}\sum_{l''m''}(-1)^{m+m'+m''}
  {\cal G}_{l l' l''}^{-m m' m''}\\
 \label{eq:almlens}
  & &\times
  \frac{l'(l'+1)-l(l+1)+l''(l''+1)}2
  a_{l'-m'}^{P*}\Theta^*_{l''-m''}
  +a_{lm}^{\rm SZ},
\end{eqnarray}
%%%%%%%%%%%%%%%%%%%%%%%%%%%%%%%%%%%%%%%%%%%%%%%%%%%%%%%%%%%%%%%%%%
where ${\cal G}_{l_1l_2l_3}^{m_1m_2m_3}$ is the Gaunt integral
(Eq.(\ref{eq:gaunt})).
Substituting equation (\ref{eq:almlens}) into (\ref{eq:blllmmm}),
and using the identity, ${\cal G}_{l_1l_2l_3}^{-m_1-m_2-m_3}
={\cal G}_{l_1l_2l_3}^{m_1m_2m_3}$, we obtain the bispectrum,
%%%%%%%%%%%%%%%%%%%%%%%%%%%%%%%%%%%%%%%%%%%%%%%%%%%%%%%%%%%%%%%%%%
\begin{eqnarray}
 \nonumber
  B_{l_1l_2l_3}^{m_1m_2m_3}
  &=&{\cal G}_{l_1 l_2 l_3}^{m_1 m_2 m_3}
  \left[
  \frac{l_1(l_1+1)-l_2(l_2+1)+l_3(l_3+1)}2
  C_{l_1}^{\rm P} \left<\Theta^*_{l_3m_3} a_{l_3m_3}^{\rm SZ}\right>\right.\\
  \label{eq:szlensbispec}
   & &
   \left. + \mbox{5 permutations}\right].
\end{eqnarray}
%%%%%%%%%%%%%%%%%%%%%%%%%%%%%%%%%%%%%%%%%%%%%%%%%%%%%%%%%%%%%%%%%%
The form of equation (\ref{eq:func}) is confirmed;
the reduced bispectrum $b_{l_1l_2l_3}^{\rm sz-lens}$ 
includes the terms in the square bracket.

\begin{figure}
\begin{center}
    \leavevmode\epsfxsize=9cm \epsfbox{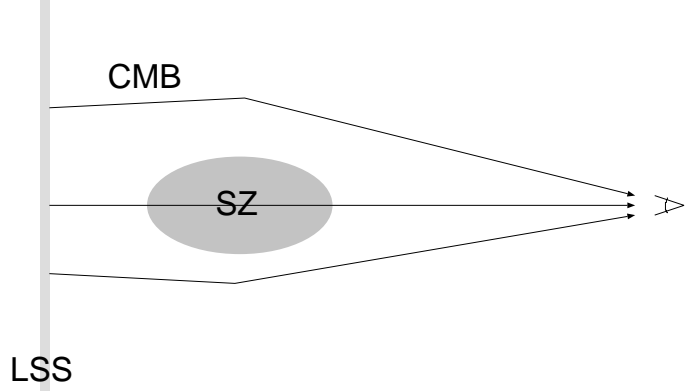}
\end{center}
 \caption{SZ--lensing Coupling}
 \mycaption{A schematic view of the SZ--lensing coupling bispectrum.
 One of the three CMB photons, which are decoupled at the last
 scattering surface (LSS), penetrates through a SZ cluster, 
 changing its temperature, and coming toward us.
 As the other two photons pass through near the SZ cluster, 
 they are deflected by the lensing effect, changing their propagation 
 directions, and coming toward us.
 As a result, the three photons are correlated, generating three-point
 correlation, the bispectrum.}
\label{fig:szlens}
\end{figure}

While equation~(\ref{eq:szlensbispec}) is complicated, we can understand
the physical effect producing the SZ--lensing bispectrum intuitively.
Figure~\ref{fig:szlens} shows how the SZ--lensing coupling 
produces the three-point correlation. 
Suppose that there are three CMB photons decoupled at the last 
scattering surface (LSS), and
one of these photons penetrates through a SZ cluster between LSS and us;
the energy of the photon changes because of the SZ effect.
When the other two photons pass through near the SZ cluster, they are
deflected by the gravitational lensing effect, changing their
propagation directions, and coming toward us.
What do we see after all?
We see that the three CMB photons are correlated; we measure
non-zero angular bispectrum.
The cross-correlation strength between the SZ and lensing
effects, $\left<\Theta^*_{l_3m_3} a_{l_3m_3}^{\rm SZ}\right>$,
thus determines the bispectrum amplitude, as indicated by 
equation~(\ref{eq:szlensbispec}).

%%%%%%%%%%%%%%%%%%%%%%%%%%%%%%%%%%%%%%%%%%%%%%%%%%%%%%%%%%%%%%%%%%
% Figure 3.5
%%%%%%%%%%%%%%%%%%%%%%%%%%%%%%%%%%%%%%%%%%%%%%%%%%%%%%%%%%%%%%%%%%

\citet{GS99} have derived $\left<\Theta^*_{lm} a_{lm}^{\rm SZ}\right>$, 
assuming the linear pressure bias model \citep{PSCO95},
$T_\rho=\overline{T}_\rho b_{\rm gas} \delta$,
and the mean temperature evolution,
$\overline{T}_\rho\simeq \overline{T}_{\rho0}(1+z)^{-1}$, 
for $z<2$, which is roughly suggested by recent hydrodynamic simulations
\citep{CO99,RKSP00,SWH01}.
They have obtained
%%%%%%%%%%%%%%%%%%%%%%%%%%%%%%%%%%%%%%%%%%%%%%%%%%%%%%%%%%%%%%%%%%
\begin{equation}
  \label{eq:blsz}
  \left<\Theta^*_{lm} a_{lm}^{\rm SZ}\right>
  \simeq
  -j_\nu\frac{4y_0b_{\rm gas}l^2}{3\Omega_{\rm m}H_0^2}
  \int_0^{z_*} dz \frac{dr}{dz}D^2(z)(1+z)^2
  \frac{r_*-r(z)}{r_*^2r^5(z)}
  P_\Phi\left(k=\frac{l}{r(z)}\right),
\end{equation}
%%%%%%%%%%%%%%%%%%%%%%%%%%%%%%%%%%%%%%%%%%%%%%%%%%%%%%%%%%%%%%%%%%
where $D(z)$ is the linear growth factor.
Simulations without non-gravitational heating \citep{RKSP00,SWH01} 
suggest that $\overline{T}_{\rho0}\sim 0.2-0.4~{\rm keV}$ and 
$b_{\rm gas}\sim 5-10$; analytic estimations give similar numbers 
\citep{RKSP00,ZP01}.
In the pressure bias model, free parameters except cosmological 
parameters are $\overline{T}_{\rho0}$ and $b_{\rm gas}$;  
however, both actually depend upon cosmological models \citep{RKSP00}.
Since 
$l^3\left<\Theta^*_{lm}a_{lm}^{\rm SZ}\right>\sim 
2\times 10^{-10}j_\nu \overline{T}_{\rho0}b_{\rm gas}$
 \citep{GS99,CH00} and $l^2C_l^{\rm P}\sim 6\times 10^{-10}$, we have
%%%%%%%%%%%%%%%%%%%%%%%%%%%%%%%%%%%%%%%%%%%%%%%%%%%%%%%%%%%%%%%%%%
\begin{equation}
  \label{eq:orderestszlens}
  b_{lll}^{\rm sz-lens}\sim 
  l^{-3}
  \left[
  \left(l^2C_l^{\rm P}\right)
  \left(l^3\left<\Theta^*_{lm} a_{lm}^{\rm SZ}\right>\right)\times 5/2\right]
  \sim 
  l^{-3}\times
  3\times 10^{-19}j_\nu\overline{T}_{\rho0}b_{\rm gas},
\end{equation}
%%%%%%%%%%%%%%%%%%%%%%%%%%%%%%%%%%%%%%%%%%%%%%%%%%%%%%%%%%%%%%%%%%
where $\overline{T}_{\rho0}$ is in units of 1~keV, and
$b_{l_1l_2l_3}=
B_{l_1l_2l_3}^{m_1m_2m_3}\left({\cal
G}_{l_1l_2l_3}^{m_1m_2m_3}\right)^{-1}$  
is the reduced bispectrum (Eq.(\ref{eq:func})).
Comparing this with equation (\ref{eq:orderest}), we obtain
%%%%%%%%%%%%%%%%%%%%%%%%%%%%%%%%%%%%%%%%%%%%%%%%%%%%%%%%%%%%%%%%%%
\begin{equation}
  \label{eq:ordercomp}
  \frac{b_{lll}^{\rm primary}}{b_{lll}^{\rm sz-lens}}  
  \sim l^{-1}\times 10
  \left(\frac{f_{\rm NL}}{j_\nu\overline{T}_{\rho0}b_{\rm gas}}\right).
\end{equation}
%%%%%%%%%%%%%%%%%%%%%%%%%%%%%%%%%%%%%%%%%%%%%%%%%%%%%%%%%%%%%%%%%%  
This estimate suggests that the SZ--lensing bispectrum overwhelms
the primary bispectrum in small angular scales.
This is why we have to separate the primary from the SZ--lensing effect.

While the pressure bias model gives a rough estimate of the SZ 
power spectrum, more accurate predictions exist.
Several authors have predicted the SZ power spectrum analytically using the 
Press-Schechter approach \citep{CK88,MS93,AM99,KK99,Coo00}
or the hyper-extended perturbation theory \citep{ZP01}.
The predictions agree with hydrodynamic simulations well
\citep{RKSP00,SBP01,SWH01,RT00,Kom01a}.
While a big uncertainty in the predictions has been lying in
a phenomenological model which describes the SZ surface brightness profile
of halos, \citet{KSel01} have proposed universal gas and temperature profiles, 
and predicted the SZ profile relying on the physical basis.
Hence, using those universal profiles, one can improve
the analytic prediction for the SZ power spectrum.
We expect that the universal profiles describe the SZ profile
in the average sense; on the individual halo-to-halo basis, there could be 
significant deviation from the universal profile, owing to
substructures in halos \citep{Kom01c}.

\subsection{Extragalactic radio and infrared sources}

The bispectrum from extragalactic radio and infrared sources 
whose fluxes, $F$, are smaller than a certain detection threshold, 
$F_{\rm d}$, is simple to estimate, when we assume the Poisson distribution.
\citet{tof98} have shown that the Poisson distribution is good 
approximation.
The Poisson distribution has the white noise spectrum; thus, 
the reduced bispectrum (Eq.(\ref{eq:func})) is constant,
$b_{l_1l_2l_3}^{\rm ps}=b^{\rm ps}={\rm constant}$, and we obtain
%%%%%%%%%%%%%%%%%%%%%%%%%%%%%%%%%%%%%%%%%%%%%%%%%%%%%%%%%%%%%%%%%%
\begin{equation}
  \label{eq:pointsource}
  B_{l_1l_2l_3}^{m_1m_2m_3}
  ={\cal G}_{l_l1_2l_3}^{m_1m_2m_3} b^{\rm ps},
\end{equation}
%%%%%%%%%%%%%%%%%%%%%%%%%%%%%%%%%%%%%%%%%%%%%%%%%%%%%%%%%%%%%%%%%%
where 
%%%%%%%%%%%%%%%%%%%%%%%%%%%%%%%%%%%%%%%%%%%%%%%%%%%%%%%%%%%%%%%%%%
\begin{equation}
  \label{eq:Bps}
  b^{\rm ps}(<F_{\rm d})
  \equiv g^3(x)
  \int_0^{F_{\rm d}} dF F^3 \frac{dn}{dF}
  =
  g^3(x)\frac{n(>F_{\rm d})}{3-\beta}F_{\rm d}^3.
\end{equation}
%%%%%%%%%%%%%%%%%%%%%%%%%%%%%%%%%%%%%%%%%%%%%%%%%%%%%%%%%%%%%%%%%%
Here, $dn/dF$ is the differential source count per unit solid angle, and 
$n(>F_{\rm d})\equiv \int_{F_{\rm d}}^\infty dF (dn/dF)$. 
We have assumed a power-law count, $dn/dF\propto F^{-\beta-1}$, 
for $\beta<2$.
The other symbols mean $x\equiv h\nu/k_{\rm B} T \simeq 
(\nu/56.80~{\rm GHz})(T/2.726~{\rm K})^{-1}$, and
%%%%%%%%%%%%%%%%%%%%%%%%%%%%%%%%%%%%%%%%%%%%%%%%%%%%%%%%%%%%%%%%%%
\begin{equation}
  \label{eq:gx}
  g(x)\equiv
  2\frac{(hc)^2}{(k_{\rm B} T)^3}\left(\frac{\sinh x/2}{x^2}\right)^2
  \simeq
  \frac1{67.55~{\rm MJy~sr^{-1}}}\left(\frac{T}{2.726~{\rm K}}\right)^{-3}
  \left(\frac{\sinh x/2}{x^2}\right)^2.
\end{equation}
%%%%%%%%%%%%%%%%%%%%%%%%%%%%%%%%%%%%%%%%%%%%%%%%%%%%%%%%%%%%%%%%%%

Using the Poisson angular power spectrum, $C^{\rm ps}$, given by
%%%%%%%%%%%%%%%%%%%%%%%%%%%%%%%%%%%%%%%%%%%%%%%%%%%%%%%%%%%%%%%%%%
\begin{equation}
  \label{eq:Cps}
  C^{\rm ps}(<F_{\rm d})
  \equiv g^2(x)
  \int_0^{F_{\rm d}} dF F^2 \frac{dn}{dF}
  =
  g^2(x)\frac{n(>F_{\rm d})}{2-\beta}F_{\rm d}^2,
\end{equation}
%%%%%%%%%%%%%%%%%%%%%%%%%%%%%%%%%%%%%%%%%%%%%%%%%%%%%%%%%%%%%%%%%%
we can rewrite $b^{\rm ps}$ into a different form, 
%%%%%%%%%%%%%%%%%%%%%%%%%%%%%%%%%%%%%%%%%%%%%%%%%%%%%%%%%%%%%%%%%%
\begin{equation}
  \label{eq:Bps*}
  b^{\rm ps}(<F_{\rm d})
  = \frac{(2-\beta)^{3/2}}{3-\beta}\left[n(>F_{\rm d})\right]^{-1/2}
  \left[C^{\rm ps}(<F_{\rm d})\right]^{3/2}.
\end{equation}
%%%%%%%%%%%%%%%%%%%%%%%%%%%%%%%%%%%%%%%%%%%%%%%%%%%%%%%%%%%%%%%%%%

\citet{tof98} have estimated $n(>F_{\rm d})\sim 300~{\rm sr^{-1}}$ 
for $F_{\rm d}\sim 0.2~{\rm Jy}$ at 217~GHz. 
This $F_{\rm d}$ corresponds to $5\sigma$ detection
threshold for the {\it Planck} experiment at 217~GHz.
\citet{RSH00} have extrapolated their estimation to 94~GHz, finding 
$n(>F_{\rm d})\sim 7~{\rm sr^{-1}}$ for $F_{\rm d}\sim 2~{\rm Jy}$,
which corresponds to {\it MAP} $5\sigma$ threshold.
These values yield
%%%%%%%%%%%%%%%%%%%%%%%%%%%%%%%%%%%%%%%%%%%%%%%%%%%%%%%%%%%%%%%%%%
\begin{eqnarray}
  \label{eq:cl90ghz}
  C^{\rm ps}(90~{\rm GHz},<2~{\rm Jy})&\sim& 2\times 10^{-16},\\
  \label{eq:cl217ghz}
  C^{\rm ps}(217~{\rm GHz},<0.2~{\rm Jy})&\sim& 1\times 10^{-17}.
\end{eqnarray}
%%%%%%%%%%%%%%%%%%%%%%%%%%%%%%%%%%%%%%%%%%%%%%%%%%%%%%%%%%%%%%%%%%
Thus, rough estimates for $b^{\rm ps}$ are
%%%%%%%%%%%%%%%%%%%%%%%%%%%%%%%%%%%%%%%%%%%%%%%%%%%%%%%%%%%%%%%%%%
\begin{eqnarray}
  \label{eq:bl90ghz}
  b^{\rm ps}(90~{\rm GHz},<2~{\rm Jy})&\sim& 2\times 10^{-25},\\
  \label{eq:bl217ghz}
  b^{\rm ps}(217~{\rm GHz},<0.2~{\rm Jy})&\sim& 5\times 10^{-28}.
\end{eqnarray}
%%%%%%%%%%%%%%%%%%%%%%%%%%%%%%%%%%%%%%%%%%%%%%%%%%%%%%%%%%%%%%%%%%
While we have assumed the Euclidean source count ($\beta=3/2$) for 
definiteness, this assumption does not affect an order of magnitude
estimate here.

As the primary reduced bispectrum is $\propto l^{-4}$ (Eq.(\ref{eq:orderest})),
and the SZ--lensing reduced bispectrum is $\propto l^{-3}$
(Eq.(\ref{eq:orderestszlens})), the point-source bispectrum rapidly
becomes to dominate the total bispectrum in small angular scales:
%%%%%%%%%%%%%%%%%%%%%%%%%%%%%%%%%%%%%%%%%%%%%%%%%%%%%%%%%%%%%%%%%%
\begin{eqnarray}
  \label{eq:ordercomp*}
  \frac{b_{lll}^{\rm primary}}{b^{\rm ps}}
  &\sim& l^{-4}\times 10^7
  \left(\frac{f_{\rm NL}}{b^{\rm ps}/10^{-25}}\right),\\
  \label{eq:ordercomp**}
  \frac{b_{lll}^{\rm sz-lens}}
       {b^{\rm ps}}
  &\sim& l^{-3}\times 10^6
  \left(\frac{j_\nu\overline{T}_{\rho0}b_{\rm gas}}
             {b^{\rm ps}/10^{-25}}\right).
\end{eqnarray}
%%%%%%%%%%%%%%%%%%%%%%%%%%%%%%%%%%%%%%%%%%%%%%%%%%%%%%%%%%%%%%%%%%  
For example, the point sources overwhelm the SZ--lensing bispectrum 
measured by {\it MAP} at $l\simgt 100$.

\begin{figure}
 \plotone{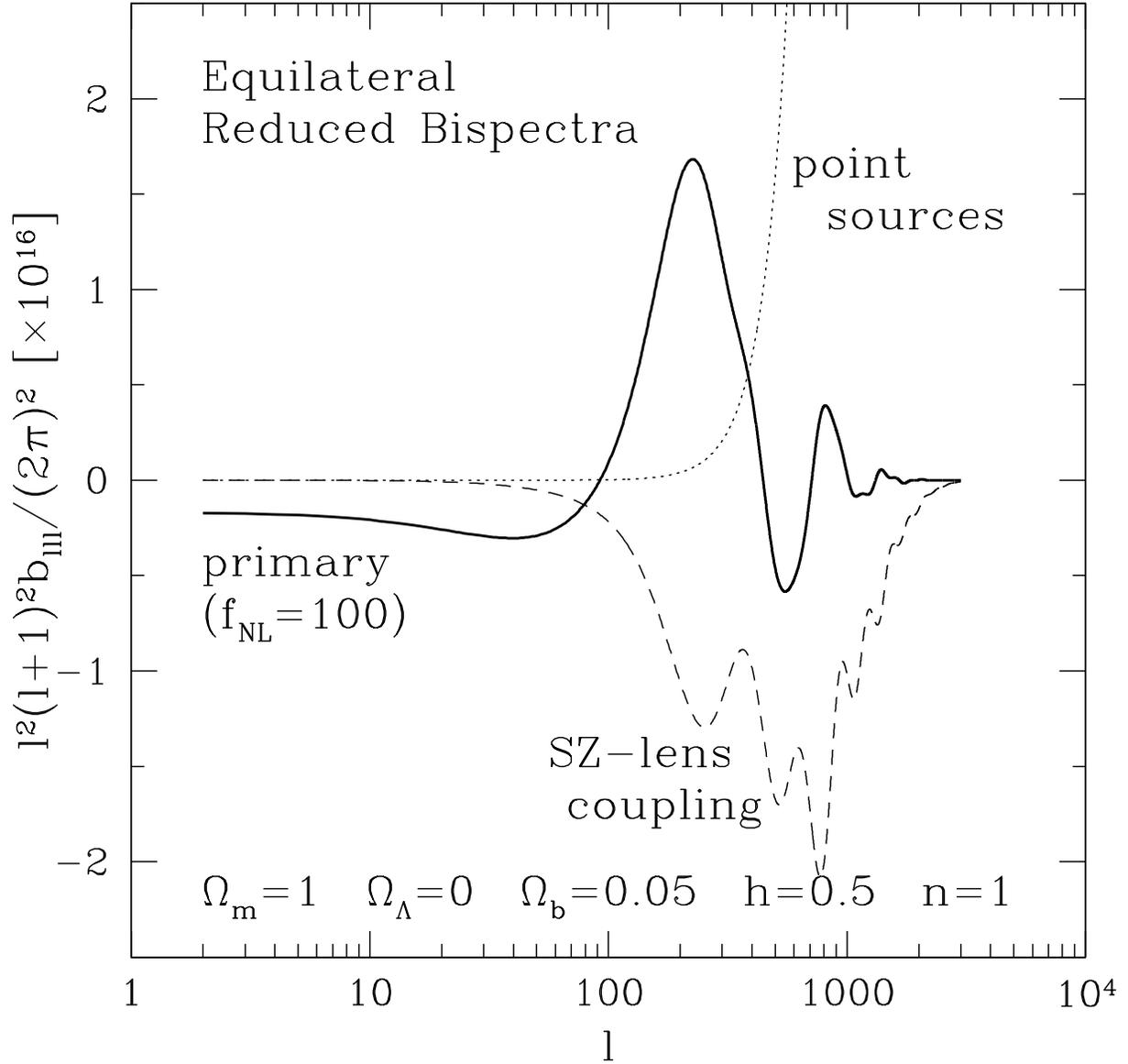}
 \caption{Equilateral Reduced Bispectra}
 \mycaption{Comparison between the primary (solid line), 
 the SZ--lensing (dashed line), and the point-source (dotted line) 
 reduced bispectra for the equilateral configurations, $l\equiv l_1=l_2=l_3$.
 We have plotted $\left[l^2(l+1)^2b_{lll}/(2\pi)^2\right]\times 10^{16}$,
 which makes the Sachs--Wolfe plateau of the primary reduced bispectrum
 on large angular scales, $l\simlt 10$, easily seen.}
\label{fig:reducedb}
\end{figure}

What the SZ--lensing bispectrum and the point-source bispectrum
look like?
Figure~\ref{fig:reducedb} plots the primary, the SZ--lensing,
and the point-source reduced bispecta for the equilateral configurations,
$l\equiv l_1=l_2=l_3$.
We have plotted $l^2(l+1)^2b_{lll}/(2\pi)^2$.
We find that these bispecra are very different from each other in shape 
on small angular scales.
It thus suggests that we can separate these three contributions
on the basis of shape difference. 
We study this point in the next section.

%%%%%%%%%%%%%%%%%%%%%%%%%%%%%%%%%%%%%%%%%%%%%%%%%%%%%%%%%%%%%%%%%%
%%%%%%%%%%%%%%%%%%%%%%%%%%%%%%%%%%%%%%%%%%%%%%%%%%%%%%%%%%%%%%%%%%

%%%%%%%%%%%%%%%%%%%%%%%%%%%%%%%%%%%%%%%%%%%%%%%%%%%%%%%%%%%%%%%%%%
\section{Measuring Bispectra}
\label{sec:measure}

\subsection{Fisher matrix}

In this section, we study how well we can measure the 
primary bispectrum, and how well
we can separate it from the secondary bispectra.
Suppose that we fit the observed bispectrum, $B_{l_1l_2l_3}^{\rm obs}$,
by theoretically calculated bispectra, which include both the primary 
and secondary sources.
We minimize $\chi^2$ defined by
%%%%%%%%%%%%%%%%%%%%%%%%%%%%%%%%%%%%%%%%%%%%%%%%%%%%%%%%%%%%%%%%%%
\begin{equation}
%  \label{eq:chisq}
  \chi^2
  \equiv 
  \sum_{2\leq l_1\leq l_2\leq l_3}
  \frac{\left(B_{l_1l_2l_3}^{\rm obs}
	     -\sum_i A_i B^{(i)}_{l_1l_2l_3}\right)^2}
  {\sigma^2_{l_1l_2l_3}},
\end{equation}
%%%%%%%%%%%%%%%%%%%%%%%%%%%%%%%%%%%%%%%%%%%%%%%%%%%%%%%%%%%%%%%%%%
where $i$ denotes a component such as the primary, the SZ and 
lensing effects, extragalactic sources, and so on. 
We have removed unobservable modes, $l=0$ and $1$.

As we have shown in chapter~\ref{chap:spectrum}, the variance of the 
bispectrum, $\sigma^2_{l_1l_2l_3}$, is the six-point function of 
$a_{lm}$ \citep{Luo94,heavens98}.
When non-Gaussianity is weak, 
we calculate it as \citep{SG99,GM00}
%%%%%%%%%%%%%%%%%%%%%%%%%%%%%%%%%%%%%%%%%%%%%%%%%%%%%%%%%%%%%%%%%%
\begin{equation}
  \sigma^2_{l_1l_2l_3}
  \equiv \left<B_{l_1l_2l_3}^2\right>-\left<B_{l_1l_2l_3}\right>^2
  \approx
  {\cal C}_{l_1}{\cal C}_{l_2}{\cal C}_{l_3}\Delta_{l_1l_2l_3},
\end{equation}
%%%%%%%%%%%%%%%%%%%%%%%%%%%%%%%%%%%%%%%%%%%%%%%%%%%%%%%%%%%%%%%%%% 
where $\Delta_{l_1l_2l_3}$ takes values 1, 2, or 6 when
all $l$'s are different, two are same, or all are same, respectively. 
${\cal C}_l\equiv C_l+C_l^{\rm N}$ is the total CMB angular power spectrum,
which includes the power spectrum of the detector noise, $C_l^{\rm N}$.
We calculate $C_l^{\rm N}$ analytically following 
\citet{Knox95} with the noise characteristics of relevant experiments.
We do not include $C_l$ from secondary sources, as they are subdominant 
compared with the primary $C_l$ and $C_l^{\rm N}$ for relevant experiments.
Including $C_l$ from extragalactic sources (Eqs.(\ref{eq:cl90ghz}) 
or (\ref{eq:cl217ghz})) changes our results less than 10\%.

Taking $\partial\chi^2/\partial A_i=0$, we obtain normal equation,
%%%%%%%%%%%%%%%%%%%%%%%%%%%%%%%%%%%%%%%%%%%%%%%%%%%%%%%%%%%%%%%%%%
\begin{equation}
  \label{eq:fiseq}
  \sum_j
  \left[\sum_{2\leq l_1\leq l_2\leq l_3}
        \frac{B_{l_1l_2l_3}^{(i)}B_{l_1l_2l_3}^{(j)}}{\sigma_{l_1l_2l_3}^2}
  \right]A_j
  =
  \sum_{2\leq l_1\leq l_2\leq l_3}
     \frac{B_{l_1l_2l_3}^{\rm obs}B_{l_1l_2l_3}^{(i)}}{\sigma_{l_1l_2l_3}^2}.
\end{equation}
%%%%%%%%%%%%%%%%%%%%%%%%%%%%%%%%%%%%%%%%%%%%%%%%%%%%%%%%%%%%%%%%%%
We then define the Fisher matrix, $F_{ij}$, as
%%%%%%%%%%%%%%%%%%%%%%%%%%%%%%%%%%%%%%%%%%%%%%%%%%%%%%%%%%%%%%%%%%
\begin{eqnarray}
 \nonumber
  F_{ij}&\equiv& 
  \sum_{2\leq l_1\leq l_2\leq l_3}
  \frac{B_{l_1l_2l_3}^{(i)}B_{l_1l_2l_3}^{(j)}}{\sigma_{l_1l_2l_3}^2}\\
 \label{eq:fis}
  &=&
  \frac{2}{\pi}\sum_{2\leq l_1\leq l_2\leq l_3}
  \left(l_1+\frac12\right)\left(l_2+\frac12\right)\left(l_3+\frac12\right)
   \left(\begin{array}{ccc}l_1&l_2&l_3\\0&0&0\end{array}\right)^2
  \frac{b_{l_1l_2l_3}^{(i)}b_{l_1l_2l_3}^{(j)}}{\sigma_{l_1l_2l_3}^2},
\end{eqnarray}
%%%%%%%%%%%%%%%%%%%%%%%%%%%%%%%%%%%%%%%%%%%%%%%%%%%%%%%%%%%%%%%%%%
where we have used equation (\ref{eq:wigner*}) to replace $B_{l_1l_2l_3}$ by
the reduced bispectrum, $b_{l_1l_2l_3}$ (see Eq.(\ref{eq:func}) for
definition). 
Since the covariance matrix of $A_i$ is $F_{ij}^{-1}$,
we define the signal-to-noise ratio, $(S/N)_i$, for a component $i$, 
the correlation coefficient, $r_{ij}$, between different components 
$i$ and $j$, and the degradation parameter, $d_i$, of $(S/N)_i$ 
due to $r_{ij}$, as
%%%%%%%%%%%%%%%%%%%%%%%%%%%%%%%%%%%%%%%%%%%%%%%%%%%%%%%%%%%%%%%%%%
\begin{eqnarray}
  \label{eq:sn}
  \left(\frac{S}{N}\right)_i &\equiv& \frac1{\sqrt{F_{ii}^{-1}}},\\
  \label{eq:r}
  r_{ij}&\equiv& 
  \frac{F_{ij}^{-1}}{\sqrt{F^{-1}_{ii}F^{-1}_{jj}}},\\
  \label{eq:d}
  d_i&\equiv& F_{ii}F_{ii}^{-1}.
\end{eqnarray}
%%%%%%%%%%%%%%%%%%%%%%%%%%%%%%%%%%%%%%%%%%%%%%%%%%%%%%%%%%%%%%%%%%
Note that $r_{ij}$ does not depend upon the amplitude of bispectra,
but the shape. 
We have defined $d_i$ so as $d_i=1$ for zero degradation,
while $d_i>1$ for degraded $(S/N)_i$.
\citet{SG99} and \citet{CH00} have considered the diagonal component 
of $F_{ij}^{-1}$.
We study all the components to study the separatability 
between various bispectra.

We estimate an order of magnitude of $S/N$ as a function of 
a certain angular resolution, $l$, as follows.
Since the number of modes contributing to $S/N$ increases as 
$l^{3/2}$, and 
$l^3\left(\begin{array}{ccc}l&l&l\\0&0&0\end{array}\right)^2
\sim 0.36\times l$, we estimate $(S/N)_i\sim (F_{ii})^{1/2}$ as
%%%%%%%%%%%%%%%%%%%%%%%%%%%%%%%%%%%%%%%%%%%%%%%%%%%%%%%%%%%%%%%%%%
\begin{equation}
  \label{eq:snorder}
  \left(\frac{S}{N}\right)_i
  \sim 
  \frac1{3\pi} l^{3/2}
  \times l^{3/2}
  \left|
  \left(\begin{array}{ccc}l&l&l\\0&0&0\end{array}\right)
  \right|\times
   \frac{l^3b_{lll}^{(i)}}
        {(l^2 C_l)^{3/2}}
  \sim l^5b_{lll}^{(i)}\times 4\times 10^{12},
\end{equation}
%%%%%%%%%%%%%%%%%%%%%%%%%%%%%%%%%%%%%%%%%%%%%%%%%%%%%%%%%%%%%%%%%%
where we have used $l^2C_l\sim 6\times 10^{-10}$.

Table~\ref{tab:fis} tabulates $F_{ij}$, while table~\ref{tab:invfis}
tabulates $F_{ij}^{-1}$; table~\ref{tab:sn} tabulates $(S/N)_i$, while 
table~\ref{tab:corr} tabulates $d_i$ in the diagonal, and $r_{ij}$ 
in the off-diagonal parts.

%%%%% Table I: F_ij %%%%%
%%%%%%%%%%%%%%%%%%%%%%%%%%%%%%%%%%%%%%%%%%%%%%%%%%%%%%%%%%%%%%%%%%%%%%%%%%%
% Table 1
\begin{table}
 \caption{Fisher Matrix}
 \mycaption{Fisher matrix, $F_{ij}$ (Eq.(\ref{eq:fis})).
 $i$ denotes a component in the first row; 
 $j$ denotes a component in the first column.
 $\overline{T}_{\rho0}$ is in units of 1 keV, 
 $b^{\rm ps}_{25}\equiv b^{\rm ps}/10^{-25}$, and
 $b^{\rm ps}_{27}\equiv b^{\rm ps}/10^{-27}$.}
\label{tab:fis}
\begin{center} 
  \begin{tabular}{cccc}\hline\hline
  {\it COBE} & primary & SZ--lensing & point sources \\
  \hline 
  primary &
  $4.2\times 10^{-6}~f_{\rm NL}^2$ & 
  $-4.0\times 10^{-7}~f_{\rm NL} j_\nu \overline{T}_{\rho0} b_{\rm gas}$ &
  $-1.0\times 10^{-9}~f_{\rm NL} b^{\rm ps}_{25}$ \\
  SZ--lensing &
  & 
  $1.3\times 10^{-7}~(j_\nu \overline{T}_{\rho0} b_{\rm gas})^2$ &
  $3.1\times 10^{-10}~j_\nu \overline{T}_{\rho0} b_{\rm gas} b^{\rm ps}_{25}$ \\
  point sources&
  & &
  $1.1\times 10^{-12}~(b^{\rm ps}_{25})^2$ \\
  \hline
  {\it MAP} & & & \\
  \hline 
  primary &
  $3.4\times 10^{-3}~f_{\rm NL}^2$ & 
  $2.6\times 10^{-3}~f_{\rm NL} j_\nu \overline{T}_{\rho0} b_{\rm gas}$ &
  $2.4\times 10^{-3}~f_{\rm NL} b^{\rm ps}_{25}$ \\
  SZ--lensing &
  & 
  $0.14~(j_\nu \overline{T}_{\rho0} b_{\rm gas})^2$ &
  $0.31~j_\nu \overline{T}_{\rho0} b_{\rm gas} b^{\rm ps}_{25}$ \\
  point sources&
  & &
  $5.6~(b^{\rm ps}_{25})^2$ \\
  \hline
  {\it Planck} & & & \\
  \hline 
  primary &
  $3.8\times 10^{-2}~f_{\rm NL}^2$ & 
  $7.2\times 10^{-2}~f_{\rm NL} j_\nu \overline{T}_{\rho0} b_{\rm gas}$ &
  $1.6\times 10^{-2}~f_{\rm NL} b^{\rm ps}_{27}$ \\
  SZ--lensing &
  & 
  $39~(j_\nu \overline{T}_{\rho0} b_{\rm gas})^2$ &
  $5.7~j_\nu \overline{T}_{\rho0} b_{\rm gas} b^{\rm ps}_{27}$ \\
  point sources&
  & &
  $2.7\times 10^3~(b^{\rm ps}_{27})^2$\\
   \hline\hline
  \end{tabular}
\end{center}
\end{table}
%%%%%%%%%%%%%%%%%%%%%%%%%%%%%%%%%%%%%%%%%%%%%%%%%%%%%%%%%%%%%%%%%%%%%%%%%%

%%%%% Table II: F_ij^-1 %%%%%
%%%%%%%%%%%%%%%%%%%%%%%%%%%%%%%%%%%%%%%%%%%%%%%%%%%%%%%%%%%%%%%%%%%%%%%%%%%
% Table 2
\begin{table}
 \caption{Inverted Fisher Matrix}
 \mycaption{Inverted Fisher matrix, $F_{ij}^{-1}$.
 The meaning of the symbols is the same as in table~\ref{tab:fis}.}
\label{tab:invfis}
\begin{center} 
 \begin{tabular}{cccc}\hline\hline
  {\it COBE} & primary & SZ--lensing & point sources \\
  \hline 
  primary &
  $3.5\times 10^{5}~f_{\rm NL}^{-2}$ & 
  $1.1\times 10^{6}~(f_{\rm NL} j_\nu \overline{T}_{\rho0} b_{\rm gas})^{-1}$ &
  $1.3\times 10^7~(f_{\rm NL} b^{\rm ps}_{25})^{-1}$ \\
  SZ--lensing &
  & 
  $3.1\times 10^7~(j_\nu \overline{T}_{\rho0} b_{\rm gas})^{-2}$ &
  $-7.8\times 10^9~(j_\nu \overline{T}_{\rho0} b_{\rm gas} b^{\rm ps}_{25})^{-1}$ \\
  point sources&
  & &
  $3.1\times 10^{12}~(b^{\rm ps}_{25})^{-2}$ \\
  \hline
  {\it MAP} & & & \\
  \hline 
  primary &
  $3.0\times 10^2~f_{\rm NL}^{-2}$ & 
  $-6.1~(f_{\rm NL} j_\nu \overline{T}_{\rho0} b_{\rm gas})^{-1}$ &
  $0.21~(f_{\rm NL} b^{\rm ps}_{25})^{-1}$ \\
  SZ--lensing &
  & 
  $8.4~(j_\nu \overline{T}_{\rho0} b_{\rm gas})^{-2}$ &
  $-0.46~(j_\nu \overline{T}_{\rho0} b_{\rm gas} b^{\rm ps}_{25})^{-1}$ \\
  point sources&
  & &
  $0.21~(b^{\rm ps}_{25})^{-2}$ \\
  \hline
  {\it Planck} & & & \\
  \hline 
  primary &
  $26~f_{\rm NL}^{-2}$ & 
  $-4.9\times 10^{-2}~(f_{\rm NL} j_\nu \overline{T}_{\rho0} b_{\rm gas})^{-1}$ &
  $-5.7\times 10^{-5}~(f_{\rm NL} b^{\rm ps}_{27})^{-1}$ \\
  SZ--lensing &
  & 
  $2.6\times 10^{-2}~(j_\nu \overline{T}_{\rho0} b_{\rm gas})^{-2}$ &
  $-5.4\times 10^{-5}~(j_\nu \overline{T}_{\rho0} b_{\rm gas} b^{\rm ps}_{27})^{-1}$ \\
  point sources&
  & &
  $3.7\times 10^{-4}~(b^{\rm ps}_{27})^{-2}$\\
  \hline\hline
 \end{tabular}
\end{center}
\end{table}
%%%%%%%%%%%%%%%%%%%%%%%%%%%%%%%%%%%%%%%%%%%%%%%%%%%%%%%%%%%%%%%%%%%%%%%%%%

%%%%% Table III: (S/N)_i %%%%%
%%%%%%%%%%%%%%%%%%%%%%%%%%%%%%%%%%%%%%%%%%%%%%%%%%%%%%%%%%%%%%%%%%%%%%%%%%%
% Table 3
\begin{table}
 \caption{Signal-to-noise Ratio}
 \mycaption{Signal-to-noise ratio, $(S/N)_i$ (Eq.(\ref{eq:sn})), of 
 detecting the bispectrum.
 $i$ denotes a component in the first row.
 The meaning of the symbols is the same as in table~\ref{tab:fis}.}
\label{tab:sn}
\begin{center} 
  \begin{tabular}{cccc}\hline\hline
       & primary & SZ--lensing & point sources \\
  \hline 
  {\it COBE} & $1.7\times 10^{-3}~f_{\rm NL}$ 
       & $1.8\times 10^{-4}~\left|j_\nu\right| \overline{T}_{\rho0}b_{\rm gas}$
       & $5.7\times 10^{-7}~b_{25}^{\rm ps}$ \\
  {\it MAP}  & $5.8\times 10^{-2}~f_{\rm NL}$ 
       & $0.34~\left|j_\nu\right| \overline{T}_{\rho0}b_{\rm gas}$
       & $2.2~b_{25}^{\rm ps}$ \\
  {\it Planck} & $0.19~f_{\rm NL}$ 
         & $6.2~\left|j_\nu\right| \overline{T}_{\rho0}b_{\rm gas}$
         & $52~b_{27}^{\rm ps}$\\
   \hline\hline
  \end{tabular}
\end{center}
\end{table}
%%%%%%%%%%%%%%%%%%%%%%%%%%%%%%%%%%%%%%%%%%%%%%%%%%%%%%%%%%%%%%%%%%%%%%%%%%%

%%%%% Table IV: d_i and r_ij %%%%%
%%%%%%%%%%%%%%%%%%%%%%%%%%%%%%%%%%%%%%%%%%%%%%%%%%%%%%%%%%%%%%%%%%%%%%%%%%%
% Table 4
\begin{table}
\caption{Signal Degradation and Correlation Matrix}
 \mycaption{Signal degradation parameter, $d_i$ (Eq.(\ref{eq:d})), and 
 correlation coefficient, $r_{ij}$ (Eq.(\ref{eq:r})), matrix. 
 $i$ denotes a component in the first row; 
 $j$ denotes a component in the first column.
 $d_i$ for $i=j$, while $r_{ij}$ for $i\neq j$.}
\label{tab:corr}
\begin{center} 
  \begin{tabular}{cccc}\hline\hline
  {\it COBE} & primary & SZ--lensing & point sources \\
  \hline 
  primary &
  $1.46$ & 
  $0.33~{\rm sgn}(j_\nu)$ &
  $1.6\times 10^{-2}$ \\
  SZ--lensing &
  & 
  $3.89$ &
  $-0.79~{\rm sgn}(j_\nu)$ \\
  point sources&
  & &
  $3.45$ \\
  \hline
  {\it MAP} & & & \\
  \hline 
  primary &
  $1.01$ & 
  $-0.12~{\rm sgn}(j_\nu)$ &
  $2.7\times 10^{-2}$ \\
  SZ--lensing &
  & 
  $1.16$ &
  $-0.35~{\rm sgn}(j_\nu)$ \\
  point sources&
  & &
  $1.14$ \\
  \hline
  {\it Planck} & & & \\
  \hline 
  primary &
  $1.00$ & 
  $-5.9\times 10^{-2}~{\rm sgn}(j_\nu)$ &
  $-5.8\times 10^{-4}$ \\
  SZ--lensing &
  & 
  $1.00$ &
  $-1.8\times 10^{-2}~{\rm sgn}(j_\nu)$ \\
  point sources&
  & &
  $1.00$\\
   \hline\hline
  \end{tabular}
\end{center}
\end{table}
%%%%%%%%%%%%%%%%%%%%%%%%%%%%%%%%%%%%%%%%%%%%%%%%%%%%%%%%%%%%%%%%%%%%%%%%%%

\subsection{Measuring primary bispectrum}

\begin{figure}
 \plotone{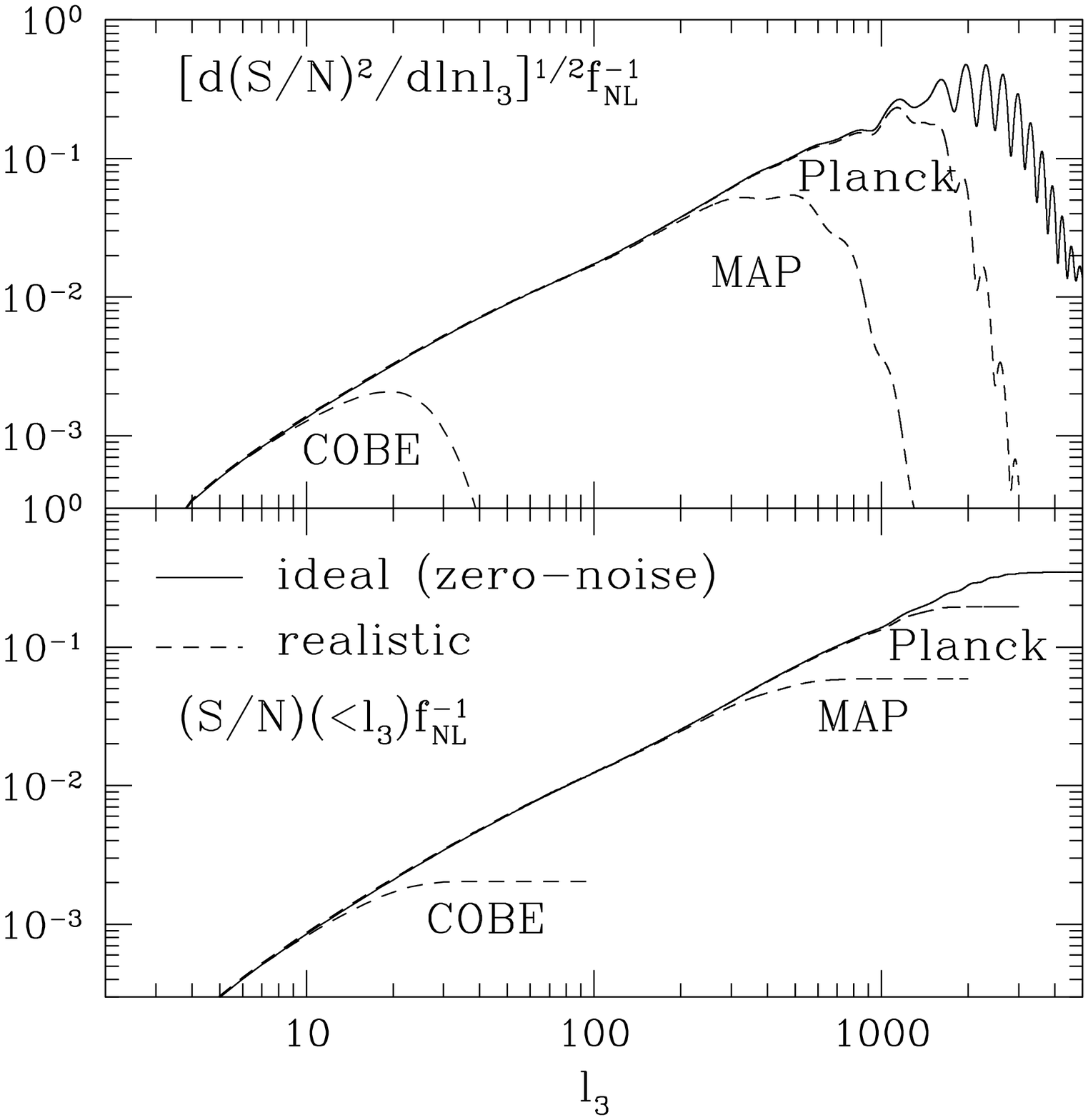}
 \caption{Signal-to-noise Ratio}
 \mycaption{The predictions of the signal-to-noise ratio, $S/N$, 
 for {\it COBE}, {\it MAP}, and {\it Planck} experiments 
 (see Eq.(\ref{eq:sn})).
 The differential $S/N$ at $\ln l_3$ interval is shown in the upper
 panel, while the cumulative $S/N$ up to a certain $l_3$ is shown
 in the bottom panel. 
 Both are in units of $f_{\rm NL}$.
 Solid line represents the zero-noise ideal experiment, while dotted 
 lines show the realistic experiments mentioned above.
 The total $(S/N)f^{-1}_{\rm NL}$ 
 are $1.7\times 10^{-3}$, $5.8\times 10^{-2}$, and $0.19$
 for {\it COBE}, {\it MAP}, and {\it Planck} experiments, respectively.}
\label{fig:fis11}
\end{figure}

Figure~\ref{fig:fis11} shows the signal-to-noise ratio, $S/N$.
The top panel shows the differential $S/N$ for the primary bispectrum 
at $\ln l_3$ interval, 
$\left[d(S/N)^2/d\ln l_3\right]^{1/2}f_{\rm NL}^{-1}$, and the bottom
panel shows the cumulative $S/N$, $(S/N)(<l_3)f_{\rm NL}^{-1}$, which is 
$S/N$ summed up to a certain $l_3$. 
We have computed the detector noise power spectrum, $C_l^{\rm N}$, for 
{\it COBE} four-year map \citep{Ben96}, {\it MAP} 90 GHz channel, 
and {\it Planck} 217 GHz channel, and assumed full sky coverage.
Figure~\ref{fig:fis11} also shows the ideal experiment 
with no noise: $C_l^{\rm N}=0$.
Both $\left[d(S/N)^2/d\ln l_3\right]^{1/2}$ and 
$(S/N)(<l_3)$ increase monotonically with $l_3$, roughly $\propto l_3$, 
up to $l_3\sim 2000$ for the ideal experiment.

%%%%% Figure 4: signal-to-noise %%%%%
%%%%%%%%%%%%%%%%%%%%%%%%%%%%%%%%%%%%%%%%%%%%%%%%%%%%%%%%%%%%%%%%%%
% Figure 4
%%%%%%%%%%%%%%%%%%%%%%%%%%%%%%%%%%%%%%%%%%%%%%%%%%%%%%%%%%%%%%%%%%

Beyond $l_3\sim 2000$, an enhancement of the damping tail in $C_l$ 
because of the weak lensing effect \citep{Seljak96} stops 
$\left[d(S/N)^2/d\ln l_3\right]^{1/2}$, and hence $(S/N)(<l_3)$,
increasing. 
This leads to an important constraint on the observation; 
even for the ideal noise-free, infinitesimally thin beam experiment, 
there is an upper limit on the value of $S/N\simlt 0.3f_{\rm NL}$.
For a given realistic experiment, 
$\left[d(S/N)^2/d\ln l_3\right]^{1/2}$ has a maximum at 
a scale near the beam size.

For {\it COBE}, {\it MAP} and {\it Planck} experiments,
the total $(S/N)f^{-1}_{\rm NL}$ are $1.7\times 10^{-3}$, 
$5.8\times 10^{-2}$, and $0.19$, respectively (see table~\ref{tab:sn}).
To obtain $S/N>1$, we need $f_{\rm NL}>600$, 20, and $5$, while the 
ideal experiment requires $f_{\rm NL}>3$ (see table~\ref{tab:fnl}).
We can also roughly obtain these values by substituting equation 
(\ref{eq:orderest}) into (\ref{eq:snorder}),
%%%%%%%%%%%%%%%%%%%%%%%%%%%%%%%%%%%%%%%%%%%%%%%%%%%%%%%%%%%%%%%%%%
\begin{equation}
  \label{eq:snorderprim}
  \left(\frac{S}{N}\right)_{\rm primary}
  \sim 
  l \times 10^{-4}f_{\rm NL}.
\end{equation}
%%%%%%%%%%%%%%%%%%%%%%%%%%%%%%%%%%%%%%%%%%%%%%%%%%%%%%%%%%%%%%%%%%
 
The degradation parameters, $d_{\rm primary}$,  
are 1.46, 1.01, and 1.00 for {\it COBE}, {\it MAP}, and {\it Planck} 
experiments, respectively (see table~\ref{tab:corr}),
suggesting that {\it MAP} and {\it Planck} experiments will separate the 
primary bispectrum from the others with 1\% or better accuracy;
however, {\it COBE} cannot discriminate between them very well, as 
the primary and the secondary sources change monotonically on 
the {\it COBE} angular scales.
On the {\it MAP} and {\it Planck} scales, the primary bispectrum
starts oscillating around zero, being well separated in shape 
from the secondaries that do not oscillate.
This is good news for the forthcoming high angular
resolution CMB experiments.

\subsection{Measuring secondary bispectra}

Signal-to-noise ratios of detecting the SZ--lensing bispectrum, 
$(S/N)_{\rm sz-lens}$, in units of 
$\left|j_\nu\right| \overline{T}_{\rho0}b_{\rm gas}$ are 
$1.8\times 10^{-4}$, 0.34, and 6.2 for {\it COBE}, {\it MAP}, and {\it Planck}
experiments, respectively (see table~\ref{tab:sn}), where 
$\overline{T}_{\rho0}$ is in units of 1~keV.
Using equations (\ref{eq:snorder}) and (\ref{eq:orderestszlens}), 
we estimate $(S/N)_{\rm sz-lens}$ roughly as
%%%%%%%%%%%%%%%%%%%%%%%%%%%%%%%%%%%%%%%%%%%%%%%%%%%%%%%%%%%%%%%%%%
\begin{equation}
  \label{eq:snorderszlens}
  \left(\frac{S}{N}\right)_{\rm sz-lens}
  \sim 
  l^2 \times 10^{-6}\left|j_\nu\right| \overline{T}_{\rho0}b_{\rm gas}.
\end{equation}
%%%%%%%%%%%%%%%%%%%%%%%%%%%%%%%%%%%%%%%%%%%%%%%%%%%%%%%%%%%%%%%%%%
Hence, $(S/N)_{\rm sz-lens}$ increases with the angular resolution
more rapidly than the primary bispectrum (see Eq.(\ref{eq:snorderprim})).
Since $\left|j_\nu\right| \overline{T}_{\rho0}b_{\rm gas}$ should be
of order unity, {\it COBE} and {\it MAP} cannot detect the 
SZ--lensing bispectrum; however, {\it Planck} is sensitive enough 
to detect, depending on the frequency, i.e., a value of $j_\nu$.
For example, 217~GHz is insensitive to the SZ effect as
$j_\nu\sim 0$, while $j_\nu=-2$ in the Rayleigh--Jeans regime.

The degradation parameters, $d_{\rm sz-lens}$, are
3.89, 1.16, and 1.00 for {\it COBE}, {\it MAP}, and {\it Planck} experiments,
respectively (see table~\ref{tab:corr}); thus, 
{\it Planck} will separate the SZ--lensing bispectrum from the other effects.
Note that $(S/N)_{\rm sz-lens}$ values must be an order of magnitude 
estimation, for our cosmological model is the {\it COBE} normalized SCDM 
that yields $\sigma_8=1.2$, which is a factor of 2 greater than the cluster
normalization for $\Omega_{\rm m}=1$, and $20\%$ greater than
the normalization for $\Omega_{\rm m}=0.3$ \citep{KS97}.
Hence, this factor tends to overestimate 
$\left<\Theta^*_{lm} a_{lm}^{\rm SZ}\right>$ (Eq.(\ref{eq:blsz})) by a 
factor of less than 10; on the other hand, using the linear
$P_\Phi(k)$ power spectrum rather than the non-linear power spectrum 
tends to underestimate the effect by a factor of less than 10 at 
$l\sim 3000$ \citep{CH00}.
Yet, our main goal is to discriminate between 
the shapes of various bispectra, not to determine the amplitude, 
so that this factor does not affect our conclusion on the 
degradation parameters, $d_i$.

For the extragalactic radio and infrared sources, we estimate 
the signal-to-noise ratios as $5.7\times 10^{-7}(b^{\rm ps}/10^{-25})$, 
$2.2(b^{\rm ps}/10^{-25})$, and $52(b^{\rm ps}/10^{-27})$ for 
{\it COBE}, {\it MAP}, and {\it Planck} experiments, respectively (see
table~\ref{tab:sn}), and the degradation parameters, $d_{\rm ps}$, as
3.45, 1.14, and 1.00 (see table~\ref{tab:corr}).
Our estimate is consistent with \citet{RSH00}. 
From equation (\ref{eq:snorder}), we find 
%%%%%%%%%%%%%%%%%%%%%%%%%%%%%%%%%%%%%%%%%%%%%%%%%%%%%%%%%%%%%%%%%%
\begin{equation}
  \left(\frac{S}{N}\right)_{\rm ps}
  \sim l^5\times 10^{-13}\left(\frac{b^{\rm ps}}{10^{-25}}\right);
\end{equation}
%%%%%%%%%%%%%%%%%%%%%%%%%%%%%%%%%%%%%%%%%%%%%%%%%%%%%%%%%%%%%%%%%%
thus, $S/N$ of the point-source bispectrum increases 
very rapidly with the angular resolution.

Although {\it MAP} cannot separate the Poisson bispectrum from 
the SZ--lensing bispectrum very well (see 
$r_{ij}$ in table~\ref{tab:corr}), the SZ--lensing bispectrum is too 
small to be measured by {\it MAP} anyway.
{\it Planck} will do an excellent job on separating all kinds of bispectra,
at least including the primary signal, SZ--lensing coupling, 
and extragalactic point sources, on the basis of the shape difference.

\subsection{Measuring primary skewness}

For the skewness, we define $S/N$ as
%%%%%%%%%%%%%%%%%%%%%%%%%%%%%%%%%%%%%%%%%%%%%%%%%%%%%%%%%%%%%%%%%%%%%%%%%%%
\begin{equation}
  \label{eq:skew_sn}
  \left(\frac{S}{N}\right)^2\equiv
  \frac{S_3^2}{\sigma^2_{S_3}},
\end{equation}
%%%%%%%%%%%%%%%%%%%%%%%%%%%%%%%%%%%%%%%%%%%%%%%%%%%%%%%%%%%%%%%%%%%%%%%%%%%
where the variance is \citep{Srednicki93}
%%%%%%%%%%%%%%%%%%%%%%%%%%%%%%%%%%%%%%%%%%%%%%%%%%%%%%%%%%%%%%%%%%%%%%%%%%%
\begin{eqnarray}
  \nonumber
  \sigma_{S_3}^2
  &\equiv& \left<\left(S_3\right)^2\right> =
  6\int_{-1}^{1}\frac{d\cos\theta}2 \left[{\cal C}(\theta)\right]^3\\
  \nonumber
  &=&
  6\sum_{l_1l_2l_3}
  \frac{\left(2l_1+1\right)\left(2l_2+1\right)\left(2l_3+1\right)}
        {(4\pi)^3}
  \left(
  \begin{array}{ccc}
  l_1 & l_2 & l_3 \\ 0 & 0 & 0 
  \end{array}
  \right)^2
   {\cal C}_{l_1}{\cal C}_{l_2}{\cal C}_{l_3}
   W^2_{l_1}W^2_{l_2}W^2_{l_3}\\
 \nonumber
  &=&
  \frac{9}{2\pi^3}\sum_{2\leq l_1\leq l_2\leq l_3}
  \left(l_1+\frac12\right)\left(l_2+\frac12\right)\left(l_3+\frac12\right)
  \left(
  \begin{array}{ccc}
  l_1 & l_2 & l_3 \\ 0 & 0 & 0 
  \end{array}
  \right)^2\\
 \label{eq:skewvar}
  & &\times
   {\cal C}_{l_1}{\cal C}_{l_2}{\cal C}_{l_3}
   W^2_{l_1}W^2_{l_2}W^2_{l_3}.
\end{eqnarray}
%%%%%%%%%%%%%%%%%%%%%%%%%%%%%%%%%%%%%%%%%%%%%%%%%%%%%%%%%%%%%%%%%%%%%%%%%%%
In the last equality, we have used symmetry of the summed 
quantity with respect to indices (Eq.(\ref{eq:sumchange})), 
and removed unobservable modes, $l=0$ and $1$.
Typically $\sigma_{S_3}\sim 10^{-15}$, as $\sigma_{S_3}\sim
\left[{\cal C}(0)\right]^{3/2}\sim 10^{-15}$, 
where ${\cal C}(\theta)$ is the temperature auto correlation function 
including noise.

The bottom panel of figure~\ref{fig:skewness} plots $\sigma_{S_3}(<l_3)$,  
which is $\sigma_{S_3}$ summed up to a certain $l_3$, for 
{\it COBE}, {\it MAP}, and {\it Planck} experiments as well as for the 
ideal experiment.
Since ${\cal C}_{l}W^2_l= C_l e^{-l(l+1)\sigma^2_{\rm b}} + w^{-1}$,
where $w^{-1}$ is the white noise power spectrum of the detector
noise \citep{Knox95}, $w^{-1}$ keeps $\sigma_{S_3}(<l_3)$ 
slightly increasing with $l_3$ beyond the experimental angular resolution 
scale, $l\sim \sigma_{\rm b}^{-1}$.
In contrast, $S_3(<l_3)$ becomes constant beyond $l\sim \sigma_{\rm b}^{-1}$
(see the top panel of figure~\ref{fig:skewness}). 
As a result, $S/N$ starts slightly decreasing beyond the resolution.
We use the maximum $S/N$ for calculating the minimum value of 
$f_{\rm NL}$ above which the primary $S_3$ is detectable; 
we find that 
$f_{\rm NL}> 800$, 80, 70, and 60 for {\it COBE}, {\it MAP}, 
{\it Planck}, and the ideal experiments, respectively, assuming 
full sky coverage.

These $f_{\rm NL}$ values are systematically larger than those for
detecting $B_{l_1l_2l_3}$ by a factor of 1.3, 4, 14, and 20, respectively
(see table~\ref{tab:fnl}).
The higher the angular resolution is, the less sensitive the primary
$S_3$ is to non-Gaussianity than $B_{l_1l_2l_3}$.
This is because the cancellation effect on smaller angular scales
because of the oscillation of $B_{l_1l_2l_3}$ damps $S_3$.

%%%%%%%%%%%%%%%%%%%%%%%%%%%%%%%%%%%%%%%%%%%%%%%%%%%%%%%%%%%%%%%%%%
\section{Discussion and Conclusions}
\label{sec:discussion}

In this chapter, using the full radiation transfer function, 
we have numerically computed the primary cosmic microwave background 
bispectrum (Eq.(\ref{eq:almspec})) and skewness 
(Eq.(\ref{eq:skewness})) down to arcminutes angular scales.
As the primary bispectrum oscillates around zero
(figure~\ref{fig:bispectrum}), the primary skewness saturates at 
the {\it MAP} angular resolution scale, $l\sim 500$ 
(figure~\ref{fig:skewness}).
We have introduced the {\it reduced} bispectrum, $b_{l_1l_2l_3}$, 
defined by equation (\ref{eq:func}), and found that 
this quantity is more useful to describe physical properties of 
the bispectrum than the angular averaged bispectrum,
$B_{l_1l_2l_3}$ (Eq.(\ref{eq:blll})).

\begin{figure}
 \plotone{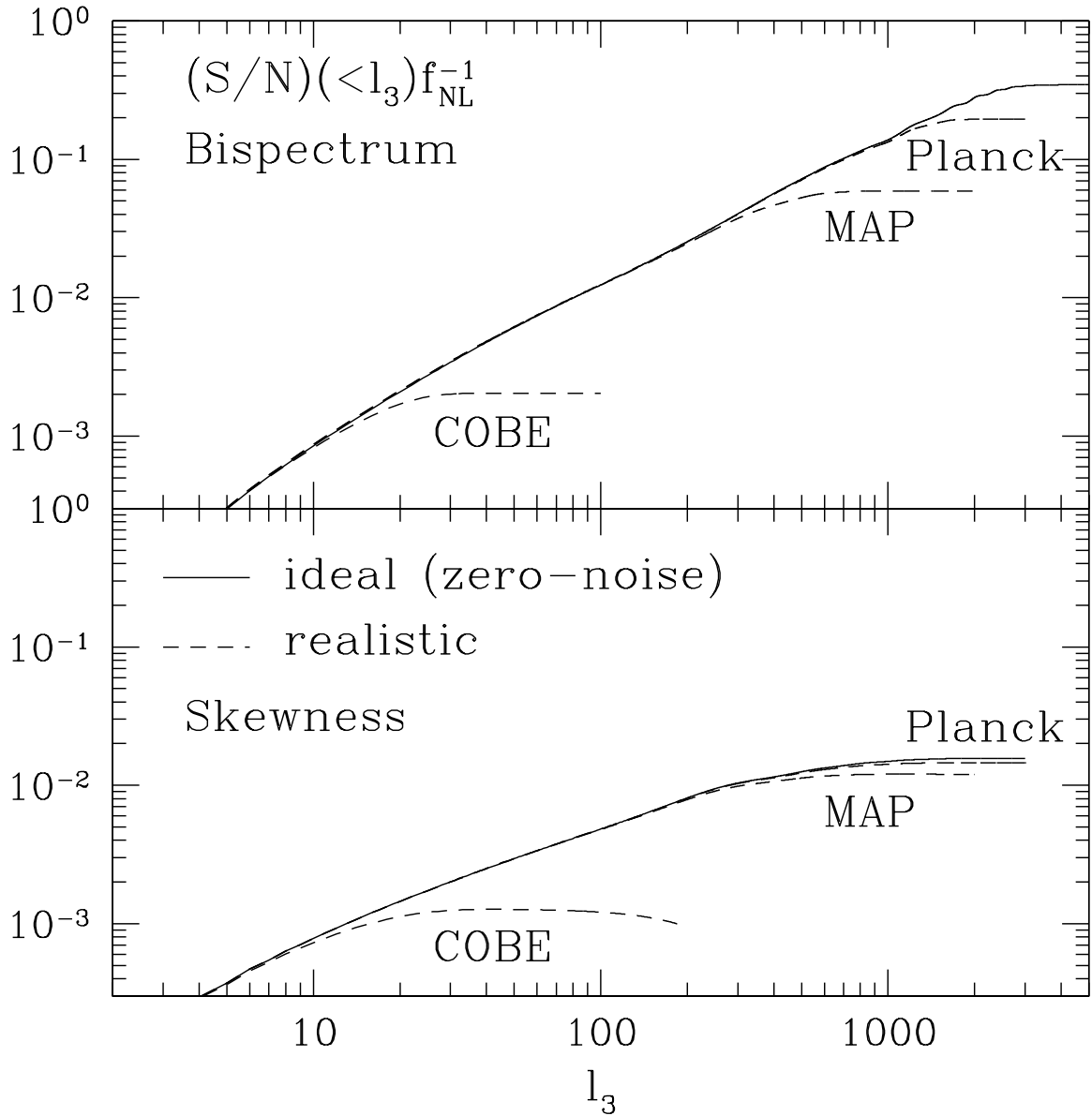}
 \caption{Bispectrum vs Skewness}
 \mycaption{Comparison of the signal-to-noise ratio summed up to a certain
 $l_3$, $S/N(<l_3)$, for the bispectrum (top panel; Eq.(\ref{eq:sn})) 
 and the skewness (bottom panel; Eq.(\ref{eq:skew_sn})).
 $S/N(<l_3)$ is in units of $f_{\rm NL}$.
 The dotted lines show {\it COBE}, {\it MAP}, and {\it Planck} 
 experiments (dotted lines), while the solid line shows the ideal experiment.
 See table~\ref{tab:fnl} for $f_{\rm NL}$ to obtain $S/N>1$.}
\label{fig:sn}
\end{figure}

Figure~\ref{fig:sn} compares the expected signal-to-noise ratio 
of detecting the primary non-Gaussianity based on the bispectrum
(Eq.(\ref{eq:sn})) with that based on the skewness (Eq.(\ref{eq:skew_sn})). 
It shows that the bispectrum is almost an order of magnitude more 
sensitive to the non-Gaussianity than the skewness.
We conclude that when we can compute the predicted form of the
bispectrum, it becomes a ``matched filter'' for detecting the 
non-Gaussianity in data, and thus much more powerful tool than the 
skewness.
Table~\ref{tab:fnl} summarizes the minimum $f_{\rm NL}$ for 
detecting the primary non-Gaussianity using the bispectrum or the skewness 
for {\it COBE}, {\it MAP}, {\it Planck}, and the ideal experiments. 
This shows that even the ideal experiment 
needs $f_{\rm NL}>3$ to detect the primary bispectrum.

%%%%% Figure 5: signal-to-noise; comparison %%%%%
%%%%%%%%%%%%%%%%%%%%%%%%%%%%%%%%%%%%%%%%%%%%%%%%%%%%%%%%%%%%%%%%%%
% Figure 5
%%%%%%%%%%%%%%%%%%%%%%%%%%%%%%%%%%%%%%%%%%%%%%%%%%%%%%%%%%%%%%%%%%

%%%%% Table V: summary of f_NL %%%%%
%%%%%%%%%%%%%%%%%%%%%%%%%%%%%%%%%%%%%%%%%%%%%%%%%%%%%%%%%%%%%%%%%%%%%%%%%%%
% Table 5
\begin{table} 
 \caption{Detection Limit for the Non-linear Coupling Parameter}
 \mycaption{The minimum non-linear coupling parameter, $f_{\rm NL}$, needed
 for detecting the primary non-Gaussianity by the bispectrum or the skewness
 with the signal-to-noise ratio greater than 1. 
 These estimates include the effects of cosmic variance, detector noise,
 and foreground sources.}
\label{tab:fnl}
\begin{center} 
  \begin{tabular}{ccc}\hline\hline
  Experiments & $f_{\rm NL}$ (Bispectrum) & $f_{\rm NL}$ (Skewness) \\
  \hline 
  {\it COBE}   & 600 & 800 \\
  {\it MAP}    & 20  & 80 \\
  {\it Planck} & 5   & 70 \\
  Ideal  & 3   & 60 \\
   \hline\hline
  \end{tabular}
\end{center}
\end{table}
%%%%%%%%%%%%%%%%%%%%%%%%%%%%%%%%%%%%%%%%%%%%%%%%%%%%%%%%%%%%%%%%%%%%%%%%%%

We have calculated the secondary bispectra from the coupling between the 
SZ effect and the weak lensing effect, and from 
the extragalactic radio and infrared sources.
Only {\it Planck} will detect the SZ--lensing bispectrum, 
while both {\it MAP} and {\it Planck} will detect the 
 extragalactic point-source bispectrum (table~\ref{tab:sn}).

We have also studied how well we can discriminate between the primary,
the SZ--lensing coupling, and the extragalactic point-source bispectra.
We have found that {\it MAP} and {\it Planck} will separate the 
primary from the other secondary sources with 1\% or better accuracy. 
This conclusion is due to the acoustic oscillation in the primary
bispectrum that does not appear in the secondary bispectra. 
The SZ--lensing coupling and the extragalactic sources are 
well separately measured by {\it Planck} experiment, although {\it COBE} and 
{\it MAP} cannot discriminate between them (table~\ref{tab:corr}).

Our arguments on the ability to discriminate between various bispectra
have been based upon the shape difference, and thus have not taken into 
account the spectral difference in frequency space.
\citet{TE96} and \citet{CHT00} have shown that the multi-band observation is 
efficient for discriminating between the primary signal and the other 
foreground sources in the CMB power spectrum.
Their scheme should be effective on the bispectrum as well, and
will improve the accuracy of the foreground removal further. 
We thus expect that {\it MAP} and {\it Planck} will measure the primary
bispectrum separately from the foregrounds.

Simple slowly-rolling single-field inflation models predict 
$f_{\rm NL}\sim {\cal O}(10^{-2})$ \citep{SB90,SB91,Gan94}, while
the second order perturbation theory predicts 
$f_{\rm NL}\sim {\cal O}(1)$ \citep{PC96};
thus, significant detection of the primary bispectrum or the
skewness with any experiments means that these inflation models need to 
be modified.
According to our results, if the reported detection \citep{FMG98,Mag00} 
of the bispectrum on the {\it COBE} map were cosmological in origin, then
{\it MAP} and {\it Planck} would detect the primary bispectrum much more 
significantly.
While Banday, Zaroubi and G\'orski \citep{BZG00} have shown that
the one of those detections \citep{FMG98} is accounted for by 
the experimental systematic effects,
the other \citep{Mag00} is significant even after removing 
such the systematics.

Although we have not discussed so far, spatial distribution of 
interstellar dust emissions is a potential source of microwave 
non-Gaussianity. 
While it is very hard to estimate the bispectrum
analytically, we can use the dust template map compiled by 
\citet{SFD98} to estimate the dust bispectrum. 
For example, we have found that the dimensionless 
skewness parameter,
$\left<(\Delta T)^3\right>/\left<(\Delta T)^2\right>^{3/2}$,
is as large as 51 on the template map.
We have used the publicly available 
HEALPix-formatted \citep{GHW98} $100~\mu{\rm m}$ map, 
which contains 12,582,912 pixels without sky-cut. 
The mean intensity in the map is $14.8~{\rm MJy~sr^{-1}}$.
Of course, this skewness is largely an overestimate for the 
real CMB measurement; we need to cut a fraction of the sky that
contains the Galactic plane, and this will greatly reduce the non-Gaussianity.
Nevertheless, residual non-Gaussianity is still a source of the 
microwave bispectrum, and has to be taken into account.
Moreover, the form of the measured bispectrum on the dust map reflects
the physics of interstellar dust, which is highly uncertain at present;
thus, studying the interstellar dust bispectrum is a challenging field.

%%%%%%%%%%%%%%%%%%%%%%%%%%%%%%%%%%%%%%%%%%%%%%%%%%%%%%%%%%%%%%%%%%%
%
%  Measurement of Bispectrum on the COBE DMR sky maps
%
%     1st draft:  06/20/2001
%     revision:   07/17/2001
%     final:      08/05/2001
%
%%%%%%%%%%%%%%%%%%%%%%%%%%%%%%%%%%%%%%%%%%%%%%%%%%%%%%%%%%%%%%%%%%%
\chapter{Measurement of Bispectrum on the {\it COBE} DMR sky maps}
\label{chap:obs_bl}

%%%%% why study non-Gaussianity? %%%%%

Several authors have attempted to measure non-Gaussianity 
in CMB using various statistical techniques 
(e.g., Kogut et al. 1996b); as yet no conclusive detection has been 
reported except for measurement of several 
modes of the normalized CMB bispectrum on the {\it COBE} Differential 
Microwave Radiometer (DMR) sky maps \citep{FMG98,Mag00}.
The existence of non-Gaussianity in the DMR data is controversial.
If the CMB sky were non-Gaussian, this would challenge our simplest 
inflationary model.

%%%%% previous work on COBE bispectrum %%%%%

The angular bispectrum, $B_{l_1l_2l_3}$, is the harmonic transform of 
the three-point correlation function.
We carefully distinguish the normalized bispectrum, 
$B_{l_1l_2l_3}/\left(C_{l_1}C_{l_2}C_{l_3}\right)^{1/2}$, from 
the bispectrum, $B_{l_1l_2l_3}$.
\citet{FMG98} have measured 9 equilateral ($l_1=l_2=l_3$) modes of the 
normalized bispectrum, 
$B_{l_1l_2l_3}/\left(C_{l_1}C_{l_2}C_{l_3}\right)^{1/2}$,
on the DMR map, claiming detection at $l_1=l_2=l_3=16$.
Their result has been under extensive efforts to confirm 
its significance and origin.
\citet{BT99} claim that a few individual pixels in the DMR map 
are responsible for the most of the signal.
\citet{BZG00} have proposed an eclipse effect by the Earth against 
the {\it COBE} satellite as a possible source of the signal.
\citet{Mag00} has measured other 8 inter-$l$ modes of the normalized 
bispectrum such as $B_{l-1ll+1}/\left(C_{l-1}C_{l}C_{l+1}\right)^{1/2}$,
and claims that scatter of the normalized bispectrum among 8 modes 
is too small to be consistent with Gaussian. 
\citet{SM00} further report measurement of 24 other inter-$l$ modes for
different lags in $l$, and conclude they are consistent with Gaussian.

%%%%% our work %%%%%

Thus, until now 41 modes of the normalized bispectrum have been 
measured on the DMR map.
Here, we simply ask: ``how many modes are available in the DMR map for 
the bispectrum?''
The answer is 466, up to a maximum multipole of 20 that corresponds to 
the DMR beam size; thus, it is conceivable that the claimed detection of the 
normalized bispectrum at $l_1=l_2=l_3=16$ would be explained by 
a statistical fluctuation, as 9 modes are expected to have statistical 
significance above 98\% out of 466 independent modes even 
if CMB is exactly Gaussian.
In this chapter, we measure 466 modes of the CMB bispectrum on the 
{\it COBE} DMR sky maps, testing the claimed detection of the 
bispectrum and non-Gaussianity.
We take into account the covariance between these modes due to the
Galactic cut, which has not been done in the previous work.

%%%%% theory %%%%%

On the theoretical side, several predictions for the CMB bispectrum exist.
Several authors \citep{FRS93,LS93,Gan94} have predicted 
the primary bispectrum (or equivalently three-point correlation
function) on the DMR angular scales from slow-roll inflation models. 
In chapter~\ref{chap:theory_bl}, we have extended the prediction 
down to arcminutes scales using the full radiation transfer function.

In addition to the primary one, secondary sources in the low-redshift 
universe and foreground sources produce the bispectrum through their 
non-linearity. 
\citet{LS93} and \citet{SG99} have calculated the secondary bispectrum 
arising from non-linear evolution of gravitational potential;
\citet{GS99} and \citet{CH00} have calculated the one from the gravitational 
lensing effect coupled with various secondary anisotropy sources.
In chapter~\ref{chap:theory_bl}, we have calculated the foreground 
bispectrum from extragalactic radio and infrared point sources.
While the bispectrum is not the best tool for detecting the signature
of rare highly non-linear events, e.g., textures \citep{PK01},
it is sensitive to weakly non-linear effects.

%%%%% fit the data (primary) %%%%%

Having theoretical predictions is a great advantage in extracting
physical information from measurement; one can fit a predicted 
bispectrum to the data so as to constrain parameters in a theory. 
Since the DMR beam size is large enough to minimize contribution 
from the secondary and the extragalactic foreground sources, 
the only relevant source would be the primary one.
In this chapter, we fit a theoretical primary bispectrum 
\citep{KS01a} to the data.

%%%%% fit the data (foreground) %%%%%

The Galactic plane contains strong microwave emissions from interstellar 
sources.
The emissions are highly non-Gaussian, and distributed on fairly large 
angular scales.
Unfortunately, predicting the CMB bispectrum from interstellar sources
is very difficult; thus, we excise the galactic plane from the DMR data.
We model the residual foreground bispectrum at high
galactic latitude using foreground template maps.
By simultaneously fitting the foreground bispectrum and the primary 
bispectrum to the DMR data for three different Galactic cuts, 
we quantify the importance of the interstellar emissions in our analysis.

%%%%% structure of the chapter %%%%%

This chapter is organized as follows.
In \S~\ref{sec:bispectrum*}, we define the angular bispectrum, and 
show how to compute it efficiently from observational data. 
In \S~\ref{sec:measurement}, we study statistical properties of the 
bispectrum and the normalized bispectrum.
We then measure the normalized bispectrum from the {\it COBE} DMR 
four-year sky maps \citep{Ben96}, testing Gaussianity of the DMR map.
In \S~\ref{sec:fit}, we fit predicted bispectra to the DMR data, 
constraining parameters in the predictions. 
The predictions include the primary bispectrum from inflation and the 
foreground bispectrum from interstellar Galactic emissions.
Finally, \S~\ref{sec:discussion_bl} concludes.

%%%%%%%%%%%%%%%%%%%%%%%%%%%%%%%%%%%%%%%%%%%%%%%%%%%%%%%%%%%%%%%%%
\section{Angular Bispectrum}
\label{sec:bispectrum*}

The CMB angular bispectrum consists of a product of three 
harmonic transforms of the CMB temperature field.
For Gaussian fields, expectation value of the bispectrum is exactly zero.
Given statistical isotropy of the universe, the angular averaged 
bispectrum, $B_{l_1l_2l_3}$, is given by
%%%%%%%%%%%%%%%%%%%%%%%%%%%%%%%%%%%%%%%%%%%%%%%%%%%%%%%%%%%%%%%%%%
\begin{equation}
  \label{eq:blll**}
  B_{l_1l_2l_3}= \sum_{{\rm all}~m}
  \left(
  \begin{array}{ccc}
  l_1&l_2&l_3\\
  m_1&m_2&m_3
  \end{array}
  \right)
  a_{l_1m_1}a_{l_2m_2}a_{l_3m_3},
\end{equation}
%%%%%%%%%%%%%%%%%%%%%%%%%%%%%%%%%%%%%%%%%%%%%%%%%%%%%%%%%%%%%%%%%%
where the matrix denotes the Wigner-$3j$ symbol, and harmonic
coefficients, $a_{lm}$, are given by
%%%%%%%%%%%%%%%%%%%%%%%%%%%%%%%%%%%%%%%%%%%%%%%%%%%%%%%%%%%%%%%%%%
\begin{equation}
 \label{eq:alm*}
  a_{lm}= 
  \int_{\Omega_{\rm obs}} 
  d^2\hat{\mathbf n}~\frac{\Delta T\left(\hat{\mathbf n}\right)}T
  Y_{lm}^*\left(\hat{\mathbf n}\right).
\end{equation}
%%%%%%%%%%%%%%%%%%%%%%%%%%%%%%%%%%%%%%%%%%%%%%%%%%%%%%%%%%%%%%%%%%
$\Omega_{\rm obs}$ denotes a solid angle of the observed sky. 
$B_{l_1l_2l_3}$ satisfies the triangle condition,
$\left|l_i-l_j\right|\leq l_k \leq l_i+l_j$ for all permutations of
indices, and parity invariance, $l_1+l_2+l_3={\rm even}$.

We can rewrite equation~(\ref{eq:blll**}) into a more useful form.
Using the identity,
%%%%%%%%%%%%%%%%%%%%%%%%%%%%%%%%%%%%%%%%%%%%%%%%%%%%%%%%%%%%%%%%%%
\begin{eqnarray}
 \nonumber
  \left(
   \begin{array}{ccc}
    l_1 & l_2 & l_3 \\ m_1 & m_2 & m_3 
   \end{array}
 \right)
  &=&
  \left(
   \begin{array}{ccc}
    l_1 & l_2 & l_3 \\ 0 & 0 & 0 
   \end{array}
 \right)^{-1}
  \sqrt{
  \frac{(4\pi)^3}{\left(2l_1+1\right)\left(2l_2+1\right)\left(2l_3+1\right)}
  }\\
% \label{eq:gaunt**}
 & &\times
  \int \frac{d^2\hat{\mathbf n}}{4\pi}~
  Y_{l_1m_1}(\hat{\mathbf n})
  Y_{l_2m_2}(\hat{\mathbf n})
  Y_{l_3m_3}(\hat{\mathbf n}),
\end{eqnarray}
%%%%%%%%%%%%%%%%%%%%%%%%%%%%%%%%%%%%%%%%%%%%%%%%%%%%%%%%%%%%%%%%%%
we rewrite equation~(\ref{eq:blll**}) as
%%%%%%%%%%%%%%%%%%%%%%%%%%%%%%%%%%%%%%%%%%%%%%%%%%%%%%%%%%%%%%%%%%
\begin{equation}
  \label{eq:bobs}
  B_{l_1l_2l_3}=
  \left(
   \begin{array}{ccc}
    l_1 & l_2 & l_3 \\ 0 & 0 & 0 
   \end{array}
 \right)^{-1}
  \int \frac{d^2\hat{\mathbf n}}{4\pi}~
  e_{l_1}(\hat{\mathbf n})
  e_{l_2}(\hat{\mathbf n})
  e_{l_3}(\hat{\mathbf n}),
\end{equation}
%%%%%%%%%%%%%%%%%%%%%%%%%%%%%%%%%%%%%%%%%%%%%%%%%%%%%%%%%%%%%%%%%%
where the integral is not over $\Omega_{\rm obs}$, but over the 
whole sky; $e_l(\hat{\mathbf n})$ already encapsulates 
the information of incomplete sky coverage through $a_{lm}$.
Following chapter~\ref{chap:spectrum}, 
we used the azimuthally averaged harmonic transform of the 
CMB temperature field, $e_l(\hat{\mathbf n})$,
%%%%%%%%%%%%%%%%%%%%%%%%%%%%%%%%%%%%%%%%%%%%%%%%%%%%%%%%%%%%%%%%%%
\begin{equation}
  \label{eq:el**}
   e_{l}(\hat{\mathbf n})
   =
   \sqrt{\frac{4\pi}{2l+1}}
   \sum_m a_{lm} Y_{lm}(\hat{\mathbf n}).
\end{equation}
%%%%%%%%%%%%%%%%%%%%%%%%%%%%%%%%%%%%%%%%%%%%%%%%%%%%%%%%%%%%%%%%%%
Similarly, we write the angular power spectrum, $C_l$, as
%%%%%%%%%%%%%%%%%%%%%%%%%%%%%%%%%%%%%%%%%%%%%%%%%%%%%%%%%%%%%%%%%%
\begin{equation}
  \label{eq:cl**}
   C_l=   
   \int \frac{d^2\hat{\mathbf n}}{4\pi}~
   e^2_{l}(\hat{\mathbf n}).
\end{equation}
%%%%%%%%%%%%%%%%%%%%%%%%%%%%%%%%%%%%%%%%%%%%%%%%%%%%%%%%%%%%%%%%%%
Thus, $e_l(\hat{\mathbf n})$ is a square-root of $C_l$ 
at a given position of the sky.
Equation~(\ref{eq:bobs}) is computationally efficient, as we can calculate 
$e_l(\hat{\mathbf n})$ quickly with the spherical harmonic transform 
for a given $l$. 
Since the HEALPix pixels have the equal area \citep{GHW98},
the average over the whole sky, $\int d^2\hat{\mathbf n}/(4\pi)$,
is done by the sum over all pixels divided by the total number 
of pixels, $N^{-1}\sum_i^{N}$.

%%%%%%%%%%%%%%%%%%%%%%%%%%%%%%%%%%%%%%%%%%%%%%%%%%%%%%%%%%%%%%%%%
\section{Measurement of Bispectrum on the DMR Sky Maps}
\label{sec:measurement} 

%%%%%%%%%%%%%%%%%%%%%%%%%%%%%%%%%%%%%%%%%%%%%%%%%%%%%%%%%%%%%%%%%
\subsection{The data}

We use the HEALPix-formatted \citep{GHW98} {\it COBE} DMR four-year sky
map, which contains 12,288 pixels in Galactic coordinate
with a pixel size $1.\hspace{-4pt}^\circ 83$.
We obtain the most sensitive sky map to CMB by combining 53~GHz map with 
90~GHz map, after coadding the channels A and B at each frequency.
We do not subtract eclipse season time-ordered data;
while \citet{BZG00} ascribe the reported non-Gaussianity to this data,
we will argue in this paper that the claimed detection of the normalized
bispectrum at $l_1=l_2=l_3=16$ \citep{FMG98} can also be explained 
in terms of a statistical fluctuation.

We reduce interstellar Galactic emissions by using three different
Galactic cuts: $20^\circ$, $25^\circ$, and $30^\circ$ in Galactic latitude.
Since we want to see how the different Galactic cuts affect the measured
bispectrum, we use the three different Galactic cuts instead of the extended
Galactic cut \citep{Ban97}, which is commonly used for analyzing 
the DMR sky maps.
Then, we subtract the monopole and the dipole from each cut map,
minimizing contaminations from these two multipoles to
higher order multipoles through the mode-mode coupling.
The coupling arises from incomplete sky coverage.
This is very important to do, for the leakage of power from 
the monopole and the dipole to the higher order multipoles is rather big.
We use the least-squares fit weighted by the pixel noise variance
to measure the monopole and the dipole on each cut map.
Table~\ref{tab:skycut} shows the amplitude of the subtracted
monopole and the dipole for the different Galactic cuts.

%%%%%%%%%%%%%%%%%%%%%%%%%%%%%%%%%%%%%%%%%%%%%%%%%%%%%%%%%%%%%%%%%%%%%%%%%%%
% Table 2
\begin{table}
 \caption{Monopole and Dipole Subtraction}
 \mycaption{The monopole, $T_0$, and the dipole, $T_1$, anisotropies
 that have been subtracted from the DMR sky maps.
 We show the subtracted values for zero, 
 $20^\circ$, $25^\circ$, and $30^\circ$ cuts.
 The rightmost column shows the directions of the subtracted dipole
 in Galactic coordinate.}
\begin{center} 
  \begin{tabular}{cccc}\hline\hline
   $\left|b_{\rm cut}\right|$ & ${T}_0$ [$\mu$K] &
  ${T}_1$ [$\mu$K] & 
  ${\mathbf T}_1/{T}_1$ ($l,b$) \\
   \hline
  $0^\circ$ & 1.40 & 63.1 & 
  ($28.\hspace{-2.5pt}^\circ 07$, $2.\hspace{-2.5pt}^\circ 12$) \\
  $20^\circ$ & $-70.3$ & 26.2 & 
  ($89.\hspace{-2.5pt}^\circ 23$, $-4.\hspace{-2.5pt}^\circ 17$) \\
  $25^\circ$ & $-72.4$ & 26.9 & 
  ($84.\hspace{-2.5pt}^\circ 89$, $-5.\hspace{-2.5pt}^\circ 13$) \\
  $30^\circ$ & $-73.4$ & 28.6 & 
  ($97.\hspace{-2.5pt}^\circ 32$, $-6.\hspace{-2.5pt}^\circ 31$) \\
   \hline\hline
  \end{tabular}
\end{center}
 \label{tab:skycut}
\end{table}
%%%%%%%%%%%%%%%%%%%%%%%%%%%%%%%%%%%%%%%%%%%%%%%%%%%%%%%%%%%%%%%%%%%%%%%%%%

We measure the bispectrum, $B_{l_1l_2l_3}$, on the DMR sky maps as follows.
First, we measure $a_{lm}$ using equation~(\ref{eq:alm*}).
Then, we transform $a_{lm}$ for $-l\le m\le l$ into 
$e_l(\hat{\mathbf n})$ through equation~(\ref{eq:el**}).
Finally, we obtain $B_{l_1l_2l_3}$ from equation~(\ref{eq:bobs}),
arranging $l_1$, $l_2$, and $l_3$ in order of $l_1\le l_2\le l_3$, where
the maximum $l_3$ is set to be 20.
In total, we have 466 non-zero modes after taking into account
$\left|l_i-l_j\right|\leq l_k \leq l_i+l_j$ and $l_1+l_2+l_3={\rm even}$.
Measurement of 466 modes takes about 1 second of CPU time on 
a Pentium-III single processor personal computer.

\subsection{Monte--Carlo Simulations}

We use Monte--Carlo simulations to estimate the covariance matrix of the 
measured bispectrum.
Our simulation includes 
(a) a Gaussian random realization of the primary CMB anisotropy field 
drawn from the {\it COBE}-normalized ${\Lambda}$CDM power spectrum,
and (b) a Gaussian random realization of the instrumental noise drawn from
diagonal terms of the {\it COBE} DMR noise covariance matrix \citep{Lin94}.
For computational efficiency, we do not use off-diagonal terms 
as they are smaller than 1\% of the diagonal terms \citep{Lin94}.

We generate the input power spectrum, $C_l$, using the {\sf CMBFAST} code 
\citep{SZ96} with cosmological parameters fixed at 
$\Omega_{\rm cdm}=0.25$, $\Omega_\Lambda=0.7$, $\Omega_{\rm b}=0.05$, 
$h=0.7$, and $n=1$; the {\sf CMBFAST} code uses the \citet{BW97} normalization.

In each realization, we generate $a_{lm}$ from the power spectrum, 
multiply it by the harmonic-transformed DMR beam, $G_l$
\citep{Wri94}, transform $G_l a_{lm}$ back to a sky map, and add an
instrumental noise realization to the map.
Finally, we measure 466 modes of the bispectrum from each realization. 
We generate 50,000 realizations for one simulation;
processing one realization takes about 1 second, so that 
one simulation takes about 16 hours of CPU time on a Pentium-III single 
processor personal computer.

\subsection{Normalized bispectrum}

The input power spectrum determines the variance of the bispectrum.
Off-diagonal terms in the covariance matrix arise from incomplete 
sky coverage.
When non-Gaussianity is weak, the variance is given by 
\citep{Luo94,heavens98,SG99,GM00}
%%%%%%%%%%%%%%%%%%%%%%%%%%%%%%%%%%%%%%%%%%%%%%%%%%%%%%%%%%%%%%%%%
\begin{equation}
 \label{eq:var}
  \left<B_{l_1l_2l_3}^2\right>
  = \left<C_{l_1}\right>\left<C_{l_2}\right>\left<C_{l_3}\right>
  \Delta_{l_1l_2l_3},
\end{equation}
%%%%%%%%%%%%%%%%%%%%%%%%%%%%%%%%%%%%%%%%%%%%%%%%%%%%%%%%%%%%%%%%%% 
where $\Delta_{l_1l_2l_3}$ takes values 1, 2, or 6 
for all $l$'s are different, two are same, or all are same, respectively.
The brackets denote the ensemble average.

The variance is undesirably sensitive to the input power spectrum;
even if the input power spectrum were slightly different 
from the true power spectrum on the DMR map, the estimated variance 
from simulations would be significantly wrong,
and we would erroneously conclude that the DMR map is inconsistent 
with Gaussian.
It is thus not a robust test of Gaussianity to  
compare the measured bispectrum with the Monte--Carlo simulations.

The {\it normalized} bispectrum, 
$B_{l_1l_2l_3}/\left(C_{l_1}C_{l_2}C_{l_3}\right)^{1/2}$,
is more sensible quantity than the bare bispectrum.
\citet{Mag95} shows that the normalized bispectrum is 
a rotationally invariant spectrum independent of the power spectrum,
as it factors out fluctuation amplitude in $a_{lm}$, which is 
measured by $C_l^{1/2}$.
By construction, the variance of the normalized bispectrum is 
insensitive to the power spectrum, approximately given by 
$\Delta_{l_1l_2l_3}$.

\begin{figure}
 \plotone{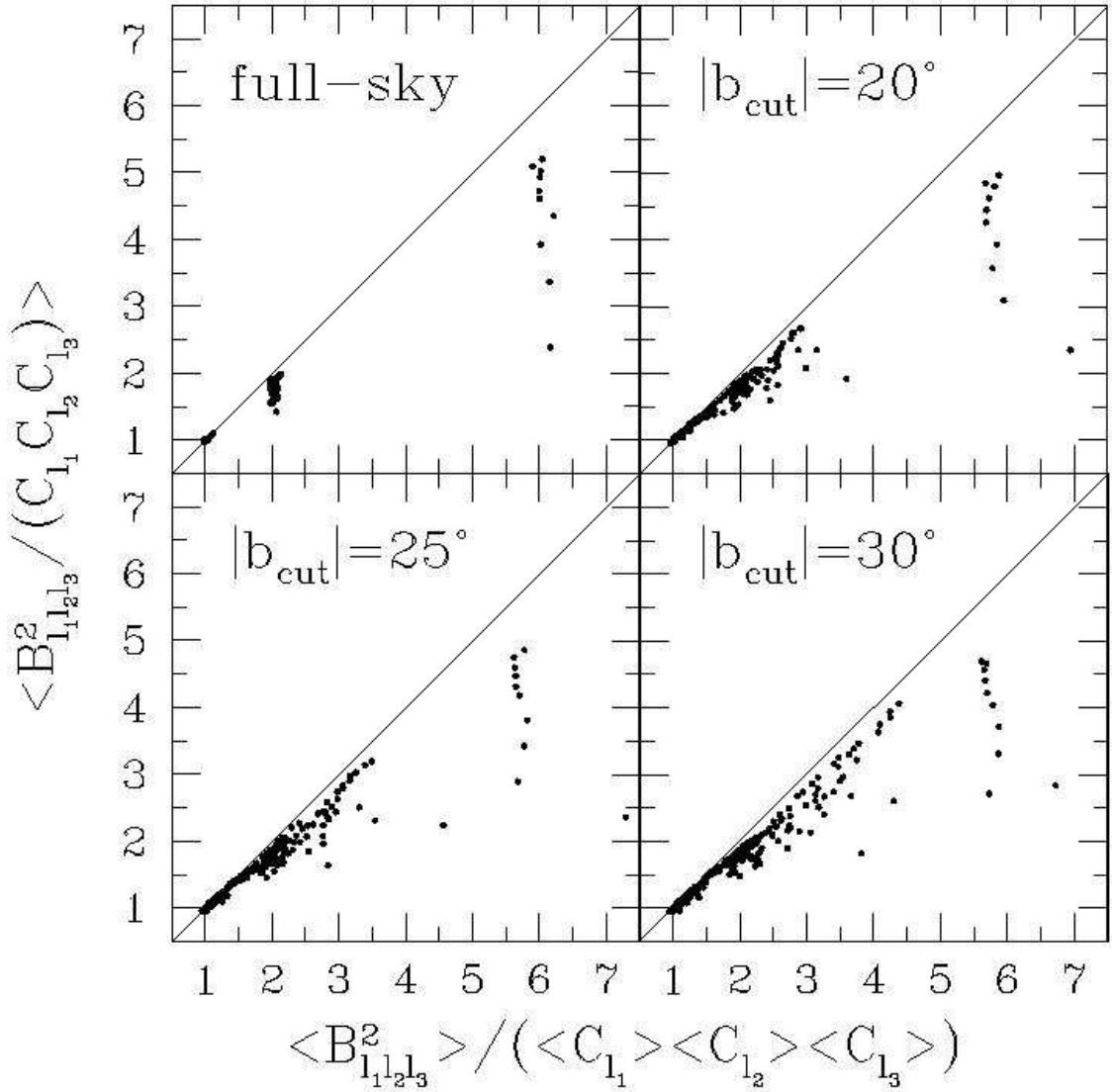}
 \caption{Variance of Normalized Bispectrum and Bare Bispectrum}
 \mycaption
 {Variance of the normalized bispectrum, 
 $\left<B_{l_1l_2l_3}^2/\left(C_{l_1}C_{l_2}C_{l_3}\right)\right>$,
 in comparison with variance of the bare bispectrum,
 $\left<B_{l_1l_2l_3}^2\right>/
 \left(\left<C_{l_1}\right>\left<C_{l_2}\right>\left<C_{l_3}\right>\right)$.
 These are derived from simulated realizations of a Gaussian sky.
 The top-left panel shows the case of full sky coverage, while
 the rest of panels show the cases of incomplete sky coverage.
 The top-right, bottom-left, and bottom-right panels use
 the $20^\circ$, $25^\circ$, and $30^\circ$ Galactic cuts, respectively.}
\label{fig:variance}
\end{figure}

One might wonder if the normalized bispectrum is too noisy to be
useful, as the power spectrum in the denominator is also uncertain
to some extent; however, we find that the variance is actually
slightly smaller than $\Delta_{l_1l_2l_3}$.
Figure~\ref{fig:variance} compares the variance of the normalized
bispectrum, 
$\left<B_{l_1l_2l_3}^2/\left(C_{l_1}C_{l_2}C_{l_3}\right)\right>$,
with that of the bispectrum,
$\left<B_{l_1l_2l_3}^2\right>/
\left(\left<C_{l_1}\right>\left<C_{l_2}\right>\left<C_{l_3}\right>\right)$.
The top-left panel shows the case of full sky coverage. 
We find that the variance of the normalized bispectrum is precisely 1 
when all $l$'s are different, while it is slightly smaller than 2 or 6 
when two $l$'s are same or all $l$'s are same, respectively.
This arises due to correlation between the uncertainties in the 
bispectrum and the power spectrum, and this correlation tends to reduce 
the total variance of the normalized bispectrum.
The rest of panels show the cases of incomplete sky coverage.
While the variance becomes more scattered than the case of 
full sky coverage, the variance of the normalized bispectrum is 
still systematically smaller than that of the bare bispectrum.
Thus, the normalized bispectrum is reasonably sensitive to
non-Gaussianity, yet it is not sensitive to the overall normalization
of power spectrum.

%%%%%%%%%%%%%%%%%%%%%%%%%%%%%%%%%%%%%%%%%%%%%%%%%%%%%%%%%%%%%%%%%%%%%%
%%%%%%%%%%%%%%%%%%%%%%%%%%%%%%%%%%%%%%%%%%%%%%%%%%%%%%%%%%%%%%%%%%%%%%

\begin{figure}
 \plotone{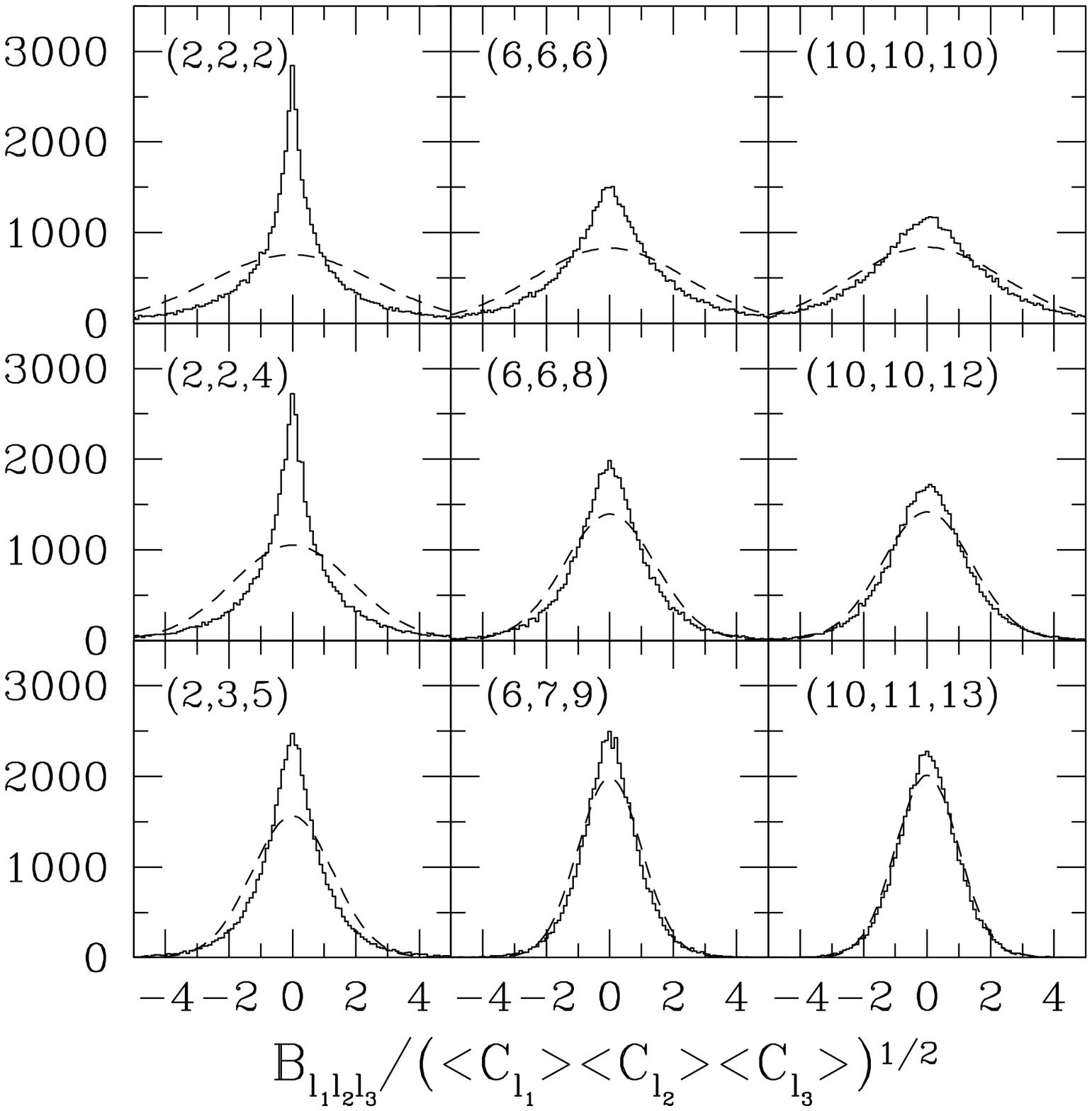}
 \caption{Distribution of Bispectrum}
 \mycaption
 {Distribution of the bare bispectrum drawn from the Monte--Carlo
 simulations (solid lines).
 The $20^\circ$ Galactic cut is used.
 $B_{l_1l_2l_3}/\left(\left<C_{l_1}\right>
 \left<C_{l_2}\right>\left<C_{l_3}\right>\right)^{1/2}$ is plotted,
 where the brackets denote the ensemble average over realizations
 from the Monte--Carlo simulations.
 The dashed lines plot Gaussian distributions calculated from 
 r.m.s. values.
 Each panel represents a certain mode of $(l_1,l_2,l_3)$ as quoted in 
 the panels.}
\label{fig:dist_bare}
\end{figure}

\begin{figure}
 \plotone{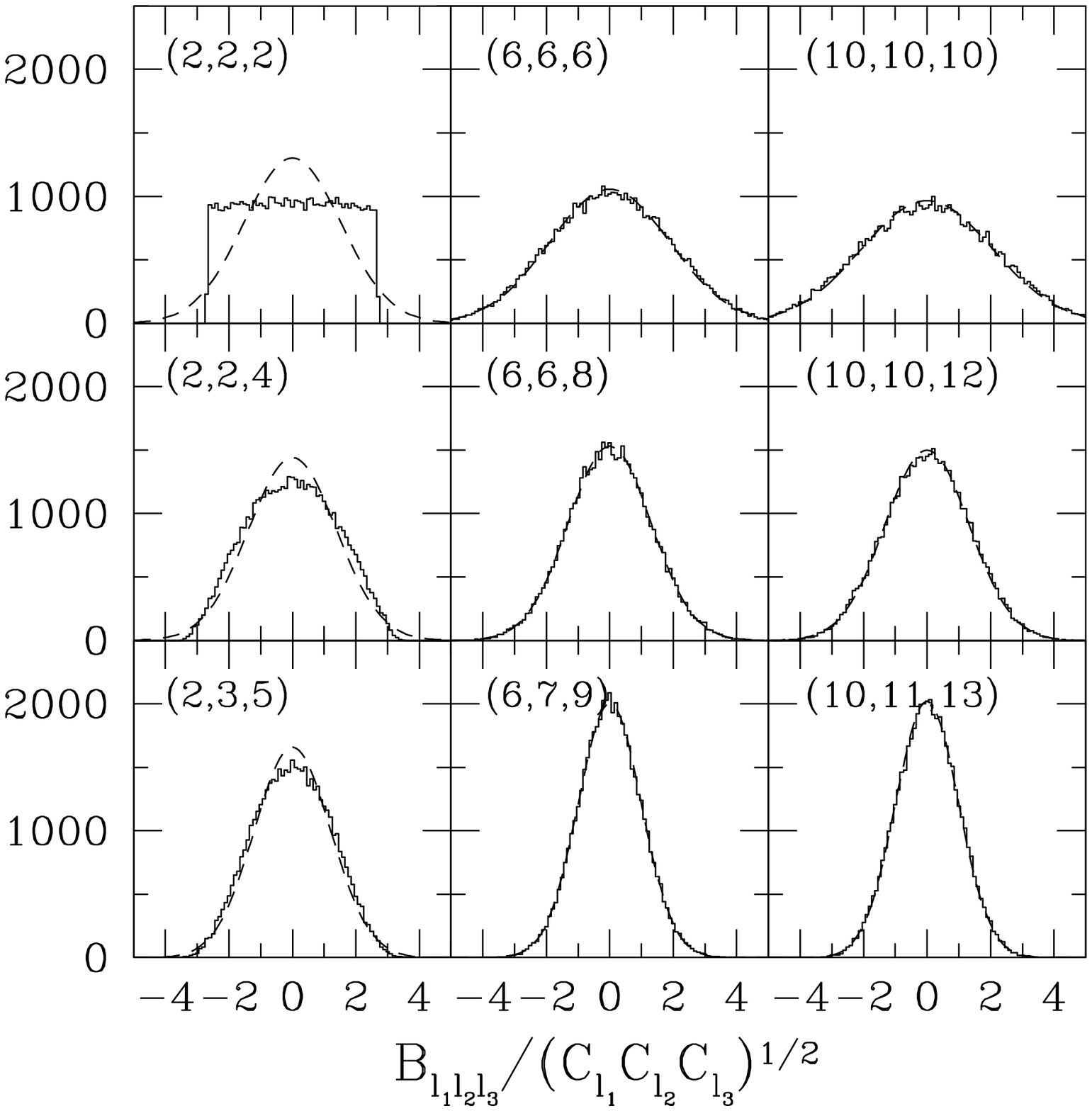}
 \caption{Distribution of Normalized Bispectrum}
 \mycaption
 {Distribution of the normalized bispectrum (solid lines),
 $B_{l_1l_2l_3}/\left(C_{l_1}C_{l_2}C_{l_3}\right)^{1/2}$,
 in comparison with Gaussian distributions calculated from r.m.s. values
 (dashed lines).
 The $20^\circ$ Galactic cut is used.
 %drawn from the Monte--Carlo simulations for 
 The meaning of the panels is the same as in figure~\ref{fig:dist_bare}.}
\label{fig:dist_norm}
\end{figure}

What distribution does the normalized bispectrum obey for a Gaussian field?
First, even for a Gaussian field, the probability distribution 
of a single mode of $B_{l_1l_2l_3}$ is non-Gaussian,
characterized by a large kurtosis.
Figure~\ref{fig:dist_bare} plots the distributions of 9 modes of  
$B_{l_1l_2l_3}$ drawn from the Monte--Carlo simulations 
(solid lines) in comparison with Gaussian distributions calculated from 
r.m.s. values (dashed lines).
We find that the distribution does not fit the Gaussian very well.
Then, we examine distribution of the normalized bispectrum, 
$B_{l_1l_2l_3}/\left(C_{l_1}C_{l_2}C_{l_3}\right)^{1/2}$.
We find that the distribution is very much Gaussian except 
for $l_1=l_2=l_3=2$.
Figure~\ref{fig:dist_norm} plots the distributions of the 9 modes 
of the normalized bispectrum (solid lines) in comparison with 
Gaussian distributions calculated from r.m.s. values (dashed lines).
The distribution fits the Gaussian remarkably well; this
motivates our using standard statistical methods developed for 
Gaussian fields to analyze the normalized bispectrum.
We could not make this simplification if we were analyzing the bare bispectrum.
Furthermore, the central limit theorem implies that when we combine 466 modes
the deviation of the distribution from Gaussianity becomes even smaller.

%%%%%%%%%%%%%%%%%%%%%%%%%%%%%%%%%%%%%%%%%%%%%%%%%%%%%%%%%%%%%%%%%%%%%%
%%%%%%%%%%%%%%%%%%%%%%%%%%%%%%%%%%%%%%%%%%%%%%%%%%%%%%%%%%%%%%%%%%%%%%

%%%%%%%%%%%%%%%%%%%%%%%%%%%%%%%%%%%%%%%%%%%%%%%%%%%%%%%%%%%%%%%%%%%%%%
%%%%%%%%%%%%%%%%%%%%%%%%%%%%%%%%%%%%%%%%%%%%%%%%%%%%%%%%%%%%%%%%%%%%%%

\citet{FMG98} claim detection of the normalized bispectrum 
at $l_1=l_2=l_3=16$; 
\citet{Mag00} claims that the scatter of the 
normalized bispectrum for $l_1=l_2-1$ and $l_3=l_2+1$ is 
too small to be consistent with Gaussian.
The former has analyzed 9 modes, while the latter has analyzed 8 modes.
In the next section, we analyze 466 modes, testing the statistical 
significance of the non-Gaussianity with much more samples than the 
previous work.
We calculate $C_l$ from equation~(\ref{eq:cl**}), and then
divide $B_{l_1l_2l_3}$ by $\left(C_{l_1}C_{l_2}C_{l_3}\right)^{1/2}$
to obtain the normalized bispectrum.

\subsection{Testing Gaussianity of the DMR map}

We characterize statistical significance of the normalized bispectrum
as probability of the measured normalized bispectrum being greater than 
those drawn from the Monte--Carlo simulations.
We define the probability $P$ as 
%%%%%%%%%%%%%%%%%%%%%%%%%%%%%%%%%%%%%%%%%%%%%%%%%%%%%%%%%%%%%%%%%
\begin{equation}
 \label{eq:significance}
  P_\alpha\equiv 
  \frac{N\left(\left|I_{\alpha}^{\rm DMR}\right|
	  >\left|I_{\alpha}^{\rm MC}\right|\right)}
  {N_{\rm total}}
  =
  \int_{-\left|I_\alpha^{\rm DMR}\right|}^{\left|I_\alpha^{\rm DMR}\right|}
  dx~F_\alpha^{\rm MC}(x),
\end{equation}
%%%%%%%%%%%%%%%%%%%%%%%%%%%%%%%%%%%%%%%%%%%%%%%%%%%%%%%%%%%%%%%%%  
where $I_\alpha$ is the normalized bispectrum,
$N_{\rm total}=50,000$ is the total number of simulated realizations,
and $\alpha=$ 1, 2, 3, 4,\dots, 466 represent
$(l_1,l_2,l_3)=$
(2,2,2), (2,3,3), (2,2,4), (3,3,4),\dots, (20,20,20), respectively,
with satisfying $l_1\le l_2\le l_3$, 
$\left|l_i-l_j\right|\leq l_k \leq l_i+l_j$, and $l_1+l_2+l_3={\rm even}$.
$F_\alpha^{\rm MC}(x)$ is the probability density distribution function
(p.d.f) of the simulated 
realizations for the normalized bispectrum, $x=I_{\alpha}^{\rm MC}$.
The p.d.f is normalized to unity: 
$\int_{-\infty}^\infty dx F_\alpha^{\rm MC}(x) =1$; thus,
$P_\alpha$ lies in $0\le P_\alpha\le 1$.

By construction, the distribution of $P_\alpha$ is uniform, if
the DMR map is consistent with the simulated realizations, i.e.,
Gaussian.
We give the proof as follows.
By rewriting equation~(\ref{eq:significance}) as 
$P_\alpha=f(\left|I_\alpha^{\rm DMR}\right|)$, we calculate the p.d.f 
of $P_\alpha$, $G(P_\alpha)$, as
%%%%%%%%%%%%%%%%%%%%%%%%%%%%%%%%%%%%%%%%%%%%%%%%%%%%%%%%%%%%%%%%%
\begin{eqnarray}
 \nonumber
 G(P_\alpha)&=& 
  \int_{-\infty}^\infty dy~\delta\left[P_\alpha=f(\left|y\right|)\right]
  F_\alpha^{\rm DMR}(y)\\
  &=&
   \nonumber
   \int_{0}^\infty dy~\delta\left[P_\alpha=f(y)\right]
   \left[F_\alpha^{\rm DMR}(y)+F_\alpha^{\rm DMR}(-y)\right]\\
 &=&
  \nonumber
   \int_{0}^\infty dy~
   \frac{\delta\left[y=f^{-1}(P_\alpha)\right]}{df/dy}
   \left[F_\alpha^{\rm DMR}(y)+F_\alpha^{\rm DMR}(-y)\right]\\
 &=&
  \label{eq:uniformity}
  \int_{0}^\infty dy~\delta\left[y=f^{-1}(P_\alpha)\right]
  \frac{F_\alpha^{\rm DMR}(y)+F_\alpha^{\rm DMR}(-y)}
  {F^{\rm MC}_\alpha(y)+F^{\rm MC}_\alpha(-y)},
\end{eqnarray}
%%%%%%%%%%%%%%%%%%%%%%%%%%%%%%%%%%%%%%%%%%%%%%%%%%%%%%%%%%%%%%%%%
where $F_\alpha^{\rm DMR}(y)$ is the p.d.f of the measured normalized
bispectrum on the DMR map, $y=I_\alpha^{\rm DMR}$.
Our goal is to see if $F_\alpha^{\rm DMR}(y)$ is consistent with
the DMR data being Gaussian.
It follows from equation~(\ref{eq:uniformity}) 
that $G(P_\alpha)\equiv 1$, when 
$F_\alpha^{\rm DMR}(y)\equiv F^{\rm MC}_\alpha(y)$, regardless of 
the functional form of $F^{\rm MC}_\alpha(y)$.
In other words, the distribution of $P_\alpha$ is uniform, if
the distribution of the measured normalized bispectrum is the same as
the simulated realizations.
Since our simulation assumes the DMR map Gaussian, the $P$ 
distribution, $G(P)$, tests the Gaussianity of the DMR map.
If the $P$ distribution is not uniform, then we conclude the DMR data 
to be non-Gaussian.

A Gaussian field gives equal number of modes 
in each bin of $P$. 
For example, it gives 46.6 modes in $\Delta P=10\%$ bin:
$466\times G(P)\Delta P=466\times 0.1= 46.6$.
If we detect the normalized bispectrum significantly, then we find that
$G(P)$ is not uniform, but increases rapidly as $P$ increases.

\begin{figure}
 \plotone{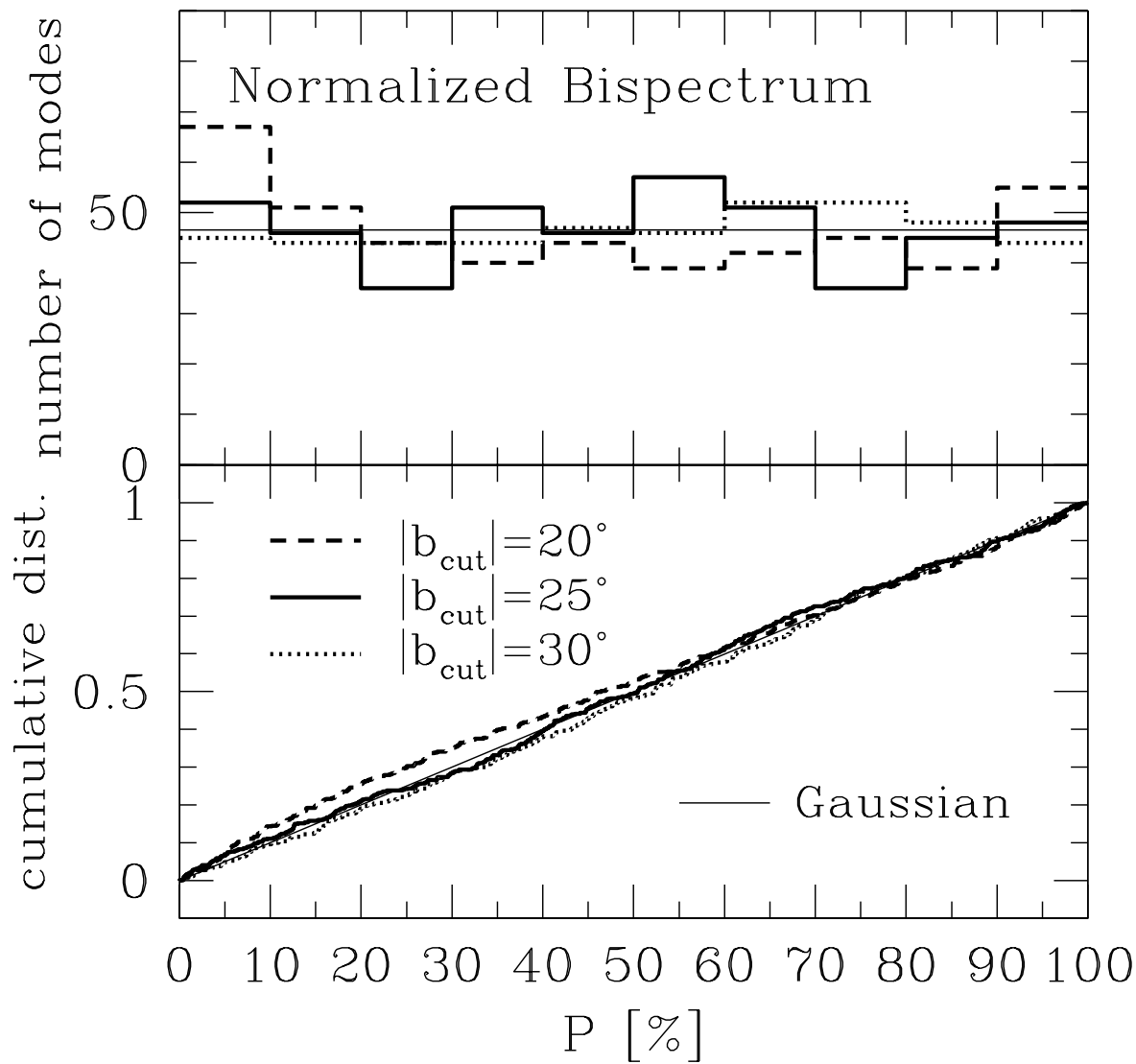}
 \caption{KS Test for Gaussianity with Bispectrum}
 \mycaption
 {$P$ distribution (Eq.(\ref{eq:significance})).
 $P$ is the probability of the CMB normalized bispectrum, 
 $B_{l_1l_2l_3}/\left(C_{l_1}C_{l_2}C_{l_3}\right)^{1/2}$, 
 measured on the {\it COBE} DMR $53+90~{\rm GHz}$ sky map, 
 being larger than those drawn from the Monte--Carlo simulations.
 There are 466 modes in total.
 The thick dashed, solid, and dotted lines represent 
 the three different Galactic cuts as quoted in the figure.
 The thin solid line shows the expectation value for a Gaussian field.
 The top panel shows the $P$ distribution, while the bottom panel
 shows the cumulative $P$ distribution,
 for which we calculate the KS statistic.
 The KS statistic gives the probability of the distribution being 
 consistent with the expectation for Gaussianity as
 6.7\%, 77\%, and 52\% for the three Galactic cuts, respectively.}
\label{fig:KS_norm}
\end{figure}

The top panel of figure~\ref{fig:KS_norm} plots the $P$ distribution
for the three different Galactic cuts.
We find that the distribution is uniform, and the number of modes
in the bin ($\Delta P=10\%$) is consistent with the expectation value 
for Gaussian fluctuations (46.6).

%%%%%%%%%%%%%%%%%%%%%%%%%%%%%%%%%%%%%%%%%%%%%%%%%%%%%%%%%%%%%%%%%%%%%%
%%%%%%%%%%%%%%%%%%%%%%%%%%%%%%%%%%%%%%%%%%%%%%%%%%%%%%%%%%%%%%%%%%%%%%

To further quantify how well it is uniform, we calculate 
the Kolmogorov--Smirnov (KS) statistic for the $P$ distribution in 
comparison with the uniform distribution.
The bottom panel of figure~\ref{fig:KS_norm} plots the cumulative 
$P$ distribution, for which we calculate the KS statistic.
The probability of the distribution being uniform is 6.7\%, 77\%, and 52\% 
for the three Galactic cuts, respectively.
While we have confirmed that the normalized bispectrum at $l_1=l_2=l_3=16$ 
has $P=97.8\%$ for the $20^\circ$ cut and $P=99.3\%$ for the $25^\circ$
cut as similar to \citet{FMG98}, 
our result shows that it is well within statistical fluctuations.
Thus, the properties of the normalized bispectrum of the DMR map are
consistent with CMB being a Gaussian field.

%%%%%%%%%%%%%%%%%%%%%%%%%%%%%%%%%%%%%%%%%%%%%%%%%%%%%%%%%%%%%%%%%
\section{Model Fitting}
\label{sec:fit}

In this section, we fit predicted CMB bispectra to the measured 
normalized bispectrum.
The predictions include the primary bispectrum from inflation
and the interstellar foreground bispectrum from the Galactic emissions.
Then, we constrain a parameter characterizing the primary bispectrum.

\subsection{Primary bispectrum}

For the primary bispectrum from inflation, we consider weakly 
non-Gaussian adiabatic perturbations generated through non-linearity 
in slow-roll inflation. 
The simplest weak non-linear coupling gives
%%%%%%%%%%%%%%%%%%%%%%%%%%%%%%%%%%%%%%%%%%%%%%%%%%%%%%%%%%%%%%%%%%
\begin{equation}
%  \label{eq:modelreal}
  \Phi({\mathbf x})
 =\Phi_{\rm L}({\mathbf x})
 +f_{\rm NL}\left[
              \Phi^2_{\rm L}({\mathbf x})-
	      \left<\Phi^2_{\rm L}({\mathbf x})\right>
        \right],
\end{equation}
%%%%%%%%%%%%%%%%%%%%%%%%%%%%%%%%%%%%%%%%%%%%%%%%%%%%%%%%%%%%%%%%%%
where the square bracket denotes the volume average, and 
$\Phi_{\rm L}({\mathbf x})$ is a linear Gaussian part of 
curvature perturbations. 
We call $f_{\rm NL}$ the non-linear coupling parameter.

\citet{SB90,SB91} and \citet{Gan94} show that
slow-roll inflation gives this coupling; 
\citet{PC96} shows that the second-order general relativistic 
perturbation theory gives this.  
The former predicts $f_{\rm NL}$ as a certain combination of 
slope and curvature of a inflaton potential
($\Phi_3=-2f_{\rm NL}$ in \citet{Gan94}).
The latter predicts $f_{\rm NL}\sim{\cal O}(1)$.
In chapter~\ref{chap:theory_bl}, we have given the exact form of 
$B_{l_1l_2l_3}$ for this model; we do not repeat it here.
$f_{\rm NL}$ is the parameter that we try to constrain by measuring
the CMB bispectrum.

Since the theoretical bispectrum assumes full sky coverage,
we must correct it for the bias arising from incomplete sky
coverage.
We use an approximate correction factor for the bias, 
$\Omega_{\rm obs}/4\pi$, which we have derived in chapter~\ref{chap:spectrum}.
Moreover, the theoretical bispectrum must also be convolved with the 
DMR beam.
We use the harmonic transform of the DMR beam, $G_l$, given in \citet{Wri94}.
Hence, the observed bispectrum is related to the theoretical one
through
%%%%%%%%%%%%%%%%%%%%%%%%%%%%%%%%%%%%%%%%%%%%%%%%%%%%%%%%%%%%%%%%%%
\begin{equation}
 \label{eq:theory2obs}
  B_{l_1l_2l_3}^{\rm obs}
  =\frac{\Omega_{\rm obs}}{4\pi}
  B_{l_1l_2l_3}^{\rm theory}
  G_{l_1}G_{l_2}G_{l_3}.
\end{equation}
%%%%%%%%%%%%%%%%%%%%%%%%%%%%%%%%%%%%%%%%%%%%%%%%%%%%%%%%%%%%%%%%%%
Note that $\Omega_{\rm obs}/4\pi= 1-\sin \left|b_{\rm cut}\right|$
for an azimuthally symmetric cut within certain latitude $b_{\rm cut}$;
$\Omega_{\rm obs}/4\pi= 0.658$, 0.577, and 0.5 for 
$\left|b_{\rm cut}\right|=20^\circ$, $25^\circ$, and $30^\circ$, 
respectively.

\subsection{Foreground bispectra from interstellar emissions}

Although we cut a fraction of the sky to reduce interstellar 
emissions from the Galactic plane, there should be some residuals
at high Galactic latitude.
\citet{Kog96a} have found significant correlation between {\it COBE} DMR
maps at high Galactic latitude and {\it COBE} Diffuse Infrared 
Background Experiment (DIRBE) maps which mainly trace dust emission 
from the Galactic plane.

The interstellar emissions are highly non-Gaussian.
For example, the one-point p.d.f of the all-sky dust template map 
\citep{SFD98} is highly skewed.
We find the normalized skewness, 
$\left<(\Delta T)^3\right>/\left<(\Delta T)^2\right>^{3/2}\sim 51$.
Since these non-Gaussian emissions would confuse the parameter 
estimation of the primary CMB bispectrum, we take the effect into account.

We estimate the foreground bispectra from interstellar sources
by using two foreground template maps.
One is the dust template map of \citet{SFD98}; the other is the 
synchrotron map of \citet{Has81}.
Both maps are in the HEALPix format \citep{GHW98}.

We extrapolate the dust map to 53~GHz and 90~GHz with taking into account
spatial variations of dust temperatures across the sky \citep{Fin99}.
We then cross-correlate the extrapolated maps with the DMR maps
to confirm that the extrapolation is reasonable.
We find that while the dust-correlated emission in the DMR 90~GHz map 
is consistent with the extrapolated dust emission,
that in the DMR 53~GHz map is much larger than the extrapolated one.
This is consistent with the anomalous microwave emission of \citet{Kog96a}. 
To take the excess emission into account, we multiply our
extrapolated 53~GHz maps by factors of 3.66, 2.59, and 2.45
for the $20^\circ$ cut, the $25^\circ$ cut, and the $30^\circ$ cut, 
respectively.
Note that we do not essentially need the correction for the excess emission,
as it does not alter spatial distribution of the emission.
Nevertheless, we do it for convenience of subsequent analyses.

We also extrapolate the synchrotron map to these two bands, assuming
the spectrum of the source, $T(\nu)\propto \nu^{-2.9}$. 
We do not need the extrapolation of the synchrotron template map either, 
as the extrapolation does not alter spatial distribution of 
the emission in contrast to the dust template map in which the 
extrapolation does alter it.
We find no significant correlation between the DMR maps and the 
extrapolated synchrotron maps at both 53~GHz and 90~GHz.

After coadding the extrapolated 53 and 90~GHz maps with the same 
weight as used for the DMR maps, we measure $B_{l_1l_2l_3}$ from the
maps for the three different Galactic cuts, multiplying it by 
$G_{l_1}G_{l_2}G_{l_3}$ to take into account the DMR beam.

\subsection{Constraints on non-linearity in inflation}

We simultaneously fit the primary, dust, and synchrotron bispectra to 
the measured bispectrum on the DMR map.
We use the least-squares method based on a $\chi^2$ statistic defined by
%%%%%%%%%%%%%%%%%%%%%%%%%%%%%%%%%%%%%%%%%%%%%%%%%%%%%%%%%%%%%%%%%%%%
\begin{equation}
  \label{eq:chisq}
  \chi^2(f_j)\equiv 
  \sum_{\alpha\alpha'}
  \left(I^{\rm DMR}_\alpha-\sum_jf_jI_\alpha^{j}\right)
  \left(C^{-1}\right)_{\alpha\alpha'}
  \left(I^{\rm DMR}_{\alpha'}-\sum_jf_jI_{\alpha'}^{j}\right).
\end{equation}
%%%%%%%%%%%%%%%%%%%%%%%%%%%%%%%%%%%%%%%%%%%%%%%%%%%%%%%%%%%%%%%%%%%%
$I_\alpha^{j}$ is a model bispectrum divided by 
$\left(\left<C_{l_1}^{\rm MC}\right>
\left<C_{l_2}^{\rm MC}\right>\left<C_{l_3}^{\rm MC}\right>\right)^{1/2}$, 
where $j$ represents a certain component such as the primary, dust, 
and synchrotron. 
$f_j$ is a fitting parameter for a component $j$,
where $f_{\rm primary}\equiv f_{\rm NL}$ is the non-linear coupling
parameter (Eq.(\ref{eq:modelreal})). 
$f_{\rm dust}$ and $f_{\rm sync}$ characterize amplitude of the 
foreground bispectra.

$C_{\alpha\alpha'}$ is the covariance matrix of the 
normalized bispectrum calculated from the Monte--Carlo simulations:
%%%%%%%%%%%%%%%%%%%%%%%%%%%%%%%%%%%%%%%%%%%%%%%%%%%%%%%%%%%%%%%%%%%%
\begin{equation}
  \label{eq:cov}
  C_{\alpha\alpha'}\equiv
  \frac1{N-1} 
  \sum_{i=1}^{N}
  \left(I^{{\rm MC}(i)}_\alpha -\left<I^{\rm MC}_\alpha \right>\right)
  \left(I^{{\rm MC}(i)}_{\alpha'}-\left<I^{\rm MC}_{\alpha'}\right>\right),
\end{equation}
%%%%%%%%%%%%%%%%%%%%%%%%%%%%%%%%%%%%%%%%%%%%%%%%%%%%%%%%%%%%%%%%%%%% 
where $N=50,000$ is the number of realizations. 
The bracket denotes an average over all realizations,
$\left<I^{\rm MC}_\alpha \right>\equiv N^{-1} \sum_i^N 
I_\alpha^{{\rm MC}(i)}$.  
Here, we have implicitly assumed the non-Gaussianity weak, so that
we calculate the covariance matrix from Gaussian realizations.

As we have observed in the previous section, distribution of
$I_\alpha$ is very much Gaussian; thus, $\chi^2(f_j)$ should obey 
the $\chi^2$ distribution to good accuracy.
Hence, minimizing $\chi^2(f_j)$ with respect to $f_j$ gives 
the maximum-likelihood value of $f_j$ as a solution to the normal equation:
%%%%%%%%%%%%%%%%%%%%%%%%%%%%%%%%%%%%%%%%%%%%%%%%%%%%%%%%%%%%%%%%%%%%
\begin{equation}
 \label{eq:normal}
  f_j = \sum_i\left(F^{-1}\right)_{ji}
  \left[\sum_{\alpha\alpha'}
  I_\alpha^i\left(C^{-1}\right)_{\alpha\alpha'}I_{\alpha'}^{\rm DMR}
 \right],
\end{equation}
%%%%%%%%%%%%%%%%%%%%%%%%%%%%%%%%%%%%%%%%%%%%%%%%%%%%%%%%%%%%%%%%%%%%
where
%%%%%%%%%%%%%%%%%%%%%%%%%%%%%%%%%%%%%%%%%%%%%%%%%%%%%%%%%%%%%%%%%%%%
\begin{equation}
 \label{eq:fisher}
  F_{ij}\equiv
  \sum_{\alpha\alpha'}
  I_\alpha^i\left(C^{-1}\right)_{\alpha\alpha'}I_{\alpha'}^j.
\end{equation}
%%%%%%%%%%%%%%%%%%%%%%%%%%%%%%%%%%%%%%%%%%%%%%%%%%%%%%%%%%%%%%%%%%%%
We estimate statistical uncertainties of the parameters using 
the Monte--Carlo simulations; we obtain parameter realizations
by substituting $I_\alpha^{\rm MC}$ for $I_\alpha^{\rm DMR}$ 
in equation~(\ref{eq:normal}).

\begin{figure}
 \plotone{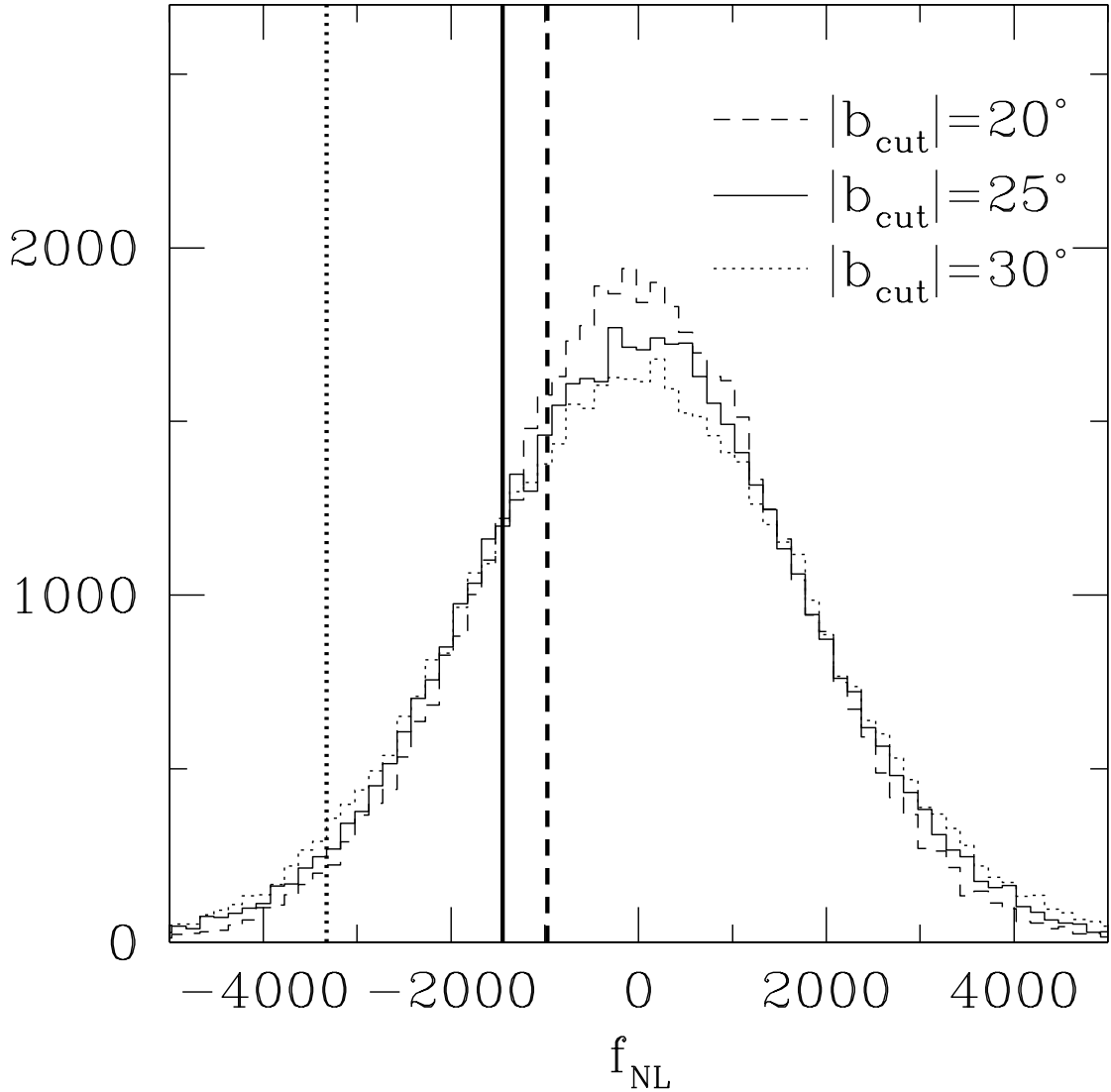}
 \caption{Constraint on Non-linearity in Inflation}
 \mycaption
 {DMR constraint on the non-linear coupling parameter, $f_{\rm NL}$, which
 characterizes non-linearity in inflation (Eq.(\ref{eq:modelreal})).
 The dashed, solid, and dotted lines represent the three different 
 Galactic cuts as quoted in the figure.
 The thick vertical lines plot the measured values of $f_{\rm NL}$
 from the {\it COBE} DMR maps, while the histograms plot those drawn
 from the Monte--Carlo simulations for each cut.
 68\% confidence limits on $f_{\rm NL}$ are 
 $\left|f_{\rm NL}\right|<1.6\times 10^3$, $1.7\times 10^3$, and
 $2.0\times 10^3$ for the three Galactic cuts, respectively.}
\label{fig:fNL}
\end{figure}

Figure~\ref{fig:fNL} plots the measured values of the non-linear
coupling parameter, $f_{\rm NL}$, as well as the simulated realizations, 
for the three different Galactic cuts.
The measured values are well within the cosmic variance: we place 
68\% confidence limits on $f_{\rm NL}$ as
$\left|f_{\rm NL}\right|<1.6\times 10^3$, $1.7\times 10^3$, and
$2.0\times 10^3$, for the $20^\circ$, $25^\circ$, and $30^\circ$ cuts,
respectively.

%%%%%%%%%%%%%%%%%%%%%%%%%%%%%%%%%%%%%%%%%%%%%%%%%%%%%%%%%%%%%%%%%%%%%%
%%%%%%%%%%%%%%%%%%%%%%%%%%%%%%%%%%%%%%%%%%%%%%%%%%%%%%%%%%%%%%%%%%%%%%

\begin{figure}
 \plotone{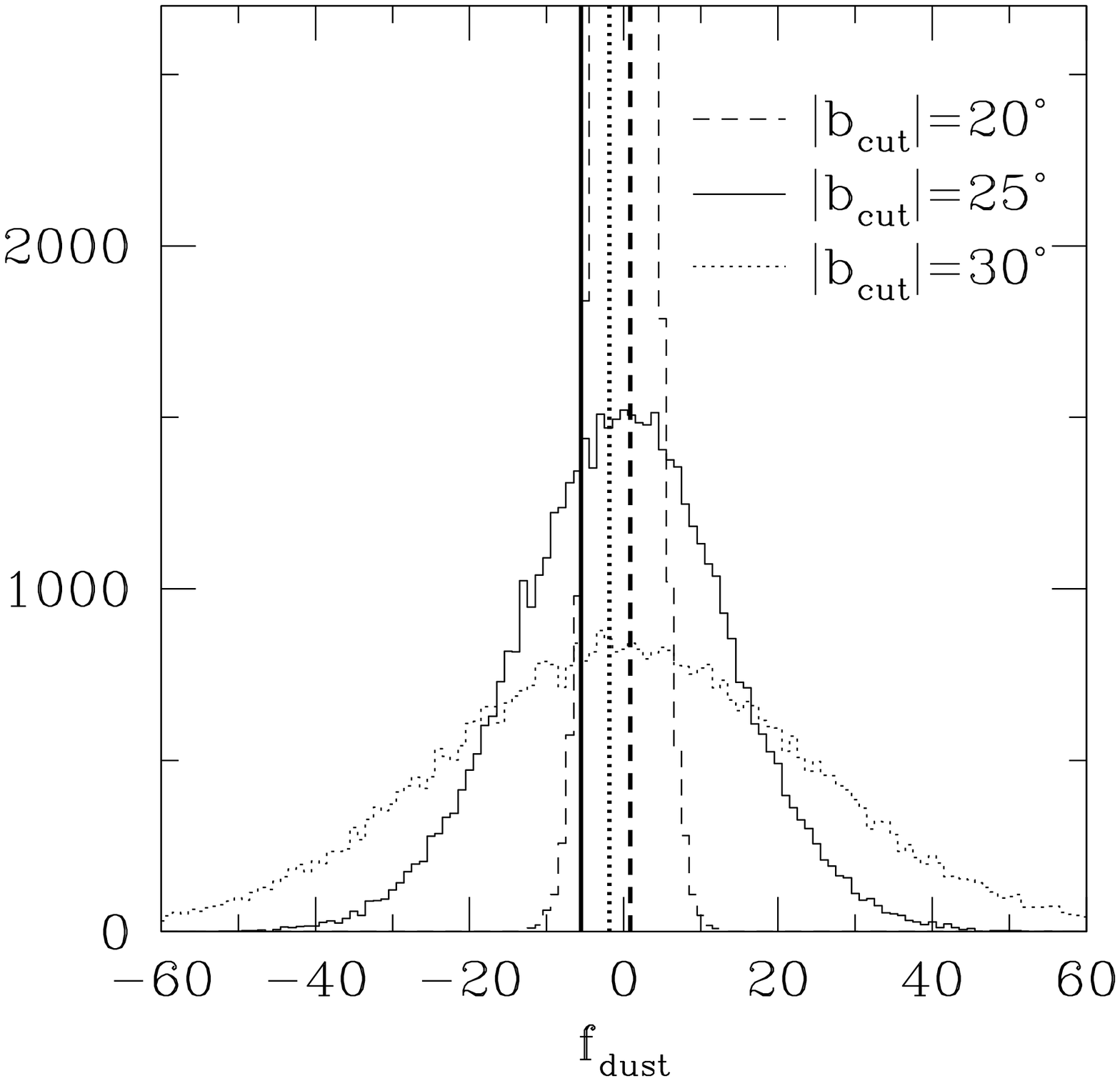}
 \caption{Constraint on Dust Bispectrum}
 \mycaption
 {DMR constraint on the amplitude of the interstellar dust bispectrum
 at high Galactic latitude, $f_{\rm dust}$.
 The meaning of the lines is the same as in figure~\ref{fig:fNL}.}
\label{fig:fdust}
\end{figure}

\begin{figure}
 \plotone{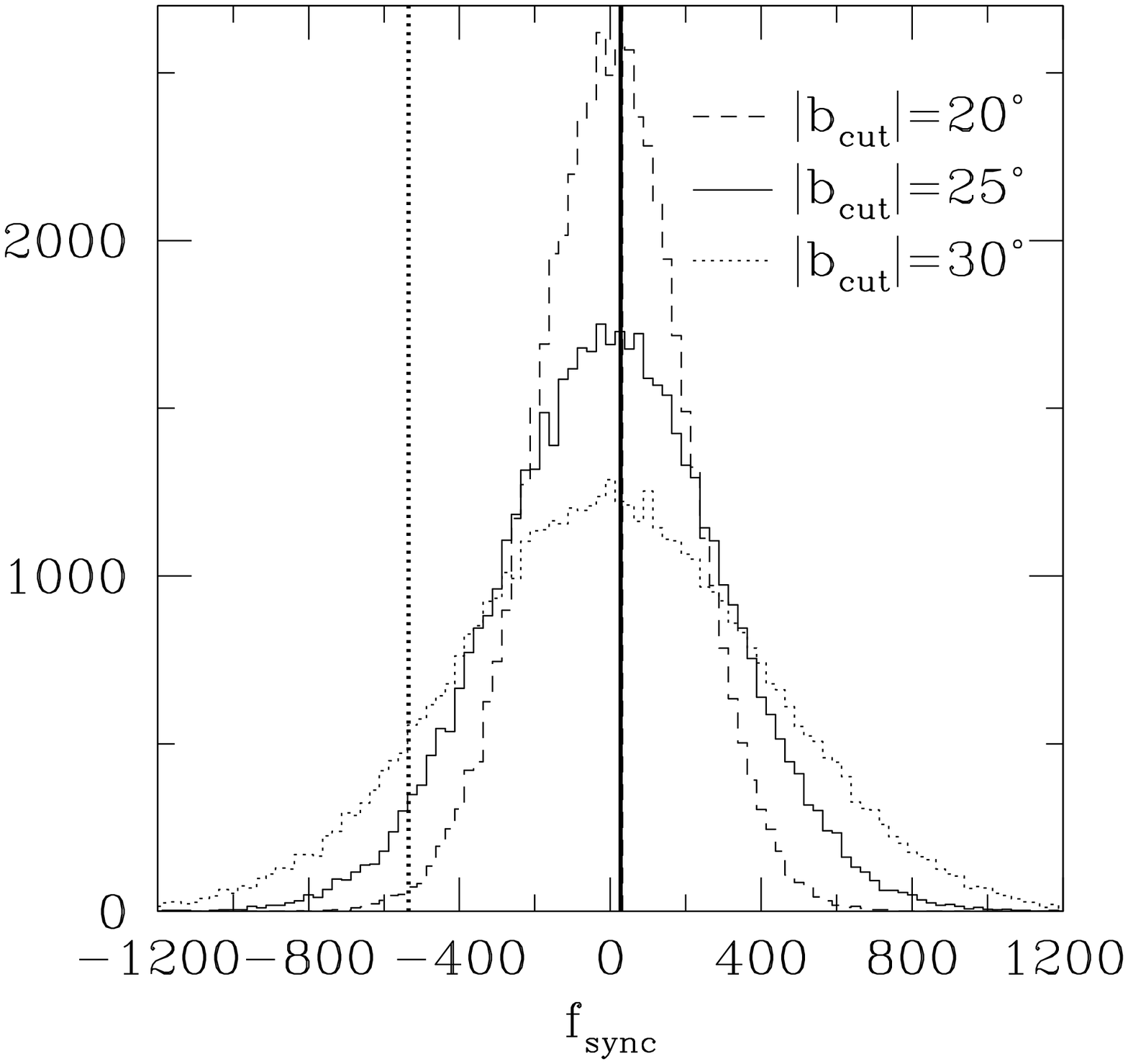}
 \caption{Constraint on Synchrotron Bispectrum}
 \mycaption
 {DMR constraint on the amplitude of the interstellar synchrotron bispectrum
  at high Galactic latitude,  $f_{\rm sync}$.
 The meaning of the lines is the same as in figure~\ref{fig:fNL}.}
\label{fig:fsync}
\end{figure}

Figures~\ref{fig:fdust} and \ref{fig:fsync} plot
constraints on $f_{\rm dust}$ and $f_{\rm sync}$, respectively.
There is no indication of either component contributing
to the measured bispectrum significantly.
The constraints are rapidly weakened as the Galactic cut is widened.

%%%%%%%%%%%%%%%%%%%%%%%%%%%%%%%%%%%%%%%%%%%%%%%%%%%%%%%%%%%%%%%%%%%%%%
%%%%%%%%%%%%%%%%%%%%%%%%%%%%%%%%%%%%%%%%%%%%%%%%%%%%%%%%%%%%%%%%%%%%%%

%%%%%%%%%%%%%%%%%%%%%%%%%%%%%%%%%%%%%%%%%%%%%%%%%%%%%%%%%%%%%%%%%%%%%%
%%%%%%%%%%%%%%%%%%%%%%%%%%%%%%%%%%%%%%%%%%%%%%%%%%%%%%%%%%%%%%%%%%%%%%

\subsection{Null test of the normalized bispectrum}

Using $\chi^2$ defined by equation~(\ref{eq:chisq}), we can test
Gaussianity of the DMR map. 
While the minimization of $\chi^2(f_j)$ gives constraints on 
the parameters, a value of $\chi^2(f_j)$ tells us goodness-of-fit;
$\chi^2(0)$ tests a hypothesis of the bispectrum being zero.
When $\chi^2(0)$ is either significantly greater or smaller than 
those drawn from the simulations, we conclude that the 
DMR map is inconsistent with zero bispectrum.

$\chi^2(0)$ is similar to what several authors have used
for quantifying statistical significance of non-Gaussianity in 
the DMR map \citep{FMG98,Mag00,SM00}.
They use only diagonal terms of the covariance matrix; however,
the matrix is diagonal only on the full sky.
As lack of sky coverage correlates one mode to the others, 
we should include off-diagonal terms as well.
We did so in equation~(\ref{eq:chisq}).

\begin{figure} 
 \plotone{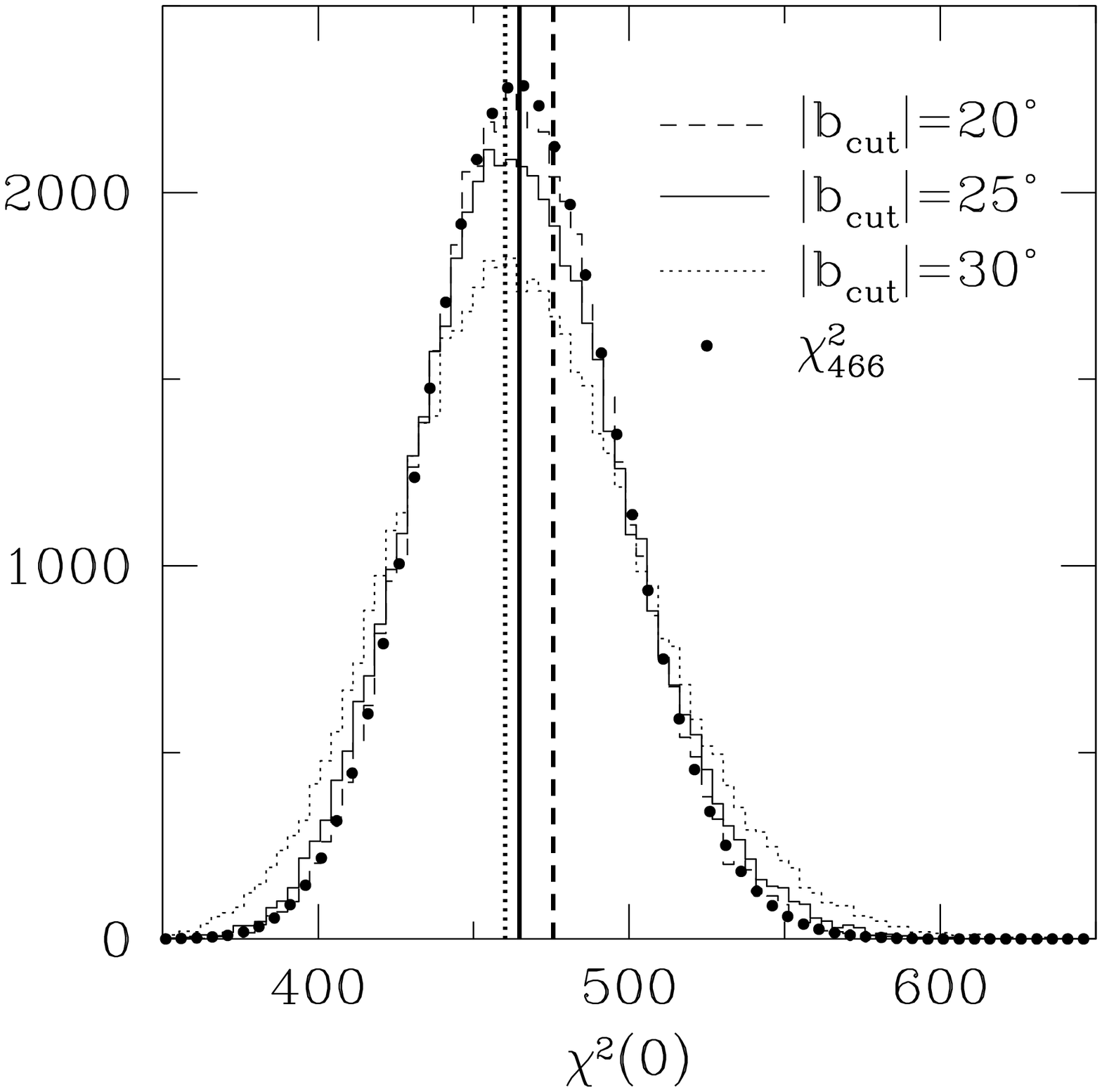}
 \caption{Null Test of Normalized Bispectrum}
 \mycaption
 {Testing hypothesis of the normalized bispectrum, 
 $B_{l_1l_2l_3}/\left(C_{l_1}C_{l_2}C_{l_3}\right)^{1/2}$, being zero
 in the {\it COBE} DMR four-year $53+90$~GHz sky map.
 The dashed, solid, and dotted lines represent 
 the $20^\circ$, $25^\circ$, and $30^\circ$ cuts, respectively.
 The thick vertical lines plot the measured $\chi^2(0)$, while the 
 histograms plot those drawn from the Monte--Carlo simulations.
 The filled circles plot the $\chi^2$ distribution for 466 degrees of 
 freedom.}
\label{fig:null_norm}
\end{figure}

Figure~\ref{fig:null_norm} compares $\chi^2_{\rm DMR}(0)$ with
$\chi^2_{\rm MC}(0)$ for the different Galactic cuts.
The measured values are $\chi^2_{\rm DMR}(0)=475.6$, 464.8, and 460.2 for 
the corresponding cuts, respectively, while 
$\left<\chi^2_{\rm MC}(0)\right>=466$.
We find the probability of $\chi^2_{\rm MC}(0)$ being larger than 
$\chi^2_{\rm DMR}(0)$ to be 
$P\left(\chi^2_{\rm MC}>\chi^2_{\rm DMR}\right)=36.9\%$, 49.7\%, and 
54.2\%, respectively.

%%%%%%%%%%%%%%%%%%%%%%%%%%%%%%%%%%%%%%%%%%%%%%%%%%%%%%%%%%%%%%%%%%%%%%
%%%%%%%%%%%%%%%%%%%%%%%%%%%%%%%%%%%%%%%%%%%%%%%%%%%%%%%%%%%%%%%%%%%%%%

We conclude that the DMR map is comfortably consistent with zero 
normalized bispectrum.
We explain the claimed detection \citep{FMG98} by a statistical 
fluctuation without invoking the eclipse effect of \citet{BZG00}.

There is no evidence that the scatter of
the normalized bispectrum is too small to be consistent with
Gaussian, in contrast to the claim of \citet{Mag00} based on 
$\chi^2(0)$ derived from 8 modes.
To clarify, our analysis does not reject the possibility that the CMB sky
is non-Gaussian for only a small number of modes; however, in the absence
of a theoretical motivation for limiting the analysis to a specific set of
modes, we choose to treat all the bispectrum modes on an equal footing.
\citet{SM00} claim that the non-Gaussianity found by \cite{Mag00}
does not spread to other modes. 
This is consistent with our result.

Incidentally, we plot in the figure~\ref{fig:null_norm} 
the $\chi^2$ distribution for 466 degrees of freedom, 
$\chi^2_{466}$, in filled circles; we find that the distribution of 
$\chi^2_{\rm MC}(0)$ is very similar to the $\chi^2_{466}$ distribution 
for a smaller cut as expected, while it becomes 
slightly broader for a larger cut for which the distribution of the 
normalized bispectrum at some modes deviates from Gaussian appreciably. 
Yet, we find that the $20^\circ-25^\circ$ cuts reasonably 
retain the Gaussianity of the distribution of the normalized bispectrum.

%%%%%%%%%%%%%%%%%%%%%%%%%%%%%%%%%%%%%%%%%%%%%%%%%%%%%%%%%%%%%%%%%
\section{Discussion and Conclusions}
\label{sec:discussion_bl}

In this chapter, we have measured all independent 
configurations of the angular bispectrum on the {\it COBE} DMR map,
down to the DMR beam size.
Using the most sensitive sky map to CMB, which combines the maps at 
53 and 90~GHz, we test the Gaussianity of the DMR map.

We find that the normalized bispectrum, 
$B_{l_1l_2l_3}/\left(C_{l_1}C_{l_2}C_{l_3}\right)^{1/2}$, 
gives more robust test of Gaussianity than the bare bispectrum, 
$B_{l_1l_2l_3}$. 
We compare the measured data with the simulated realizations,
finding the DMR map comfortably consistent with Gaussian.
We explain the reported detection of the normalized bispectrum
at $l_1=l_2=l_3=16$ \citep{FMG98} by a statistical fluctuation
as an alternative to the "eclipse effect" proposition 
made in \citet{BZG00}.

We fit the predicted bispectra to the data, constraining the parameters 
in the predictions, which include the primary bispectrum from inflation
and the foreground bispectra from interstellar dust and synchrotron
emissions.
We find that neither dust nor synchrotron emissions contribute to
the bispectrum significantly.

We have obtained a weak constraint on the non-linear coupling parameter,
$f_{\rm NL}$, that characterizes non-linearity in inflation. 
We interpret the constraint in terms of a single-field inflation as follows. 
According to the analysis of non-linear perturbations on super horizon 
scales \citep{SB90}, we can explicitly calculate $f_{\rm NL}$ as
%%%%%%%%%%%%%%%%%%%%%%%%%%%%%%%%%%%%%%%%%%%%%%%%%%%%%%%%%%%%%%%%%
\begin{equation}
 \label{eq:SB91}
  f_{\rm NL}= 
  -\frac{5}{24\pi G}
  \left(\frac{\partial^2\ln H}{\partial\phi^2}\right),
\end{equation}
%%%%%%%%%%%%%%%%%%%%%%%%%%%%%%%%%%%%%%%%%%%%%%%%%%%%%%%%%%%%%%%%%
where $H$ is the Hubble parameter during inflation.
When applying the slow-roll conditions to an inflaton potential 
$V(\phi)$, we have $\partial\ln H/\partial\phi\approx (d\ln V/d\phi)/2$;
thus, $f_{\rm NL}$ is on the order of curvature of a slow-roll potential,
implying that $\left|f_{\rm NL}\right|$ should be smaller than 1 in
slow-roll inflation.
Therefore, the obtained constraint, 
$\left|f_{\rm NL}\right|<1.6\times 10^3$, seems too weak to be 
interesting; however, any deviation from slow-roll could yield larger 
$\left|f_{\rm NL}\right|$, bigger non-Gaussianity.

The next generation satellite experiments, {\it MAP} and {\it Planck}, 
should be able to put more
stringent constraints on $f_{\rm NL}$. 
In chapter~\ref{chap:theory_bl}, we have shown 
that {\it MAP} and {\it Planck} should be sensitive down to 
$\left|f_{\rm NL}\right|\sim 20$ and 5, respectively.
We find that the actual constraint from {\it COBE} (figure~\ref{fig:fNL}) is 
much worse than the estimate.
This is partly due to different cosmology used for the model, but mainly 
due to incomplete sky coverage; the statistical power of the 
bispectrum at low multipoles is significantly weakened by the Galactic cut.
Since {\it MAP} and {\it Planck} probe much smaller angular scales, 
and their better angular resolution makes an extent of the Galactic cut 
smaller,
the degradation of sensitivity should be minimal. 
Moreover, the improved frequency coverage of future experiments 
will aid in extracting more usable CMB pixels from the data.
At this level of sensitivity, any deviation from slow-roll could give 
an interesting amount of the bispectrum, and {\it MAP} and {\it Planck}
will put severe constraints on any substantial deviation from slow-roll.

While we have explored adiabatic generation of the 
bispectrum only, isocurvature perturbations from inflation also generate
non-Gaussianity \citep{LM97,P97,BZ97}. 
They are in general more non-Gaussian than the adiabatic
perturbations; it is worth constraining these models by the same
strategy as we have done in this chapter.

%%%%%%%%%%%%%%%%%%%%%%%%%%%%%%%%%%%%%%%%%%%%%%%%%%%%%%%%%%%%%%%%%%%
%
%  In Pursuit of Angular Trispectrum
%
%     1st draft:  07/12/2001
%     revision:   07/17/2001
%     final:      08/05/2001
%
%%%%%%%%%%%%%%%%%%%%%%%%%%%%%%%%%%%%%%%%%%%%%%%%%%%%%%%%%%%%%%%%%%%
\chapter{In Pursuit of Angular Trispectrum}
\label{chap:obs_tl}

%%%%% Why study trispectrum? %%%%%

The angular trispectrum, the harmonic transform of the angular 
four-point correlation function, carries cosmological information
which is independent of the power spectrum and the bispectrum.

Several authors show that the weak lensing effect produces 
non-trivial connected CMB angular trispectrum or four-point correlation
function on small angular scales \citep{Ber97,ZS99,Zal00,Hu01}.
On large angular scales, \citet{Ino01a,Ino01b} shows that 
the connected trispectrum is produced if topology of our 
universe is closed flat or closed hyperbolic.
These effects do not produce the bispectrum, the angular three-point
harmonic spectrum, but the trispectrum.
Hence, while we have not found significant bispectrum on the DMR map in 
chapter~\ref{chap:obs_bl}, we could find the trispectrum.

%%%%% power spectrum covariance %%%%%

The connected angular trispectrum contributes to the power spectrum
covariance matrix.
It increases the variance, and produces non-zero off-diagonal terms in
the covariance matrix.
Measuring the connected trispectrum thus
constrains how much the connected trispectrum contributes to 
the power spectrum covariance.
This is important to do.
We have to know the power spectrum covariance matrix as accurate as 
possible, when we measure the power spectrum with better than 
1\% accuracy, and determine
many of cosmological parameters with better than 10\% accuracy.

%%%%% First measurement %%%%%

So far, there has been no attempt to measure the angular trispectrum 
of the CMB anisotropy.
In this chapter, we present the first measurement of 
the CMB trispectrum on the DMR map.
We measure all independent terms of the angular trispectrum down to
the DMR beam size, and test Gaussianity of the DMR data.

%%%%% structure of the chapter %%%%%

This chapter is organized as follows.
In \S~\ref{sec:trispectrum*}, we give our method of measuring 
the angular trispectrum from CMB sky maps. 
In \S~\ref{sec:norm_tl}, we study statistical properties of 
the normalized trispectrum.
In \S~\ref{sec:test}, applying the method to the DMR data, we measure 
the angular trispectrum.
We then test Gaussianity of the DMR data with the normalized trispectrum.
Finally, \S~\ref{sec:conclusion_obstl} concludes this chapter,

%%%%%%%%%%%%%%%%%%%%%%%%%%%%%%%%%%%%%%%%%%%%%%%%%%%%%%%%%%%%%%%%%
\section{Angular Trispectrum}
\label{sec:trispectrum*}

The angular trispectrum comprises four harmonic transforms of the 
CMB temperature anisotropy field, 
$a_{l_1m_1}a_{l_2m_2}a_{l_3m_3}a_{l_4m_4}$, where 
%%%%%%%%%%%%%%%%%%%%%%%%%%%%%%%%%%%%%%%%%%%%%%%%%%%%%%%%%%%%%%%%%%
\begin{equation}
 \label{eq:alm**}
  a_{lm}= 
  \int_{\Omega_{\rm obs}} 
  d^2\hat{\mathbf n}~\frac{\Delta T\left(\hat{\mathbf n}\right)}T
  Y_{lm}^*\left(\hat{\mathbf n}\right).
\end{equation}
%%%%%%%%%%%%%%%%%%%%%%%%%%%%%%%%%%%%%%%%%%%%%%%%%%%%%%%%%%%%%%%%%%
$\Omega_{\rm obs}$ denotes a solid angle of the observed sky. 
A rotationally invariant angular averaged trispectrum,
$T^{l_1l_2}_{l_3l_4}(L)$, takes the form \citep{Hu01}
%%%%%%%%%%%%%%%%%%%%%%%%%%%%%%%%%%%%%%%%%%%%%%%%%%%%%%%%%%%%%%%%%%%%%%
\begin{eqnarray}
 \nonumber
  T^{l_1l_2}_{l_3l_4}(L)
  &=& (2L+1)\sum_{{\rm all}~m}\sum_M(-1)^M
  \left(\begin{array}{ccc}l_1&l_2&L\\m_1&m_2&M\end{array}\right)
  \left(\begin{array}{ccc}l_3&l_4&L\\m_3&m_4&-M\end{array}\right)\\
 \label{eq:test*}
 & &\times
  a_{l_1m_1}a_{l_2m_2}a_{l_3m_3}a_{l_4m_4},
\end{eqnarray}
%%%%%%%%%%%%%%%%%%%%%%%%%%%%%%%%%%%%%%%%%%%%%%%%%%%%%%%%%%%%%%%%%%%%%%
where the Matrices denote the Wigner-3$j$ symbol.
By construction, $l_1$, $l_2$, and $L$ form one triangle,
while $l_3$, $l_4$, and $L$ form the other triangle in a quadrilateral
with sides of $l_1$, $l_2$, $l_3$, and $l_4$.
$L$ represents a diagonal of the quadrilateral.
Figure~\ref{fig:quad} sketches a configuration of the angular 
trispectrum.
When we arrange $l_1$, $l_2$, $l_3$, and $l_4$ in order of 
$l_1\le l_2\le l_3\le l_4$, $L$ lies in 
$\max(l_2-l_1,l_4-l_3)\le L\le \min(l_1+l_2,l_3+l_4)$.
Hence, $l_1=l_2$ and $l_3=l_4$ for $L=0$.
Parity invariance of the angular four-point correlation function 
demands $l_1+l_2+L=\mbox{even}$ and $l_3+l_4+L=\mbox{even}$.

The angular trispectrum generically consists of two parts.
One part is the unconnected part, the contribution from Gaussian fields,
which is given by the angular power spectra \citep{Hu01},
%%%%%%%%%%%%%%%%%%%%%%%%%%%%%%%%%%%%%%%%%%%%%%%%%%%%%%%%%%%%%%%%%%
\begin{eqnarray}
 \nonumber
  \left<T^{l_1l_2}_{l_3l_4}(L)\right>_{\rm unconnected}
  &=&
  (-1)^{l_1+l_3}\sqrt{(2l_1+1)(2l_3+1)}\left<C_{l_1}\right>\left<C_{l_3}\right>
  \delta_{l_1l_2}\delta_{l_3l_4}\delta_{L0} \\
 \label{eq:unconnected}
  & & 
  + (2L+1)\left<C_{l_1}\right>\left<C_{l_2}\right>\left[(-1)^{l_2+l_3+L}
		       \delta_{l_1l_3}\delta_{l_2l_4}
		     +\delta_{l_1l_4}\delta_{l_2l_3}\right].
\end{eqnarray}
%%%%%%%%%%%%%%%%%%%%%%%%%%%%%%%%%%%%%%%%%%%%%%%%%%%%%%%%%%%%%%%%%%
For $l_1\le l_2\le l_3\le l_4$, the unconnected terms are 
non-zero only when $L=0$ or $l_1=l_2=l_3=l_4$.
We have numerically confirmed that our estimator given below 
(Eq.(\ref{eq:tobs})) accurately reproduces the unconnected terms
(Eq.(\ref{eq:unconnected})) on a simulated Gaussian sky.

The other part is the connected part whose expectation value is 
exactly zero for Gaussian fields; thus, the connected part is sensitive 
to non-Gaussianity.
When none of $l$'s are same in $T^{l_1l_2}_{l_3l_4}(L)$,
one might expect the trispectrum to comprise the connected part only;
however, it is true only on the full sky.
The unconnected terms on the incomplete sky, which are often much bigger 
than the connected terms, leak the power to the other modes for which
all $l$'s are different.
We should take this effect into account in the analysis.

Using the azimuthally averaged harmonic transform, 
$e_{l}(\hat{\mathbf n})$ (Eq.(\ref{eq:el})), 
we rewrite equation~(\ref{eq:test*}) into a much more computationally 
efficient form,
%%%%%%%%%%%%%%%%%%%%%%%%%%%%%%%%%%%%%%%%%%%%%%%%%%%%%%%%%%%%%%%%%%
\begin{equation}
  \label{eq:tobs}
   T^{l_1l_2}_{l_3l_4}(L)
   =
   \frac1{2L+1} \sum_{M=-L}^L t_{LM}^{l_1l_2*} t_{LM}^{l_3l_4},
\end{equation}
%%%%%%%%%%%%%%%%%%%%%%%%%%%%%%%%%%%%%%%%%%%%%%%%%%%%%%%%%%%%%%%%%%
where
%%%%%%%%%%%%%%%%%%%%%%%%%%%%%%%%%%%%%%%%%%%%%%%%%%%%%%%%%%%%%%%%%%
\begin{equation}
% \label{eq:tLM*}
  t_{LM}^{l_1l_2}
  \equiv
  \sqrt{\frac{2L+1}{4\pi}}  
  \left(
   \begin{array}{ccc}
    l_1 & l_2 & L \\ 0 & 0 & 0 
   \end{array}
 \right)^{-1} 
  \int d^2\hat{\mathbf n}
  \left[e_{l_1}(\hat{\mathbf n}) e_{l_2}(\hat{\mathbf n})\right]
  Y_{LM}^*(\hat{\mathbf n}).
\end{equation}
%%%%%%%%%%%%%%%%%%%%%%%%%%%%%%%%%%%%%%%%%%%%%%%%%%%%%%%%%%%%%%%%%%
Since $t_{LM}^{l_1l_2}$ is the harmonic transform on the full sky, 
we can calculate it quickly.
This method makes measurement of the angular trispectrum computationally
feasible even for the {\it MAP} data in which we have more than millions of
pixels; thus, the methods developed here can be applied not only to the 
{\it COBE} DMR data, but also to the {\it MAP} data.

For the DMR data for which the maximum $l_4$ is 20, 
we have 21,012 non-zero trispectrum modes after taking into account 
the triangle conditions in a quadrilateral and parity invariance.
Measurement of 21,012 modes takes about 5 second of CPU time on 
a Pentium-III single processor personal computer.

We use Monte-Carlo simulations to estimate the covariance matrix
of the angular trispectrum.
We have described details of our simulations in 
chapter~\ref{chap:obs_bl}.
We generate 5,000 realizations for one simulation; processing one 
realization takes about 5 second, so that 
one simulation takes about 7 hours of CPU time on a Pentium-III single 
processor personal computer.
In total, we run 6 simulations, generating 30,000 realizations.

%%%%%%%%%%%%%%%%%%%%%%%%%%%%%%%%%%%%%%%%%%%%%%%%%%%%%%%%%%%%%%%%%%
\section{Normalized Trispectrum}
\label{sec:norm_tl}

In general, analytic forms of the angular trispectrum covariance,
$\left<T^{l_1l_2}_{l_3l_4}(L)T^{l_1'l_2'}_{l_3'l_4'}(L')\right>$,
are highly complicated \citep{Hu00}; however, one can reduce them to rather
simplified forms for $l_1\leq l_2< l_3\leq l_4$ and $L\neq 0$
(chapter~\ref{chap:spectrum}).
The covariance of these terms on the full sky is diagonal, and 
the variance is given by equation~(\ref{eq:variance4}),
%%%%%%%%%%%%%%%%%%%%%%%%%%%%%%%%%%%%%%%%%%%%%%%%%%%%%%%%%%%%%%%%%
\begin{equation}
 \label{eq:variance4*}
  {\left<\left[T^{l_1l_2}_{l_3l_4}(L)\right]^2\right>} 
  = {(2L+1)\left<C_{l_1}\right>\left<C_{l_2}\right>\left<C_{l_3}\right>
  \left<C_{l_4}\right>}
  \left(
   1 + \delta_{l_1l_2} + \delta_{l_3l_4} + \delta_{l_1l_2}\delta_{l_3l_4}
   \right).
\end{equation}
%%%%%%%%%%%%%%%%%%%%%%%%%%%%%%%%%%%%%%%%%%%%%%%%%%%%%%%%%%%%%%%%%% 
One finds that the variance of the trispectrum is very sensitive to the 
power spectrum normalization.
A slight difference in the power spectrum normalization alters 
the variance of the trispectrum substantially.
This makes a test for Gaussianity with the ``bare'' trispectrum very
difficult, as it requires precise determination of the power spectrum
normalization.

We overcome the difficulty by normalizing the trispectrum as
%%%%%%%%%%%%%%%%%%%%%%%%%%%%%%%%%%%%%%%%%%%%%%%%%%%%%%%%%%%%%%%%%
\begin{equation}
 \frac{T^{l_1l_2}_{l_3l_4}(L)}
  {\left[(2L+1)C_{l_1}C_{l_2}C_{l_3}C_{l_4}\right]^{1/2}}.
\end{equation}
%%%%%%%%%%%%%%%%%%%%%%%%%%%%%%%%%%%%%%%%%%%%%%%%%%%%%%%%%%%%%%%%%
This statistic, the normalized trispectrum, is analogous to the 
normalized bispectrum that we have used in chapter~\ref{chap:obs_bl}.
As similar to the normalized bispectrum, the variance of the 
normalized trispectrum is insensitive to the power spectrum
normalization, and systematically smaller than that of the bare
trispectrum.
The normalized trispectrum is thus reasonably sensitive to non-Gaussianity.

\begin{figure}
 \plotone{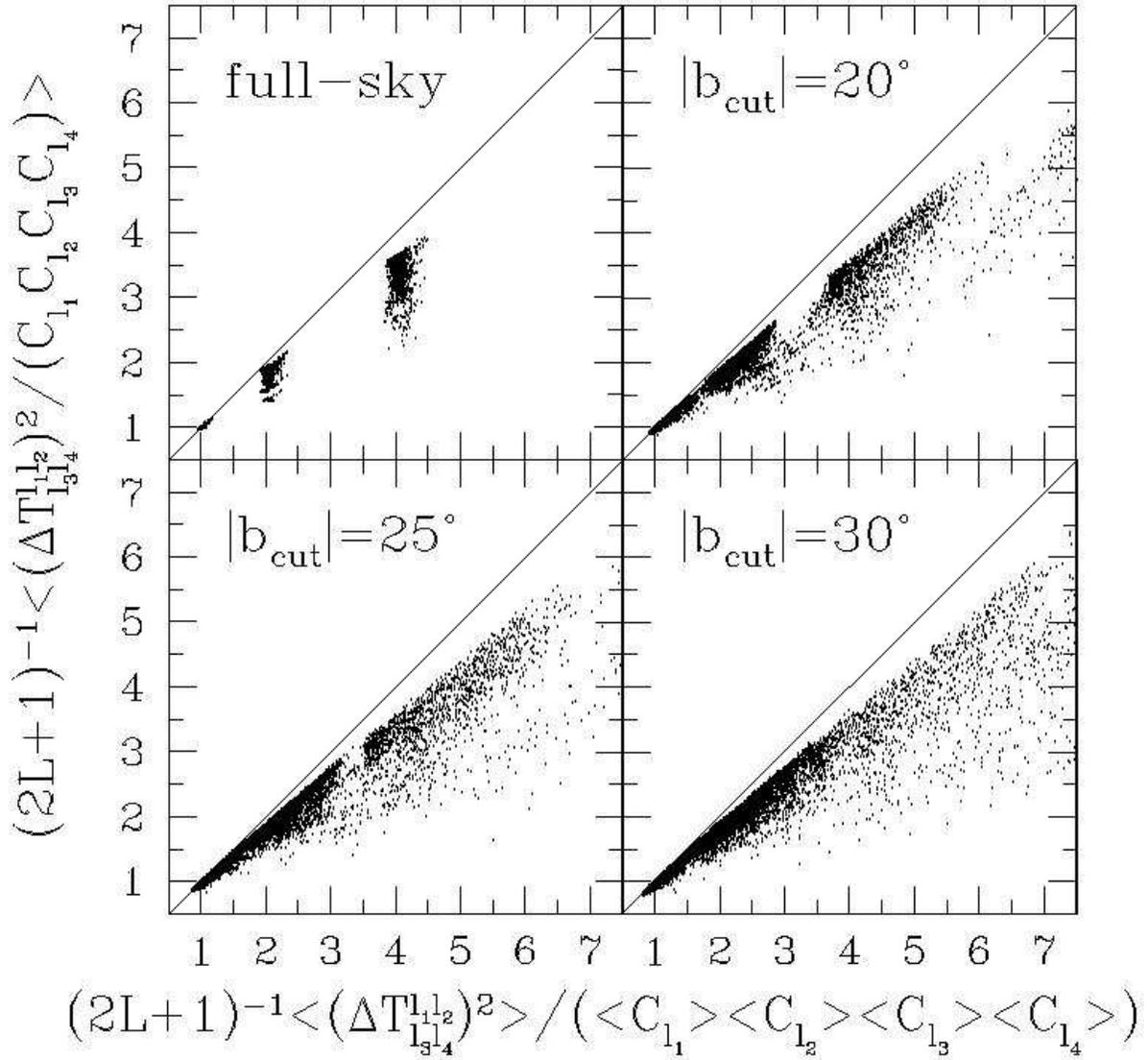}
 \caption
 {Variance of Normalized Trispectrum and Bare Trispectrum} 
 \mycaption{Comparison of the variance of the normalized trispectrum
 with that of the bare trispectrum for
 $l_1\le l_2< l_3\le l_4$ and $L\neq 0$ (group~(a)).
 These are derived from simulated realizations of a Gaussian sky.
 The top-left panel shows the case of full sky coverage, while
 the rest of panels show the cases of incomplete sky coverage.
 The top-right, bottom-left, and bottom-right panels use
 the $20^\circ$, $25^\circ$, and $30^\circ$ Galactic cuts, respectively.}
\label{fig:variance_tl}
\end{figure}

Figure~\ref{fig:variance_tl} compares the variance of the normalized 
trispectrum with that of the bare trispectrum, for full sky coverage
as well as for incomplete sky coverage.
We have used $l_1\leq l_2< l_3\leq l_4$ and $L\neq 0$ terms, and 
calculated the variance from simulated realizations of a Gaussian sky.
We confirm that the variance of the normalized trispectrum is 
systematically smaller than that of the bare trispectrum, and 
that the variance distribution becomes more scattered on 
the incomplete sky.

%%%%%%%%%%%%%%%%%%%%%%%%%%%%%%%%%%%%%%%%%%%%%%%%%%%%%%%%%%%%%%%%%%%%%%
%%%%%%%%%%%%%%%%%%%%%%%%%%%%%%%%%%%%%%%%%%%%%%%%%%%%%%%%%%%%%%%%%%%%%%

%%%%%%%%%%%%%%%%%%%%%%%%%%%%%%%%%%%%%%%%%%%%%%%%%%%%%%%%%%%%%%%%%
\section{Testing Gaussianity of the DMR Map}
\label{sec:test}

In this section, we test Gaussianity of the DMR data using the
normalized trispectrum.
Before performing the analysis, we should recall that even if CMB is 
exactly Gaussian, there are significant, non-zero unconnected 
trispectrum terms for $L=0$ or $l_1=l_2=l_3=l_4$.
We should analyze these configurations separately from the others
for which the unconnected terms vanish.

\subsection{Classification of trispectrum configurations}

We divide the measured 21,012 modes into four groups:
%%%%%%%%%%%%%%%%%%%%%%%%%%%%%%%%%%%%%%%%%%%%%%%%%%%%%%%%%%%%%%%%%
\begin{itemize}
 \item [(a)] $l_2\neq l_3$, and $L\ne 0$ (16,554 modes)
 \item [(b)] $l_2=l_3$, $l_1\neq l_4$, and $L\neq 0$ (4,059 modes)
 \item [(c)] $l_1=l_2=l_3=l_4$, and $L\neq 0$ (209 modes)
 \item [(d)] $L=0$ (190 modes)
\end{itemize}
%%%%%%%%%%%%%%%%%%%%%%%%%%%%%%%%%%%%%%%%%%%%%%%%%%%%%%%%%%%%%%%%%
Note that all the groups satisfy $l_1\leq l_2\leq l_3\leq l_4$,
$\max(l_2-l_1,l_4-l_3)\le L\le \min(l_1+l_2,l_3+l_4)$, 
$l_1+l_2+L=\mbox{even}$, and $l_3+l_4+L=\mbox{even}$.

The groups (a) and (b) are most sensitive to non-Gaussian
signals, as for which the unconnected terms~(Eq.(\ref{eq:unconnected})) 
vanish on the full sky. 
On the other hand, the groups (c) and (d)
are dominated by the unconnected terms, and thus
less sensitive to non-Gaussianity.
These statements should, however, depend upon what kind of
non-Gaussianity exists in the data.
If non-Gaussian signals are subject to the groups (c) and (d) only, then
the groups (a) and (b) become the least sensitive modes to the non-Gaussianity.

On the incomplete sky, the unconnected terms also contaminate 
the groups (a) and (b) through the mode-mode coupling.
We take this effect into account by using the Monte-Carlo simulations
that use the same Galactic cut as the data.

We discriminate between the group (a) and the group (b) in terms of the 
covariance matrix: the covariance matrix of the group (a) is diagonal, 
while that of the group (b) is not diagonal in $L$.
We discriminate between the group (c) and the group (d) in terms
of the statistical power: 
the group (d) has no statistical power of testing Gaussianity.
The reason is as follows.
For the group (d), the estimator given by equation~(\ref{eq:test*})
becomes $T^{l_1l_1}_{l_3l_3}(0)=(-1)^{l_1+l_3}
\sqrt{(2l_1+1)(2l_3+1)}C_{l_1}C_{l_3}$.
The normalized bispectrum for the group (d) is thus just a pure number,
%%%%%%%%%%%%%%%%%%%%%%%%%%%%%%%%%%%%%%%%%%%%%%%%%%%%%%%%%%%%%%%%%
\begin{equation}
 \label{eq:remarkable}
 \frac{T^{l_1l_1}_{l_3l_3}(0)}{C_{l_1}C_{l_3}}
  = (-1)^{l_1+l_3}\sqrt{(2l_1+1)(2l_3+1)}.
\end{equation}
%%%%%%%%%%%%%%%%%%%%%%%%%%%%%%%%%%%%%%%%%%%%%%%%%%%%%%%%%%%%%%%%%
This property holds regardless of Gaussianity.
Even strongly non-Gaussian fields give exactly the same number.
It thus follows from this result that we cannot measure the connected 
trispectrum for $L=0$.

This is unfortunate.
It is the connected part of $T^{l_1l_1}_{l_3l_3}(0)$
that contributes to the covariance matrix of the power spectrum,
%%%%%%%%%%%%%%%%%%%%%%%%%%%%%%%%%%%%%%%%%%%%%%%%%%%%%%%%%%%%%%%%%
\begin{equation}
 \label{eq:covpowerspec}
  \left<C_lC_{l'}\right> - \left<C_l\right>\left<C_{l'}\right>
   =
  \frac{2\left<C_l\right>^2}{2l+1}\delta_{ll'}
   + \frac{(-1)^{l+l'}}{\sqrt{(2l+1)(2l'+1)}}
   \left<T^{ll}_{l'l'}(0)\right>_{\rm c},
\end{equation}
%%%%%%%%%%%%%%%%%%%%%%%%%%%%%%%%%%%%%%%%%%%%%%%%%%%%%%%%%%%%%%%%%
where $\left<T^{ll}_{l'l'}(0)\right>_{\rm c}$ is the ensemble average
of the connected $T^{ll}_{l'l'}(0)$.
Even if we find the groups (a)--(c) consistent with Gaussianity, 
we can conclude nothing about the power spectrum covariance, unless 
we have a model for the connected trispectrum.
We will discuss this point in \S~\ref{sec:conclusion_obstl}.

Henceforth, we analyze the groups (a)--(c) only, while
we have used the group (d) to see if our code works properly.
Our code reproduces equation~(\ref{eq:remarkable}) very well, and
the numerical error is at most of order $10^{-4}$.

\subsection{Gaussianity test}

To quantify statistical significance of the measured trispectrum, 
we use a statistic $P$, which is the probability of the measured normalized 
trispectrum being greater than those drawn from the Monte--Carlo simulations:
%%%%%%%%%%%%%%%%%%%%%%%%%%%%%%%%%%%%%%%%%%%%%%%%%%%%%%%%%%%%%%%%%
\begin{equation}
 \label{eq:significance_tl}
  P_\alpha\equiv
  \frac{N\left(\left|J_{\alpha}^{\rm DMR}\right|
	  >\left|J_{\alpha}^{\rm MC}\right|\right)}
  {N_{\rm total}},
\end{equation}
%%%%%%%%%%%%%%%%%%%%%%%%%%%%%%%%%%%%%%%%%%%%%%%%%%%%%%%%%%%%%%%%%  
where $J_\alpha$ denotes the normalized trispectrum,
$N_{\rm total}=30,000$ is the total number of the simulated realizations,
and $\alpha$ represents a set of $(l_1,l_2,l_3,l_4,L)$.
The distribution of $P_\alpha$ is uniform if the DMR map is consistent with 
Gaussian, for which there are equal number of modes in each bin of $P$.
For example, when we calculate $P_\alpha$ for all 21,012 $\alpha$'s,  
we expect a Gaussian field to give 210.12 modes in $\Delta P=1\%$ bin.
If we detect the normalized trispectrum significantly, then the number 
of modes having higher $P$ is much larger than the expectation value 
for a Gaussian field.

In chapter~\ref{chap:obs_bl}, we have proven the $P$ distribution 
uniform, if the DMR data are consistent with the simulated realizations
(see Eq.(\ref{eq:uniformity})).
We have also shown that this property holds regardless of the
distribution function of $J_{\alpha}^{\rm MC}$.

\begin{figure}
 \plotone{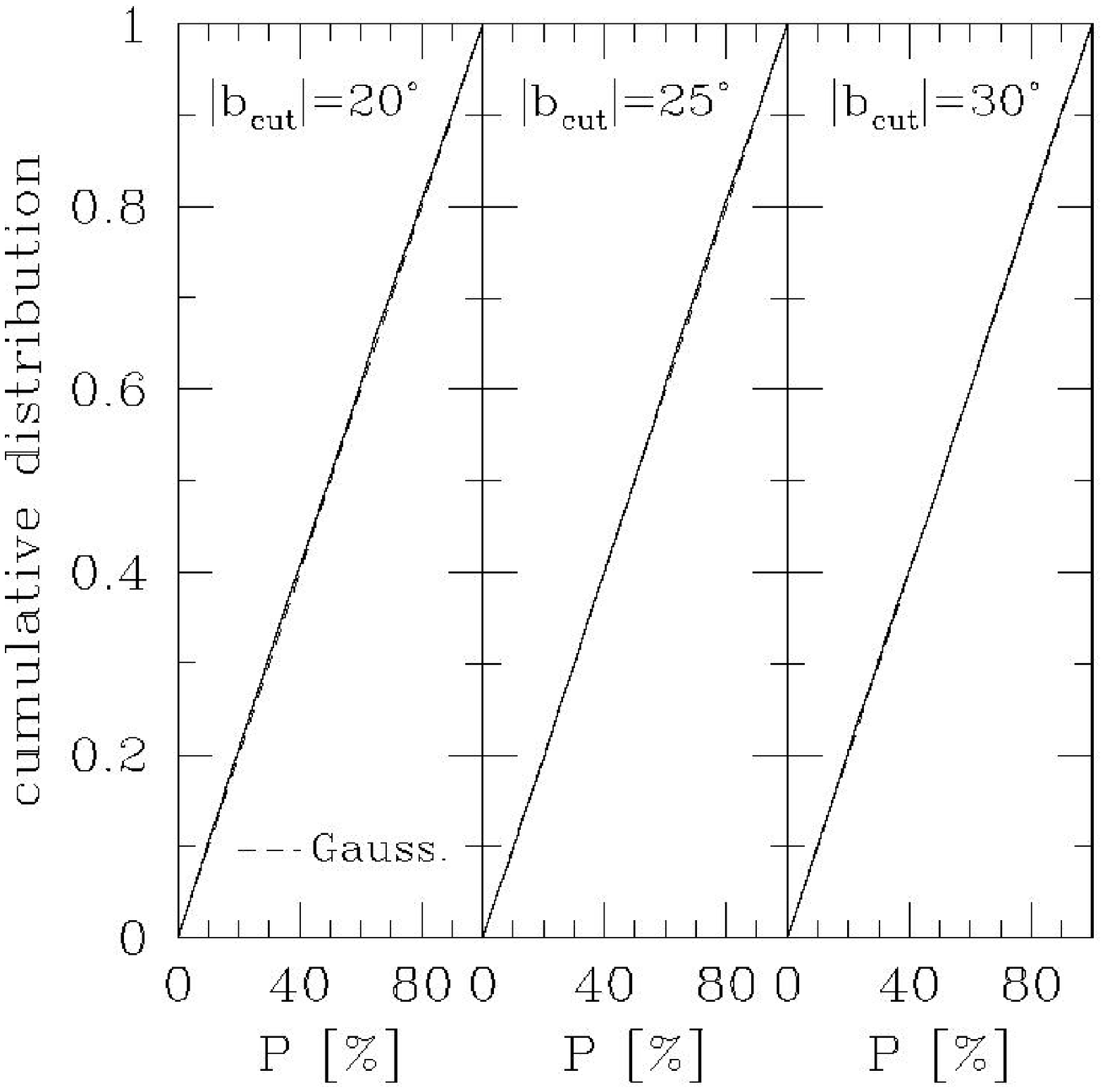}
 \caption{KS Test for Gaussianity with Trispectrum I}
 \mycaption{Cumulative $P$ Distribution (Eq.(\ref{eq:significance_tl})),
 for which we calculate the KS statistic.
 $P$ is the probability of the normalized trispectrum, 
 $T^{l_1l_2}_{l_3l_4}(L)
 /\left[(2L+1)C_{l_1}C_{l_2}C_{l_3}C_{l_4}\right]^{1/2}$,
 for the group (a) ($L\neq 0$ and $l_1\le l_2<l_3\le l_4$), 
 measured on the {\it COBE} DMR $53+90~{\rm GHz}$ sky map, 
 being larger than those drawn from the Monte--Carlo simulations.
 There are 16,654 modes.
 From left to right panels, we use the
 $20^\circ$, $25^\circ$, and $30^\circ$ Galactic cuts, respectively.
 The dashed lines show the expectation value for a Gaussian field.
 The KS statistic gives the probability of the distribution being 
 consistent with Gaussianity as
 5.4\%, 12\%, and 48\% for the three Galactic cuts, respectively.}
\label{fig:KS_norm_tl_c1}
\end{figure}

\begin{figure}
 \plotone{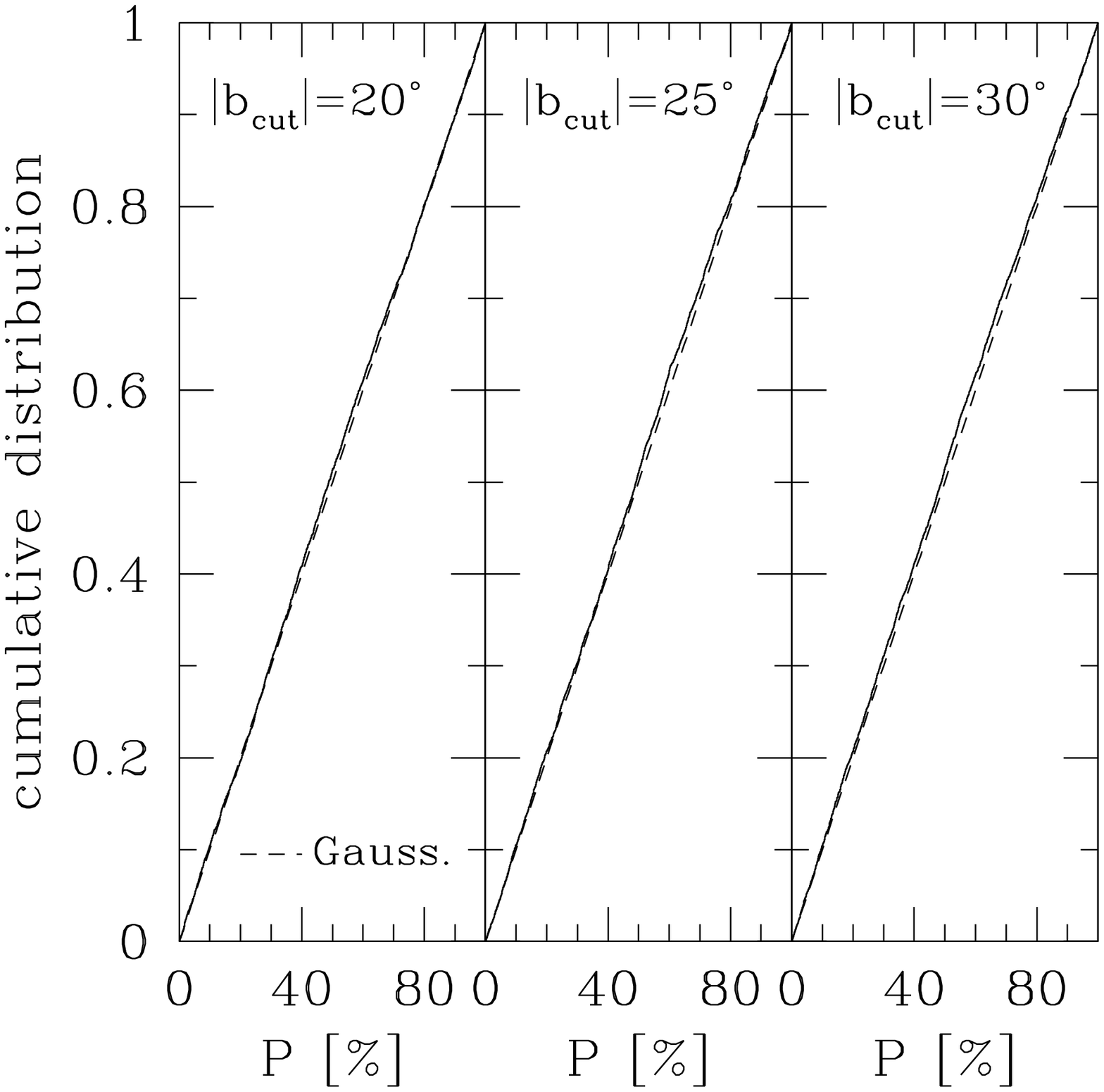}
 \caption{KS Test for Gaussianity with Trispectrum II}
 \mycaption{The same as figure~\ref{fig:KS_norm_tl_c1} but for the group (b)
 ($l_2=l_3$, $l_1\neq l_4$, and $L\neq 0$).
 There are 4,059 modes.
 The KS statistic gives the probability of the distribution being 
 consistent with Gaussianity as
 38\%, 2.5\%, and 5.2\% for the three Galactic cuts, respectively.}
\label{fig:KS_norm_tl_c2}
\end{figure}

\begin{figure} 
 \plotone{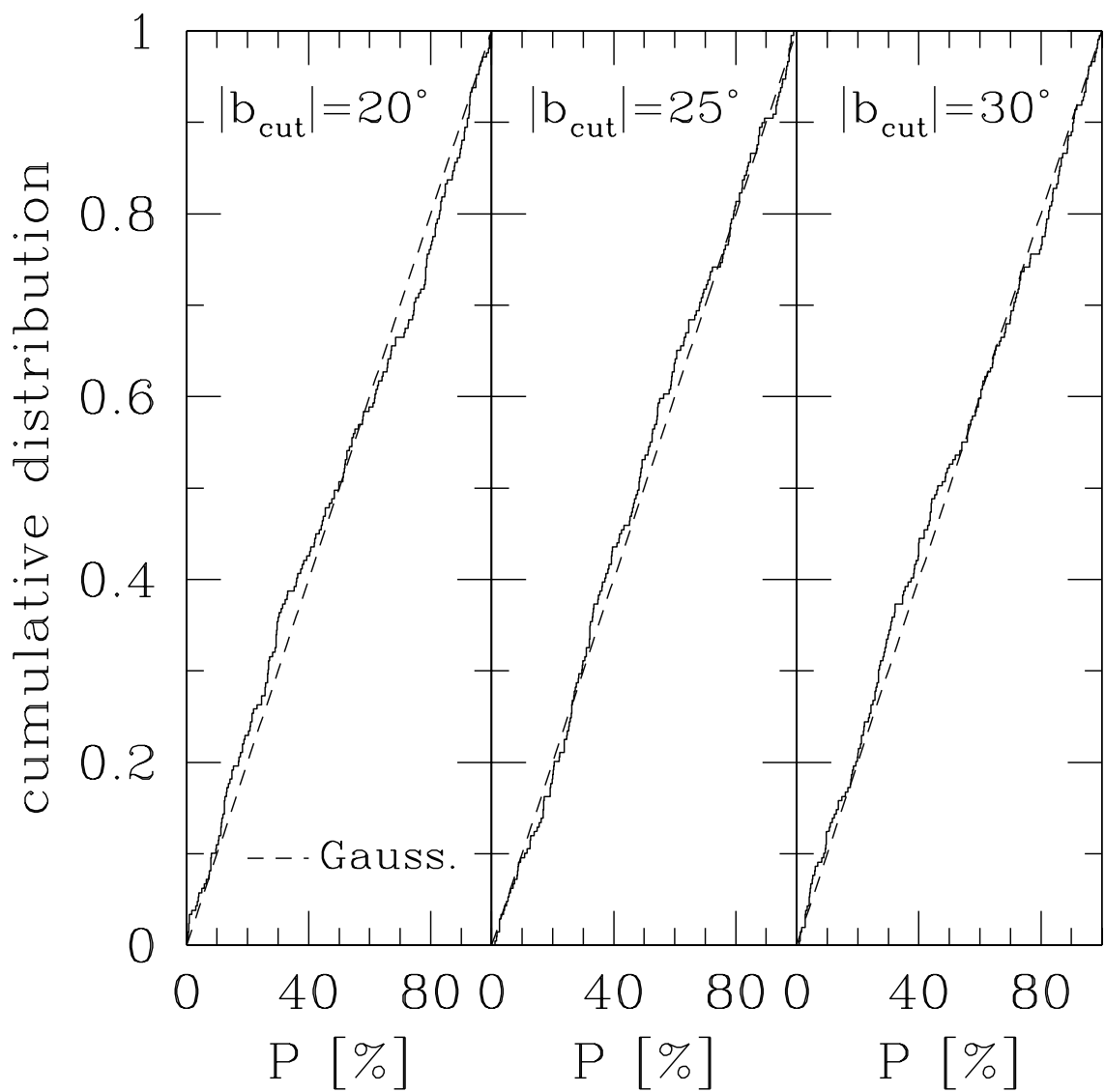}
 \caption{KS Test for Gaussianity with Trispectrum III}
 \mycaption{The same as figure~\ref{fig:KS_norm_tl_c1} but for the group (c)
 ($l_1=l_2=l_3=l_4$ and $L\neq 0$).
 There are 209 modes.
 The KS statistic gives the probability of the distribution being 
 consistent with Gaussianity as
 41\%, 73\%, and 63\% for the three Galactic cuts, respectively.}
\label{fig:KS_norm_tl_unc2}
\end{figure}

We calculate the KS statistic for the $P$ distribution
in comparison with the uniform distribution, to quantify
how well the $P$ distribution is uniform.
We calculate the KS statistic for the groups (a)--(c) separately. 
Table~\ref{tab:KS} summarizes the KS-test results, and
figures~\ref{fig:KS_norm_tl_c1}--\ref{fig:KS_norm_tl_unc2} plot
the cumulative $P$ distribution, for which we have calculated the KS statistic.
We find that the measured trispectrum is comfortably consistent with Gaussianity
for all of the analyzed groups, (a)--(c).

Since the groups (a) and (b) are
zero for Gaussian fields, these groups provide the strongest constraint 
on generic non-Gaussian fluctuations.
As we have done in chapter~\ref{chap:obs_bl} for the angular bispectrum,
if we have predictions for the CMB angular trispectrum 
(e.g., appendix~\ref{app:CH}), then our measurement tests those predictions.

%%%%%%%%%%%%%%%%%%%%%%%%%%%%%%%%%%%%%%%%%%%%%%%%%%%%%%%%%%%%%%%%%%%%%%%%%%%
% Table 1
\begin{table}
 \caption{Gaussianity Test with Normalized Trispectrum}
 \mycaption{Probability of the measured trispectrum being consistent 
 with Gaussianity (the rightmost column) for the three different Galactic 
 cuts.
 The probability is derived from the KS test for the $P$ distribution
 in comparison with the uniform distribution.
 The group (a) comprises the modes of $l_2\neq l_3$, and $L\ne 0$ 
 (16,554 modes), the group (b) of $l_2=l_3$, $l_1\neq l_4$, and 
 $L\neq 0$ (4,059 modes), and the group (c) of $l_1=l_2=l_3=l_4$, and 
 $L\neq 0$ (209 modes).}
\begin{center} 
  \begin{tabular}{cccc}\hline\hline
  group & \# of modes & Galactic cut & probability [\%]\\
  \hline 
   (a)  & 16,554 & $20^\circ$ & 5.4 \\
        &        & $25^\circ$ & 12  \\
        &        & $30^\circ$ & 48  \\
   (b)  & 4,059  & $20^\circ$ & 38  \\
        &        & $25^\circ$ & 2.5 \\
        &        & $30^\circ$ & 5.2 \\
   (c)  & 209    & $20^\circ$ & 41 \\
        &        & $25^\circ$ & 71 \\
        &        & $30^\circ$ & 63 \\
   \hline\hline
  \end{tabular}
\end{center}
 \label{tab:KS}
\end{table}
%%%%%%%%%%%%%%%%%%%%%%%%%%%%%%%%%%%%%%%%%%%%%%%%%%%%%%%%%%%%%%%%%%%%%%%%%%

%%%%%%%%%%%%%%%%%%%%%%%%%%%%%%%%%%%%%%%%%%%%%%%%%%%%%%%%%%%%%%%%%%%%%%
%%%%%%%%%%%%%%%%%%%%%%%%%%%%%%%%%%%%%%%%%%%%%%%%%%%%%%%%%%%%%%%%%%%%%%

%%%%%%%%%%%%%%%%%%%%%%%%%%%%%%%%%%%%%%%%%%%%%%%%%%%%%%%%%%%%%%%%%%%%%%
%%%%%%%%%%%%%%%%%%%%%%%%%%%%%%%%%%%%%%%%%%%%%%%%%%%%%%%%%%%%%%%%%%%%%%

%%%%%%%%%%%%%%%%%%%%%%%%%%%%%%%%%%%%%%%%%%%%%%%%%%%%%%%%%%%%%%%%%%%%%%
%%%%%%%%%%%%%%%%%%%%%%%%%%%%%%%%%%%%%%%%%%%%%%%%%%%%%%%%%%%%%%%%%%%%%%

%%%%%%%%%%%%%%%%%%%%%%%%%%%%%%%%%%%%%%%%%%%%%%%%%%%%%%%%%%%%%%%%%%%%%%
\section{Discussion and Conclusions}
\label{sec:conclusion_obstl}

In this chapter, we have presented the first measurement of the 
CMB angular trispectrum on the {\it COBE} DMR sky maps.
We have measured all the trispectrum terms, 21,012 terms, down to the 
DMR beam size.
Since 190 $L=0$ modes have no statistical power of testing Gaussianity, 
we have used 20,822 $L\neq 0$ modes to test Gaussianity of the DMR
data, and found that the DMR map is comfortably consistent with Gaussianity.

Our results do not directly constrain the connected trispectrum
for $L=0$, $T^{ll}_{l'l'}(0)$, which contributes to the power spectrum 
covariance through equation~(\ref{eq:covpowerspec}).
We can thus conclude nothing as to whether the covariance matrix is 
diagonal on the DMR angular scales.
Moreover, $T^{ll}_{ll}(0)$ increases the power spectrum variance;
we have no idea how much the contribution is.
We need to use other statistics than the angular trispectrum to investigate
the power spectrum covariance.
Otherwise, we have to have a model for the connected trispectrum,
and constrain $T^{ll}_{l'l'}(0)$ by measuring the other trispectrum
configurations.

One example for a trispectrum model is the one produced in a closed 
hyperbolic universe.
\citet{Ino01b} suggests that the closed hyperbolic geometry produces 
non-zero connected trispectrum.
In appendix~\ref{app:CH}, we have derived an analytic prediction for the 
connected trispectrum produced in a closed hyperbolic universe
(Eq.(\ref{eq:CHprediction})).
For $L=0$, we reduce the prediction to
%%%%%%%%%%%%%%%%%%%%%%%%%%%%%%%%%%%%%%%%%%%%%%%%%%%%%%%%%%%%%%%%%%
\begin{equation}
  \left<T^{l_1l_1}_{l_3l_3}(0)\right>_{\rm c}
  =
  (-1)^{l_1+l_3}\sqrt{(2l_1+1)(2l_3+1)}\left(2F_{l_1l_3}\right)
  + 4 F_{l_1l_1}\delta_{l_1l_3}, 
\end{equation}
%%%%%%%%%%%%%%%%%%%%%%%%%%%%%%%%%%%%%%%%%%%%%%%%%%%%%%%%%%%%%%%%%%
where 
%%%%%%%%%%%%%%%%%%%%%%%%%%%%%%%%%%%%%%%%%%%%%%%%%%%%%%%%%%%%%%%%%%%%%%
\begin{equation}
 F_{l_1l_3}=
  \sum_{\nu}P_\Phi^2(\nu)g_{{\rm T}l_1}^2(\nu)g_{{\rm T}l_3}^2(\nu)
  \left<\left|\xi_{l_1m_1}(\nu)\right|^2\right>
  \left<\left|\xi_{l_3m_3}(\nu)\right|^2\right>,
\end{equation}
%%%%%%%%%%%%%%%%%%%%%%%%%%%%%%%%%%%%%%%%%%%%%%%%%%%%%%%%%%%%%%%%%%%%%%
and $\nu=\sqrt{k^2-1}$ is a discrete wavenumber,
$\Phi(\nu)$ is the primordial curvature perturbation,
$g_{{\rm T}l}(\nu)$ is the radiation transfer function, and
$\xi_{lm}(\nu)$ describes eigenmodes in a closed hyperbolic
geometry.
Hence, a single function, $F_{ll'}$, determines the connected
trispectrum completely.

$F_{ll'}$ contributes to the power spectrum covariance directly.
Using the prediction, we find the power spectrum covariance matrix
in a closed hyperbolic universe (Eq.(\ref{eq:CHcov})),
%%%%%%%%%%%%%%%%%%%%%%%%%%%%%%%%%%%%%%%%%%%%%%%%%%%%%%%%%%%%%%%%%
\begin{equation}
  \left<C_lC_{l'}\right> - \left<C_l\right>\left<C_{l'}\right>
   =
  \frac{2}{2l+1}\left(\left<C_l\right>^2+2F_{ll}\right)\delta_{ll'}
  +2F_{ll'}.
\end{equation}
%%%%%%%%%%%%%%%%%%%%%%%%%%%%%%%%%%%%%%%%%%%%%%%%%%%%%%%%%%%%%%%%%
The power spectrum is given by
%%%%%%%%%%%%%%%%%%%%%%%%%%%%%%%%%%%%%%%%%%%%%%%%%%%%%%%%%%%%%%%%%%%%%%
\begin{equation}
 \left<C_l\right>=
  \sum_{\nu}P_\Phi(\nu)g_{{\rm T}l}^2(\nu)
  \left<\left|\xi_{lm}(\nu)\right|^2\right>.
\end{equation}
%%%%%%%%%%%%%%%%%%%%%%%%%%%%%%%%%%%%%%%%%%%%%%%%%%%%%%%%%%%%%%%%%%%%%%
Note that $\left<C_l\right>\left<C_{l'}\right>\ge F_{ll'}$.
It follows from this equation that to constrain the contribution of 
the connected $T^{ll}_{l'l'}(0)$ to the power spectrum covariance, 
we need to constrain $F_{ll'}$.

In addition to $L=0$ modes, fortunately, a closed hyperbolic universe also 
produces the connected trispectrum for the group (c), 
$l_1=l_2=l_3=l_4\equiv l$ and $L\neq 0$.
The prediction is 
$\left<T^{ll}_{ll}(L)\right>_{\rm c}=4(2L+1)F_{ll}$.
We may constrain this term with our measurement of the group (c).
The group (c) is, however, the most noisy group, and the constraint
from the DMR measurement is too weak to be useful yet.

The most promising way to investigate the angular trispectrum in near 
future is to use the {\it MAP} CMB sky maps.
The big advantage of {\it MAP} over {\it COBE} is the much higher
angular resolution.
The high-resolution measurement is important even on large angular 
scales, as we can minimize the effect of the Galactic cut.
Since the rather big {\it COBE} Galactic cut has been the main cause 
of numerous 
complications in the data analysis, we expect that the {\it MAP} data
will measure the angular trispectrum with much better sensitivity,
and with much smaller systematic errors than {\it COBE}.
The {\it MAP} trispectrum will put strong constraints on 
several challenging non-Gaussian models:
the non-Gaussianity induced from topology of the universe, the weak lensing
effect, and so on, which are not probed by the angular bispectrum.

%%%%%%%%%%%%%%%%%%%%%%%%%%%%%%%%%%%%%%%%%%%%%%%%%%%%%%%%%%%%%%%%%%%
%
%  Bibliography
%
%%%%%%%%%%%%%%%%%%%%%%%%%%%%%%%%%%%%%%%%%%%%%%%%%%%%%%%%%%%%%%%%%%%

%%%%% appendix %%%%%
\appendix

\chapter{Slow-roll Approximation}
\label{app:slowroll}

%%%%%%%%%%%%%%%%%%%%%%%%%%%%%%%%%%%%%%%%%%%%%%%%%%%%%%%%%%%%%%%%%%%%%%
In this appendix, we describe the {\it slow-roll approximation}, which
has played a central role in the analysis of inflation dynamics.
We then apply the approximation to the effective mass of scalar-field
fluctuations, $m^2_\chi(\tau)$, to show explicitly the approximation 
that we have used in equation~(\ref{eq:effmass**}).

A slowly-rolling scalar field on a potential, $V(\phi)$, 
is a key ingredient of a successful inflation model, for to achieve
the accelerated expansion of the universe 
the kinetic energy of $\phi$ needs to be smaller than the potential 
energy (see Eq.(\ref{eq:energycondition})),
%%%%%%%%%%%%%%%%%%%%%%%%%%%%%%%%%%%%%%%%%%%%%%%%%%%%%%%%%%%%%%%%%%%%%%
\begin{equation}
 \label{eq:slow-roll1}
  \left(\frac{d\phi}{dt}\right)^2< V(\phi).
\end{equation}
%%%%%%%%%%%%%%%%%%%%%%%%%%%%%%%%%%%%%%%%%%%%%%%%%%%%%%%%%%%%%%%%%%%%%%
It then follows from this condition that the second-order time 
derivative, $d^2\phi/dt^2$, needs to be smaller than the 
potential slope, $dV/d\phi$,
%%%%%%%%%%%%%%%%%%%%%%%%%%%%%%%%%%%%%%%%%%%%%%%%%%%%%%%%%%%%%%%%%%%%%%
\begin{equation}
 \label{eq:slow-roll2}
  2\left|\frac{d^2\phi}{dt^2}\right|< \left|\frac{dV}{d\phi}\right|.
\end{equation}
%%%%%%%%%%%%%%%%%%%%%%%%%%%%%%%%%%%%%%%%%%%%%%%%%%%%%%%%%%%%%%%%%%%%%%
These conditions provide us an useful scheme of approximation, the 
slow-roll approximation.
The approximation demands that the l.h.s's of the conditions
be much smaller than the r.h.s's.

The trace-part Einstein equation and the Friedmann equation give the 
exact relation between the time derivative of $H$ and $d\phi/dt$,
%%%%%%%%%%%%%%%%%%%%%%%%%%%%%%%%%%%%%%%%%%%%%%%%%%%%%%%%%%%%%%%%%%%%%%
\begin{equation}
 \frac{dH}{dt}= -4\pi G \left(\frac{d\phi}{dt}\right)^2.
\end{equation}
%%%%%%%%%%%%%%%%%%%%%%%%%%%%%%%%%%%%%%%%%%%%%%%%%%%%%%%%%%%%%%%%%%%%%%
Hence, we obtain another form of the slow-roll condition, 
$\left|dH/dt\right|< 4\pi G V(\phi)$.
In the slow-roll approximation, the dimensionless variables such as
$V^{-1}(d\phi/dt)^2$, $(Hd\phi/dt)^{-1}(d^2\phi/dt^2)$, $H^{-2}(dH/dt)$,
and so on, are small order parameters which control the approximation.

By the physical requirement, the accelerated expansion of the universe, 
the slow-roll conditions are defined by the time derivative with respect
to the physical time, $t$.
If we use the conformal time, $\tau$ ($d\tau=a^{-1}dt$), 
then we have $\dot{\phi}= a(d\phi/dt)$, and
%%%%%%%%%%%%%%%%%%%%%%%%%%%%%%%%%%%%%%%%%%%%%%%%%%%%%%%%%%%%%%%%%%%%%%
\begin{equation}
 \ddot{\phi}= aH\dot{\phi} + a^2\frac{d^2\phi}{dt^2},
\end{equation}
%%%%%%%%%%%%%%%%%%%%%%%%%%%%%%%%%%%%%%%%%%%%%%%%%%%%%%%%%%%%%%%%%%%%%%
where the dots denote the conformal time derivative: 
$\dot{x}\equiv d\phi/d\tau$.
In contrast to $d^2\phi/dt^2$, $\ddot{\phi}$ is not so small 
compared with $dV/d\phi$ because of the first term.
From this, we obtain a rule for the slow-roll analysis: 
{\it evaluate the second- or the higher-order time derivative with respect 
to the physical time, not the conformal time}.
Using this rule, we can use the slow-roll approximation consistently 
in the conformal time coordinate.

Using the slow-roll approximation, we derive the approximation to the 
effective mass of scalar-field fluctuations, $ m_\chi^2$, which 
we have done in chapter~\ref{chap:inflation} (Eq.~(\ref{eq:effmass**})).
To derive
%%%%%%%%%%%%%%%%%%%%%%%%%%%%%%%%%%%%%%%%%%%%%%%%%%%%%%%%%%%%%%%%%%%%%%
\begin{equation}
% \label{eq:effmass**}
  m_\chi^2
  = -\frac{H}{\dot{\phi}}\frac{d^2({\dot{\phi}}/{H})}{d\tau^2}
  \approx 
  \left(\frac{d^2V}{d\phi^2}+9\frac{dH}{dt}\right)a^2-\frac2{\tau^2},
\end{equation}
%%%%%%%%%%%%%%%%%%%%%%%%%%%%%%%%%%%%%%%%%%%%%%%%%%%%%%%%%%%%%%%%%%%%%%
we begin with
%%%%%%%%%%%%%%%%%%%%%%%%%%%%%%%%%%%%%%%%%%%%%%%%%%%%%%%%%%%%%%%%%%%%%%
\begin{equation}
 -\frac{H}{\dot{\phi}}\frac{d^2({\dot{\phi}}/{H})}{d\tau^2}
  =
  \frac{\ddot{H}}{H}-\frac{\dddot{\phi}}{\dot{\phi}}
  +2\frac{\dot{H}}{H}\left(\frac{\ddot{\phi}}{\dot{\phi}}
		    -\frac{\dot{H}}H\right),
\end{equation}
%%%%%%%%%%%%%%%%%%%%%%%%%%%%%%%%%%%%%%%%%%%%%%%%%%%%%%%%%%%%%%%%%%%%%%
where
%%%%%%%%%%%%%%%%%%%%%%%%%%%%%%%%%%%%%%%%%%%%%%%%%%%%%%%%%%%%%%%%%%%%%%
\begin{eqnarray}
 \frac{\ddot{H}}H &=& 
  - a\dot{H} - 8\pi G \frac{\dot{\phi}\ddot{\phi}}{aH}
  = a\dot{H} - 8\pi G\frac{a\dot{\phi}}{H}\left(\frac{d^2\phi}{dt^2}\right),\\
 \frac{\dddot{\phi}}{\dot{\phi}} &=&
  2a^2H^2 - 2a\dot{H} - a^2\frac{d^2V}{d\phi^2},\\
  \frac{\ddot{\phi}}{\dot{\phi}} &=&
  aH + a^2\frac{d^2\phi}{dt^2}.
\end{eqnarray}
%%%%%%%%%%%%%%%%%%%%%%%%%%%%%%%%%%%%%%%%%%%%%%%%%%%%%%%%%%%%%%%%%%%%%%
Here, to calculate $\dddot{\phi}$, we have used the Klein--Gordon 
equation for a homogeneous scalar field, 
$\ddot{\phi}+ 2aH\dot{\phi}+ a^2(dV/d\phi)=0$.
These equations are exact.

We then neglect the higher-order slow-roll terms such as
$\dot{H}^2$ and $(d^2\phi/dt^2)\dot{H}$, and obtain 
%%%%%%%%%%%%%%%%%%%%%%%%%%%%%%%%%%%%%%%%%%%%%%%%%%%%%%%%%%%%%%%%%%%%%%
\begin{equation}
 \label{eq:app1}
 -\frac{H}{\dot{\phi}}\frac{d^2({\dot{\phi}}/{H})}{d\tau^2}
  \approx
  \left(\frac{d^2V}{d\phi^2} - 2H^2 + 5\frac{dH}{dt}\right)a^2.
\end{equation}
%%%%%%%%%%%%%%%%%%%%%%%%%%%%%%%%%%%%%%%%%%%%%%%%%%%%%%%%%%%%%%%%%%%%%%
Next step is to relate $H$ to the conformal time, $\tau$, through
%%%%%%%%%%%%%%%%%%%%%%%%%%%%%%%%%%%%%%%%%%%%%%%%%%%%%%%%%%%%%%%%%%%%%%
\begin{equation}
 \tau\equiv \int\frac{dt}{a}= \int\frac{da}{a^2H}
  = -\frac1{aH}-\int da\frac{\dot{H}}{(aH)^3}.
\end{equation}
%%%%%%%%%%%%%%%%%%%%%%%%%%%%%%%%%%%%%%%%%%%%%%%%%%%%%%%%%%%%%%%%%%%%%%
By rewriting the conformal-time derivative in the second term 
with the physical-time derivative, 
$\int da (dH/dt)/(a^2 H^3)$, and neglecting the higher-order slow-roll
terms, we obtain the first-order
slow-roll correction to the conformal time, 
%%%%%%%%%%%%%%%%%%%%%%%%%%%%%%%%%%%%%%%%%%%%%%%%%%%%%%%%%%%%%%%%%%%%%%
\begin{equation}
 \tau\approx -\frac1{aH} + \frac{dH/dt}{aH^3},
\end{equation}
%%%%%%%%%%%%%%%%%%%%%%%%%%%%%%%%%%%%%%%%%%%%%%%%%%%%%%%%%%%%%%%%%%%%%%
and hence $H^2\approx (a\tau)^{-1} - 2(dH/dt)$.
Finally, substituting this for $H^2$ in equation~(\ref{eq:app1}), we obtain
%%%%%%%%%%%%%%%%%%%%%%%%%%%%%%%%%%%%%%%%%%%%%%%%%%%%%%%%%%%%%%%%%%%%%%
\begin{equation}
  -\frac{H}{\dot{\phi}}\frac{d^2({\dot{\phi}}/{H})}{d\tau^2}
  \approx 
  \left(\frac{d^2V}{d\phi^2}+9\frac{dH}{dt}\right)a^2-\frac2{\tau^2}.
\end{equation}
%%%%%%%%%%%%%%%%%%%%%%%%%%%%%%%%%%%%%%%%%%%%%%%%%%%%%%%%%%%%%%%%%%%%%%

\chapter{Wigner 3-$j$ Symbol}
\label{app:wigner}

%%%%%%%%%%%%%%%%%%%%%%%%%%%%%%%%%%%%%%%%%%%%%%%%%%%%%%%%%%%%%%%%%%%%%%
In this appendix, we summarize basic properties of the 
Wigner 3-$j$ symbol, following \citet{RBMW59}.
The Wigner 3-$j$ symbol characterizes geometric properties of 
the angular bispectrum.

\section{Triangle conditions}

The Wigner 3-$j$ symbol,
%%%%%%%%%%%%%%%%%%%%%%%%%%%%%%%%%%%%%%%%%%%%%%%%%%%%%%%%%%%%%%%%%%%%%%
\begin{equation}
 \label{eq:3j}
 \left(\begin{array}{ccc}l_1&l_2&l_3\\m_1&m_2&m_3\end{array}\right),
\end{equation}
%%%%%%%%%%%%%%%%%%%%%%%%%%%%%%%%%%%%%%%%%%%%%%%%%%%%%%%%%%%%%%%%%%%%%%
is related to the Clebsh--Gordan coefficients which describe
coupling of two angular momenta in the quantum mechanics.
In the quantum mechanics, $l$ is the eigenvalue of the angular 
momentum operator, ${\mathbf L}={\mathbf r}\times {\mathbf p}$: 
${\mathbf L}^2 Y_{lm}= l(l+1) Y_{lm}$.
$m$ is the eigenvalue of the $z$-direction component of the 
angular momentum, $L_z Y_{lm}= m Y_{lm}$.

The symbol such as  
%%%%%%%%%%%%%%%%%%%%%%%%%%%%%%%%%%%%%%%%%%%%%%%%%%%%%%%%%%%%%%%%%%%%%%
\begin{equation}
 (-1)^{m_3}\left(\begin{array}{ccc}l_1&l_2&l_3\\m_1&m_2&-m_3\end{array}\right)
\end{equation}
%%%%%%%%%%%%%%%%%%%%%%%%%%%%%%%%%%%%%%%%%%%%%%%%%%%%%%%%%%%%%%%%%%%%%%
describes coupling of two angular-momentum states,
${\mathbf L}_1$ and ${\mathbf L}_2$, forming a coupled state,
${\mathbf L}_3={\mathbf L}_1+{\mathbf L}_2$.
It follows from ${\mathbf L}_1+{\mathbf L}_2-{\mathbf L}_3=0$ that
$m_1+m_2-m_3=0$; thus, the Wigner 3-$j$ symbol~(\ref{eq:3j}) 
describes three angular momenta forming a triangle, 
${\mathbf L}_1+{\mathbf L}_2+{\mathbf L}_3=0$, and 
satisfies $m_1+m_2+m_3=0$.

Since ${\mathbf L}_1$, ${\mathbf L}_2$, and ${\mathbf L}_3$ form a 
triangle, they have to satisfy the triangle conditions, 
$\left|L_i-L_j\right|\le L_k\le L_i+L_j$,
where $L_i\equiv \left|{\mathbf L}_i\right|$.
Hence, $l_1$, $l_2$, and $l_3$ also satisfy the triangle conditions,
%%%%%%%%%%%%%%%%%%%%%%%%%%%%%%%%%%%%%%%%%%%%%%%%%%%%%%%%%%%%%%%%%%%%%%
\begin{equation}
 \left|l_i-l_j\right|\le l_k\le l_i+l_j;
\end{equation}
%%%%%%%%%%%%%%%%%%%%%%%%%%%%%%%%%%%%%%%%%%%%%%%%%%%%%%%%%%%%%%%%%%%%%%
otherwise, the Wigner 3-$j$ symbol vanishes.
The triangle conditions also include $m_1+m_2+m_3=0$.
These properties may regard ($l$, $m$) as vectors, ${\mathbf l}$,
which satisfy ${\mathbf l}_1+{\mathbf l}_2+{\mathbf l}_3=0$.
Note that, however, ${\mathbf L}\neq {\mathbf l}$.

For $l_1=l_2$ and $m_3=0$, the Wigner 3-$j$ symbol reduces to
%%%%%%%%%%%%%%%%%%%%%%%%%%%%%%%%%%%%%%%%%%%%%%%%%%%%%%%%%%%%%%%%%%%%%%
\begin{equation}
 (-1)^m\left(\begin{array}{ccc}l&l&l'\\m&-m&0\end{array}\right)
 =
  \frac{(-1)^l}{\sqrt{2l+1}}\delta_{l'0}.
\end{equation}
%%%%%%%%%%%%%%%%%%%%%%%%%%%%%%%%%%%%%%%%%%%%%%%%%%%%%%%%%%%%%%%%%%%%%%
In chapter~\ref{chap:spectrum}, we have used this relation to reduce 
the covariance matrix of the angular bispectrum and trispectrum.
We have also used this relation to reduce the angular trispectrum
for $L=0$ (see Eq.(\ref{eq:special})).

\section{Symmetry}

The Wigner 3-$j$ symbol is invariant under even permutations,
%%%%%%%%%%%%%%%%%%%%%%%%%%%%%%%%%%%%%%%%%%%%%%%%%%%%%%%%%%%%%%%%%%%%%%
\begin{equation}
 \left(\begin{array}{ccc}l_1&l_2&l_3\\m_1&m_2&m_3\end{array}\right)
 =\left(\begin{array}{ccc}l_3&l_1&l_2\\m_3&m_1&m_2\end{array}\right)
 =\left(\begin{array}{ccc}l_2&l_3&l_1\\m_2&m_3&m_1\end{array}\right),
\end{equation}
%%%%%%%%%%%%%%%%%%%%%%%%%%%%%%%%%%%%%%%%%%%%%%%%%%%%%%%%%%%%%%%%%%%%%%
while it changes the phase for odd permutations if
$l_1+l_2+l_3=\mbox{odd}$,
%%%%%%%%%%%%%%%%%%%%%%%%%%%%%%%%%%%%%%%%%%%%%%%%%%%%%%%%%%%%%%%%%%%%%%
\begin{eqnarray}
 & &
 (-1)^{l_1+l_2+l_3}
	\left(\begin{array}{ccc}l_1&l_2&l_3\\m_1&m_2&m_3\end{array}\right)\\
 &=&\left(\begin{array}{ccc}l_2&l_1&l_3\\m_2&m_1&m_3\end{array}\right)
 =\left(\begin{array}{ccc}l_1&l_3&l_2\\m_1&m_3&m_2\end{array}\right)
 =\left(\begin{array}{ccc}l_3&l_2&l_1\\m_3&m_2&m_1\end{array}\right).
\end{eqnarray}
%%%%%%%%%%%%%%%%%%%%%%%%%%%%%%%%%%%%%%%%%%%%%%%%%%%%%%%%%%%%%%%%%%%%%%

The phase also changes under the transformation of $m_1+m_2+m_3\rightarrow -(m_1+m_2+m_3)$,
if $l_1+l_2+l_3=\mbox{odd}$,
%%%%%%%%%%%%%%%%%%%%%%%%%%%%%%%%%%%%%%%%%%%%%%%%%%%%%%%%%%%%%%%%%%%%%%
\begin{equation}
 \left(\begin{array}{ccc}l_1&l_2&l_3\\m_1&m_2&m_3\end{array}\right)
 =(-1)^{l_1+l_2+l_3}
  \left(\begin{array}{ccc}l_1&l_2&l_3\\-m_1&-m_2&-m_3\end{array}\right).
\end{equation}
%%%%%%%%%%%%%%%%%%%%%%%%%%%%%%%%%%%%%%%%%%%%%%%%%%%%%%%%%%%%%%%%%%%%%%
%It follows from this property that parity invariance of the system demands 
%$l_1+l_2+l_3=\mbox{even}$.
If there is no $z$-direction component of the angular momenta 
in the system, i.e., $m_i=0$, then the Wigner 3-$j$ symbol of the system,
%%%%%%%%%%%%%%%%%%%%%%%%%%%%%%%%%%%%%%%%%%%%%%%%%%%%%%%%%%%%%%%%%%%%%%
\begin{equation}
 \left(\begin{array}{ccc}l_1&l_2&l_3\\0&0&0\end{array}\right),
\end{equation}
%%%%%%%%%%%%%%%%%%%%%%%%%%%%%%%%%%%%%%%%%%%%%%%%%%%%%%%%%%%%%%%%%%%%%%
is non-zero only if $l_1+l_2+l_3=\mbox{even}$.
%; thus, 
%this symbol demands parity invariance automatically. 
This symbol is invariant under any permutations of $l_i$.

In chapter~\ref{chap:theory_bl}, we have frequently used the Gaunt 
integral, ${\cal G}^{m_1m_2m_3}_{l_1l_2l_3}$, defined by
%%%%%%%%%%%%%%%%%%%%%%%%%%%%%%%%%%%%%%%%%%%%%%%%%%%%%%%%%%%%%%%%%%
\begin{eqnarray}
  \nonumber
  {\cal G}_{l_1l_2l_3}^{m_1m_2m_3}
  &\equiv&
  \int d^2\hat{\mathbf n}
  Y_{l_1m_1}(\hat{\mathbf n})
  Y_{l_2m_2}(\hat{\mathbf n})
  Y_{l_3m_3}(\hat{\mathbf n})\\
 \nonumber
  &=&\sqrt{
   \frac{\left(2l_1+1\right)\left(2l_2+1\right)\left(2l_3+1\right)}
        {4\pi}
        }
  \left(
  \begin{array}{ccc}
  l_1 & l_2 & l_3 \\ 0 & 0 & 0 
  \end{array}
  \right)\\
 & &\times
  \left(
  \begin{array}{ccc}
  l_1 & l_2 & l_3 \\ m_1 & m_2 & m_3 
  \end{array}
  \right),
\end{eqnarray}
%%%%%%%%%%%%%%%%%%%%%%%%%%%%%%%%%%%%%%%%%%%%%%%%%%%%%%%%%%%%%%%%%%
to calculate the angular bispectrum.
By definition, the Gaunt integral is invariant under 
both the odd and the even permutations, and non-zero only if
$l_1+l_2+l_3=\mbox{even}$, $m_1+m_2+m_3=0$, and
$\left|l_i-l_j\right|\le l_k\le l_i+l_j$.
In other words, the Gaunt integral describes fundamental geometric 
properties of the angular bispectrum such as
the triangle conditions.
%Note that the angular bispectrum is parity invariant, if we assume 
%statistical isotropy of the universe.

The Gaunt integral for $m_i=0$ gives the identity for the Legendre polynomials,
%%%%%%%%%%%%%%%%%%%%%%%%%%%%%%%%%%%%%%%%%%%%%%%%%%%%%%%%%%%%%%%%%
\begin{equation}
 \int_{-1}^{1}\frac{dx}2~
  P_{l_1}(x)P_{l_2}(x)P_{l_3}(x)
  =
  \left(\begin{array}{ccc}l_1&l_2&l_3\\0&0&0\end{array}\right)^2.
\end{equation}
%%%%%%%%%%%%%%%%%%%%%%%%%%%%%%%%%%%%%%%%%%%%%%%%%%%%%%%%%%%%%%%%%
In chapter~\ref{chap:spectrum}, we have used this identity to derive 
the bias for the angular bispectrum on the incomplete sky
(Eq.(\ref{eq:biasbl})).
Here, we have used
%%%%%%%%%%%%%%%%%%%%%%%%%%%%%%%%%%%%%%%%%%%%%%%%%%%%%%%%%%%%%%%%%
\begin{equation}
 Y_{l0}(\hat{\mathbf n})= \sqrt{\frac{4\pi}{2l+1}}P_l(\cos\theta).
\end{equation}
%%%%%%%%%%%%%%%%%%%%%%%%%%%%%%%%%%%%%%%%%%%%%%%%%%%%%%%%%%%%%%%%%

\section{Orthogonality}

The Wigner 3-$j$ symbol has the following orthogonality properties:
%%%%%%%%%%%%%%%%%%%%%%%%%%%%%%%%%%%%%%%%%%%%%%%%%%%%%%%%%%%%%%%%%
\begin{equation}
 \sum_{l_3m_3}
  (2l_3+1)
 \left(\begin{array}{ccc}l_1&l_2&l_3\\m_1&m_2&m_3\end{array}\right)
 \left(\begin{array}{ccc}l_1&l_2&l_3\\m'_1&m'_2&m_3\end{array}\right)
 =
  \delta_{m_1m_1'}\delta_{m_2m_2'},
\end{equation}
%%%%%%%%%%%%%%%%%%%%%%%%%%%%%%%%%%%%%%%%%%%%%%%%%%%%%%%%%%%%%%%%%%%%%%
and
%%%%%%%%%%%%%%%%%%%%%%%%%%%%%%%%%%%%%%%%%%%%%%%%%%%%%%%%%%%%%%%%%
\begin{equation}
 \sum_{m_1m_2}
 \left(\begin{array}{ccc}l_1&l_2&l_3\\m_1&m_2&m_3\end{array}\right)
 \left(\begin{array}{ccc}l_1&l_2&l'_3\\m_1&m_2&m'_3\end{array}\right)
 =
  \frac{\delta_{l_3l_3'}\delta_{m_3m_3'}}{2l_3+1},
\end{equation}
%%%%%%%%%%%%%%%%%%%%%%%%%%%%%%%%%%%%%%%%%%%%%%%%%%%%%%%%%%%%%%%%%%%%%%
or
%%%%%%%%%%%%%%%%%%%%%%%%%%%%%%%%%%%%%%%%%%%%%%%%%%%%%%%%%%%%%%%%%
\begin{equation}
 \label{eq:normalrelation}
 \sum_{{\rm all}~m}
 \left(\begin{array}{ccc}l_1&l_2&l_3\\m_1&m_2&m_3\end{array}\right)^2
 = 1.
\end{equation}
%%%%%%%%%%%%%%%%%%%%%%%%%%%%%%%%%%%%%%%%%%%%%%%%%%%%%%%%%%%%%%%%%%%%%%
The orthogonality properties are essential for any basic calculations 
involving the Wigner 3-$j$ symbols.
Note that these orthogonality properties are consistent with 
orthonormality of the angular-momentum eigenstate vectors,
and unitality of the Crebsh--Gordan coefficients, by definition.

\section{Rotation matrix}

A finite rotation operator for the Euler angles
$\alpha$, $\beta$, and $\gamma$, $D(\alpha,\beta,\gamma)$, 
comprises angular momentum operators,
%%%%%%%%%%%%%%%%%%%%%%%%%%%%%%%%%%%%%%%%%%%%%%%%%%%%%%%%%%%%%%%%%%
\begin{equation}
 D(\alpha,\beta,\gamma)= e^{-i\alpha L_z}e^{-i\beta L_y}e^{-i\gamma L_z}.
\end{equation}
%%%%%%%%%%%%%%%%%%%%%%%%%%%%%%%%%%%%%%%%%%%%%%%%%%%%%%%%%%%%%%%%%%
Since the Wigner 3-$j$ symbol describes coupling of two angular momenta,
it also describes coupling of two rotation operators.
Using the rotation matrix element, 
$D_{m'm}^{(l)}=\left<l,m'\left|D\right|l,m\right>$,
we have
%%%%%%%%%%%%%%%%%%%%%%%%%%%%%%%%%%%%%%%%%%%%%%%%%%%%%%%%%%%%%%%%%%
\begin{equation}
 \label{eq:rotreduce}
 D_{m'_1m_1}^{(l_1)}D_{m'_2m_2}^{(l_2)}
  =
  \sum_{l_3}(2l_3+1)
  \sum_{m_3m'_3}D_{m'_3m_3}^{(l_3)*}
  \left(\begin{array}{ccc}l_1&l_2&l_3\\m_1&m_2&m_3\end{array}\right)
  \left(\begin{array}{ccc}l_1&l_2&l_3\\m'_1&m'_2&m'_3\end{array}\right).
\end{equation}
%%%%%%%%%%%%%%%%%%%%%%%%%%%%%%%%%%%%%%%%%%%%%%%%%%%%%%%%%%%%%%%%%%
In chapter~\ref{chap:spectrum}, we have used this relation to 
evaluate rotationally invariant harmonic spectra.
Note that the rotation matrix is orthonormal, 
%%%%%%%%%%%%%%%%%%%%%%%%%%%%%%%%%%%%%%%%%%%%%%%%%%%%%%%%%%%%%%%%%%
\begin{equation}
 \sum_{m}D_{m'm}^{(l)*}D_{m''m}^{(l)}= \delta_{m'm''}.
\end{equation}
%%%%%%%%%%%%%%%%%%%%%%%%%%%%%%%%%%%%%%%%%%%%%%%%%%%%%%%%%%%%%%%%%%

\section{Wigner 6-$j$ symbol}

The Wigner 6-$j$ symbol,
%%%%%%%%%%%%%%%%%%%%%%%%%%%%%%%%%%%%%%%%%%%%%%%%%%%%%%%%%%%%%%%%%%%%%%
\begin{equation}
 \label{eq:6j}
 \left\{\begin{array}{ccc}l_1&l_2&l_3\\l'_1&l'_2&l'_3\end{array}\right\},
\end{equation}
%%%%%%%%%%%%%%%%%%%%%%%%%%%%%%%%%%%%%%%%%%%%%%%%%%%%%%%%%%%%%%%%%%%%%%
describes coupling of three angular momenta.
We often encounter the Wigner 6-$j$ symbol, when we calculate 
the angular bispectrum which has more complicated geometric structures
\citep{GS99}.
The angular trispectrum also often includes the Wigner 6-$j$ symbol
\citep{Hu01}.

The Wigner 6-$j$ symbol is related to the Wigner 3-$j$ symbols through
%%%%%%%%%%%%%%%%%%%%%%%%%%%%%%%%%%%%%%%%%%%%%%%%%%%%%%%%%%%%%%%%%%%%%%
\begin{eqnarray}
 \nonumber
  & &
  (-1)^{l_1'+l_2'+l_3'}
 \left\{\begin{array}{ccc}l_1&l_2&l_3\\l'_1&l'_2&l'_3\end{array}\right\}
 \left(\begin{array}{ccc}l_1&l_2&l_3\\m_1&m_2&m_3\end{array}\right)\\
 \nonumber
 & &=
 \sum_{{\rm all}~m'}
 (-1)^{m_1'+m_2'+m_3'}\\
 & &\times
 \left(\begin{array}{ccc}l_1&l'_2&l'_3\\m_1&m'_2&-m'_3\end{array}\right)
 \left(\begin{array}{ccc}l'_1&l_2&l'_3\\-m'_1&m_2&m'_3\end{array}\right)
 \left(\begin{array}{ccc}l'_1&l'_2&l_3\\m'_1&-m'_2&m_3\end{array}\right).
\end{eqnarray}
%%%%%%%%%%%%%%%%%%%%%%%%%%%%%%%%%%%%%%%%%%%%%%%%%%%%%%%%%%%%%%%%%%%%%%
In appendix~\ref{app:iso}, we use this relation to derive the angular
bispectrum from isocurvature fluctuations in inflation 
(Eq.(\ref{eq:isobispectrum})).
By using equation~(\ref{eq:normalrelation}), we also obtain
%%%%%%%%%%%%%%%%%%%%%%%%%%%%%%%%%%%%%%%%%%%%%%%%%%%%%%%%%%%%%%%%%%%%%%
\begin{eqnarray}
 \nonumber
  (-1)^{l_1'+l_2'+l_3'}
 \left\{\begin{array}{ccc}l_1&l_2&l_3\\l'_1&l'_2&l'_3\end{array}\right\}
 &=&
 \sum_{{\rm all}~mm'}
 (-1)^{m_1'+m_2'+m_3'}\\
 \nonumber
  & &\times
 \left(\begin{array}{ccc}l_1&l_2&l_3\\m_1&m_2&m_3\end{array}\right)
 \left(\begin{array}{ccc}l_1&l'_2&l'_3\\m_1&m'_2&-m'_3\end{array}\right)\\
 & &\times
 \left(\begin{array}{ccc}l'_1&l_2&l'_3\\-m'_1&m_2&m'_3\end{array}\right)
 \left(\begin{array}{ccc}l'_1&l'_2&l_3\\m'_1&-m'_2&m_3\end{array}\right).
\end{eqnarray}
%%%%%%%%%%%%%%%%%%%%%%%%%%%%%%%%%%%%%%%%%%%%%%%%%%%%%%%%%%%%%%%%%%%%%%

\chapter{Angular Bispectrum from Isocurvature Fluctuations}
\label{app:iso}

%%%%%%%%%%%%%%%%%%%%%%%%%%%%%%%%%%%%%%%%%%%%%%%%%%%%%%%%%%%%%%%%%%%%%%
In this appendix, we derive the angular bispectrum from
isocurvature fluctuations generated in inflation.
The mechanism of generating isocurvature fluctuations 
we consider here is that of \citet{LM97}:
a massive-free scalar field, $\sigma$, oscillating about $\sigma=0$
during inflation.

Quantum fluctuations of $\sigma$ produce Gaussian fluctuations, $\delta\sigma$.
Since there is no mean $\sigma$-field because of the 
oscillation about $\sigma=0$ in the model,
the model predicts density fluctuations which are {\it quadratic} in
$\delta\sigma$:
%%%%%%%%%%%%%%%%%%%%%%%%%%%%%%%%%%%%%%%%%%%%%%%%%%%%%%%%%%%%%%%%%%
\begin{equation}
 \delta\rho_\sigma\sim m^2\sigma\delta\sigma + m^2(\delta\sigma)^2
  = m^2(\delta\sigma)^2.
\end{equation}
%%%%%%%%%%%%%%%%%%%%%%%%%%%%%%%%%%%%%%%%%%%%%%%%%%%%%%%%%%%%%%%%%%
$\delta\rho_\sigma$ is thus non-Gaussian.
Moreover, since the energy density of $\sigma$ does not
dominate the universe during inflation, $\delta\rho_\sigma$  does not
perturb the spatial curvature, being isocurvature density fluctuations.

After inflation, $\delta\rho_\sigma$ may produce 
the spatial curvature perturbations in the Newtonian gauge, $\Phi$,
through the evolution.
If $\delta\rho_\sigma$ becomes dominant in the universe at some point,
then the linear perturbation theory gives 
$\Phi=\frac18(\delta\rho_\sigma/\rho_\sigma)(a/a_{\rm eq})$ 
in the radiation era, and 
$\Phi=\frac15(\delta\rho_\sigma/\rho_\sigma)$ in the matter era \citep{KS84}.
$\Phi$ then produces CMB fluctuations through the
Sachs--Wolfe effect, $\Delta T/T= -2\Phi$.
Since $\delta\rho_\sigma$ is non-Gaussian, $\Phi$ is also non-Gaussian,
so is $\Delta T/T$.

Our goal in this appendix is to calculate the CMB angular bispectrum 
from the $\Phi$-field bispectrum.
We start with writing $\Phi$ as a Gaussian-variable-squared in real space,
$\Phi({\mathbf x})= \eta^2({\mathbf x}) - \left<\eta^2({\mathbf x})\right>$,
where $\eta$ is Gaussian.
We have subtracted the mean from $\Phi$, ensuring 
$\left<\Phi({\mathbf x})\right>=0$.
Transforming $\Phi({\mathbf x})$ into Fourier space, we obtain
%%%%%%%%%%%%%%%%%%%%%%%%%%%%%%%%%%%%%%%%%%%%%%%%%%%%%%%%%%%%%%%%%%
\begin{equation}
  \Phi({\mathbf k})=
  \int \frac{d^3{\mathbf p}}{(2\pi)^3}~
  \eta({\mathbf k}+{\mathbf p})\eta^*({\mathbf p})
  -(2\pi)^3\delta^{(3)}({\mathbf k})\left<\eta^2({\mathbf x})\right>.
\end{equation}
%%%%%%%%%%%%%%%%%%%%%%%%%%%%%%%%%%%%%%%%%%%%%%%%%%%%%%%%%%%%%%%%%%
Using the $\eta$ power spectrum, $P_\eta(k)$, we write
the $\Phi$ power spectrum, $P_\Phi(k)$, as
%%%%%%%%%%%%%%%%%%%%%%%%%%%%%%%%%%%%%%%%%%%%%%%%%%%%%%%%%%%%%%%%%%
\begin{equation}
 P_\Phi(k) = 2\int \frac{d^3{\mathbf p}}{(2\pi)^3}~
  P_\eta(p)P_\eta\left(\left|{\mathbf k}+{\mathbf p}\right|\right).
\end{equation}
%%%%%%%%%%%%%%%%%%%%%%%%%%%%%%%%%%%%%%%%%%%%%%%%%%%%%%%%%%%%%%%%%%
If we use a conventional power-law spectrum, 
$P_\Phi(k)\propto k^{n-4}$, then we find $P_\eta(k)\propto k^{(n-7)/2}$.
On the other hand, quantum fluctuations of a massive-free scalar field
give $P_\sigma(k)\propto k^{-3+2m^2/(3H^2)}$, 
where $H$ is the Hubble parameter during inflation 
(chapter~\ref{chap:inflation}).
It then follows from $\Phi\propto \delta\rho_\sigma$
that $P_\eta(k)\propto P_\sigma(k)\propto k^{-3+2m^2/(3H^2)}$, and
$n=1+4m^2/(3H^2)$; thus, the model predicts a tilted ``blue'' spectrum
\citep{LM97}.

The $\Phi$-field bispectrum is
%%%%%%%%%%%%%%%%%%%%%%%%%%%%%%%%%%%%%%%%%%%%%%%%%%%%%%%%%%%%%%%%%%
\begin{eqnarray}
 \nonumber
 \left<\Phi({\mathbf k}_1)\Phi({\mathbf k}_2)\Phi({\mathbf k}_3)\right>
  &=& (2\pi)^3\delta^{(3)}({\mathbf k}_1+{\mathbf k}_2+{\mathbf k}_3)
  \frac83\int\frac{d^3{\mathbf p}}{(2\pi)^3}~P_\eta(p)\\
 \nonumber
 & &\times
  \left[P_\eta\left(\left|{\mathbf k}_1+{\mathbf p}\right|\right)
   P_\eta\left(\left|{\mathbf k}_2-{\mathbf p}\right|\right)\right.\\
 \nonumber
  & &
  \left.+
  P_\eta\left(\left|{\mathbf k}_2+{\mathbf p}\right|\right)
  P_\eta\left(\left|{\mathbf k}_3-{\mathbf p}\right|\right)\right.\\
 \label{eq:phichisq}
 & &
 \left.+
  P_\eta\left(\left|{\mathbf k}_3+{\mathbf p}\right|\right)
  P_\eta\left(\left|{\mathbf k}_1-{\mathbf p}\right|\right)\right].
\end{eqnarray}
%%%%%%%%%%%%%%%%%%%%%%%%%%%%%%%%%%%%%%%%%%%%%%%%%%%%%%%%%%%%%%%%%%
The delta function has appeared in the formula to satisfy the 
triangle condition,
${\mathbf k}_1+{\mathbf k}_2+{\mathbf k}_3=0$.
Since we will work in harmonic space eventually, it may be useful to expand 
$P_\eta\left(\left|{\mathbf k}\pm {\mathbf p}\right|\right)$ into
harmonic space,
%%%%%%%%%%%%%%%%%%%%%%%%%%%%%%%%%%%%%%%%%%%%%%%%%%%%%%%%%%%%%%%%%%
\begin{equation}
 P_\eta\left(\left|{\mathbf k}\pm {\mathbf p}\right|\right)
  =
  \sum_{LM}\widetilde{P}_{\eta L}^{(\pm)}(k,p)
  Y_{LM}(\hat{\mathbf k})Y^*_{LM}(\hat{\mathbf p}).
\end{equation}
%%%%%%%%%%%%%%%%%%%%%%%%%%%%%%%%%%%%%%%%%%%%%%%%%%%%%%%%%%%%%%%%%%

The next step is to project the 3-dimensional $\Phi$-field bispectrum onto
the 2-dimensional CMB angular bispectrum.
We write the harmonic coefficients of the CMB anisotropy, $a_{lm}$, 
with $\Phi$ and the radiation transfer function, $g_{{\rm T}l}(k)$, as
%%%%%%%%%%%%%%%%%%%%%%%%%%%%%%%%%%%%%%%%%%%%%%%%%%%%%%%%%%%%%%%%%%%%%%
\begin{equation}
 a_{lm}= 4\pi(-1)^l
  \int\frac{d^3{\mathbf k}}{(2\pi)^3}\Phi({\mathbf k})g_{{\rm T}l}(k)
  Y_{lm}^*(\hat{\mathbf k}).
\end{equation}
%%%%%%%%%%%%%%%%%%%%%%%%%%%%%%%%%%%%%%%%%%%%%%%%%%%%%%%%%%%%%%%%%%%%%%
On large angular scales, the Sachs--Wolfe effect gives 
$g_{{\rm T}l}(k)= -2j_l\left[k(\tau_0-\tau_{\rm dec})\right]$,
where $\tau_0$ and $\tau_{\rm dec}$ are the present-day conformal time and 
the decoupling-epoch conformal time, respectively.
On small angular scales, we need the full radiation transfer function
for isocurvature fluctuations.

We then calculate $\left<a_{l_1m_1}a_{l_2m_2}a_{l_3m_3}\right>$.
A key point on calculations is to expand the delta function in 
equation~(\ref{eq:phichisq}) into harmonic space with the Rayleigh's formula.
After lengthy calculations, we obtain the CMB angular bispectrum
from isocurvature fluctuations in inflation
%%%%%%%%%%%%%%%%%%%%%%%%%%%%%%%%%%%%%%%%%%%%%%%%%%%%%%%%%%%%%%%%%%%%%%
\begin{eqnarray}
 \nonumber
 \left<a_{l_1m_1}a_{l_2m_2}a_{l_3m_3}\right>
  &=&
  {\cal G}_{l_1l_2l_3}^{m_1m_2m_3} \frac83
  \int_0^\infty r^2 dr
  \int_0^\infty p^2 dp~P_\eta(p)
  \left[\sum_{l_1'l_2'L} {\cal F}_{l_2'l_1'L}^{l_1l_2l_3}\right. \\
 \nonumber
 & &\left.\times
 \frac{2}{\pi}\int_0^\infty k_1^2 dk_1~\widetilde{P}_{\eta L}^{(+)}(k_1,p)
 g_{{\rm T}l_1}(k_1)j_{l_1'}(k_1r)(-i)^{l_1-l_1'}\right.\\
 \nonumber
 & &\left.\times
 \frac{2}{\pi}\int_0^\infty k_2^2 dk_2~\widetilde{P}_{\eta L}^{(-)}(k_2,p)
 g_{{\rm T}l_2}(k_2)j_{l_2'}(k_2r)(-i)^{l_2-l_2'}\right.\\
 \nonumber
 & &\left.\times
 \frac{2}{\pi}\int_0^\infty k_3^2 dk_3~
 g_{{\rm T}l_3}(k_3)j_{l_3}(k_3r)\right.\\
 \label{eq:isobispectrum}
 & & \left. +~ (1 \leftrightarrow 2, 2 \leftrightarrow 3) + 
      (1 \leftrightarrow 3, 2 \leftrightarrow 1)\right],
\end{eqnarray}
%%%%%%%%%%%%%%%%%%%%%%%%%%%%%%%%%%%%%%%%%%%%%%%%%%%%%%%%%%%%%%%%%%%%%%
where ${\cal G}_{l_1l_2l_3}^{m_1m_2m_3}$ and 
${\cal F}_{l_2'l_1'L}^{l_1l_2l_3}$ represent geometric structures
of the bispectrum.
They are written with the Wigner 3-$j$ and 6-$j$ symbols, 
%%%%%%%%%%%%%%%%%%%%%%%%%%%%%%%%%%%%%%%%%%%%%%%%%%%%%%%%%%%%%%%%%%%%%%
\begin{equation}
  {\cal G}_{l_1l_2l_3}^{m_1m_2m_3}
  \equiv \sqrt{
   \frac{\left(2l_1+1\right)\left(2l_2+1\right)\left(2l_3+1\right)}
        {4\pi}
        }
  \left(
  \begin{array}{ccc}
  l_1 & l_2 & l_3 \\ 0 & 0 & 0 
  \end{array}
  \right)
  \left(
  \begin{array}{ccc}
  l_1 & l_2 & l_3 \\ m_1 & m_2 & m_3 
  \end{array}
  \right),
\end{equation}
%%%%%%%%%%%%%%%%%%%%%%%%%%%%%%%%%%%%%%%%%%%%%%%%%%%%%%%%%%%%%%%%%%%%%%
%%%%%%%%%%%%%%%%%%%%%%%%%%%%%%%%%%%%%%%%%%%%%%%%%%%%%%%%%%%%%%%%%%%%%%
\begin{eqnarray}
 \nonumber
 {\cal F}_{l_2'l_1'L}^{l_1l_2l_3}
  &\equiv&
  \frac{(2l_1'+1)(2l_2'+1)(2L+1)}{4\pi}
  \left(\begin{array}{ccc}l_1&l_2&l_3\\0&0&0\end{array}\right)^{-1}
  \left\{\begin{array}{ccc}l_1&l_2&l_3\\l_2'&l_1'&L\end{array}\right\}\\
 & &\times
  \left(\begin{array}{ccc}l_1'&l_2'&l_3\\0&0&0\end{array}\right)
  \left(\begin{array}{ccc}l_1&l_1'&L\\0&0&0\end{array}\right)
  \left(\begin{array}{ccc}l_2&l_2'&L\\0&0&0\end{array}\right)
  (-1)^{l_1'+l_2'+L}.
\end{eqnarray}
%%%%%%%%%%%%%%%%%%%%%%%%%%%%%%%%%%%%%%%%%%%%%%%%%%%%%%%%%%%%%%%%%%%%%%

\chapter{Angular Trispectrum in Closed Hyperbolic Universe}
\label{app:CH}

%%%%%%%%%%%%%%%%%%%%%%%%%%%%%%%%%%%%%%%%%%%%%%%%%%%%%%%%%%%%%%%%%%%%%%
In this appendix, we derive the angular connected trispectrum
in a closed hyperbolic universe.
If topology of our universe is infinite flat, then 
we can expand the harmonic coefficients of the CMB anisotropy 
into Fourier series,
%%%%%%%%%%%%%%%%%%%%%%%%%%%%%%%%%%%%%%%%%%%%%%%%%%%%%%%%%%%%%%%%%%%%%%
\begin{equation}
 a_{lm}= 4\pi(-1)^l
  \int\frac{d^3{\mathbf k}}{(2\pi)^3}\Phi({\mathbf k})g_{{\rm T}l}(k)
  Y_{lm}^*(\hat{\mathbf k}),
\end{equation}
%%%%%%%%%%%%%%%%%%%%%%%%%%%%%%%%%%%%%%%%%%%%%%%%%%%%%%%%%%%%%%%%%%%%%%
where  $\Phi({\mathbf k})$ is the primordial curvature perturbation,
and $g_{{\rm T}l}(k)$ is the radiation transfer function.
Hence, statistical properties of $a_{lm}$ take over 
statistical properties of $\Phi$ directly;
if $\Phi$ is Gaussian, then $a_{lm}$ is also Gaussian.

If topology of our universe is closed hyperbolic, then the 
relation between $a_{lm}$ and $\Phi$ becomes \citep{Ino01b}
%%%%%%%%%%%%%%%%%%%%%%%%%%%%%%%%%%%%%%%%%%%%%%%%%%%%%%%%%%%%%%%%%%%%%%
\begin{equation}
 \label{eq:CH}
  a_{lm}= \sum_{\nu}\Phi(\nu)g_{{\rm T}l}(\nu)\xi_{lm}(\nu),
\end{equation}
%%%%%%%%%%%%%%%%%%%%%%%%%%%%%%%%%%%%%%%%%%%%%%%%%%%%%%%%%%%%%%%%%%%%%%
where $\nu=\sqrt{k^2-1}$ is a discrete wavenumber, and
new expansion coefficients, $\xi_{lm}(\nu)$, describe
eigenmodes in a closed hyperbolic geometry. 
The complex conjugate of $\xi_{lm}$ is given by $\xi^*_{lm}(\nu)=(-1)^m\xi_{l-m}(\nu)$.
Since a closed hyperbolic universe is globally inhomogeneous,
$\xi_{lm}(\nu)$ is also a function of observer's positions and
orientations in the universe.
\citet{Ino01b} have shown that this property makes 
$\xi_{lm}(\nu)$ behave as if it were a Gaussian random number with 
the covariance matrix diagonal,
%%%%%%%%%%%%%%%%%%%%%%%%%%%%%%%%%%%%%%%%%%%%%%%%%%%%%%%%%%%%%%%%%%%%%%
\begin{equation}
 \left<\xi_{lm}(\nu)\xi_{l'm'}(\nu')\right>
  = \left<\left|\xi_{lm}(\nu)\right|^2\right>
  (-1)^{m'}\delta_{ll'}\delta_{m-m'}\delta_{\nu\nu'},
\end{equation}
%%%%%%%%%%%%%%%%%%%%%%%%%%%%%%%%%%%%%%%%%%%%%%%%%%%%%%%%%%%%%%%%%%%%%%
and $\left<\left|\xi_{lm}(\nu)\right|^2\right>\propto \nu^{-2}$,
where the bracket denotes the ensemble average over observer's positions
and orientations in the universe.
Hence, in addition to $\Phi$, statistical properties of $\xi_{lm}$ also affect 
statistical properties of $a_{lm}$. 
If $\Phi$ is Gaussian, then it follows from the relation among 
$a_{lm}$, $\Phi$, and $\xi_{lm}$ (Eq.(\ref{eq:CH})) that 
$a_{lm}$ comprises two independent Gaussian random numbers.

The angular power spectrum, $C_l$, is given by
%%%%%%%%%%%%%%%%%%%%%%%%%%%%%%%%%%%%%%%%%%%%%%%%%%%%%%%%%%%%%%%%%%%%%%
\begin{equation}
 \left<C_l\right>= \left<\left|a_{lm}\right|^2\right>
  =\sum_{\nu}P_\Phi(\nu)g_{{\rm T}l}^2(\nu)
  \left<\left|\xi_{lm}(\nu)\right|^2\right>,
\end{equation}
%%%%%%%%%%%%%%%%%%%%%%%%%%%%%%%%%%%%%%%%%%%%%%%%%%%%%%%%%%%%%%%%%%%%%%
where $P_\Phi(\nu)\delta_{\nu\nu'}= 
\left<\Phi(\nu)\Phi^*(\nu')\right>$.
The ensemble average for $\Phi$ is taken over initial conditions.
We should keep in mind that the ensemble average for $\xi_{lm}$
is taken over all possible observer's positions and orientations in 
the universe.
Hence, there exist those observers who measure $C_l$  
different from $\left<C_l\right>$ substantially.
In other words, there exists theoretical uncertainty in 
$\left<C_l\right>$ arising from uncertainty as to where we are.

We calculate the angular trispectrum from equation~(\ref{eq:CH}) as
follows.
We begin with
%%%%%%%%%%%%%%%%%%%%%%%%%%%%%%%%%%%%%%%%%%%%%%%%%%%%%%%%%%%%%%%%%%%%%%
\begin{eqnarray}
 \nonumber
 \left<a_{l_1m_1}a_{l_2m_2}a_{l_3m_3}a_{l_4m_4}\right>
  &=&
  \sum_{{\rm all}~\nu}
  g_{{\rm T}l_1}(\nu_1)g_{{\rm T}l_2}(\nu_2)
  g_{{\rm T}l_3}(\nu_3)g_{{\rm T}l_4}(\nu_4)\\
\nonumber
 & &\times 
  \left<\Phi(\nu_1)\Phi(\nu_2)\Phi(\nu_3)\Phi(\nu_4)\right>\\
 & &\times 
  \left<\xi_{l_1m_1}(\nu_1)\xi_{l_2m_2}(\nu_2)
   \xi_{l_3m_3}(\nu_3)\xi_{l_4m_4}(\nu_4)\right>.
\end{eqnarray}
%%%%%%%%%%%%%%%%%%%%%%%%%%%%%%%%%%%%%%%%%%%%%%%%%%%%%%%%%%%%%%%%%%%%%%
If $\Phi$ is Gaussian, then this equation yields 9 terms.
First, we reduce the second line to
%%%%%%%%%%%%%%%%%%%%%%%%%%%%%%%%%%%%%%%%%%%%%%%%%%%%%%%%%%%%%%%%%%%%%%
\begin{eqnarray}
 \nonumber
 \left<\Phi(\nu_1)\Phi(\nu_2)\Phi(\nu_3)\Phi(\nu_4)\right>
  &=&
  P_\Phi({\nu_1})P_\Phi({\nu_3})\delta_{\nu_1\nu_2}\delta_{\nu_3\nu_4}\\
 & &
  +
  P_\Phi({\nu_1})P_\Phi({\nu_2})
  \left(\delta_{\nu_1\nu_3}\delta_{\nu_2\nu_4}
 +\delta_{\nu_1\nu_4}\delta_{\nu_2\nu_3}\right).
\end{eqnarray}
%%%%%%%%%%%%%%%%%%%%%%%%%%%%%%%%%%%%%%%%%%%%%%%%%%%%%%%%%%%%%%%%%%%%%%
We then evaluate the first three terms in the trispectrum:
%%%%%%%%%%%%%%%%%%%%%%%%%%%%%%%%%%%%%%%%%%%%%%%%%%%%%%%%%%%%%%%%%%%%%%
\begin{eqnarray}
 \nonumber
  & &
  \sum_{\nu_1\nu_3}
  g_{{\rm T}l_1}(\nu_1)g_{{\rm T}l_2}(\nu_1)
  g_{{\rm T}l_3}(\nu_3)g_{{\rm T}l_4}(\nu_3)
  P_\Phi({\nu_1})P_\Phi({\nu_3})\\
 \nonumber
 & &\times
  \left[
   \left<\xi_{l_1m_1}(\nu_1)\xi_{l_2m_2}(\nu_1)\right>
   \left<\xi_{l_3m_3}(\nu_3)\xi_{l_4m_4}(\nu_3)\right>
   \right.\\
 \nonumber
  & &
  \left.
  +
  \left<\xi_{l_1m_1}(\nu_1)\xi_{l_3m_3}(\nu_3)\right>
   \left<\xi_{l_2m_2}(\nu_1)\xi_{l_4m_4}(\nu_3)\right>\right.\\
 \nonumber
 & &
 \left.
  +
  \left<\xi_{l_1m_1}(\nu_1)\xi_{l_4m_4}(\nu_3)\right>
   \left<\xi_{l_2m_2}(\nu_1)\xi_{l_3m_3}(\nu_3)\right>
 \right]\\
 \nonumber
 &=&
 \left<C_{l_1}\right>\left<C_{l_3}\right>
 (-1)^{m_2+m_4}
 \delta_{l_1l_2}\delta_{l_3l_4}\delta_{m_1-m_2}\delta_{m_3-m_4}\\
 \label{eq:CHfull}
 & &+ F_{l_1l_2}
  (-1)^{m_3+m_4}
 \left(\delta_{l_1l_3}\delta_{l_2l_4}\delta_{m_1-m_3}\delta_{m_2-m_4}
  + \delta_{l_1l_4}\delta_{l_2l_3}\delta_{m_1-m_4}\delta_{m_2-m_3}\right),
\end{eqnarray}
%%%%%%%%%%%%%%%%%%%%%%%%%%%%%%%%%%%%%%%%%%%%%%%%%%%%%%%%%%%%%%%%%%%%%%
where 
%%%%%%%%%%%%%%%%%%%%%%%%%%%%%%%%%%%%%%%%%%%%%%%%%%%%%%%%%%%%%%%%%%%%%%
\begin{equation}
 F_{l_il_j}\equiv
  \sum_{\nu}P_\Phi^2(\nu)g_{{\rm T}l_i}^2(\nu)g_{{\rm T}l_j}^2(\nu)
  \left<\left|\xi_{l_im_i}(\nu)\right|^2\right>
  \left<\left|\xi_{l_jm_j}(\nu)\right|^2\right>.
\end{equation}
%%%%%%%%%%%%%%%%%%%%%%%%%%%%%%%%%%%%%%%%%%%%%%%%%%%%%%%%%%%%%%%%%%%%%%
The first term in the r.h.s. of equation~(\ref{eq:CHfull}) is 
the unconnected term,
while the second and third terms are the connected ones.
By collecting all the connected terms, we obtain 
%%%%%%%%%%%%%%%%%%%%%%%%%%%%%%%%%%%%%%%%%%%%%%%%%%%%%%%%%%%%%%%%%%%%%%
\begin{eqnarray}
 \nonumber
  & &
 \left<a_{l_1m_1}a_{l_2m_2}a_{l_3m_3}a_{l_4m_4}\right>_{\rm c}\\
 \nonumber
 &=&
 F_{l_1l_2}
  (-1)^{m_3+m_4}
 \left(\delta_{l_1l_3}\delta_{l_2l_4}\delta_{m_1-m_3}\delta_{m_2-m_4}
  + \delta_{l_1l_4}\delta_{l_2l_3}\delta_{m_1-m_4}\delta_{m_2-m_3}\right)\\
 \nonumber
 & & + (2\leftrightarrow 3) + (2\leftrightarrow 4).
\end{eqnarray}
%%%%%%%%%%%%%%%%%%%%%%%%%%%%%%%%%%%%%%%%%%%%%%%%%%%%%%%%%%%%%%%%%%%%%%
Actually, the functional form of this equation is exactly the same
as the unconnected terms, if we replace $2F_{l_il_j}$ with 
$\left<C_{l_i}\right>\left<C_{l_j}\right>$.
Hence, by substituting $2F_{l_il_j}$
for $\left<C_{l_i}\right>\left<C_{l_j}\right>$  in the formula of 
the unconnected trispectrum (Eq.(\ref{eq:unconnected*})), 
we find the angular connected trispectrum in a closed 
hyperbolic universe
%%%%%%%%%%%%%%%%%%%%%%%%%%%%%%%%%%%%%%%%%%%%%%%%%%%%%%%%%%%%%%%%%%%%%%
\begin{eqnarray}
 \nonumber
  \left<T^{l_1l_2}_{l_3l_4}(L)\right>_{\rm c}
 &=&
 (-1)^{l_1+l_3}\sqrt{(2l_1+1)(2l_3+1)}\left(2F_{l_1l_3}\right)
  \delta_{l_1l_2}\delta_{l_3l_4}\delta_{L0}\\
 \label{eq:CHprediction}
 & & +
  (2L+1)\left(2F_{l_1l_2}\right)\left[(-1)^{l_2+l_3+L}
		   \delta_{l_1l_3}\delta_{l_2l_4}
		   +\delta_{l_1l_4}\delta_{l_2l_3}\right].
\end{eqnarray}
%%%%%%%%%%%%%%%%%%%%%%%%%%%%%%%%%%%%%%%%%%%%%%%%%%%%%%%%%%%%%%%%%%%%%%

Unfortunately, the unconnected terms always contaminate 
the connected terms produced in a closed hyperbolic universe: 
there is no non-Gaussian signal in the modes for which the unconnected 
terms vanish.
Since what we measure is the sum of the unconnected terms and the connected 
terms, $\left<C_{l_i}\right>\left<C_{l_j}\right>+2F_{l_il_j}$, 
we have to subtract $\left<C_{l_i}\right>\left<C_{l_j}\right>$ 
from measurement to obtain $2F_{l_il_j}$.
Although $F_{l_il_j}$ is difficult to measure because of the contamination,
if we detect non-Gaussian signals that exist only in these limited
modes ($L=0$, or $l_1=l_2=l_3=l_4$),
then it may suggest that the universe is closed hyperbolic.

The connected trispectrum for $L=0$ contributes to the power spectrum
covariance matrix (Eq.(\ref{eq:error2*})).
Using our prediction for the connected trispectrum 
(Eq.(\ref{eq:CHprediction})) for $L=0$, we obtain
the power spectrum covariance matrix in a closed hyperbolic universe,
%%%%%%%%%%%%%%%%%%%%%%%%%%%%%%%%%%%%%%%%%%%%%%%%%%%%%%%%%%%%%%%%%
\begin{equation}
 \label{eq:CHcov}
  \left<C_lC_{l'}\right> - \left<C_l\right>\left<C_{l'}\right>
   =
  \frac{2}{2l+1}\left(\left<C_l\right>^2+2F_{ll}\right)\delta_{ll'}
  +2F_{ll'}.
\end{equation}
%%%%%%%%%%%%%%%%%%%%%%%%%%%%%%%%%%%%%%%%%%%%%%%%%%%%%%%%%%%%%%%%%
The variance is amplified, and the off-diagonal terms appear in the 
covariance matrix; these properties may be used to search for 
a signature of closed hyperbolic geometry in the CMB angular power
spectrum covariance.

%%%%%%%%%%%%%%%%%%%%%%%%%%%%%%%%%%%%%%%%%%%%%%%%%%%%%%%%%%%%%%%%%%%
\end{document}